%% file: thesPhD.tex
\documentclass[a4paper,12pt]{report}

\usepackage{amsthm}
\usepackage{graphicx}
\usepackage{amsmath}
\usepackage{amssymb}
\usepackage{amsfonts}
\usepackage[english]{babel}
\usepackage[]{graphics}
\usepackage{slashed} 
\usepackage{fancyhdr} %richiama il pacchetto fancyhdr 
\theoremstyle{plain} 
\newtheorem{prop}{Proposition}

\newtheorem{thm}{Theorem}
\newcommand{\di}{\genfrac{}{}{0pt}{}}
\title{}
\author{Bernardino Spisso}
\begin{document}
\include{intestazione}

\include{abstr}
\tableofcontents
\include{thanks}
\addcontentsline{toc}{chapter}{Introdution}
\include{intro}
\fancyhead[RO]{\bfseries \leftmark}             
\fancyhead[LE]{\bfseries \rightmark}

\include{cap1} %Intoduction C-L Model
\include{cap2} %8-dim spectral triple
\include{cap3} %Statistical FT
\include{cap4} %Preparing for numericals
\include{cap5} %simulation 
\addcontentsline{toc}{chapter}{Conclusion}
\include{concl}

\appendix
\include{appt}%Il app teo
\include{appn}%Il app numerica

\end{document}

%% file: intestazione.tex
\thispagestyle{empty}
\begin{center}
\small

  \Large{Mathematik}\\
  \vspace*{\stretch{3}}
  \huge{A numerical approach to harmonic non-commutative spectral field theory}\\

  \vspace*{\stretch{6}}

    \Large{Inaugural-Dissertation\\
    zur Erlangung des Doktorgrades\\
    der Naturwissenschaften im Fachbereich\\
    Mathematik und Informatik\\
    der Mathematisch-Naturwissenschaftlichen Fakult\"{a}t\\
    der Westf\"{a}lischen Wilhelms-Universit\"{a}t M\"{u}nster}

  \vspace*{\stretch{4}}%
   \Large{ vorgelegt von\\
    {Bernardino Spisso\\
    aus Napoli(Italien)\\
    -- 2011 --}}
\end{center}

\clearpage
\thispagestyle{empty}

\renewcommand{\arraystretch}{1.7}
\vspace*{\fill}
{\normalsize
\begin{tabular}{ll}
Dekan:                      & Prof.~Dr.~Matthias L\"owe\\
Erster Gutachter:           & Prof.~Dr.~Raimar Wulkenhaar\\
Zweiter Gutachter:          & Prof.~Dr.~Denjoe O'Connor\\
Tag der m\"undlichen Pr\"ufung:\hspace{.5cm}  & \dotfill \\
Tag der Promotion:          & \dotfill \\
\end{tabular}
}

\clearpage

%% file: abstr.tex
\noindent
{\large\bf Zusammenfassung}\\
\\
Gegenstand der Arbeit ist die numerische Untersuchung einer \"uber
das
Spektralwirkungsprinzip definierten nichtkommutativen Feldtheorie.\\
Ausgangspunkt dieser Konstruktion ist ein (als harmonisch
bezeichnetes) spektrales Tripel
$(\mathcal{A}_4,\mathcal{H}_4,\mathcal{D}_4)$. Dabei ist
$\mathcal{A}_4$ die 4-dimensionale nichtkommutative Moyal-Algebra und
$\mathcal{D}_4$ ein selbstadjungierter (Dirac-)Operator auf dem
Hilbert-Raum $\mathcal{H}_4$, so dass $\mathcal{D}_4^2$ der
Schr\"odinger-Operator des 4-dimensionalen harmonischen Oszillators
wird. Die Konstruktion dieser Daten basiert auf einer 8-dimensionalen
Clifford-Algebra. F\"ur das Produkt aus dem Tripel
$(\mathcal{A}_4,\mathcal{H}_4,\mathcal{D}_4)$ mit einem matrixwertigen
spektralen Tripel wird analog zum Standardverfahren der
nichtkommutativen Geometrie die Spektralwirkung \\ berechnet.

Die Renormierungstheorie assoziiert zur Spektralwirkung ein (z.T.\ nur
formales) Wahrscheinlichkeitsma\ss{}, deren zugeh\"orige
Korrelationsfunktionen eine Feldtheorie definieren. In der
st\"orungstheoretischen Variante wird das
\\Wahrscheinlichkeitsma\ss{} als
formale Potenzreihe konstruiert. Voraussetzung daf\"ur ist die
explizite Kenntnis der L\"osungen der Euler-Lagrange-Gleichungen
zur
Spektralwirkung. F\"ur das betrachtete Modell erweist es sich als
unm\"oglich, diese L\"osungen (als Vakuum bezeichnet) zu
gewinnen.

Ein alternatives Verfahren besteht in der Diskretisierung aller
Variablen und der numerischen Untersuchung des Verhaltens der
Korrelationsfunktionen bei Verfeinerung der Diskretisierung. Die
Diskretisierung des Modells wird durch die Matrix-Basis des
Moyal-Raums und Beschr\"ankung auf endliche Matrizen erreicht.
Durch
Monte-Carlo-Simulationen werden wichtige Korrelationsfunktionen wie
die Energiedichte, die spezifische W\"arme sowie einige
Ordnungsparameter untersucht, sowohl in Abh\"angigkeit von der
Gr\"o\ss{}e der Matrizen als auch von den unabh\"angigen
Parametern
des Modells. Dabei werden trotz der gro\ss{}en Komplexit\"at der
approximierten Spektralwirkung verl\"a\ss{}liche numerische
Resultate
erzielt, die zeigen, da\ss{} eine numerische Behandlung dieser Art von
Modellen in der Matrix-Moyal-Basis m\"oglich ist.

\newpage

\noindent
{\large\bf Abstract}\\
\\
The object of this work is the numerical investigation of a
non-commutative field theory defined via the spectral action
principle. Starting point of this construction is a spectral triple
$(\mathcal{A}_4,\mathcal{H}_4,\mathcal{D}_4)$ referred to as harmonic.
Here, $\mathcal{A}_4$ is the 4-dimensional (noncommutative) Moyal
algebra, and $\mathcal{D}_4$ is a selfadjoint (Dirac-)operator on the
Hilbert space $\mathcal{H}_4$ such that $\mathcal{D}_4^2$ is the
Schr\"odinger operator of the 4-dimensional harmonic oscillator.
The
construction of these data relies on an 8-dimensional Clifford
algebra. In analogy to the standard procedure of non-commutative
geometry, the spectral action is computed for the product of the triple
$(\mathcal{A}_4,\mathcal{H}_4,\mathcal{D}_4)$ with a matrix-valued
spectral triple.

Renormalization theory associates to the spectral action a (often only
formal) probability measure. Its associated correlation functions
define then a field theory. In the perturbative approach
this measure is constructed as a formal power series. This requires
explicit knowledge of the solutions of the Euler-Lagrange equations
for the spectral action. For the model under consideration, it turns
out impossible to obtain these solutions.

An alternative approach consists in a discretization of all variables
and a numerical investigation of the behavior of the correlation
functions when the discretization becomes finer. For the model under
consideration, the discretization is achieved in the matrix basis of
the Moyal algebra restricted to finite matrices. By Monte Carlo
simulation we study several important correlation functions such as
the energy density, the specific heat and some order parameters, in
dependence both of the matrix size and of the independent parameters
of the model. Despite the complexity of the approximated spectral
action, some reliable numerical results are obtained, showing that a
numerical treatment of this kind of models in the Moyal matrix basis
is possible. 

%% file: thanks.tex
\chapter*{Acknowledgements}
This work has been supported by the Marie Curie Research Training Network MRTN-CT-2006-031962 in Noncommutative Geometry, EU-NCG.
I wish here to acknowledge all those who contributed and help me making this work possible. In particular
to the Dublin Institute for Advanced Studies where  I had a very useful discussion with  Thomas Kaltenbrunner and Martin Vachovski about numerical 
simulation on matrix models.
To my advisor Prof. Raimar Wulkenhaar for all the guidance, and support, I  am very grateful for his
careful review of this thesis and his very useful comments on my work.

%% file: intro.tex
\chapter*{Introduction}

The main object of this work is a particular non-commutative field theory  which is  derived using the spectral 
action principle and then treated numerically. Non-commutativity can be found in many fields of physics 
like quantum field theories, string theory \cite{nc-string}, condensed matter physics. The first application of  non-commutativity 
into physicss is  dated from the middle of the last century inspired by the ideas of quantum mechanics, where  starting from classical 
mechanics,  the commutative algebra of functions on the phase space is replaced by a non-commutative operator algebra on a Hilbert space.
The duality between ordinary spaces $M$ and proper commutative
algebras is expressed by the Gel'fand-Naimark theorem  which states the
fact that the algebra of all continuous functions on $M$ is the only possible
type of commutative $C^*$-algebra. Additionally, given a commutative $C^*$-algebra $C$, it is
possible to reconstruct a Hausdorff topological space $M$ in order to obtain that $C$ is the algebra
of continuous functions on $M$. The study of commutative
$C^*$-algebras is equivalent to the study of topological Hausdorff spaces.
The previous duality has inspired the identification, in non-commutative geometry, of  
some algebraical objects as a category of non-commutative topological spaces.
Alain Connes \cite{Connes}, one of the founders of non-commutative geometry, has proposed a candidate for the objects of such  category,
the spectral triples $(\mathcal{A},\mathcal{H}, \mathcal{D})$ \cite{Connes-1}, composed by an algebra $\mathcal{A}$, an Hilbert
space $\mathcal{H}$ on which $\mathcal{A}$ is represented and an  selfadjoint operator $\mathcal{D}$.
In fact, every compact oriented Riemannian manifold can be used to define a spectral triple, 
this kind of manifold $M$ characterizes a symmetric Dirac type operators on self-adjoint Clifford module bundles
over $M$. Connes, after a conjecture in 1996 \cite{Connes3}
and some considerable attempts of Rennie and Varilly \cite{Rennie-vari},
 proved the so called reconstruction theorem \cite{ConnesRec} for commutative
spectral triples satisfying various axioms, showing that 
exists a compact oriented smooth manifold $X$ such
that $A = C^\infty(X)$ is the algebra of smooth functions on $X$ and
every compact oriented smooth manifold emerges in this way.
Pushed by the aim of reformulating the standard model of particles in a non-commutative  way \cite{Connes2,Connes3},  
Connes has introduced the almost-commutative spectral triple extending the axioms of the reconstruction theorem to a non-commutative algebra. The almost-commutative spectral triples are
defined as the non-commutative Cartesian product of a commutative spectral
triple of a compact spin manifold, with a spectral triple where the Hilbert space is finite-dimensional, this triples are often labeled as finite spectral triple.

A field theory can be interpreted \cite{field-theory} as a theory concerning maps $\phi$ (usually referred as fields) between the space-time $M^\prime$ and a target space $M$. On this spaces are defined some structures, they can be Riemannian or pseudo-Riemannian manifold depending on the signature of the space-time. A fundamental role in the field theory is played by the action $S(\phi)$ which is a  functional of the elementary fields in the theory. In the classical theory the aim is to study the extrema of the action functional $\delta S(\phi_0)=0$ in order to obtain the solution of the equations of motion. Quoting the intuitive Feynman formulation of quantum field theory \cite{path1}, the action is used to study the functional integrals:
\begin{equation*}
\int_\phi D\phi F(\phi)e^{-\frac{1}{\hslash} S(\phi)} 
\end{equation*}
where $F(\phi)$ is a functional of $\phi$ and $D\phi$ is the functional measure. The previous expression in general is  not well defined and should be considered as an approximate expression. However, the functional integral approach is very useful in studying the quantum field theory connection with the expectation values in statistical mechanics; all field  configurations contributes in the estimation of the previous functional each whit probability amplitude $P(\phi)\sim e^{-\frac{1}{\hslash}S(\phi)}$. More formally the action functional is used to compute the correlation functions which are one the main physical output of a quantum field theory with a close connection to the experimental measurements. In the path integral formalism \cite{path} the correlation function of the fields $\phi_1(x_1)\cdots\phi_n(x_n)$ is given, for the euclidean case, by the functional integral:
\begin{equation*}
\int D\Phi\phi_1(x_1)\cdots\phi_n(x_n) e^{-S(\phi_i)}
\end{equation*} 
where is the path integral measure $D\Phi$ on the space of configuration of the elementary fields. In the Lorentz signature the last term is replaced by $e^{iS}$. A crucial point in the quantum field theory is the formalization of the previous integral in order to obtain all correlation functions of the theory well defined. This procedure is called renormalization and is achieved using a perturbative approach. A safe perturbative analysis can only be conducted after expanding   the action around its vacuum, of course this requirement needs the explicit expression for the vacuum usually obtained minimizing the action. It is clear that for any quantum field theory  the explicit determination of vacuum is a indispensable step before the perturbative study of its renormalizability can be done.  

In order to define an action for a non-commutative geometry theory  Connes and Chamsedinne introduced a general formalism for spectral triples, the spectral action principle \cite{nc-S}. The term  spectral come from the fact that it depends only on the spectrum of the Dirac operator and it takes the form
\begin{equation*}
\textrm{Tr}(\chi(\mathcal{D}/\Lambda))
\end{equation*}
where  $\Lambda$  is a real parameter and it fixes the energy scale. The  $\chi$ is a differentiable function 
$\chi:\mathbb{R}_+ \to \mathbb{R}_+$ of sufficiently fast decrease such that the spectral action converges.

Using this approach the standard model of particles  has been reformulated, including 
a Riemannian formulation of gravity. The algebra of the spectral triple used is defined by the
tensor product of $C^\infty(M)$, the regular functions on a manifold $M$, times a matrix algebra
of finite dimension. Formally the natural group of invariance of the standard model, including gravity, is 
the semidirect product:
\begin{equation*}
G = \mathcal{U}\rtimes \textrm{Diff}(M)
\end{equation*}
Were  $\mathcal{U}= C^\infty (M, U(1)\times SU(2)\times SU (3))$ is the group of local gauge transformations.
The total group of invariance admits $\mathcal{U}$ as a normal subgroup. 
A.Connes has showed, defining $\mathcal{A} = C^\infty(M) \otimes (C \oplus \mathbb{H} \oplus M_3 (C))$,
that the automorphisms group Aut($\mathcal{A}$) of the non-commutative
algebra $\mathcal{A}$ admits the inner group Int($\mathcal{A}$) as a normal subgroup and Aut$(A) \approx G$. 
The algebra $\mathcal{A}_F=(C \oplus \mathbb{H} \oplus M_3 (C))$ is finite and $\mathbb{H}$ is the algebra of quaternions.
The rest of the spectral geometry is defined by the action of $\mathcal{A}$ on $\mathcal{H}$ and by suitable Dirac operator operator $\mathcal{D}$. In \cite{nc-S} is showed that the total spectral triple is given by:
\begin{equation*}
(\mathcal{A},\mathcal{H},\mathcal{D})=(C^\infty(M),L^2(M,S),\ \slashed{\partial}_M)\otimes (\mathcal{A}_F,\mathcal{H}_F,\mathcal{D}_F)
\end{equation*}
with
\begin{equation*}
\mathcal{H}= L^2(M,\mathcal{S})\otimes\mathcal{H}_F, \  \mathcal{D}=\slashed{\partial}_M\otimes 1 +\gamma_5\otimes \mathcal{D}_F
\end{equation*}
The algebra $\mathcal{A}_F$ is finite dimensional so the dimension of the corresponding Hilbert space $\mathcal{H}_F$ must be finite dimensional \cite{Schu1}. 
The final step, in order to obtain a field theory, is to define an action, this task is achieved using the spectral action 	principle and the invariance under the symmetry group is implement fluctuating the Dirac operator.
The subgroup Int$(\mathcal{A})$ of inner automorphisms is a normal subgroup and the group Aut$(\mathcal{A})$ of diffeomorphisms falls in equivalence classes under Int$(\mathcal{A})$. This induces a natural foliation into equivalence classes in the space of metrics. The internal fluctuations \cite{C-L} of a given metric are given by the formula:
\begin{equation*}
\mathcal{D} = \mathcal{D}_0 + A + JAJ^{-1},\ \  A = \sum a_i [D_0 , b_i ] \ a_i , b_i \in \mathcal{A}
\end{equation*}
where $J$ is the real structure. In this way starting from $(\mathcal{A}, \mathcal{H}, \mathcal{D}_0)$, instead modify the representation of  $\mathcal{A}$ in  $\mathcal{H}$, it is changed the operator $\mathcal{D}_0$  where $A$ is
a self-adjoint operator in $\mathcal{H}$ of the form $A = \sum a_i [\mathcal{D}_0 , b_i ]$. 
The spectral action principle  applied to inner fluctuations reproduces the bosonic part
of the model, the gravity is naturally present in the model, while the other interactions are 
encoded in the matrix algebra of the total spectral triple. The non-commutativity of matrices corresponds to the non-abelianity of the gauge theory defining a so called Yang-Mills theory. 

The previous example of application of non-commutative geometry to a field theory is founded on the use of a spectral triple  where the algebra is almost-commutative, a very interesting task is to formulate a field theory for a truly  non-commutative algebra. A first attempt was obtained replacing in the usual field theory action the point-wise multiplication of the fields with a non-commutative one, namely a $\star$-product.
The fields now belongs to  $\mathbb{R}^4_\Theta$,  a vector space defined by an enough regular class functions on $\mathbb{R}^4_\Theta$ equipped with the Moyal product:
\begin{equation*}
(f \star g)(x) = \int \int d^4 y \frac{d^4 k}{(2\pi)^4} f(x+\frac{1}{2}\Theta \cdot k) g(x+y) e^{i\langle k,y\rangle} 
\end{equation*}
Where $\Theta$ is a skew-symmetric matrix. Unfortunately  all the attempts to renormalize such quantum field theories on the non-commutative $\mathbb{R}^4$ failed 
and these models show a phenomenon called UV/IR-mixing \cite{one-loop-nc3,UV-mix,UV-mix2}. 
A great step towards the non-commutative field theory was made when H.Grosse and R.Wulkenhaar \cite{phi4-non}, found a non-commutative $\varphi^4$-theory renormalizable action which develops additional marginal coupling, corresponding to an harmonic oscillator potential for the real-valued free field $\varphi$ on $\mathbb{R}^4_\Theta$ :
\begin{equation}
S[\varphi]=\int d^4x \left(\frac{1}{2} \varphi \star (-\Delta+\Omega^2 \tilde{x}^2 + \mu^2)\star\varphi + \frac{\lambda}{4}\varphi \star \varphi \star \varphi \star \varphi\right)(x) \nonumber
\end{equation}
Where $x = 2\Theta^{-1} \cdot x, \ \lambda \in \mathbb{R} \ \Omega \in [0,1]$, and $\mu$ is a real parameter.
Using the Moyal matrix base, which turns the  $\star$-product into a standard (infinite) matrix product, H.Grosse and R.Wulkenhaar  were able to prove the perturbative renormalizability of the theory \cite{matrix-renorm}.
Afterward, R.Wulkenhaar et al. \cite{nc-renorm} found an alternative simpler normalization proof using  multi-scale analysis in matrix base, showing the equivalence of various renormalization schemes. A last, but useful, renormalization proof was formulated  using Symanzik type hyperbolic polynomials \cite{nc-renorm2}. It is  worthwhile  to mention  \cite{nc-renorm2,nc-renorm3}.

The non-commutative model treated in this thesis is a sort of extension, via spectral action principle, of the scalar W-G model, in which we are interested to formulate a  Yang-Mills theory in renormalizable way on Moyal space. From the previous discussions  we can expect that usual Yang-Mills theory on Moyal space without modifications of the action by something similar to an oscillator potential, to be not renormalizable \cite{UV-mix3}. 
Additionally, the Moyal space with usual Dirac operator is a spectral triple, the corresponding spectral action was computed in \cite{nc-S}, with the result that it is the usual not renormalizable action on Moyal plane. 
In \cite{8-dim} H.Grosse and R.Wulkenhaar, in order to obtain a gauge theory with an oscillator potential via the spectral action principle, used a Dirac operator constructed  using the statement $ \mathcal{D}^2= H$  where the four dimensional Laplacian is substituted  by the four dimensional oscillator Hamiltonian  $H = -\Delta+\Omega^2 ||x||^2$.  
The idea  behind is that the spectral dimension is defined through the Dirac operator so the spectral dimension  defined
by such Dirac operator is related to the harmonic oscillator phase space dimension. It turns out that to write down an Dirac operator, so that its square equals the 4D harmonic oscillator Hamiltonian, is an easy task using eight dimension Clifford algebra. In addition, can be shown that using this Dirac operator on 4D-Moyal space, is possible define  an eight-dimensional spectral triple. In this thesis is used the  approach described in \cite{non-com-tri} in which the Dirac operator is constructed using $n$-dimensional bosonic and fermionic creation and annihilation operators. After defined the Dirac operator with the desired spectrum it is considered the total spectral triple as the tensor product
of the "oscillating" spectral triple $(\mathcal{A}_4, \mathcal{H}_4, \mathcal{D}_4 )$ with an almost-commutative triple
and then is perform the previous described procedure of non-commutative geometry to compute
the spectral action. We notice that matrix algebra introduces an extension of the standard  potential in the commutative case, in fact the  scalar field $\phi$ and the $X_\mu$ fields are present together in a potential of the form\footnote{Einstein notation on repeated indices is used.} $(\alpha X_\mu \star X^\mu  + \beta \bar{\varphi} \star \varphi - 1)^2$, with $\alpha, \beta \in \mathbb{R}^+$ and $X_\mu(x) = (\Theta^{-1} )_{\mu}^\nu x_\nu + A_\mu(x)$ is a covariant coordinate. These two additional terms, the integral over
$X_\mu \star X^\mu$ and over its square, were conjectured in \cite{wul-hab}. 

The high  non-triviality of the vacuum makes very difficult to explicit the vacuum configuration of the system in \cite{Goursac1} A. de Goursac, J.C. Wallet, and R. Wulkenhaar, using the matrix base formalism, have found an expressions from vacuum solutions deriving them from the relevant solutions of the equations of motion.
Although, the  complexity  of the vacuum configuration makes the perturbative approach very complicated, in order to conduct some investigations in this thesis will be considered a non-perturbative scheme  using a discretized matrix model of the action in which the fields become matrices, the star product become the matrix multiplication and the integral turns in a matrix trace.

Now comes in to play the numerical treatment, the standard method is to approximate the space by discrete points, for example using a lattice approximation  and then calculate the observables over that set of points \cite{lattice}. 
Since an approximation in the position space is not suitable due to the oscillator factor of the Moyal product, instead  the lattice approximation, will be used the matrix Moyal base, which was already used in the first renormalization proof of $\varphi^4$-model restricted to finite matrices.  Hence, will be performed a Monte Carlo simulation studying some statistical quantity such the energy density and specific heat varying the parameters $\Omega, \frac{\chi_{-1}}{\chi_0}, \alpha$ and gathering some informations on the various contributions of the fields to the action. The simulations are quite cumbersome due the  complexity of the action and the number of independent matrices to handle but using  particular algorithm we are able to get an acceptable balance between the  computation precision and the computation time. For the simulations is applied a standard Metropolis-Monte Carlo algorithm \cite{Metro} with 
various estimators for the error (see appendix B) and for the autocorrelation time of the samples.
The initial conditions of the Markov chain are chosen randomly in the phase space,  they  can be of two
types; hot initial conditions, which are configurations far from the minimum, and cold start conditions, which correspond to configurations  close to the  minimum of the
action. In general we chose the range of parameters in order to avoid problems with the thermalization process, obtaining numerical simulations where is enough to wait a relative small number  of Monte Carlo steps  to compute  independent results  from the initial conditions.
For the model under study, we are interested on the continuous limit that correspond
to matrices of infinite size. We will consider various size of the matrices expecting  a stabilization of the values of observables like the energy density,  increasing the matrix size. In order to find same possible phase transitions will be used the specific heat which is a measure of the dispersion of the energy. The  phase transitions are registered as peaks of the specific heat,  increasing the matrices size. Beside, we define others quantities, such as $\langle \textrm{Tr}[\varphi]^2\rangle $ and $\langle \sum_{n=0}^N|\varphi_{nn} |^2 \rangle$ which are used as order parameters \cite{order-par}, we have also  defined their susceptibilities but are not collected due the high number of samples required to obtain a sufficient precision

\section*{Plan of the work}

This work is structured in order to introduce the reader to the model treated and to the numerical analysis, using an heuristic approach. The first chapter and the third one are introductory   chapters in which are explained respectively the fundamentals non-commutative geometry and the main concepts of Monte Carlo  numerical simulation. 
In particular in the first chapter we will introduce the mathematical tools and the machinery required to formulate our spectral model. In the first part of chapter one will be described the basics of non-commutative geometry, Weyl-Wigner map. Further will be introduced some notions of spectral geometry and in the end will presented a simple example of  application of the spectral action principle. The second chapter is devoted to construct the W-G spectral action model. Will be described how  to construct  the harmonic Dirac operator starting from the one-dimensional case and then generalizing it using  annihilations and creations operators for the bosonic and fermionic sectors. Having the 4D-Dirac harmonic operator with harmonic oscillator spectrum and extending it to an spectral triple,  will be considered the tensor product of the non-commutative triple with a finite Connes-Lott type spectral triple \cite{con-lot}. Following the previous described standard procedure of non-commutative geometry, to obtain a "gauged" Dirac operator, we will fluctuate the total Dirac operator, thus  we  will proceed to compute the spectral action in which are present two U(1)-Moyal Yang-Mills fields unified with a complex Higgs field. In the end of the chapter will be discussed some aspects of the resulting action in particular of vacuum and the needs for numerical treatment. 
The third chapter is an introduction to Monte Carlo analysis, will be set up all the ingredients required to conduct a numerical simulation of the model treated in the previous chapter.  Will be explained the basics of the Monte Carlo simulation, focusing ourself on the application in the field theory. In the first sections of chapter three  will be briefly discussed the path integral formulation, this formulation  is essential to connect the field theories to  statistical systems in which the Monte Carlo methods are born.  Then will be introduced the Metropolis algorithm used to produce a Markov chain. In order to resume all the previous concepts and to show an example of phase analysis will be presented an application of numerical simulation on the well know Ising model. Right before the presentation of the numerical results, in the chapter 4, will be showed how the Monte Carlo it is implemented in our case starting from the discretization scheme introducing the Moyal base. In the second part, in order to define the  observables of the upcoming Monte Carlo simulation,  will be designated the expectation values, statistical quantities like energy, specific heat and the order parameters. Finally, in the last chapter will be showed the numerical results for the previous defined quantities for the full 4-dimensional model and for the 2-dimensional one.

%% file: cap1.tex
\chapter{Introduction to Connes-Lott models}

\emph{In this chapter we introduce some mathematical tools taken from \-non\--com\-mu\-ta\-ti\-ve geometry in order to define a spectral
triple, a spectral action and Connes-Lott models. In the first part of this chapter we describe the fundamentals of
non-commutative geometry in particular the Weyl-Wigner map, further will be introduced 
some notions of spectral geometry.}

\

\section{Non-commutative geometry and \\ Weyl-Wigner map}

In this section we will discuss, as an example 
of non-commutative space, the fate of the classical phase space under the process of quantization. 
This is a very large topic and we limit ourselves to some
aspects. In particular we wish to introduce a way to quantize the phase space 
(and in general an $L^2(\mathbb{R}^n)$ space) in which the quantized space can be seen as 
either a set of operators on an Hilbert space or as a deformation of the product of functions 
on the classical space.
At least in the simple cases, the non-commutative algebra of a quantum phase space can be taken as the one
generated by the position and momentum operators acting on a separable
Hilbert space. 
We now establish a connection between the classical and quantum phase
spaces, at least for systems with a well defined classical counterpart. We
want to associate to each classical observable an bounded operator on $\mathcal{H}$. 
This will be done by the Weyl map \cite{Weyl}. Starting from the correspondence principle, an 
immediate problem is to solve an ordering ambiguity, beside $p$ and $q$ are unbounded operators, 
additionally arises problems about the definition of their domains.
To solve these problems Weyl has suggested to introduce a base in the operator space on
$L^2(\mathbb{R})$ using the unitary operators:
\begin{align}
U(q)=e^{\frac{i}{\hslash}qP}\\
V(p)=e^{\frac{i}{\hslash}pQ}
\end{align}
Where $P$ and $Q$ are respectively the momentum operator and the position one.
Being bounded, their domains can be extended to the whole Hilbert space 
$L^2(\mathbb{R})$. In general it is convenient to use the base defined by the following
unitary operators:
\begin{equation}
W(\alpha,\beta) =e^{\frac{i}{\hslash}(\alpha P + \beta{Q})}
\label{w}
\end{equation}
Where $\alpha$ and $\beta$ are two real parameters. A generic
operator  using this base can be expanded like:
\begin{equation}
F(P,Q)=\frac{1}{2\pi\hslash}\int_{\mathbb{R}^2}g(\alpha,\beta)e
^{\frac{i}{\hslash}(\alpha P + \beta{Q})}d\alpha d\beta \label{A}
\end{equation}
Beside this expansion let us considering the base of $L^2(\mathbb{R}^2)$:
\begin{equation}
w(\alpha,\beta) =e^{\frac{i}{\hslash}(\alpha p + \beta q)}
\end{equation}
using this base any function can be written as
\begin{equation}
f(p,q)=\frac{1}{2\pi\hslash}\int_{\mathbb{R}^2}
\tilde{f}(\alpha,\beta)e ^{\frac{i}{\hslash}(\alpha p + \beta
q)}d\alpha d\beta \label{f}
\end{equation}
where $\tilde{f}(\alpha,\beta)$ is the Fourier transform
of $f(p,q)$.
Now, thanks to  the expansion \eqref{f} and \eqref{A}, it is possible
to build an application which maps each function in
$L^2(\mathbb{R}^{2})$ to an bounded operator. In order to construct this map 
 we use the function  $\tilde{f}(\alpha,\beta)$ in the
 operatorial expansion \eqref{A}. In such a way the ordering problem   
 is solved, in fact the association is no more ambiguous.

Formally the Weyl map  is the  application:
$$\Omega:L^2(\mathbb{R}^{2})\to\mathcal{B}(\mathcal{H})$$
defined by
\begin{equation}
f(p,q)\rightarrow\Omega(f)(P,Q)=\frac{1}{2\pi\hslash}\int_{\mathbb{R}^2}\tilde{f}(\alpha,\beta)
W(\alpha,\beta)\mbox{d}\alpha\mbox{d}\beta\label{Weyl map}
\end{equation}
with
\begin{equation}
\tilde{f}(\alpha,\beta)=\frac{1}{2\pi\hslash}\int_{\mathbb{R}^2}f(p,q)
e^{-\frac{i}{\hslash}(\alpha p + \beta q)}\mbox{d}p\mbox{d}q
\end{equation}

This particular Weyl map selects a certain ordering: the symmetric ordering or the so called Weyl's ordering.
Other ordering are  possible, for example normal ordering or Wick ordering,
introducing into \eqref{Weyl map} a weight function \cite{ordering}.
In addition, Weyl map is linear and satisfies the condition:
\begin{equation}
\Omega(\overline{f})=\Omega(f)^\dag
\end{equation}
In particular, if $f$ is
real, then
\begin{equation}
\Omega(f)=\Omega(f)^\dag
\end{equation}
in other words $\Omega(f)$ is symmetric.
The inverse of Weyl map, the so called Wigner map \cite{Wigner}, is
the map which associates to each operator a function of $p$ and $q$, 
it is defined by
\begin{equation}
      \Omega^{-1}(F)(p,q)\equiv\frac{1}{(2\pi\hslash)^2}\int_{\mathbb{R}^2}
e^{-\frac{i}{\hslash}(\alpha p + \beta q)}\operatorname{Tr}(F
W^\dagger(\alpha,\beta))\mbox{d}\alpha\mbox{d}\beta \label{inversa
Weyl}
\end{equation}
with $F\in\mathcal{B}(\mathcal{H})$. In fact, combining
\eqref{Weyl map} and  \eqref{inversa Weyl} we obtain:
\small
\begin{eqnarray}
\Omega^{-1}(\Omega(f))(p,q)&=&\frac{1}{(2\pi\hslash)^2}\int_{\mathbb{R}^2}
\mbox{d}\alpha\mbox{d}\beta e^{-\frac{i}{\hslash}(\alpha p + \beta q)}\operatorname{Tr}\left(
\Omega(f)W^\dagger(\alpha,\beta)\right)
\label{omega omega} \\
\operatorname{Tr}\left(
\Omega(f)W(\alpha,\beta)\right)&=&\operatorname{Tr}\left(\int_{\mathbb{R}^2}\frac{1}{2\pi\hslash}\tilde{f}(\xi,\eta)
W(\xi,\eta)\mbox{d}\xi\mbox{d}\eta W^\dagger(\alpha,\beta)\right)
\end{eqnarray}
\normalsize
The integral of the previous relation  is in d$\xi$
d$\eta$, assuming that the trace and the  $W^\dagger(\alpha,\beta)$ can be switched with the integral 
and using the identity :
\begin{equation}
      \operatorname{Tr}e^{\frac{i}{\hslash}(\alpha P +
      \beta{Q})}=\int_{\mathbb{R}^2} e^{\frac{i}{\hslash}(\alpha p +
      \beta{q})}\mbox{d}p\mbox{d}q=(2\pi\hslash)^2\delta(\alpha,\beta)
\end{equation}
we obtain
\begin{equation}
    \operatorname{Tr}\left(
\Omega(f)W(\alpha,\beta)\right)=2\pi\hslash\int_{\mathbb{R}^2}\tilde{f}(\xi,\eta)
\delta(\xi-\alpha,\eta-\beta)\mbox{d}\xi\mbox{d}\eta
\end{equation}
Therefore  \eqref{omega omega} becomes:
\begin{equation}
      \Omega^{-1}(\Omega(f))(p,q)=\frac{1}{2\pi\hslash}\int_{\mathbb{R}^2} \tilde{f}(\alpha,\beta)e^{\frac{i}{\hslash}(\alpha p + \beta
q)}\mbox{d}\alpha\mbox{d}\beta
\end{equation}
The left side is exactly the inverse Fourier transform 
of $f(\alpha,\beta)$. Finally we obtain:
\begin{equation}
      \Omega^{-1}(\Omega(f))(p,q)=f(p,q)
\end{equation}
this shows that the Wigner map  is  actually
the inverse application Weyl map \eqref{inversa Weyl}.

For simplicity, this construction used
functions  defined on $\mathbb{R}^2$ but can be extended to  $\mathbb{R}^n$ without  
 difficulties. In general using an $n$-dimensional
base is  possible to  define a one to one correspondence
between a function  of $\vec{p}$ and $\vec{q}$ with an operator
using their expansion in each space \cite{pool,bondia0}.

\section{Moyal product}
Using Weyl map it is possible to define a new product 
in order to describe a  quantum system, the Moyal product or
star product. This new product  is no longer commutative and is obtained
combining Weyl map 	applied on two elements of the continuum space,
with the Winger map applied on the product of the previous transformed elements. \\
Formally the Moyal product \cite{star} is defined by
\begin{equation}
f\star g=\Omega^{-1}(\Omega(f)\Omega(g))
\end{equation}
This star product is associative but non-commutative and satisfies the condition
\begin{equation}
(\alpha f)\star g=\alpha(f\star g)=f\star(\alpha g)
\end{equation}
with $\alpha\in\mathbb{C}$. In addition, it satisfies the condition
\begin{equation}
\overline{f\star g}=\overline{g}\star\overline{f},
\end{equation}
the Leibniz rule
\begin{equation}
\frac{\partial}{\partial x^i}(f\star g)=\frac{\partial f}{\partial
x^i}\star g+f\star\frac{\partial g}{\partial x^i}
\end{equation}
with $i\in\{1,2,\ldots,n\}$ and the condition
\begin{equation}
\int_{\mathbb{R}^{n}}f(x)\star
g(x)\mbox{d}x=\int_{\mathbb{R}^{n}}f(x)g(x)\mbox{d}x
\end{equation}
from which follows that the integral has trace property: 
\begin{equation}
\int_{\mathbb{R}^{n}}f(x)\star
g(x)\mbox{d}x=\int_{\mathbb{R}^{n}}g(x)\star f(x)\mbox{d}x
\end{equation}
It is possible to show \cite{bondia2} that, for $f$ and $g$ analytic, the star product can be written as   
\begin{equation}
(f\star g)(q,p)=f(q,p)e^{\frac{i\hslash}{2}
\left(\frac{\stackrel{\leftarrow}{\partial }}{\partial q^i}
\frac{\stackrel{\rightarrow}{\partial }}{\partial p_i}-
\frac{\stackrel{\rightarrow}{\partial }}{\partial q^i}
\frac{\stackrel{\leftarrow}{\partial }}{\partial
p_i}\right)}g(q,p)\label{Moyal}
\end{equation}
or equivalently
\begin{equation}
(f\star g)(q,p)= \sum_{n=0}^\infty
f(q,p)\frac{1}{n!}\left[\frac{i\hslash}{2}\left(
\frac{\stackrel{\leftarrow}{\partial }}{\partial q^i}
\frac{\stackrel{\rightarrow}{\partial }}{\partial p_i}-
\frac{\stackrel{\rightarrow}{\partial }}{\partial q^i}
\frac{\stackrel{\leftarrow}{\partial }}{\partial p_i}
\right)\right]^ng(q,p)
\end{equation}
where the arrows point the direction in which the partial derivate acts.\\
The expression is well defined only on a quite small set of functions, however  
can be obtained some integral expressions with larger domains \cite{bondia2}:

\begin{equation}
(f\star g)(u) := \frac{1}{(\pi\hslash)^{-n} }\int_{\mathbb{R}^{n}}\int_{\mathbb{R}^{n}} f(v)g(w)e^{\frac{2i}{\hslash}(u-v)\cdot\Theta^{-1}\cdot(u-w)}\mbox{d}v\mbox{d}w
\end{equation}
Equivalent formulas \ are
\begin{equation}
(f\star g)(u) := \frac{1}{(\pi\hslash)^{-n} }\int_{\mathbb{R}^{n}}\int_{\mathbb{R}^{n}} f(u+v)g(u+w)e^{\frac{2i}{\hslash}v\cdot\Theta^{-1}\cdot w}\mbox{d}v\mbox{d}w
\end{equation}
or
\begin{equation}
(f\star g)(u) := \frac{1}{(\pi\hslash)^{-n} }\int_{\mathbb{R}^{n}}\int_{\mathbb{R}^{n}} f(u+\frac{\hslash}{2}\Theta\cdot v)g(u+w)e^{i v\cdot w}\mbox{d}v\mbox{d}w
\end{equation}

A very useful remark is that Moyal product becomes the usual one 
in the limit $\hslash\to0$, in other words Moyal product can be viewed like
a deformation usual product in the deformation parameter  $\hslash$.
Therefore, we can interpret the quantum mechanic phase space as a deformation of
classical mechanic phase space obtained substituting the point wise product 
with Moyal one. This product will be used in the next chapter to construct a 
non-commutative $\mathbb{R}^4_\Theta$ vector space the so called Moyal plane.

It is possible to give some matrix basis for the algebra of functions on a two dimensional 
phase space with the star product. Defined the functions (taking $\hslash=1$)

\begin{equation}
\begin{array}{cc}
 a=\frac{1}{\sqrt{2}}(q+iq)  & \bar{a}=\frac{1}{\sqrt{2}}(q-ip)\\
\end{array}
\end{equation}
for a generic function  results
\begin{equation}
\begin{array}{cc}
 a\star f=af+\frac{\partial f}{\partial \bar{a} } & f \star a=af-\frac{\partial f}{\partial \bar{a} } \\ 
\end{array} \label{mbase1}
\end{equation}
and similar relations for $\bar{a}$. The function
\begin{equation}
\varphi_0=2e^{\frac{q^2+p^2}{2}}
\end{equation}
is Gaussian and has the useful property that
\begin{equation}
\varphi_0\star\varphi_0=\varphi_0 \label{mbase2}
\end{equation}
Defining the functions
\begin{equation}
\varphi_{mn}=\frac{1}{\sqrt{2^{n+m}n!m!}}\bar{a}^n\star\varphi_0 \star a^m
\end{equation}
and using the relations \eqref{mbase1} and \eqref{mbase2} it is easy to show that
\begin{equation}
\varphi_{mn}\star\varphi_{kl}=\delta_{nk}\varphi_{ml}
\end{equation}
Therefore the $\varphi$'s can be used as a base  \cite{bondia2} for the deformed algebra and the star multiplication
becomes the usual multiplication of (infinite) matrices.

\section{Spectral geometry}

\emph{In this section we introduce the 	argument of Connes' spectral geometry 
which is the non-commutative generalization of a usual geometry
on a manifold. In particular we will introduce some basic concepts like non-commutative infinitesimals and the  Dixmier trace  as  algebraic generalization of the usual infinitesimals and of the integral \cite{bondia,Connes,Madore}. A fundamental role will be taken by the generalized Dirac operator,  with which we will define the metric of a non-commutative space and  the spectral action of a Connes-Lott Model \cite{Schu,Schu2}}.

\subsection{Non-commutative infinitesimals}

To define the Diximier trace we need  some facts about compact operators \cite{Compact}.
Let us recall that an operator $A$  belonging $C^{*}$-algebras of  bounded operators over an Hilbert space $\mathcal{H}$, namely $A\in\mathcal{B}(\mathcal{H})$, is called of finite rank if the orthogonal complement of its kernel is finite
dimensional. Roughly speaking, even if the Hilbert space is infinite dimensional, such operators are finite dimensional matrices. Beside an operator $A$ on $\mathcal{H}$ is compact if it can be approximated in norm by finite rank operators. An equivalent way to characterize a compact operator $A$ is that $\forall\epsilon>0$ exists a subspace of finite dimension $E\subset \mathcal{H}$ which satisfies $\parallel A\mid_{E^\perp}\parallel<\epsilon$.\\ 
Taking in account the previous characterization of compact operators they can be seen in some sense small, so they are
good candidates to be infinitesimals. Calling $\mathcal{K}(\mathcal{H})\subset \mathcal{B}(\mathcal{H})$ the subset of the compact operators,  the size of the infinitesimal $A \in \mathcal{K}(\mathcal{H})$ can be defined by the rate of decay of a sequence $a_n(A)$ when $ n\rightarrow\infty$. Where $\{a_n (A)\}$ are the non vanishing 
eigenvalues of the operator $|A|$ arranged with repeated multiplicity.\\
The infinitesimals of order $\alpha$ are  all the operators $A \in \mathcal{K}(\mathcal{H})$ such that it is possible to to construct a sequence satisfying:
\begin{equation}
\begin{array}{l}
a_n(A) = O(n^{-\alpha}) \ \textrm{as} \  n\rightarrow\infty, \label{inf} \\ 
\\
\exists \ C<\infty \mid a_n(A)\leq Cn^{-\alpha}, \forall n\geq 1 . 
\end{array}
\end{equation}

\footnotesize Remark: 

The algebra $\mathcal{K}(\mathcal{H})$  is the only norm closed and two-sided ideal when $\mathcal{H}$ is separable and
it is essential, therefore is the largest two-sided ideal in the $C^{*}$-algebra $\mathcal{B}(\mathcal{H})$. Since the identity operator I on an infinite dimensional Hilbert space is not compact the algebra is not unital. Beside  the defining
representation of $K(\mathcal{H})$  is the only, up to equivalence, irreducible
representation of $K(\mathcal{H})$, in fact it is Morita equivalent to the algebras of finite rank matrices  of
complex numbers.

\normalsize

\subsection{The Dixmier Trace}
The Dixmier trace is a trace on a space of linear operators on an Hilbert space larger than the space of trace class
operators. This trace is defined  in order to have  the infinitesimals of order
$1$ in its domain but the higher order infinitesimals have vanishing trace. With this requirements the standard trace, with
domain in the two-sided ideal $ \mathcal{L}^1$ of trace class operators, is not suitable. The usual trace, defined as
$\operatorname{Tr}A := \sum_n \left\langle A\xi_n,\xi_n\right\rangle$, for any $A \in \mathcal{L}^1$,  is independent of the
orthonormal basis $ \xi_n,\ n\in N$ of $\mathcal{H}$. In the case of a positive and compact
operator $A$, the eigenvalues becomes positive and the definition turn to be $\operatorname{Tr}A := \sum_{n=0}^\infty a_n(A)$.
In general, an infinitesimal of order $1$ is not in $\mathcal{L}^1$, since we can not ensure the convergence  of its
characteristic values, we can just say that $a_n(A)\leq Cn^{-1}$ for some positive constant $C$. But we are sure  that 
$ \mathcal{L}^1$ contains infinitesimals of order higher than $1$, additionally for positive infinitesimals of order 1,
the standard trace is at most logarithmically divergent since $\operatorname{Tr}A = \sum_{n=0}^{N-1}a_n(A)\leq C\textrm{ln}N$.\\
The aim of the Dixmier trace \cite{Dixmier} is to extract the coefficient of the logarithmic divergence, enlarging the domain of the usual trace. Let $ \mathcal{L}^{(1,\infty)}$ indicate the ideal of infinitesimal of order $1$. As a first try, if $A\in\mathcal{L}^{(1,\infty)}$ is positive, we could 	think to define a positive functional using the limit
\begin{equation}
\displaystyle\lim_{N\to\infty}\frac{1}{\textrm{ln}N}\displaystyle\sum_{n=0}^{N-1}a_n(A)\label{dix1}
\end{equation}
But this definition is affected by two big problems: it is not linear and the lack of convergence. 
Dixmier proved \cite{Dixmier} that  exists an infinite number of scale invariant linear forms 
$\lim_\omega$ on the space $l^\infty(\mathbb{N})$ of bounded sequences.
For each such form $\lim_\omega$ there is a positive trace on the positive part of
$ \mathcal{L}^{(1,\infty)}$ defined by
\begin{equation}
\textrm{Tr} _\omega(A)=\textrm{lim}_\omega\frac{1}{\textrm{ln}N}\displaystyle\sum_{n=0}^{N-1}a_n(A), \ \ \forall A\in\mathcal{L}^{(1,\infty)}, \  A\geq 0
\end{equation}
This trace is called the Dixmier trace, it is invariant under unitary transformations since the eigenvalues $a_n(A)$ and
the sequence $\{\gamma_N \}$ are invariant as well. It satisfy the usual trace properties \cite{bondia} and
$$
\textrm{Tr} _\omega(A) = 0, \ \textrm{if} \ A \ \mbox {is of order higher than} \ 1
$$
This last statement follows from the fact that the space of all infinitesimals
of order higher than 1 forms a two-sided ideal and its elements satisfy

\begin{eqnarray}
a_n(A) = o\left(\frac{1}{n}\right), &\textrm{or}& na_n(T)\to 0 \ \textrm{as} \ n\to\infty
\end{eqnarray}
The higher order sequence $\{\gamma_N \}$ converges to zero vanishing the Dixmier trace.
In many practical examples in physic, like gravity and Yang-Mills theories, the sequence $\{\gamma_N\}$ converges itself. In these cases, from the property of Dixmier trace, the trace is given by \eqref{dix1} and does not depends on the choice of linear form $\omega$, this type of operators are called measurable \cite{bondia,landi}.

\subsection{Spectral triples}

We now define the spectral triple, it will be the core object of the Connes-models.
A spectral triple $(\mathcal{A},\mathcal{H},\mathcal{D})$ is defined by an involutive algebra $\mathcal{A}$ represented
as an algebra of bounded operators on the Hilbert space $\mathcal{H}$ and with a
self-adjoint operator $\mathcal{D} = \mathcal{D}^{\dagger}$ on $\mathcal{H}$ with the properties:
\begin{enumerate}
\item The resolvent $(\mathcal{D}-\lambda)^{-1}, \ \lambda\in \mathbb{R}$, is a compact operator on $\mathcal{H}$;
\item $[\mathcal{D},a] = \mathcal{D}a-a\mathcal{D} \in\mathcal{B}(\mathcal{H})$, for any $a\in\mathcal{A}$.
\end{enumerate}

The first statement tell us that the self-adjoint operator $\mathcal{D}$ has a real
discrete spectrum of eigenvalues $\lambda_n$ and each eigenvalue has a finite multiplicity.
Additionally, $|\lambda_n|\to\infty$ as $n\to\infty$; since $(\mathcal{D}-\lambda)^{-1}$
is compact, it has characteristic values $a_n(\mathcal{D}-\lambda)^{-1}\to 0$, from which
$|\lambda_n|=a_n(|\mathcal{D}|) \to\infty$. The second condition can be relaxed to be satisfied only for
a dense sub-algebra of $\mathcal{A}$. 
A spectral triple $(\mathcal{A},\mathcal{H},\mathcal{D})$ is said to be of dimension $n>0$ 
if $|\mathcal{D}|^{-1}$ is an infinitesimal (in the sense of \eqref{inf}) of order $1/n$ or, in other words, 
if $|\mathcal{D}|^{-n}$ is an infinitesimal of order 1.\\
The dimension of spacetime is a local property and it can be equivalently found from the asymptotic 
behavior of the spectrum of the Dirac operator for large eigenvalues. Ordering the eigenvalues,$ \cdots\lambda_{n−1}\leq
\lambda_{n-1}\leq\lambda_{n+1}\cdots$, the Weyl’s spectral theorem states that the eigenvalues grow asymptotically 
as $n^{1/\textrm{dim}M}$. The  local property of spacetime  are encoded
in the high energy part of the spectrum. This agreement with our intuition from
quantum mechanics and motivates the name spectral triple.
In following pages many regularity conditions on elements of $\mathcal{A}$ will be
defined \cite{Connes} using only the operator $\mathcal{D}$ and its modulus $|\mathcal{D}|$ and $\mathcal{D}$
will be a generalization of the usual Dirac operator on an ordinary spin manifold,  for simplicity
we will call it just Dirac operator.\\
In order to find the analogue of the measure integral,  the Dirac operator will play a role for definition of
the volume.
With such notion  we can define the integral of any $b\in\mathcal{A}$, for a 
$n$-dimensional spectral triple, by the following formula:
\begin{equation}
\int b :=\operatorname{Tr}_\omega(b|\mathcal{D}|^{-n})\label{nc-int}
\end{equation}
Where the operator $|\mathcal{D}|^{-n}$ is just used to bring the bounded operator $b$ into 
$\mathcal{L}^{(1,\infty)}$, in this way  the Dixmier trace is well defined. 
In the previous definition operator the $|\mathcal{D}|^{-n}$ is the analogue of the volume form of the space and can be proved that the integral \eqref{nc-int} is a non-negative normalized trace on $\mathcal{A}$ given a general spectral triple $(\mathcal{A},\mathcal{H},\mathcal{D})$ \cite{Connes}.

Now let us introduce a more structured spectral triple: the even spectral triple. The object composed by five items $(\mathcal{A},\mathcal{H},\mathcal{D},J,\chi)$ form what Connes calls \cite{Connes2} an even real spectral
triple. Again $\mathcal{A}$ is a real  associative involution algebra with unit, represented faithfully by bounded operators on the Hilbert space $\mathcal{H}$, $\mathcal{D}§$ is an unbounded self adjoint operator on $\mathcal{H}$.
In addition, $J$ is an anti-unitary operator and $\chi$ a unitary one.
They fulfill the following properties:
\begin{enumerate}
\item $J^2=-1$ in four dimensions ($J^2=1$ in zero dimensions)
\item $ \left[\rho(a),\mathcal{D}\rho(\tilde{a})\mathcal{D}^{-1}\right] =0$ for all $a,\tilde{a}\in\mathcal{A}$
\item $DJ = JD$
\item $\left[\mathcal{D},\rho(a))\right]$ is bounded for all $a\in\mathcal{A}$ and $\left[\left[\mathcal{D},\rho(a)\right], \mathcal{D}\rho(\tilde{a} )\mathcal{D}^{-1}\right] = 0$ for all  $a,\tilde{a}\in\mathcal{A}$
\item $\chi^2=1$ and $\left[\chi,\rho(a)\right] =0$ for all $a\in\mathcal{A}$ 
\item  $J\chi=\chi J$
\item  $\mathcal{D}\chi=-\chi \mathcal{D}$
\end{enumerate}
Furthermore, can be required some other properties: Poincar\'e duality, regularity, which grands that 
our functions $a \in \mathcal{A}$ are differentiable and orientability, which connects
the volume form to the chirality.
The fourth property is called first order condition because it grants that the Dirac operator is a first order differential
operator and property 5 allows the decomposition $\mathcal{H}=\mathcal{H}_L\otimes \mathcal{H}_R$.
\\
These properties were promoted to  axioms by Connes defining an even real spectral triple justified by his Reconstruction Theorem \cite{Connes3}. Considering an even spectral triple  $(\mathcal{A},\mathcal{H},\mathcal{D},J,\chi)$, with commutative algebra $\mathcal{A}$, then exists a compact Riemannian spin manifold $M$ of
even dimensions whose spectral triple $(\mathcal{C}^\infty(M),L^2(S),\slashed{\partial},C,\gamma_5)$ coincides
exactly with $(\mathcal{A},\mathcal{H},\mathcal{\mathcal{D}},J,\chi)$. Where $\mathcal{C}^\infty(M)$ is the algebra of the infinity
 derivable functions on $M$, $ L^2(S)$ is the space of square integrable functions on the spinor space $S$,
$\slashed{\partial}=\gamma^i\partial_i$ is the standard Dirac operator, $C$ is the charge conjugation operator and $\gamma_5$ is the usual
chirality operator $\gamma_5=\gamma_1\gamma_2\gamma_3\gamma_4\cdots\gamma_n$. \\

In this theorem are contained a lot of informations about the role of the Dirac operator.  
Beside, describing the dynamics of the spinor field, the dimension of spacetime and its integration,  
the Dirac operator $\slashed{\partial}$ encodes its Riemannian metric and its differential forms.
The metric can be reconstructed from the  spectral triple by Connes distance formula; 
a point $x \in M$ is reconstructed 
as the pure state and the general definition of a pure state of course does not use the
commutativity. A state $\delta$ of the algebra $\mathcal{A}$ is a linear form on $\mathcal{A}$, that is normalized 
$\delta(1) = 1$ and positive $\delta(a^{*}a)\geq 0$ for all $a\in\mathcal{A}$. A state is called pure if it cannot be
written as a linear combination of two states. In the commutative case, there
is a bijective correspondence between points $x\in M$ and pure states $\delta x$ defined
by the Dirac distribution, $\delta_x(a):=a(x)=\int_M\delta_x(y)a(y)\textrm{d}^4y$. The geodesic distance
between two points $x$ and $y$ is reconstructed by the Connes distance formula:
\begin{equation}
\textrm{sup}\{|\delta_x(a)-\delta_y(a)|; \  a\in C^\infty(M) \ \mid \ \parallel[\mathcal{\mathcal{D}},\rho(a)]\parallel\leq1\}\label{condis}
\end{equation}
For a general even spectral triple the operator norm $\parallel[\mathcal{\mathcal{D}},\rho(a)]\parallel$, in the distance formula, 
is bounded by axiom.\\
For example consider the circle $M=S^1$ of circumference $2\pi$ with Dirac operator $\partial=i\textrm{d}/\textrm{d}x$.
A function $a\in C^\infty(S^1)$ is represented faithfully on the space of square integrable functions $\psi\in L^2(S^1)$
by pointwise multiplication, $(\rho(a)\psi)(x) = a(x)\psi(x)$. The commutator $[\partial,\rho(a)]=
i\rho(a^\prime)$ is bounded and we have already seen it in quantum mechanic. The operator norm in this case is 
\begin{equation}
\parallel[\partial,\rho(a)]\parallel:=2\pi  \  \textrm{sup}_\psi|[\partial,\rho(a)]\psi|/|\psi|=\displaystyle\textrm{sup}_x|a^\prime(x)| 
\end{equation}
Where $|\psi|^2= \int_0^{2\pi}\bar{\psi}(x)\psi(x)\textrm{d}x$. 
Using \eqref{condis} we recover the standard distance between two points $x$ and $y$ on the circle:
\begin{equation}
\textrm{sup}\{|a(x)-a(y)|\mid \textrm{sup}_x|a^\prime(x)|\leq 1\}=|x-y|.
\end{equation}
It is important to note that Connes distance formula works for non-connected manifolds too, 
even for discrete spaces of dimension zero or collection of points.
The last ingredient that we need are the differential forms and again they can be recovered using the Dirac operator by an
analogy with quantum mechanic. The differential form of degree one, like $\textrm{d}a$, for a function $a\in\mathcal{A}$ are
reconstructed as $-i[\mathcal{\mathcal{D}},\rho(a)]$. For example a 1+1 dimensional spacetime, d$a$ is just the time derivative of the observable $a$ and is associated with the commutator of the Hamilton operator with $a$. Higher degree differential forms are obtained by multiple application of the commutator with $\mathcal{D}$.\\ 
We define a non-commutative geometry by a real spectral
triple, which contains all the geometric informations, with non-commutative algebra $\mathcal{A}$ .

\subsection{Spectral Action}

\cite{S-A} The axioms of the spectral triple allow us a change of point of view. A quite suited analogy is the Fourier transform, in which the points $x\in M$ of a geometric space are replaced by the spectrum $\Sigma\subset\mathbb{R}$ of the operator $\mathcal{\mathcal{D}}$.
In fact, we can forget about the algebra $\mathcal{A}$ in the spectral triple $(\mathcal{A},\mathcal{H},\mathcal{D})$ and focus ourself 
only on the operators $\mathcal{D}$, $\gamma$ and $J$ acting in $\mathcal{H}$ and we are able to characterize this data by the spectrum
$\Sigma$ of $\mathcal{D}$ which, for the even case $\Sigma =-\Sigma$, is a discrete subset with multiplicity of $\mathbb{R}$. So we can argue that all the physical informations, therefore the physical action, only depends the spectrum of $\mathcal{D}$. The next step is to look for the  existence of Riemannian manifolds which  have the same $\Sigma$, namely isospectral and in general not isometric. The previous hypothesis is much stronger than the invariance of the action under diffeomorphisms of general relativity. This approach has the virtue of not require the commutativity of the algebra $\mathcal{A}$ in order to apply this principle to a physical action. Indeed, the spectral triple axioms require just the much weaker condition between the algebra and the opposite algebra:
\begin{equation}
[a,b^0]=0 \  \forall a,b\in\mathcal{A} \  \textrm{with} \ b^0=Jb^{*}J^{-1} \label{opaxi}
\end{equation}\\
Analyzing, in the Riemannian case $\mathcal{A}=C^\infty(M)$, it is  possible to construct an isomorphism between the group  of diffeomorphism of the manifold Diff$(M)$ and the group automorphisms of the algebra Aut($\mathcal{A}$). To each diffeomorphism $\phi \in \textrm{Diff}(M)$ it is associated the algebra preserving map $\pi_\phi : \mathcal{A}\to \mathcal{A}$ given by:
\begin{equation}
\pi_\phi(f)=f \circ\phi^{-1} \ \ \forall f \in \mathcal{A}=C^\infty(M)
\end{equation}
This association is in general true, the group Aut$(\mathcal{A})$ of automorphisms of the algebra $\mathcal{A}$ is the generalization of the diffeomorphisms to the non-commutative geometry $(\mathcal{A},\mathcal{H},\mathcal{D})$. 
It is important to notice that in the general case there is a non trivial normal subgroup of the group Aut$(\mathcal{A})$
\begin{equation}
\textrm{Int}(\mathcal{A})\subset\textrm{Aut}(\mathcal{A})
\end{equation}
where $\textrm{Int}(\mathcal{A})$ is the group of inner automorphisms; $\pi$ is called inner if exists a unitary operator $u \in \mathcal{A}$ such that $\pi(a) = uau^{*}$ \ \ $\forall a \in \mathcal{A}.$\\
The subgroup Int$(\mathcal{A})$ of inner automorphisms is a normal subgroup and the group Aut$(\mathcal{A})$ of diffeomorphisms falls in equivalence classes under Int$(\mathcal{A})$.
This induces a natural foliation into equivalence classes in the space of metrics. \\
The internal fluctuations of a given metric are given by the formula,
\begin{equation}
 \mathcal{D} = \mathcal{D}_0 + A + JAJ^{-1},\ \   A\in\Omega^1(\mathcal{A}) \label{fluc}
\end{equation}
where 
\begin{equation}
\Omega^1(\mathcal{A}):= \left\{A=\Sigma_i a_i [\mathcal{D}_0 ,b_i ], \ \mid \  a_i,b_i\in \mathcal{A} \ \  \textrm{and} \ A=A^{*}\right\}
\end{equation}
Applying the previous formula to $(\mathcal{A},\mathcal{H},\mathcal{D}_0)$, where $\mathcal{D}_0$ is the un-fluctuated $\mathcal{D}$, the fluctuations does not
affect the representation of $\mathcal{A}$ in $\mathcal{H}$ but perturbs the operator $\mathcal{D}_0$ by \eqref{fluc} where $A$ is an arbitrary self-adjoint operator in $\mathcal{H}$ of the form $A=\Sigma a_i[\mathcal{D}_0,b_i] \ \ a_i,b_i \in \mathcal{A}$. 
The fluctuated Dirac operator continues to satisfy the axioms and in the commutative case (i.e. Riemann) the group of inner
automorphisms Int$(\mathcal{A}) = \{1\}$ is trivial, as a consequence the fluctuations are trivial too $\mathcal{D} = \mathcal{D}_0$ \\
The action of $\tilde{\textrm{Int}}(A)$ (where the tilde stands for taking into account
the action of automorphisms on the Hilbert space $\mathcal{H}$) on the space of metrics is restricted on the
above equivalence classes and is given by: 
\begin{equation}
\xi \in \mathcal{H} \to u\xi u^{*} = uu^{*0} \xi, \ A \to u[\mathcal{D}, u^{*}] + uAu^{*} 
\end{equation} \\
From the previous described properties of a general real spectral triple follows that it can be used to define a gauge
theory. The gauge fields are recovered from the inner fluctuations of the Dirac operator 
and the gauge group is given by the unitary elements in the algebra,  we still need to define an physical action. Let $(\mathcal{A},\mathcal{H},\mathcal{D},J)$ be a real spectral triple, given the fluctuated operator $\mathcal{D}$, a positive even function $\chi$ and a cut-off scale $\Lambda$, it is possible to define a gauge invariant spectral action for the bosonic sector:
\begin{equation}
S_b[\mathcal{A}]:=\operatorname{Tr}\chi\left(\frac{\mathcal{D}}{\Lambda}\right)
\end{equation}
The cut-off parameter $\Lambda$ is used to obtain an asymptotic series for the spectral action, in this way the
physically relevant terms will appear as the coefficients of an expansion in positive power of $\Lambda$. For the fermionic sector,  we can define a fermionic action in terms of $\psi \in H$ and $A \in \Omega^1(\mathcal{A})$:
\begin{equation}
S_f[A,\psi] := \left\langle\psi,\mathcal{D}\psi\right\rangle 
\end{equation}

In the last part of the chapter we sketch a physical application of the spectral action, treating the case of standard model plus gravity described by the action functional
\begin{equation}
S= S_E + S_{SM} \label{SM}
\end{equation}
where $S_E = \frac{1}{16\pi\mathcal{G}}\int R\sqrt{g} \ \textrm{d}^4x$ is the Einstein action and  $S_{SM}$ is the standard model action. It involves, beside the metric, other fields: bosons of spin 0 such as the Higgs, bosons of spin 1 like $\gamma$, $W^\pm$ and $Z$, the eight gluons, fermions, quarks and leptons. The two parts of the action  have a priori a very different origin; the interaction of $S_{SM}$, which  is governed  by a gauge invariance group, is a priori
very different from the interaction of the Einstein action which is governed by invariance under the diffeomorphism group. Formally the natural group of invariance of the functional \eqref{SM} is
the semidirect product,
\begin{equation}
G = \mathcal{U}\rtimes \textrm{Diff}(M)
\end{equation}
Where  $\mathcal{U}= C^\infty (M, U(1)\times SU(2)\times SU (3))$ is the group of local gauge transformations.\\
It is very useful to compare the exact sequence of endomorphisms   groups of $\mathcal{A}$ ,
\begin{equation}
1\to\textrm{Int}(\mathcal{A})\to\textrm{Aut}(\mathcal{A})\to\textrm{Out}(\mathcal{A})\to 1 \label{ex1}
\end{equation}
with the exact sequence which describes the structure of the symmetry group $G$ of the action functional \eqref{SM}.
\begin{equation}
1 \to\mathcal{U}\to G\to\textrm{Diff}(M)\to 1 \label{ex2} 
\end{equation}
The \eqref{ex1} and \eqref{ex2} look very similar and a natural question arises: to find an algebra  $\mathcal{A}$ which satisfy $\textrm{Aut}(\mathcal{A})=G$ condition. Taking into account the action of inner automorphisms 
of $\mathcal{A}$ in $\mathcal{H}$ given by
\begin{equation}
\xi\to u(u^{*})^0\xi = u\xi u^{*}
\end{equation}
this algebra turns out to be: 
\begin{equation}
 \mathcal{A}= C^\infty(M)\otimes\mathcal{A}_F
\end{equation}
With
\begin{equation}
 \mathcal{A}_F=C \otimes\mathbb{H}\otimes M_3(\mathbb{C}) \label{tesAlg}
\end{equation}
The algebra $\mathcal{A}_F$ is finite dimensional and $\mathbb{H}\subset M_2(\mathbb{C})$ is the algebra of quaternions,
\begin{equation}
\begin{pmatrix}
\alpha & \beta \\
-\bar{\beta} & \bar{\alpha} \\
\end{pmatrix} \  \ \alpha,\beta \in \mathbb{C}
\end{equation}
Next to the algebra $\mathcal{A}$, we
need the action of $\mathcal{A}$ in $\mathcal{H}$ and a suitable Dirac operator operator $\mathcal{D}$. 
The algebra $\mathcal{A}$ \eqref{tesAlg} is a tensor product of two algebras which corresponds to a product of spectral triples given by:
\begin{equation}
(\mathcal{A},\mathcal{H},\mathcal{D})=(C^\infty(M),L^2(M,S),\ \slashed{\partial}_M)\otimes (\mathcal{A}_F,\mathcal{H}_F,\mathcal{D}_F)
\end{equation}
with
\begin{equation}
\mathcal{H}= L^2(M,\mathcal{S})\otimes\mathcal{H}_F, \  \mathcal{D}=\slashed{\partial}_M\otimes 1 +\gamma_5\otimes \mathcal{D}_F
\end{equation}
The algebra $\mathcal{A}_F$ is finite dimensional so the dimension of the corresponding space is 0 and 
$\mathcal{H}_F$ must be finite dimensional \cite{Schu1}. The elementary fermions provide a
natural candidate for $\mathcal{H}_F$ and  a finite Hilbert base can be labeled
by elementary leptons and quarks. For instance, for the first generation of leptons we get
$e_L$, $e_R$, $\nu_L$, $e_L$, $e_R$, $\nu_L$  as the corresponding basis. The helicity operator $\gamma_F$ is given by the usual $\gamma_5$
and distinguishes the left handed particles and right handed ones. For quarks
one has an additional color index, $g, r, b$. The  real structure $J$ is just such that $Jf = \bar{f}$
for any $f$ in the basis. Additionally, the algebra $\mathcal{A}_F$ has a natural representation in $\mathcal{H}_F$ and:
\begin{equation}
 ab^0 = b^0a \ \ \forall  a,b \in  \mathcal{A}_F, \ \ b^0 = Jb^{*}J^{-1} .
\end{equation}
The finite Dirac operator acting in the finite dimensional Hilbert space $\mathcal{A}_F$ which fulfill the spectral triple axioms \eqref{opaxi} is:
\begin{equation}
\mathcal{D}_F =\begin{pmatrix} Y & 0 \\
                     0 & \bar{Y}
\end{pmatrix} 
\end{equation}
where $Y$ is the Yukawa coupling matrix.\\
Now we are able to determine the internal fluctuations using \eqref{fluc}, computing the internal fluctuations of the above product geometry $M \times F$, skipping the calculus for brevity, we recover the bosons \cite{Witt,Schu}. In fact, the internal fluctuations are parametrized exactly by the bosons $\gamma$, $W^\pm$ and $Z$ the eight gluons and the Higgs field $H$ of the standard model.
The last step is to compute the spectral action for the fluctuated $\mathcal{D}$, the calculus are quite cumbersome but it is possible 
to prove \cite{Witt,S-A,Schu} that for any smooth cut-off function $\chi$, $\chi(\lambda) = 1$ for $|\lambda| \leq  1$, we have:
$$
\operatorname{Tr}\chi\left(\frac{\mathcal{D}}{\Lambda}\right)= S_E + S_G + S_{GH} + S_H + S_C + O(\Lambda^{-\infty} )
$$
The last missing contributions, the femionic sector, in terms of the operator $\mathcal{D}$ alone are given by the equality:
$$
\langle\psi,\mathcal{D}\psi\rangle=\int_M (\mathcal{L}_{Gf} + \mathcal{L}_{Hf})\sqrt{g}\textrm{d}^4x 
$$
In \cite{nc-sm} this construction was improved to include massive neutrinos and  to solve some technical issues \cite{fermion-dub} at the same time.\\
For completeness, we end this chapter with some mathematical tools 	useful to compute (in particular conditions) the spectral action using the heat kernel expansions and Seeley-De Witt coefficients \cite{Seeley}. For  a vector bundle $V$ on a compact Riemannian manifold $(M,g)$ and  a second-order elliptic differential operator $P:C^\infty(V)\to C^\infty(V)$ of the form
\begin{equation}
P=-(g^{\mu\nu} \partial_\mu\partial_\nu + K^\mu\partial_\mu +L)\label{see}
\end{equation}
with $K^\mu , L \in \Gamma(\textrm{End}(V))$, it is possible to find a unique connection $\nabla$ and an endomorphism $E$ on $V$  satisfying:  
\begin{equation}
P=\nabla\nabla^{*}-E.
\end{equation}
Or locally $\nabla_\mu = \partial_\mu + \omega^\prime_\mu$, where
\begin{equation}
\omega^\prime_\mu =\frac{1}{2}(g_{\mu\nu}K^\nu + g_{\mu\nu}g^{\rho\sigma}\Gamma^\nu_{\rho\sigma})
\end{equation}
Using this $\omega^\prime_\mu$ and $L$ we find $E\in\Gamma(\textrm{End}(V))$ and we can compute the curvature $\Omega_{\mu\nu}$ of $\nabla$:
\begin{eqnarray}
E&:=& L-g^{\mu\nu}\partial_\nu(\omega^\prime_\mu)-g^{\mu\nu} \omega^\prime_\mu \omega^\prime_\nu +g^{\mu\nu} \omega^\prime_\rho\Gamma^\rho_{\mu\nu} \\
\Omega_{\mu\nu}&:=& \partial_\mu(\omega^\prime_\nu)- \partial_\nu(\omega^\prime_\mu)-\left[\omega^\prime_\mu,\omega^\prime_\nu\right]
\end{eqnarray}
Now it is convenient to make an asymptotic expansion (as $t \to 0$) of the trace of the operator
$e^{-tP}$ in powers of $t$:
\begin{eqnarray}
\operatorname{Tr}e^{-tP}\sim\sum_{n\geq 0} t^{(n-m)/2}a_n(P ),& & a_n(P):=\int_{M} a_n(x,P)\sqrt{g}\textrm{d}^m x \label{hker}
\end{eqnarray}
The coefficients $a_n(x,P)$ are called the Seeley-DeWitt coefficients and $m$ is the dimension of $M$. Can be proved \cite{Seeley} that $a_n(x,P) = 0$ for $n$ odd and  the first three even coefficients are given by
\begin{eqnarray} 
a_0(x,P)&=&(4\pi)^{-m/2} \operatorname{Tr}(\textrm{id}) \\
a_2(x,P)&=&(4\pi)^{-m/2} \operatorname{Tr}(-R/6 \ \textrm{id} + E)\\
a_4(x,P)&=&(4\pi)^{-m/2}\frac{1}{360}\operatorname{Tr}(-12R_{;\mu}^\mu + 5R^2-2R_{\mu\nu}R^{\mu\nu}+ \nonumber \\ 
& &2R_{\mu\nu\rho\sigma}R^{\mu\nu\rho\sigma}-60RE+180E^2+60E_{;\mu}^\mu +30\Omega_{\mu\nu}\Omega^{\mu\nu}) 
\end{eqnarray}
where $R_{;\mu}^\mu := \nabla^\mu\nabla_\mu R$ and $E_{;\mu}^\mu := \nabla^\mu\nabla_\mu E$. Considering  manifolds 
without boundary, the terms $E_{;\mu}^\mu$, $R_{;\mu}^\mu$ vanish due to the Stokes Theorem.
This expansion is very useful in some computations of the spectral action. Taken a fluctuated Dirac operator $\mathcal{D}$ for which
$\mathcal{D}^2$ can be written as \eqref{see} on some vector bundle $V$ on a compact Riemannian manifold $M$ and  writing $\chi$ as a Laplace transform, we obtain
\begin{equation}
\chi(\mathcal{D}_A/\Lambda)=\int_{t>0}\tilde{g}(t)e^{-tD^2_A /\Lambda^2}\textrm{d}t.
\end{equation}
Using \eqref{hker} we find that for a four-dimensional manifold the relevant terms of the expansion are
\begin{equation}
\operatorname{Tr}\chi(\mathcal{D}_A / \Lambda)=2\Lambda^4 \chi_4 a_0(\mathcal{D}^2_A)+2\Lambda^2 \chi_2 a_2(\mathcal{D}^2_A) +a_4(\mathcal{D}^2_A)\chi(0)+\mathcal{O}(\Lambda^{-2}),
\end{equation}
where  $\chi_k$ are the moments of the function $\chi$ :
\begin{equation}
\chi_k:=\int_0^\infty \chi(w)w^{k-1}\textrm{d}w; \ \ \ (k > 0).
\end{equation}

%% file: cap2.tex
\chapter{8-dim spectral action }

In this chapter will be computed a spectral action starting from a \-non\--com\-mu\-ta\-ti\-ve spectral triple. The feature of this particular triple is the choice of a 4-dimension Harmonic  Dirac  operator. The idea behind this construction \cite{8-dim} is to relate  the Dirac operator, which plays a fundamental role in a spectral triple, with the oscillator Hamiltonian operator.Roughly speaking, we  look at the Dirac operator as a generalization of the Laplace operator  so we have $\mathcal{D}^2\approx H$. 
Considering the spectrum $\omega(2n+1)$ $n \in N$ of  the one-dimensional  harmonic oscillator Hamiltonian $H$, can be deduced  that  $H^{-1}$ is a non-commutative infinitesimal of order one. According to the previous discussion  the non-commutative dimension of a spectral triple, equipped with the 4D harmonic Dirac operator $ \mathcal{D}^2_4\approx H_4$, is fixed by the order of the inverse operator $\mathcal{D}^{-1}$ which is eight not four. Due to the fact that the spectral dimension is defined by the Dirac operator, it is connected to the 
phase space dimension and not on the one of the configuration space \cite{con-phase}.
In order to construct such harmonic Dirac operator and the spectral triple we will work in the framework of the generalized n-dimensional harmonic operator. Therefore, will be studied the four-dimensional case in order to construct the non-commutative spectral triple which is starting point for the field theory we are interested in.
The first part of the chapter will be devoted to introduce the harmonic Dirac operator starting from the one-dimensional case and then generalizing it using the annihilations and creations operator for the bosonic and fermionic sectors.   
Having the 4-dimensional harmonic Dirac operator with harmonic oscillator spectrum, to implement the Higgs mechanism we will consider 
the tensor product of the non-commutative triple with a finite Connes-Lott type spectral triple \cite{con-lot}.
We will fluctuate the total Dirac operator following the previous described standard machinery \cite{S-A,Connes3} of non-commutative geometry to get "Gauged" Dirac operator. Thus  we  will proceed to compute the spectral action in which are present two U(1)-Moyal Yang-Mills fields unified with a complex Higgs field.

\section{Harmonic Dirac operators }

To introduce this subject we start to show the simple case of one dimensional harmonic oscillator, then will be introduced 
the general n-dimensional case.
In one dimension the Hamiltonian of an harmonic oscillator is well known and is defined as
\begin{equation}
H_1=-\frac{d^2}{dx^2} +\omega^2 x^2
\end{equation}
Using the Hermite polynomials is possible to construct an orthonormal base of eigenfunctions of $H_1$ in the Hilbert space $L^2(\mathbb{R})$ of square integrable functions on $\mathbb{R}$ 
\begin{eqnarray}
H_1 \phi_n=\omega(2n+1)\psi_n, & \psi_n(x)=\left(\frac{\omega}{\pi}\right)^{\frac{1}{4}}\frac{H^n(\sqrt{\omega}x)}{\sqrt{2^nn!}}
e^{- \frac{\omega x^2}{2}},& n\in \mathbb{N}
\end{eqnarray}
where $H^n(z)$ are the Hermite polynomials. From the behavior for $n \to \infty$ of the eigenvalues of $H_1$ we can infer that
the inverse operator $H_1^{-1}$ is a first order non commutative infinitesimal. Reminding previous chapter, it means that after arranging the eigenvalues in deceasing order taking in account the multiplicities, the order of the generic eigenvalues $\alpha_n$ is $\mathcal{O}(n^{-1})$.  From this evidence we can deduce that taking a Dirac operator to be 
$\mathcal{D}^2 \approx H_1$, the non-commutative order of this Dirac operator will be 2. In order to define such Dirac operator we need a Clifford algebra of dimension two which can be represented by the Pauli matrices. In this setting a choice of a two dimensions harmonic Dirac operator can be:
\begin{equation}
\mathcal{D}_1=i\sigma_1 \frac{d}{dx}+\sigma_2\omega x=\left(
\begin{array}{cc}
0& i\left(\frac{d}{dx}+\omega x \right) \\
i\left(\frac{d}{dx}-\omega x \right) & 0
\end{array}
 \right)=\left(
\begin{array}{cc}
0& \sqrt{2\omega}a \\
-\sqrt{2\omega}a^\dagger & 0
\end{array}\right) \label{dirac1}
\end{equation}

Where $a=\frac{1}{\sqrt{2\omega}}(\frac{d}{dx}+\omega x)$ and $a^\dagger=\frac{1}{\sqrt{2\omega}}(\frac{d}{dx}-\omega x)$ are the usual annihilation ad creation operators which satisfy the commutation relation $[a,a^\dagger]=1$. To complete a spectral triple data we need to define the Hilbert space on which this operator acts, an algebra and its representation on $H$. It easy to see that the Hilbert space is just $\mathcal{H}_1=L^2(\mathbb{R})\otimes \mathbb{C}^2$, about the algebra the simplest choice is the commutative algebra of the Schwartz class functions $\mathcal{A}=\mathcal{S}(\mathbb{R})$. The representation can be defined as the pointwise diagonal multiplication $\pi(f)(\psi_1,\psi_2)=(f\psi_1,f\psi_2)  $ with $\psi_1,\psi_2\in L^2(\mathbb{R})$. This spectral triple can be turned in a even spectral triple finding a suitable real structure $J_1$, using the axioms of even spectral triple  $J^2_1=\epsilon$, $J_1\mathcal{D}=\epsilon^\prime \mathcal{D}J_1$,   $J_1\chi=\epsilon^{\prime\prime} \chi J_1$ and the anti linearity, $J_1$ turns out to be $(J_1\psi)(x)=\sigma_3\overline{\psi(x)}$ with the choice $\epsilon=1$, $\epsilon^\prime=1$, $\epsilon^{\prime\prime}=1$. Can be proved \cite{non-com-tri}  that all the axioms of a even spectral triple are fulfilled, in particular the constraints about the opposite algebra  are satisfied by this choice of $J_1$ and the operator $[D,\pi(f)]=i\sigma_1\pi(f^\prime)$  is bounded.

The previous construction can be generalized in $n$-dimensions using the Clifford algebra of $\mathbb{R}^{2n}$ represented on the  Hilbert space $ \mathbb{C}^{2^n}$, in order to do it is very useful to consider $b$, $b^\dagger$ the fermionic annihilation and creation operators with the usual anti-commutation rules:
\begin{equation}
\begin{array}{ccc}
\{b,b^\dagger\}&=&\textbf{I}  \\
\{b,b\}&=&0  \\
\{b^\dagger,b^\dagger\}&=&0
 \end{array}
\end{equation}
With these operators we can redefine the previous one dimensional spectral triple as:
\begin{eqnarray}
\mathcal{D}_1=-i\sqrt{2\omega}a^\dagger\otimes b + i\sqrt{2\omega}a\otimes b^\dagger&=&i\frac{d}{dx}\otimes(b+b^\dagger)+ i\omega x\otimes(b-b^\dagger) \nonumber \\
\mathcal{H}_1=L^2(\mathbb{R})\otimes(\mathbb{C}\mid 0\rangle \oplus \mathbb{C} b^\dagger\mid 0\rangle) & & \chi_1=1\otimes(b^\dagger b- bb^\dagger)
\end{eqnarray}
Where $\mathbb{C}|0\rangle$ is the space generated by the vacuum state defined as $ b|0\rangle=0$. Now the generalization to $n$-dimensional harmonic oscillator becomes straightforward if we considering $n$-dimensional fermionic annihilation and creation operators $b_\mu$, $b_\nu^\dagger$ and  $n$-dimensional bosonic annihilation and creation operators $a_\mu$, $a^\dagger_\nu$ satisfying for $\mu,\nu=1,\cdots,n$:
\begin{eqnarray}
&[a_\mu,a_\nu]=[a^\dagger_\mu,a^\dagger_\nu]=0, & [a_\mu,a^\dagger_\nu]=\delta_{\mu\nu} \label{com1} \\ 
&\{b_\mu,b_\nu\}=\{b_\mu^\dagger ,b_\nu^\dagger\}=0, & \{b_\mu,b_\nu^\dagger\}=\delta_{\mu\nu} \label{anti1}
\end{eqnarray}
Where $a_\mu =\frac{1}{\sqrt{2\omega}}  (\omega x_\mu + \partial_\mu ), \ a_\mu^\dagger = \frac{1}{\sqrt{2\omega}}(\omega x_\mu-\partial_\mu ) $.
The generalization of the Dirac operators \eqref{dirac1} using this operator is:
\begin{equation}
\mathcal{D}_n=-i\sqrt{2\omega}\delta^{\mu\nu}a_\mu^\dagger\otimes b_\nu + i\sqrt{2\omega}\delta^{\mu\nu}a_\mu\otimes b_\nu^\dagger=i\frac{d}{dx_\mu}\otimes(b_\mu+b_\mu^\dagger)+ i\omega x^\mu\otimes(b_\mu-b^\dagger_\mu) \label{dirac2}
\end{equation}
summed over repeated index. In  analogy with the one dimensional case we can define the Hilbert space on which  the Dirac operator \eqref{dirac2} acts starting from the vacuum state by subsequent applications of  the fermionic creation operators  $b_\nu^\dagger$  on the vacuum, using the anti-commutation relations \eqref{anti1} and the definition of vacuum state 
$b|0\rangle=0$. We call this space $\Lambda(\mathbb{C}^n)$ and therefore the Hilbert space is $ \mathcal{H}_n=L^2(\mathbb{R}^n)\otimes\Lambda(\mathbb{C}^n)$. Beside, we can define a grading operator $\chi_n $ as:
 \begin{eqnarray}
\chi_n= \textbf{I}\otimes \prod^n_{\mu=1}(b_\mu b^\dagger_\mu-b_\mu b^\dagger_\mu )
\end{eqnarray}
An interesting feature of the Dirac operator \eqref{dirac2} is the preserving of the sums of excitations of bosons and fermions in the Fock space, in other words is super-symmetric. However, the super-symmetry will no longer holds for the total spectral triple due to the non-preserving behavior of algebra that will be chosen.
Using the relations \eqref{com1}-\eqref{anti1} we can compute the square the Dirac operator \eqref{dirac2} as:
\begin{equation}
 \mathcal{D}_n^2=2\omega a^\dagger_\mu a^\mu\otimes \textbf{I} -2\omega \textbf{I}\otimes b^\dagger_\mu b^\mu =2\omega N_B\otimes \textbf{I} -2\omega 1\otimes N_F
\end{equation}
Where $N_F$ and $N_B$ are the number operators. In this form it easy to see that $\mathcal{D}^2_n $, being a "difference" between fermionic and bosonic number operator, has only one zero mode corresponding to the vacuum state. For practical reasons it is convenient write $\mathcal{D}^2_n $ as:
\begin{equation}
 \mathcal{D}_n^2=\omega \delta^{\mu\nu}(a^\dagger_\mu a_\nu + a_\nu a^\dagger_\mu)\otimes \textbf{I} -2\omega \textbf{I}\otimes \delta^{\mu\nu}(b^\dagger_\mu b_\nu - b_\nu b^\dagger_\mu)=H_n\otimes \textbf{I} +\omega\otimes \Sigma_n 
\end{equation}
where in $H_n$ we can recognize the harmonic oscillator Hamiltonian and the spin operator $\Sigma_n$.
The universality property of the Clifford algebra grants the existence of an isomorphism between the 2$n$-dimensional Clifford algebra
and the Hilbert space $ \mathcal{H}_n=L^2(\mathbb{R}^n)\otimes \mathbb{C}^{2^n}$.  In this representation the Dirac operator is: 
\begin{equation}
\mathcal{D}_n=i\Gamma^\mu\partial_\mu +\omega\Gamma^{\mu+n}x_\mu  \label{Diracn}
\end{equation}
Where $\Gamma^\mu$ turns to be $\Gamma^\mu=(b_\mu+b^\dagger_\mu) $, $ \Gamma^{\mu+n}=i(b^\dagger_\mu-b_\mu)$  which satisfy the relations:
\begin{equation}
\Gamma_a\Gamma_b+\Gamma_b\Gamma_a=2\delta_{ab} \ \textrm{with} \ a,b=1,\cdots,2n \label{Gamman}
\end{equation}
Beside, the grading operator is represented as:
\begin{equation}
\chi_n=(-i)^n(-1)^{\frac{n(n-1)}{2}}\otimes \Gamma_1\cdots\Gamma_{2n}
\end{equation}

\section{An harmonic spectral triple for the Moyal plane}
In the framework of non-commutative field theories on 4-dimensional Moyal plane has been proved \cite{phi4-non,matrix-renorm} that the introduction of an harmonic oscillator term makes a $\phi^4$-model on 4-dimensional Moyal plane renormalizable. Such oscillator term can be written as:  
\begin{equation}
H_m=-\frac{\partial^2}{\partial x_\mu \partial x^\mu}+\Omega^2 \tilde{x}_\mu\tilde{x}^\mu +m^2
\end{equation}
where $\tilde{x}_\mu:=2(\Theta^{-1})_{\mu\nu} x^\nu$, $\Theta$ can be chosen as two copies of the Pauli matrix  $\Theta=i\theta\sigma$ with $\sigma = \sigma_2  \otimes \textbf{I}_2$ or explicitly:
\begin{equation}
\Theta=\left(
\begin{array}{cccc}
0& \theta &0&0 \\
-\theta &0&0&0 \\
0&0&0&\theta \\
0&0&-\theta&0
\end{array}
\right), \ \theta\in\mathbb{R}
\end{equation}
With this choice we have $\Theta^{-1} = - \frac{i\sigma}{\theta} $.
Quantum mechanics tell us that in the Hilbert space $L^2(\mathbb{R}^4)$ exists
an orthonormal basis $\psi_s ,\ s\in\mathbb{R}^4$ of eigenfunctions of $H_m$ with eigenvalues
\begin{equation}
\lambda_s(m)=\left(\frac{4\Omega}{\theta} \left(s+2+ \frac{\theta m^2}{2\Omega}\right)\right), \ s\in\mathbb{N}
\end{equation}
The inverse $H^{-1}_m$ extends to a selfadjoint compact operator on $L^2(\mathbb{R}^4)$ with eigenvalues $\lambda^{-1}_s(m)$. 
If we look  at the trace the operator $H^{-4}_m$ we find:  
\begin{equation}
\operatorname{Tr}(H^{-s}_m)=\sum_{n=0}^\infty(n + 3)(n + 2)(n + 1)(\lambda_n (m))^s
\end{equation}
Which is derived simply from the number of possibilities to express $s$ as a sum of four ordered natural numbers. 
This means that the trace exists or in other words $H^{-4}$  belongs to the Dixmier trace ideal 
$L^{(1,\infty)}(L^2(\mathbb{R}^4))$ of compact operators.
Nevertheless, the main object of non-commutative geometry, which  determine the spectral dimension,  is the Dirac operator. The relation $\mathcal{D}^2=H$, which essentially states that each degree of freedom in the configuration space contributes to the spectral dimension  twice, implies   that the 4-dimensional Moyal space has spectral dimension 8.

%In a d-dimensional space we require $|\mathcal{D}|^{-d} \in L^{(1,\infty)}(L^2(\mathbb{R}^4))$. 

From the previous section, we can define a proper Dirac operator just considering  the 4-dimensional case obtaining a Dirac operator built from a 8-dimensional Cifford algebra:
\begin{equation}
\mathcal{D}_4 = i\Gamma_\mu \partial_\mu + \Omega \Gamma_{\mu+4} \tilde{x}_\mu
\end{equation}
Here, the $\Gamma_k \in M_{16}(\mathbb{C}), k = 1, . . . , 8$ are the generators of the 8-dimensional real Clifford
algebra, satisfying
\begin{equation}
\Gamma_k \Gamma_l + \Gamma_l\Gamma_k = 2\delta_{kl} \textbf{I} 
\end{equation}
We take the Hilbert space $\mathcal{H}_4 = L^2(\mathbb{R}^4,\mathcal{S})=L^2(\mathbb{R}^4)\otimes\mathbb{C}^{16} $ of square integrable spinors over
4-dimensional euclidean space. 
Accordingly with \eqref{Gamman} for $\psi \in \mathcal{H}_4$ we obtain:
\begin{equation}
\mathcal{D}^2_4 \psi= \left((-\Delta + \Omega^2 \tilde{x}_\mu \tilde{x}^\mu )\textbf{I} + \Sigma\right)\psi \ , \ \Sigma:= -i\Omega(\Theta^{-1})^{\mu\nu} [\Gamma_\mu ,\Gamma_{\nu+4}] \label{Sigma1}
\end{equation}
with $\Delta = \partial^\mu \partial_\mu$. % Assuming a choice of the Clifford algebra where $\Sigma$ is diagonal, we obtain
%up to the 16-fold multiplicity of each level and an unimportant shift in the mass exactly
%the spectrum of the harmonic oscillator Hamiltonian $H$. In particular, $|\mathcal{D}_4|^{-8}$ belongs as
%required to the Dixmier trace ideal $\mathcal{L}^{(1,\infty)}(L^2(\mathbb{R}^4,S))$.
As algebra we chose \footnote{In the appendix A a unitalised Moyal algebra is introduced in the frame of non compact spectral triples } the Moyal algebra $\mathbb{R}^4_\Theta$:
\begin{equation}
\mathcal{A}_4 = \mathbb{R}_\Theta^4=(\mathcal{S}(\mathbb{R}^4),\star)
\end{equation}
where $(\mathcal{S}(\mathbb{R}^4),\star)$ is the algebra of the Schwartz functions on $\mathbb{R}^4$,  with the Moyal product
\begin{equation}
(f \star g)(x) = \int \int d^4 y \frac{d^4 k}{(2\pi)^4} f(x+\frac{1}{2}\Theta \cdot k) g(x+y) e^{i\langle k,y\rangle} \ ,\ f,g \in \mathcal{A}_4 \label{star}
\end{equation}
The representation of the algebra $\mathcal{A}_4$ on $\mathcal{H}_4$ is by component-wise diagonal Moyal product \cite{moyal-triple} $\star : \mathcal{A}_4 \times \mathcal{H}_4 \to \mathcal{H}_4$. The Moyal product can be  extended to constant functions using   another representation of the product with the integral representation of the Dirac distribution.
Taking in account, for smooth spinors, the identity $2x^\mu\psi=x\star\psi+\psi\star x$ and the relation 
\begin{equation}
[x^\nu,f]_\star = i\Theta^{\nu\rho} \partial_\rho \label{comxf}
\end{equation}
we compute the commutator of that action with the Dirac operator
\begin{equation}
\begin{array}{l}
\mathcal{D}_4(f\star\psi)-f\star(\mathcal{D}_4 \psi) \\
= i\Gamma^\mu ((\partial_\mu f) \star\psi + f\star \partial_\mu \psi) + \frac{1}{2} \Omega\Gamma^{\mu+4} (\tilde{x}_\mu\star(f \star \psi) + (f \star \psi) \star \tilde{x}_\mu) \\
-i\Gamma^\mu f \star \partial_\mu \psi - \frac{1}{2} \Omega\Gamma^{\mu+4}(f\star( \tilde{x}_\mu\star\psi)+f\star(\psi \star \tilde{x}_\mu))\\
= i(\Gamma^\mu + \Omega\Gamma^{\mu+4} )(\partial_\mu f ) \star \psi .
\end{array} \label{d4-com}
\end{equation}
Just the four-dimensional differential of $f$ appears, no $x$-multiplication, this differential 
is represented on $\mathcal{H}_4$ by $\pi(dx^\mu ) = \Gamma^\mu + \Omega\Gamma^{\mu+4}$. Furthermore  the previous commutator confirms that
$(\mathcal{A}_4,\mathcal{H}_4,\mathcal{D}_4)$ satisfy the main\footnote{Orientability axiom and Poincar\'e duality will be not considered} axioms of spectral triple, in fact the commutator is bounded and due to its commutation with
Moyal multiplication, order-one condition is fulfilled.
%However, the algebra generated by $[\mathcal{D}_4,A_4]$ and $A_4$ does not contain the chirality matrix 
%$\Gamma_9$ so that the orientability axiom does not hold

Now we introduce a very useful relation connected to the heat kernel  type expansion associated to a regular spectral 
triple \cite{non-com-tri}. This relation will be used later in order to  compute the spectral action. Considering a regular non-unital spectral triple $(\mathcal{A},\mathcal{H},\mathcal{D},J)$
and two pseudo-differential operator  $A_0 \in \Psi_0(\mathcal{A}) $ $A_1\in \Psi_1(\mathcal{A})$ of order respectively 0 and 1.
we consider the following decomposition: 
\begin{equation}
e^{-t(\mathcal{D}^2+A_0+A_1)}=\sum_{j=0}^4(-1)^jE_j(t)-t^5R
\end{equation}
Using Duhamel principle \cite{Duhamel}
\begin{equation}
e^{-t(A+B)}= e^{-tA}-t\int_0^1 ds \ e^{-st(A+B)}Be^{-(1-s)tA}
\end{equation}
we can identify:
\begin{eqnarray}
E_0(t)&=&e^{-t\mathcal{D}^2}\nonumber \\ \ 
E_j(t)&=&\int_{\Delta_j} d^js \ e^{-s_1t\mathcal{D}^2}(A_0+A_1)e^{-(s_2-s_1)t\mathcal{D}^2}\cdots(A_0+A_1)e^{-(1-s_j)t\mathcal{D}^2} \nonumber \\ \ 
&&
\end{eqnarray}
and
\begin{eqnarray}
&&R=\int_{\Delta_5}ds_1ds_2ds_3ds_4ds_5 e^{-s_1t(\mathcal{D}^2+A_0+A_1)}(A_0+A_1)e^{-(s_2-s_1)t\mathcal{D}^2}(A_0+A_1)    \nonumber\\
&& \times e^{-(s_3-s_2)t\mathcal{D}^2}(A_0+A_1)e^{-(s_4-s_3)t\mathcal{D}^2}(A_0+A_1)e^{-(s_5-s_4)t\mathcal{D}^2}(A_0+A_1)e^{-(1-s_5)t\mathcal{D}^2}\nonumber\\
&&
\end{eqnarray}
The domains of the integrals $\Delta_j$ are the $j$-simplex:
\begin{equation}
\Delta_j=\{s\in\mathbb{R}^j;0\leq s_1\leq s_2\leq s_1\leq\cdots\leq s_j\leq 1\} \simeq \{s\in\mathbb{R}^{j+1};s_i\geq 0,\sum_{i=0}^j s_i=1\}
\end{equation}
Taking in account the relation:
\begin{equation}
\left[ e^{-t\mathcal{D}^2},A \right]=\int_0^1 ds \frac{d}{ds}\left(e^{-ts\mathcal{D}^2}Ae^{-t(1-s)\mathcal{D}^2}\right) =-t\int_0^1 ds e^{-ts\mathcal{D}^2}\left[\mathcal{D}^2,A\right]e^{-t(1-s)\mathcal{D}^2}
\end{equation}
we can rearrange the $E_j$ collecting the heat operators as follow: 
\begin{eqnarray}
&&E_2(t) \nonumber\\
     &&=\int_{\Delta_2} e^{-s_1t\mathcal{D}^2}(A_0+A_1)e^{-(s_2-s_1)t\mathcal{D}^2}(A_0+A_1)
     e^{-(1-s_2)t\mathcal{D}^2}ds_1ds_2\nonumber\\ 
     &&=\int_{\Delta_2} e^{-s_1t\mathcal{D}^2}(A_0+A_1)^2e^{-(1-s_2)t\mathcal{D}^2}ds_1ds_2-t\int_{\Delta_2}
      \int_0^1 dr(s_2-s_1)e^{-s_1t\mathcal{D}^2}\nonumber \\
     &&\times (A_0+A_1) e^{-(s_2-s_1)rt\mathcal{D}^2}\left[\mathcal{D}^2,A_0+A_1\right]e^{-(s_2-s_1)(1-r)t
      \mathcal{D}^2}e^{-(1-s_2)t\mathcal{D}^2}ds_1ds_2\nonumber\\
     &&=\int_{\Delta_2}e^{-s_1t\mathcal{D}^2}\left\{(A_0+A_1)^2-t(s_2-s_1)
     (A_0+A_1)\left[\mathcal{D}^2,A_0+A_1\right]\right\}\nonumber \\
     &&\times e^{-(1-s_1)t\mathcal{D}^2}ds_1ds_2+t^2\int_{\Delta_2}\int_0^1 dr_1dr_2r_1(s_2-s_1)^2e^{-s_1t\mathcal{D}^2}
     (A_0 + A_1)\nonumber  \\
     &&\times e^{-(s_2-s_1)r_1r_2t\mathcal{D}^2} \left[\mathcal{D}^2,\left[\mathcal{D}^2,A_0+A_1\right]\right]
      e^{-(s_2-s_1)(1-r_1r_2)t\mathcal{D}^2}e^{-(1-s_1)t\mathcal{D}^2}ds_1ds_2\nonumber  \\
     &&=\int_{\Delta_2}e^{-s_1t\mathcal{D}^2}\big\{(A_0+A_1)^2-t(s_2-s_1)(A_0+A_1)\left[\mathcal{D}^2,A_0+A_1\right]\nonumber \\
     &&+\frac{t^2}{2}(s_2-s_1)^2(A_0+A_1)\left[\mathcal{D}^2,\left[\mathcal{D}^2,A_0+A_1\right]\right]\big\}
      e^{-(1-s_1)t\mathcal{D}^2}ds_1ds_2\nonumber \\
     &&- t^3\int_{\Delta_2}\int_0^1ds_1ds_2 r_1^2r_2(s_2-s_1)^3e^{-s_1t\mathcal{D}^2}(A_0+A_1)
     e^{-(s_2-s_1)r_1r_2r_3t\mathcal{D}^2}\nonumber \\
     &&\times\left[\mathcal{D}^2,\left[\mathcal{D}^2,\left[\mathcal{D}^2,A_0+A_1\right]\right]\right]e^{-(s_2-s_1)
     (1-r_1r_2r_3)t\mathcal{D}^2}e^{-(1-s_2)t\mathcal{D}^2}ds_1ds_2
\end{eqnarray}
The last integral is  a bounded operator which tents to zero for $t \to 0 $. \\ 
Applying the same procedure for $E_3$ we get:
\begin{eqnarray}
&&E_3(t) \nonumber \\
&&=\int_{\Delta_3} e^{-s_1t\mathcal{D}^2}\big\{(A_0+A_1)^3-t(s_3-s_2)(A_0+A_1)^2\left[\mathcal{D}^2,A_0+A_1\right]\nonumber \\
      &&-t(s_2-s_1)(A_0+A_1)\left[\mathcal{D}^2,A_0+A_1\right](A_0+A_1)\big\}e^{-(1-s_1)t\mathcal{D}^2}ds_1ds_2 ds_3 \nonumber\\ 
      &&+t^2\int_{\Delta_3}\int_0^1(s_3-s_1)^2e^{-s_1t\mathcal{D}^2}(A_0+A_1)^2e^{-(s_3-s_1)r_1r_2t\mathcal{D}^2} \nonumber \\
      &&\times \left[\mathcal{D}^2,\left[\mathcal{D}^2,A_0+A_1\right]\right]e^{-(s_3-s_1)(1-r_1r_2)t\mathcal{D}^2}
      e^{-(1-s_3)t\mathcal{D}^2}ds_1ds_2 ds_3dr_1dr_2dr_3\nonumber \\
      &&+t^2\int_{\Delta_3}\int_0^1(s_2-s_1)(s_3-s_1)e^{-s_1t\mathcal{D}^2}          
      (A_0+A_1)\left[\mathcal{D}^2,A_0+A_1\right]\nonumber\\  
      &&\times e^{-(s_3-s_1)rt\mathcal{D}^2}\left[\mathcal{D}^2,A_0+A_1\right]e^{-(s_3-s_1)(1-r)t\mathcal{D}^2}
      e^{-(1-s_3)t\mathcal{D}^2}ds_1ds_2 ds_3dr\nonumber \\
      &&+t^2\int_{\Delta_3}\int_0^1(s_2-s_1)^2e^{-s_1t\mathcal{D}^2}(A_0+A_1)e^{-(s_2-s_1)r_1r_2t\mathcal{D}^2}
      \left[\mathcal{D}^2 \left[\mathcal{D}^2,A_0+A_1\right]\right]\nonumber \\
      &&\times e^{-(s_2-s_1)(1-r_1r_2)t\mathcal{D}^2}e^{-(s_3-s_2)t\mathcal{D}^2}(A_0+A_1)e^{-(1-s_3)t\mathcal{D}^2}ds_1ds_2
      ds_3
\end{eqnarray}
Even in this case, the integrals multiplied by $t^2$ give rise to a vanishing bounded operator for $t\to 0$.  
Now using the cyclic property of the trace we can compute the traces of $E_2$ over $\Delta_2 $ and of $E_3$ over $\Delta_3$, which considering only the leading terms are: 
\begin{eqnarray}
&&\operatorname{Tr}(t^2E_2(t))\nonumber \\
	  &&=\textrm{tr}\Big(\Big\{\frac{t^2}{2}(A_0+A_1)^2-\frac{t^3}{6}(A_0+A_1)\left[\mathcal{D}^2,A_0+A_1\right]\nonumber \\
	  &&+\frac{t^4}{24}(A_0+A_1) \left[\mathcal{D}^2 \left[\mathcal{D}^2,A_0+A_1\right]\right]\Big)
	  e^{-t\mathcal{D}^2}\Big\}+\mathcal{O}(\sqrt{t})\nonumber \\
      &&=\textrm{tr}\Big(\Big\{\frac{t^2}{2}(A_0+A_1)^2-\frac{t^3}{6}
      (A_0\left[\mathcal{D}^2,A_1\right]+A_1\left[\mathcal{D}^2,A_0\right]+A_1\left[\mathcal{D}^2,A_1\right])\nonumber \\
      &&+\frac{t^4}{24}A_1\left[\mathcal{D}^2 \left[\mathcal{D}^2,A_0\right]\right]\Big)e^{-t\mathcal{D}^2}\Big\}
      +\mathcal{O} (\sqrt{t})
\end{eqnarray}
and
\begin{eqnarray}
&&\operatorname{Tr}(t^3E_3(t))\nonumber \\
     &&=\textrm{tr}\Big(\Big\{-\frac{t^3}{6}(A_0+A_1)^3+\frac{t^4}{12}(A_0+A_1)^2\left[\mathcal{D}^2,A_0+A_1\right]\nonumber \\
     &&+\frac{t^4}{24}(A_0+A_1)\left[\mathcal{D}^2,A_0+A_1\right](A_0+A_1)\Big)e^{-t\mathcal{D}^2}\Big\}+\mathcal{O}
     (\sqrt{t})\nonumber \\
     &&=\textrm{tr}\Big(\Big\{-\frac{t^3}{6}(A_0A^2_1+A_1A_0A_1+A_1^2A_0+A^3_1)\nonumber \\
     &&+\frac{t^4}{12}A^2_1\left[\mathcal{D}^2 ,A_1\right]+\frac{t^4}{12}A_1\left[\mathcal{D}^2 ,A_1\right]A_1\Big)
     e^{-t\mathcal{D}^2}\Big\}+\mathcal{O}(\sqrt{t})
\end{eqnarray}
In the exactly same way we obtain for $E_4$:
\begin{eqnarray}
\operatorname{Tr}(t^4E_4(t))&=&\textrm{tr}\left(\frac{t^4}{4}(A_0+A_1)^4e^{-t\mathcal{D}^2}\right)+\mathcal{O}(\sqrt{t}) \nonumber \\
              &=&\textrm{tr}\left(\frac{t^4}{24}A_1^4e^{-t\mathcal{D}^2}\right)+\mathcal{O}(\sqrt{t})
\end{eqnarray}
Summing all contributions together we finally arrive at:
\begin{eqnarray}
&&\operatorname{Tr}(e^{-t(\mathcal{D}^2+A_0+A_1)})  \nonumber \\
&&=\textrm{Tr}\Big(\Big\{1-t(A_0+A_1)+\frac{t^2}{2}(A^2_0+A_1A_0+A_0A_1+A^2_1) \nonumber \\
&&-\frac{t^3}{6}(A_0\left[\mathcal{D}^2,A_1\right]+A_1\left[\mathcal{D}^2,A_0\right]+A_1\left[\mathcal{D}^2,A_1\right]+A_0A^2_1+A_1A_0A_1+A_1^2A_0\nonumber \\
&&+A^3_1)+\frac{t^4}{24}(A_1\left[\mathcal{D}^2 \left[\mathcal{D}^2,A_0\right]\right]+2A^2_1\left[\mathcal{D}^2 ,A_1\right]+A_1\left[\mathcal{D}^2 ,A_1\right]A_1+A_1^4\Big)e^{-t\mathcal{D}^2}\Big\}\nonumber \\
&&+\mathcal{O}(\sqrt{t}) \label{heat1}
\end{eqnarray}

\section{8-dimensional Higgs model}
Following the Connes-Lott models, in order to implement the Higgs mechanism,  we consider the total spectral triple 
as the tensor product of the 8-dimensional spectral triple $(\mathcal{A}_4,\mathcal{H}_4,\mathcal{D}_4,\Gamma_9)$ with the two point Connes-Lott like spectral triple $(\mathbb{C}\otimes \mathbb{C},\mathbb{C}^2, M\sigma_1 )$. The total Dirac operator of the product triple is:
\begin{equation}
\mathcal{D}_T = \mathcal{D}_4 \otimes \textbf{I} + \Gamma_9 \otimes M\sigma_1
\end{equation}
Or explicitly:
\begin{equation}
\mathcal{D}_T= \left(\begin{array}{cc}
    \mathcal{D}_4& M\Gamma_9 \\
    M\Gamma_9& \mathcal{D}_4
\end{array}\right)
\end{equation}
The algebra becomes $ \mathcal{A}_T=\mathcal{A}_4 \oplus \mathcal{A}_4 $ and acts by diagonal star multiplication \eqref{star} on $\mathcal{H}_T= \mathcal{H}_4 \oplus \mathcal{H}_4$. The fluctuated Dirac operator is found  using 
$\mathcal{D}_A = \mathcal{D}_T + \Sigma_i a_i[\mathcal{D}_T,b_i]$ with $a_i,b_i\in\mathcal{A}_T $ of the form $(f,g)$, the computation of the commutator $\mathcal{D}_T$ with $(f,g)$ gives:
\begin{equation}
[\mathcal{D}_T,(f, g)] = 
\left(\begin{array}{cc}
    i(\Gamma^\mu + \Omega \Gamma^{\mu+4} )L_\star(\partial_\mu f) & M\Gamma_9 L_\star(f-g) \\
    M\Gamma_9 L_\star(g-f) & i(\Gamma^\mu + \Omega\Gamma^{\mu+4} )L_\star(\partial_\mu g)
\end{array}\right)
\end{equation}
$L_\star(f)\psi = f \star \psi$ is the left Moyal multiplication. From the commutator we deduce that the form of selfadjoint fluctuated Dirac has to be: 
\begin{equation}
\mathcal{D}_A =\left(\begin{array}{cc} 
\mathcal{D}_4 + (\Gamma_\mu + \Omega \Gamma_{\mu+4})L_\star(A^\mu )& \Gamma_9 L_\star(\varphi) \\
\Gamma_9 L_\star (\bar{\varphi}) &\mathcal{D}_4 + (\Gamma_\mu + \Omega \Gamma_{\mu+4} )L_\star (B^\mu ) \end{array}\right)
\end{equation}
Where $\phi \in \mathcal{A}_4$ is the Higgs complex field and $A_\mu, B_\mu \in \mathcal{A}_4$ are real fields. The spectral action computation needs the square of $\mathcal{D}_A$:
\begin{equation}
\mathcal{D}^2_A =\left(\begin{array}{cc}
(H^2_0 + L_\star (\phi \star \varphi))1 + \Sigma + F_A
&i(\Gamma_\mu + \Omega\Gamma_{\mu+4})\Gamma_9 L_\star (D^\mu \phi)\\
i(\Gamma_\mu + \Omega \Gamma_{\mu+4} )\Gamma_9 L_\star (\overline{D^\mu \phi})&
(H^2 + L_\star (\phi \star \phi))1 + \Sigma + F_B
\end{array}\right)
\end{equation}
with
\begin{eqnarray}
D_\mu \phi &=& \partial_\mu\phi - iA_\mu \star \phi + i\phi\star B_\mu \\ 
F_A &=& \{\mathcal{D}_4 , (\Gamma_\mu + \Omega \Gamma_{\mu+4} )L_\star (A^\mu )\} + (\Gamma_\mu + \Omega\Gamma_{\mu+4} )(\Gamma_\nu + \Omega \Gamma_{\nu+4} )L_\star (A^\mu \star A^\nu)\nonumber \\
&=& \{ L_\star(A^\mu ), i\partial_\mu + \Omega^2 M_\bullet(x_\mu)\} + (1 +\Omega^2)L_\star (A_\mu \star A^\mu )  \nonumber \\
&+& i \left(\frac{1}{4} [\Gamma_\mu, \Gamma_\nu]+ \frac{1}{4} \Omega^2 [\Gamma_{\mu+4} , \Gamma_{\nu+4} ] + \Omega\Gamma_\mu \Gamma_{\nu+4} \right) L_\star (F^{\mu\nu}_A ) ,
\end{eqnarray}
$(M_\bullet(\tilde{x}_\mu)\psi)(x) = \tilde{x}_\mu\psi(x)$ is ordinary pointwise multiplication and $F_B$  is obtained just replacing $A$ with $B$. We can recognize in  previous expression the field strength $F^A_{\mu\nu} = \partial_\mu A_\nu - \partial_\nu A_\mu -i(A_\mu\star A_\nu - A_\nu\star A_\mu )$

\subsection{Spectral action}
Recalling  the spectral action principle, the bosonic action can be defined exclusively by the
spectrum of the Dirac operator. The general form for such bosonic action is:
\begin{equation}
 S(\mathcal{D}_A) = \textrm{Tr}\chi(\mathcal{D}^2_A) \label{Action}
\end{equation}
Where $\chi$ is a regularization function $\chi: R_+\to R_+$ for which trace exists. \\
The trace in \eqref{Action}  is  defined on $\mathcal{B}(L^2(\mathbb{R}^4))$ by 
\begin{equation}
 \operatorname{Tr}(A) =\int_{\mathbb{R}^4} dx \ A(x,x)
\end{equation}
together with the matrix trace including  the Clifford algebra.
By Laplace transformation one has
\begin{equation}
S(\mathcal{D}_A) =\int_0^\infty dt \operatorname{Tr}(e^{-t\mathcal{D}^2_A})\tilde{\chi}(t)  \label{Action1}
\end{equation}
where $\tilde{\chi}$ is the inverse Laplace transform of $\chi(s)$, 
\begin{equation}
\chi(s) = \int_0^\infty dt e^{-st}\tilde{\chi}(t). 
\end{equation}
The trace in \eqref{Action1} is given by:
\begin{equation}
 \operatorname{Tr}(e^{-t\mathcal{D}^2_A}) =\int_{\mathbb{R}^4} dx \ \textrm{tr}(e^{-t\mathcal{D}^2_A})(x,x)
\end{equation}
Assuming the trace of the heat kernel $e^{-t\mathcal{D}^2_A}$ has an asymptotic expansion
\begin{equation}
 \operatorname{Tr}(e^{-t\mathcal{D}^2_A})= \sum^\infty_{n=-\delta} a_n(\mathcal{D}^2_A)t^{n} \ , \ \delta \in \mathbb{N}  \label{heat}
\end{equation}
we obtain replacing the previous expansion into \eqref{Action1}
\begin{equation}
S(\mathcal{D}_A) =\sum_{n=-\delta}^\infty  a_n(\mathcal{D}^2_A)\int_0^\infty dt \ t^{n} \tilde{\chi}(t) \label{heatd}
\end{equation}
To compute the integrals we have to consider separately the cases $ n\notin\mathbb{N}$ and $ n\in\mathbb{N}$. For $ n\in\mathbb{N} $ we have
\begin{eqnarray}
&&\int_0^\infty dt \ t^{n}\tilde{\chi}(t) \nonumber \\
 &&= \lim_{s\to 0} \int_0^\infty dt e^{-st} t^{n}\tilde{\chi}(t) 
=\lim_{s\to 0}(-1)^{n} \frac{\partial^{n}}{\partial s^{n}}  \int_0^\infty dt e^{-st}\tilde{\chi}(t)\nonumber \\
 &&= \lim_{s\to 0} (-1)^{n} \frac{\partial^{n}\chi}{\partial s^{n}}(s) = (-1)^{n} \chi^{(n)}(0) 
\end{eqnarray}
For $ n\notin\mathbb{N}$ we have
\begin{equation}
\int_0^\infty ds \ s^{-n-1}\chi(s) =
\int_0^\infty ds  \int_0^\infty dt \ e^{-st} s^{-n-1} \tilde{\chi}(t) = \Gamma(-n)\int_0^\infty dt \ t^{n} \tilde{\chi}(t) 
\end{equation}
In summary,
\begin{equation}
\chi_n =\Bigg\{
\begin{array}{ll}
\frac{1}{\Gamma(-n)}\int_0^\infty ds \ s^{-n-1} \chi(s) & \textrm{for} \ n\notin\mathbb{N}  \\
(-1)^{n-\delta} \chi^{(n)}(0) & \textrm{for} \ n\in\mathbb{N}  
\end{array}\label{chi}
\end{equation}
Due to the nature of the $\chi(t)$ function (usually one chose a characteristic function), consequently in  the expansion \eqref{heat}  we will  take in account only the finite or singular part for $t\to 0$   

Our strategy to compute the action is to use the relation \eqref{heat1}, therefore after explicitly	expressed  $A_0$ and $A_1$ we proceed to the calculus of the traces and in the end we will identify the leading part of the action comparing the result with the expansions \eqref{heat}-\eqref{heatd}. We can identify the operators $A_0$ and $A_1$ appearing in the \eqref{heat1} as follow:
\begin{equation}
A_0 =\left(\begin{array}{cc}
L_\star (V_{A,\phi})\textbf{I} + L_\star(F_A^{\mu\nu})\Gamma^\Omega_{\mu\nu}
&i(\Gamma^\mu + \Omega\Gamma^{\mu+4})\Gamma_9 L_\star (D_\mu \phi)\\
i(\Gamma^\mu + \Omega \Gamma^{\mu+4} )\Gamma_9 L_\star (\overline{D_\mu \phi})&
L_\star (V_{B,\phi})\textbf{I} + L_\star(F_B^{\mu\nu})\Gamma^\Omega_{\mu\nu}
\end{array}\right) \label{A0}
\end{equation}

\begin{equation}
A_1 =\left(\begin{array}{cc}
2i(1+\Omega^2)L_\star(A^\mu)\nabla_\mu^{(\Omega)} &0\\
0& 2i(1+\Omega^2)L_\star(B^\mu)\nabla_\mu^{(\Omega)}
\end{array}\right)\label{A1}
\end{equation}
with
\begin{equation}
 V_{A,\phi}= \phi\star\bar{\phi}+(1+\Omega^2)(i \partial_\mu A^\mu +A_\mu\star A^\mu) , \  V_{B,\phi}= \bar{\phi}\star\phi+(1+\Omega^2)(i\partial_\mu B^\mu+B_\mu\star A^\mu)
\end{equation}
$\nabla_\mu^{(\Omega)}$  are define as $\nabla_\mu^{(\Omega)}=\frac{1}{1+\Omega^2}\left(\partial_\mu-i\Omega^2 M_\bullet(\tilde{x}_\mu)\right) $ and using the \eqref{comxf} it is useful  to  compute the  commutators:
\begin{eqnarray}
\left[\nabla_\mu^{(\Omega)},L_\star(f)\right] &=& L_\star(\partial_\mu f) \label{nabla1} \\
\left[H_4,L_\star(f)\right] &=& -(1+\Omega^2)\left(L_\star(\Delta f)+2L_\star(\partial^\mu)\nabla_\mu^{(\Omega)}\right)   \\
\left[H_4,\nabla_\mu^{(\Omega)}\right]&=& \frac{8i\Omega^2}{1+\Omega^2}(\Theta^{-1})_\mu^\nu\nabla_\nu^{(1)} \label{nabla2}
\end{eqnarray} 
Referring to the \eqref{heat1} and \eqref{nabla1}-\eqref{nabla2}  for the computation of the action are required the following commutators
\begin{eqnarray}
&& [\mathcal{D}^2,2i(1+\Omega^2)L_\star(A^\mu)\nabla^{(\Omega)}_\mu]=  \nonumber \\
&&-(1+\Omega^2)^2\left(2iL_\star(\triangle A^\mu)\nabla^{(\Omega)}_\mu +4iL_\star(\partial^\nu A^\mu)\nabla^{(\Omega)}_\nu\nabla^{(\Omega)}_\mu\right) \nonumber \\
&&-16\Omega^2L_\star(A^\mu)(\Theta^{-1})_{\mu}^\nu\nabla^{(1)}_\nu \\
&& [\mathcal{D}^2,[\mathcal{D}^2,2i(1+\Omega^2)L_\star(A^\mu)\nabla^{(\Omega)}_\mu]]=  \nonumber \\
&& 8i(1+\Omega^2)^3L_\star(\partial^\rho\partial^\nu A^\mu)\nabla^{(\Omega)}_\rho\nabla^{(\Omega)}_\nu\nabla^{(\Omega)}_\mu +\cdots 
\end{eqnarray}
From the previous discussion  we are allowed to split the traces in two parts a matrix trace and the continuous one, now we focus ourself on the matrix trace contributions. Considering the matrix term $e^{-t\tilde{\Omega}\Sigma_4 1_2}$, coming from $e^{-t\tilde{\Omega}\Sigma_4}$, it contributes non trivially only for the vacuum  tr$(e^{-t\mathcal{D}^2})$ and for the first order contribution  tr$(-tA_0e^{-t\mathcal{D}^2})$, in the other cases it contributes only by its leading part $1_{32}$. 
The contribution from the vacuum is:
\begin{eqnarray}
\textrm{tr}\left(e^{-t\tilde{\Omega}\Sigma_4 1_2}\right)=2\textrm{tr}\left(e^{-t\tilde{\Omega}\Sigma_4}\right)=2\textrm{tr}\left(\prod^4_{\mu=1}e^{t\tilde{\Omega}(b^\dagger_\mu b_\mu-b_\mu b^\dagger_\mu)}\right)
\end{eqnarray}
On the base of $ \mathbb{C}^{14}$ defined by $|s_1,\cdots,s_4\rangle =(b_1^\dagger)^{s_1}\cdots(b_4^\dagger)^{s_4}|0,0,0,0\rangle, \ s_i\in\{0,1\}$ we have: 
\begin{equation}
\textrm{tr}(e^{-t\tilde{\Omega}\Sigma_41_2})=32 \cosh^4(\tilde{\Omega}t)
\end{equation}
The first order term related to the matrix trace vanishes, in fact it takes the form:
\begin{equation}
t^2\tilde{\Omega}\textrm{tr}(\Sigma_4\Gamma^\Omega_{\mu\nu})(F_A^{\mu\nu}+ F_B^{\mu\nu})
\end{equation} 
From \eqref{Sigma1} we  deduce that the matrix trace $\textrm{tr}(\Sigma_4\Gamma^\Omega_{\mu\nu})$  is proportional to $\delta_{\mu\nu}$ which contracted to the antisymmetric operator $F^{\mu\nu} $ vanishes. 
The others matrix traces required  are:
\begin{eqnarray}
&\textrm{tr}(\Gamma^\Omega_{\nu\mu}\Gamma^\Omega_{\rho\sigma})=4(1+\Omega^2)^2(\delta_{\mu\rho}\delta_{\nu\sigma}-\delta_{\mu\sigma}\delta_{\nu\rho}) \\
&\textrm{tr}(i(\Gamma_\mu +\Omega\Gamma_{\mu +4})\Gamma_9\cdot i(\Gamma_\nu +\Omega\Gamma_{\nu +4})\Gamma_9)=16(1+\Omega^2)\delta_{\mu\nu}
\end{eqnarray}
Turning our self on the calculus of the functional trace, to simplify the notation we introduce the functions :
\begin{eqnarray}
 \mathcal{T}(f)&=&\textrm{Tr}_{L^2(\mathbb{R}^4)}\left(L_\star(f)e^{-tH_4}\right) \nonumber \\
 \mathcal{T}_{\mu_1\cdots\mu_k}(f)&=&\textrm{Tr}_{L^2(\mathbb{R}^4)}\left(L_\star(f)\nabla^{(\Omega)}_{\mu_1}\cdots\nabla^{(\Omega)}_{\mu_k}e^{-tH_4}\right)  \nonumber \\
 \tilde{\mathcal{T}}_{\mu\nu}(f)&=&\textrm{Tr}_{L^2(\mathbb{R}^4)}\left(L_\star(f)\nabla^{(\Omega)}_{\mu_1}\nabla^{(1)}_{\mu_k}e^{-tH_4}\right) 
\end{eqnarray}
Now substituting into \eqref{heat1} the \eqref{A0}-\eqref{A1} and using the matrices trace computations we obtain for the $A$ field:
\begin{eqnarray}
&&\textrm{Tr}(e^{-t\mathcal{D}_A^2})\nonumber \\
&&=\Bigg\{16\cosh^4(t\Omega)\textrm{tr}(e^{-tH_4^2})-tT(16V_{A,\phi})-t\mathcal{T}_\mu(32i(1+\Omega^2)A^\mu)\nonumber \\
&&+\frac{t^2}{2}\mathcal{T}\Big(16V_{A,\phi}\star V_{A,\phi}+16(1+\Omega^2)D_\mu\phi\star\overline{D^\mu\phi}+8(1+\Omega^2)F_{\mu\nu}^A F^{\mu\nu}_A\nonumber \\
&&+ 32i(1+\Omega^2)A^\mu\star\partial_\mu V_{A,\phi}\Big)\nonumber \\
&&+ \frac{t^2}{2}\mathcal{T}_\mu\Big(32i(1+\Omega^2)A^\mu\star V_{A,\phi}+32i(1+\Omega^2)V_{A,\phi}\star A^\mu\nonumber \\
&&-64(1+\Omega^2)^2A^\nu\star\partial_\nu A^\mu\Big)+\frac{t^2}{2}\mathcal{T}_{\mu\nu}\left(-64(1+\Omega^2)^2 A^\mu\star A^\nu \right)\nonumber \\
&&- \frac{t^3}{6}\mathcal{T}_{\mu\nu}\Big(-64(1+\Omega^2)^2V_{A,\phi}\star \partial^\mu A^\nu -64i(1+\Omega^2)^2A^\nu\star\partial^\nu V_{A,\phi}\nonumber \\
&&+64(1+\Omega^2)^3A_\rho\star(\delta^{\rho\mu}\triangle A^\nu +2\partial^\rho\partial^\mu A^\nu)\nonumber \\
&&- 64(1+\Omega^2)^2(V_{A,\phi}\star A^\mu\star A^\nu + A^\mu\star V_{A,\phi}\star A^\nu +A^\mu\star A^\nu\star V_{A,\phi})\nonumber \\
&&-128i(1+\Omega^2)^3\left(A^\rho\star(\partial_\rho A^\mu)\star A^\nu +A^\rho\star A^\mu(\partial_\rho A^\nu\right)\Big)\nonumber \\
&&- \frac{t^3}{6}\mathcal{T}_{\mu\nu\rho}\left(128(1+\Omega^2)^3A^\mu\star\partial^\nu A^\rho -128i(1+\Omega^2)^3A^\nu\star A^\mu\star A^\rho\right)\nonumber \\
&&+ \frac{t^3}{6}\tilde{\mathcal{T}}_{\mu\nu}\left(512i\Omega^2(1+\Omega^2)(\Theta^{-1})^{\rho\nu}A^\mu\star A_\rho\right)\nonumber \\
&&+ \frac{t^4}{24}\mathcal{T}_{\mu\nu\rho\sigma}\Big(-256(1+\Omega^2)^4A^\mu\star\partial^\nu\partial^\rho A^\sigma +512i(1+\Omega^2)^3 A^\mu\star A^\nu\star \partial^\rho A^\sigma\nonumber \\
&&+256i(1+\Omega^2)^4 A^\mu\star(\partial^\nu A^\rho)\star A^\sigma +256(1+\Omega^2)^4 A^\mu\star A^\nu\star A^\rho\star A^\sigma\Big)\Bigg\}\nonumber \\
&& + \mbox{$B$ field contribution} \ +\mathcal{O}(\sqrt{t})
\end{eqnarray} 
The contributions for the $B$ fields are obtained operating the following substitutions:
\begin{equation}
\left\{ A_\mu \to B_\mu, \ F^A_{\mu\nu}\to F^B_{\mu\nu}, \ V_{A,\phi} \to V_{B,\phi}, \ D_\mu\phi \leftrightarrow \overline{D^\mu\phi}\right\}
\end{equation}
The next step is the computation of the $\mathcal{T},\mathcal{T}_{\mu_1\cdots\mu_k}$ \footnote{$\tilde{\mathcal{T}}_{\mu\nu}$ can be ignored.} 
the position space kernel of  $e^{-tH_4}(x,y)$ is
\begin{equation}
e^{-tH_4}(x,y)=\left(\frac{\tilde{\Omega}}{2\pi\sinh(2t\tilde{\Omega})}\right)^2e^{-\frac{\tilde{\Omega} } {4}\left(\coth(\tilde{\Omega}t)|x-y|^2+\tanh(\tilde{\Omega}t)|x+y|^2\right)}\label{Hkern}
\end{equation}
which is essentially the  four dimensional Mehler kernel \cite{Mehler} with $ \tilde{\Omega}=\frac{2\Omega}{\theta}$ and $ |x|^2=x^\mu 
x_\mu$. 
For the integral kernel  of $L_\star(f)(x,y)$  we get after a direct substitution of \eqref{star}  and a variable change:
\begin{eqnarray}
\left(L_\star(f)g\right)&=&\int d^4 y\left(\int \frac{d^4 k}{(2\pi)^2}f(x+\frac{1}{2}\Theta\cdot k)e^{i\langle k,y-x\rangle}\right)g(y)\nonumber \\
&=& \int d^4y \left(\frac{1}{(\pi\theta)^4}\int d^4z f(f)e^{2i(\langle k,\Theta^{-1}x\rangle +\langle k,\Theta^{-1}z\rangle +\langle k,\Theta^{-1}x\rangle) }\right)g(y)\nonumber \\
& &
\end{eqnarray}
From which the kernel of the operator $L_\star(f)$ is easily identified as:
\begin{equation}
L_\star(f)(x,y)=\frac{1}{(\pi\theta)^4}\int d^4z f(z) e^{i\langle x-y,\Theta^{-1}(x+y)\rangle +2i\langle z,\Theta^{-1}(x-y)\rangle}\label{Lker}
\end{equation} 
Using \eqref{Hkern} and \eqref{Lker} we compute the trace using a change of variables $u=x-y,\ v=x+y$ and performing a Gaussian integration: 
\begin{eqnarray}
&& \textrm{Tr}_{L^2(\mathbb{R}^4)}\left(L_\star(f)e^{-tH_4}\right)\nonumber \\
&&=\int d^4xd^4y e^{-tH_4}(y,x)L_\star(f)(x,y)\nonumber \\
&&=\frac{\tilde{\Omega}^2}{(2\pi\theta)^42\pi^2(1+\Omega^2)^2\sinh^2(2t\tilde{\Omega})}\int d^4z f(z)\int d^4u \ d^4v\nonumber \\
&&\times e^{-\frac{\tilde{\Omega}}{4}\left(\frac{|u|^2}{\tanh(\tilde{\Omega}t)}+\frac{|v|^2} {\coth(\tilde{\Omega}t)}\right)i\langle u,\Theta^{-1}(v-2z)\rangle}  \nonumber \\
&&=\frac{1}{\cosh^2(2t\tilde{\Omega})}\int d^4z f(z)\int d^4ve^{-\frac{\tilde{\Omega}\tanh(t\tilde{\Omega})}{4}\left(|v|^2+\frac{2}{\tilde{\Omega}^2\theta^2}|v+2z|^2\right) }\nonumber \\
&&=\frac{1}{\cosh^2(2t\tilde{\Omega})}\int d^4z f(z)\int d^4ve^{-\frac{\tanh(t\tilde{\Omega})}{2\theta\Omega}\left((1+\Omega^2)|v+\frac{2}{(1+\Omega^2)}z|^2+\frac{4\Omega^2}{(1+\Omega^2)}|z|^2\right) }\nonumber \\
&&=\frac{\tilde{\Omega}^2}{2\pi^2(1+\Omega^2)^2\sinh^2(2t\tilde{\Omega})}\int d^4z f(z)e^{-\tilde{\Omega}\frac{\tanh(\tilde{\Omega}t) }{1+\Omega^2}|z|^2 }
\end{eqnarray} 
In the end, we have:
\begin{equation}
\mathcal{T}(f)=\frac{\tilde{\Omega}^2}{2\pi^2(1+\Omega^2)^2\sinh^2(2t\tilde{\Omega})}\int d^4z f(z)e^{-\tilde{\Omega}\frac{\tanh(\tilde{\Omega}t) }{1+\Omega^2}|z|^2 }
\end{equation}
Using the same change of variables and the kernels \eqref{Hkern}-\eqref{Lker}  we can compute $\mathcal{T}_\mu$:
\begin{eqnarray}
&&\mathcal{T}_\mu(f)\nonumber \\
&&=\frac{1}{1+\Omega^2}\int d^4x d^4y L_\star(f)(x,y)\left(\frac{\partial}{\partial y^\mu}-2i(\Theta^{-1})_{\mu\nu}\Omega^2 y^\nu\right) e^{-tH_4}(y,x)\nonumber \\
&&=\frac{1}{(1+\Omega^2)(2\pi\theta)^4 4\pi^2\sinh^2(2t\tilde{\Omega})}\int d^4z \ d^4u \ d^4v \ f(z)\nonumber \\
&&\times e^{-\frac{\tilde{\Omega}}{4}\left(\frac{|u|^2}{\tanh(\tilde{\Omega}t)}+\frac{|v|^2} {\coth(\tilde{\Omega}t)}\right)i\langle u,\Theta^{-1}(v-2z)\rangle}\nonumber \\
&&\times \left(u_\mu \frac{\tilde{\Omega}}{2}\cosh(\tilde{\Omega}t)-v_\mu\frac{\tilde{\Omega}}{2}\tanh(\tilde{\Omega}t)-i\Omega^2(\Theta^{-1})_{\mu\nu}(v^\nu-u^\nu)\right)\nonumber \\
&&=\frac{1}{(1+\Omega^2)^3 4\pi^2\sinh^2(2t\tilde{\Omega})}\int d^4z f(z)e^{-\tilde{\Omega}\frac{\tanh(\tilde{\Omega}t) }{1+\Omega^2}|z|^2 }\mathfrak{D}_\mu\mathcal{E}(\xi,\eta,t)\mid_{\xi=\eta=0}\nonumber \\
&&=\frac{1}{(1+\Omega^2)^3 4\pi^2\sinh^2(2t\tilde{\Omega})}\int d^4z f(z)\mathcal{Z}_\mu e^{-\tilde{\Omega}\frac{\tanh(\tilde{\Omega}t) }{1+\Omega^2}|z|^2 }
\end{eqnarray} 
with
\begin{eqnarray}
\mathfrak{D}_\mu &=&\frac{i\Omega}{\theta\tanh(\tilde{\Omega}t)}\frac{\partial}{\partial\xi^\mu}-\frac{i\Omega\tanh(\tilde{\Omega}t)}{\theta}\frac{\partial}{\partial\eta^\mu}+\Omega^2(\Theta^{-1})_{\mu}^\nu\left(\frac{\partial}{\partial\eta_\nu}-\frac{\partial}{\partial\xi_\nu}\right) \nonumber \\
& & \\
\mathcal{E}(\xi,\eta,t)&=&e^{-\frac{\theta\Omega}{2(1+\Omega^2)}\left(\tanh(\tilde{\Omega}t)|\xi|^2+\frac{|\eta|^2}{\tanh(\tilde{\Omega}t)}-\frac{2i}{\theta\Omega}\langle\xi,\Theta\eta\rangle+2\tanh(\tilde{\Omega}t)\langle \xi,\tilde{z}\rangle  +4i\langle \eta,z\rangle\right)}\\
\mathcal{Z}_\mu &=&-\tilde{\Omega}\tanh(\tilde{\Omega}t)z_\mu -\frac{2i\Omega^2}{1+\Omega^2}\tilde{z}_\mu
\end{eqnarray}
Following the same procedure $\mathcal{T}_{\mu\nu}(f) $ turns out to be:
\begin{eqnarray}
&&\mathcal{T}_{\mu\nu}(f)\nonumber \\
&&=\frac{\tilde{\Omega}^2}{(1+\Omega^2)^4\pi^2\sin^2(\tilde{\Omega}t)}\int d^4z f(z)e^{-\tilde{\Omega}\frac{\tanh(\tilde{\Omega}t) }{1+\Omega^2}|z|^2}
\nonumber \\
&&\times \left(\mathcal{Y}_{\mu\nu}+\mathfrak{D}_\mu\mathfrak{D}_\nu\right)\mathcal{E}(\xi,\eta,t)\mid_{\xi=\eta=0}\nonumber \\
&&=\frac{\tilde{\Omega}^2}{(1+\Omega^2)^4\pi^2\sin^2(\tilde{\Omega}t)}\int d^4z f(z) \left(\mathcal{N}_{\mu\nu}+\mathcal{Z}_\mu\mathfrak{D}_\nu\right)e^{-\tilde{\Omega}\frac{\tanh(\tilde{\Omega}t) }{1+\Omega^2}|z|^2}
\end{eqnarray}
with
\begin{eqnarray}
\mathcal{Y}_{\mu\nu}&=&-\delta_{\mu\nu}\frac{\tilde{\Omega}(1+\tanh^2(\tilde{\Omega}t))}{2\tanh(\tilde{\Omega}t)}+2i\Omega(\Theta^{-1})_{\mu\nu}\\
\mathcal{N}_{\mu\nu}&=&-\delta_{\mu\nu}\frac{\Omega(1+\Omega^2)}{\theta}\left(\tanh(\tilde{\Omega}t)+\frac{1}{\tilde{\Omega}t)}\left(\frac{1-\Omega^2}{1+\Omega^2}\right)^2\right)+2i\Omega(\Theta^{-1})_{\mu\nu}\nonumber \\
&& 
\end{eqnarray}
For the $\mathcal{T}_{\mu\nu\rho}(f)$ and $\mathcal{T}_{\mu\nu\rho\sigma}(f)$ we obtain:
\begin{eqnarray}
&&\mathcal{T}_{\mu\nu\rho}(f) \nonumber \\
&&=\frac{\tilde{\Omega}^2}{(1+\Omega^2)^5\pi^2\sin^2(\tilde{\Omega}t)}\int d^4z f(z)e^{-\tilde{\Omega}\frac{\tanh(\tilde{\Omega}t) }{1+\Omega^2}|z|^2}\nonumber \\
&&\times \left(\mathcal{Y}_{\mu\nu}\mathfrak{D}_\rho +\mathcal{Y}_{\mu\rho}\mathfrak{D}_\nu +\mathcal{Y}_{\nu\rho}\mathfrak{D}_\mu \right)\mathcal{E}(\xi,\eta,t)\mid_{\xi=\eta=0} \nonumber \\
&&=\frac{\tilde{\Omega}^2}{(1+\Omega^2)^5\pi^2\sin^2(\tilde{\Omega}t)}\int d^4z f(z)e^{-\tilde{\Omega}\frac{\tanh(\tilde{\Omega}t) }{1+\Omega^2}|z|^2} \nonumber \\
&&\times \left(\mathcal{N}_{\mu\nu}\mathcal{Z}_\rho +\mathcal{N}_{\mu\rho}\mathcal{Z}_\nu + \mathcal{N}_{\nu\rho}\mathcal{Z}_\mu+ \mathcal{Z}_\mu\mathcal{Z}_\nu\mathcal{Z}_\rho\right)\nonumber \\
&&\mathcal{T}_{\mu\nu\rho\sigma}(f) \\
&&=\frac{\tilde{\Omega}^2}{(1+\Omega^2)^6\pi^2\sin^2(\tilde{\Omega}t)}\int d^4z f(z)e^{-\tilde{\Omega}\frac{\tanh(\tilde{\Omega}t) }{1+\Omega^2}|z|^2}\Big(\mathcal{N}_{\mu\nu}\mathcal{N}_{\rho\sigma}+\mathcal{N}_{\mu\rho}\mathcal{N}_{\nu\sigma}\nonumber \\
&&+\mathcal{N}_{\mu\sigma}\mathcal{N}_{\nu\rho}+\mathcal{N}_{\mu\nu}\mathcal{Z}_\rho\mathcal{Z}_\sigma +\mathcal{N}_{\mu\rho}\mathcal{Z}_\nu\mathcal{Z}_\sigma +\mathcal{N}_{\mu\sigma}\mathcal{Z}_\nu\mathcal{Z}_\rho +\mathcal{Z}_\mu\mathcal{Z}_\nu \mathcal{N}_{\rho\sigma} +\mathcal{Z}_\mu\mathcal{Z}_\rho \mathcal{N}_{\nu\sigma}\nonumber \\
&& +\mathcal{Z}_\mu\mathcal{Z}_\sigma \mathcal{N}_{\nu\rho}+\mathcal{Z}_\mu\mathcal{Z}_\nu\mathcal{Z}_\rho\mathcal{Z}_\sigma\Big)
\end{eqnarray}
Finally, we have all the ingredients required to compute the leading part of the action \eqref{Action} replacing all the traces into the \eqref{heat1}.  
Using the trace property of the star product and the identities 
\begin{eqnarray}
\tilde{z}_\mu\star f=\frac{1}{2}\{\tilde{z}_\mu,f\} +\frac{1}{2}[\tilde{z}_\mu,f]=\frac{1}{2}\{\tilde{z}_\mu,f\}+i\partial_\mu f \\
\{\tilde{z}_\mu , f\star g \}_\star=\{\tilde{z}_\mu,f \}\star g -if\star \partial_\mu g 
\end{eqnarray}
we get after some manipulations:   
\begin{eqnarray}
&&\textrm{Tr}(e^{-t\mathcal{D}^2})=2 \cosh^4(\tilde{\Omega}t) \nonumber \\
&&+\frac{1}{\pi^2(1+\Omega^2)^2} \int d^4z \Bigg\{ \frac{1}{t}\Bigg(\left(\phi\star\bar{\phi}+ \frac{4\Omega^2}{1+\Omega^2}(\tilde{X}^A_\mu\star\tilde{X}_A^\mu -\tilde{X}^0_{\mu}\star\tilde{X}_0^\mu)\right)\nonumber \\
&&+\left(\bar{\phi}\star\phi+ \frac{4\Omega^2}{1+\Omega2}(\tilde{X}^B_\mu\star\tilde{X}_B^\mu -\tilde{X}^0_{\mu}\star\tilde{X}_0^\mu)\right)\Bigg)\nonumber \\
&&+\frac{1}{2}\left( \phi\star\bar{\phi}+\frac{4\Omega^2}{1+\Omega^2}\tilde{X}^A_\mu\star\tilde{X}_A^\mu\right)^2-\frac{1}{2}\left(\frac{4\Omega^2}{1+\Omega^2}\tilde{X}^\mu_0\star\tilde{X}^0_\mu\right)^2\nonumber \\
&&+\frac{1}{2}\left( \phi\star\bar{\phi}+\frac{4\Omega^2}{1+\Omega^2}\tilde{X}^B_\mu\star\tilde{X}_B^\mu\right)^2-\frac{1}{2}\left(\frac{4\Omega^2}{1+\Omega^2}\tilde{X}^\mu_0\star\tilde{X}^0_\mu\right)^2\nonumber \\
&&+\frac{1}{2}\left(\frac{(1-\Omega^2)^2}{2}-\frac{(1-\Omega^2)^4}{6(1+\Omega^2)^2}\right)(F^A_{\mu\nu}\star F_A^{\mu\nu}+F^B_{\mu\nu}\star F_B^{\mu\nu})\nonumber \\
&&+\frac{1+\Omega^2}{2}\left(D_\mu\phi\star\overline{D^\mu\phi}+\overline{D_\mu\phi}\star D^\mu\phi\right)\Bigg\}+\mathcal{O}(\sqrt{t})
\end{eqnarray}
where 
\begin{equation}
 \tilde{X}^A_\mu=\frac{\tilde{z}_\mu}{2}+A_\mu, \ \tilde{X}^B_\mu=\frac{\tilde{z}_\mu}{2}+B_\mu, \ \tilde{X}^0_\mu=\frac{\tilde{z}_\mu}{2}
\end{equation}
Using the Laurent expansion of $ \coth^4(t^\prime)=t^{\prime-4}+ \frac{4}{3}t^{\prime-2}+\frac{26}{45}+\mathcal{O}(t^{\prime 2})$
\begin{eqnarray}
&&\textrm{Tr}(e^{-t\mathcal{D}^2})=\frac{\theta^4 t^{-4}}{8\Omega^4}+\frac{2\theta^2 t^{-2}}{3\Omega^2}+\frac{52}{45} \nonumber \\
&&+\frac{1}{\pi^2(1+\Omega^2)^2} \int d^4z \Bigg\{ -\frac{t^{-1}}{\chi_0}\Bigg(\left(\phi\star\bar{\phi}+ \frac{4\Omega^2}{1+\Omega^2}(\tilde{X}^A_\mu\star\tilde{X}_A^\mu -\tilde{X}^0_{\mu}\star\tilde{X}_0^\mu)\right)\nonumber \\
&&+\left(\bar{\phi}\star\phi+ \frac{4\Omega^2}{1+\Omega^2}(\tilde{X}^B_\mu\star\tilde{X}_B^\mu -\tilde{X}^0_{\mu}\star\tilde{X}_0^\mu)\right)\Bigg)\nonumber \\
&&+\frac{1}{2}\left( \phi\star\bar{\phi}+\frac{4\Omega^2}{1+\Omega^2}\tilde{X}^A_\mu\star\tilde{X}_A^\mu\right)^2-\frac{1}{2}\left(\frac{4\Omega^2}{1+\Omega^2}\tilde{X}^\mu_0\star\tilde{X}^0_\mu\right)^2\nonumber \\
&&+\frac{1}{2}\left( \phi\star\bar{\phi}+\frac{4\Omega^2}{1+\Omega^2}\tilde{X}^B_\mu\star\tilde{X}_B^\mu\right)^2-\frac{1}{2}\left(\frac{4\Omega^2}{1+\Omega^2}\tilde{X}^\mu_0\star\tilde{X}^0_\mu\right)^2\nonumber \\
&&+\frac{1}{2}\left(\frac{(1-\Omega^2)^2}{2}-\frac{(1-\Omega^2)^4}{6(1+\Omega^2)^2}\right)(F^A_{\mu\nu}\star F_A^{\mu\nu}+F^B_{\mu\nu}\star F_B^{\mu\nu})\nonumber \\
&&+\frac{1+\Omega^2}{2}\left(D_\mu\phi\star\overline{D^\mu\phi}+\overline{D_\mu\phi}\star D^\mu\phi\right)\Bigg\}(z)+\mathcal{O}(\sqrt{t})
\end{eqnarray}
Comparing the previous expression to the expansion \eqref{heatd} and putting  $\chi_0=\chi(0)$  we  are finally able to write the spectral action \eqref{Action} as:
\footnotesize
\begin{eqnarray}
&&S(\mathcal{D}_A)=\frac{\theta^4\chi_{-4}}{8\Omega^4}+\frac{2\theta^2\chi_{-2}}{3\Omega^2}+\frac{52\chi_{0}}{45}\nonumber \\
&&+\frac{\chi_0}{\pi^2(1+\Omega^2)^2} \int d^4z \Bigg\{ \frac{\chi_{-1}}{\chi_0}\Bigg(\left(\phi\star\bar{\phi}+ \frac{4\Omega^2}{1+\Omega^2}(\tilde{X}^A_\mu\star\tilde{X}_A^\mu -\tilde{X}^0_{\mu}\star\tilde{X}_0^\mu)\right)\nonumber \\
&&+\left(\bar{\phi}\star\phi+ \frac{4\Omega^2}{1+\Omega^2}(\tilde{X}^B_\mu\star\tilde{X}_B^\mu -\tilde{X}^0_{\mu}\star\tilde{X}_0^\mu)\right)\Bigg)\nonumber \\
&&+\frac{1}{2}\left( \phi\star\bar{\phi}+\frac{4\Omega^2}{1+\Omega^2}\tilde{X}^A_\mu\star\tilde{X}_A^\mu\right)^2-\frac{1}{2}\left(\frac{4\Omega^2}{1+\Omega^2}\tilde{X}^\mu_0\star\tilde{X}^0_\mu\right)^2\nonumber \\
&&+\frac{1}{2}\left( \phi\star\bar{\phi}+\frac{4\Omega^2}{1+\Omega^2}\tilde{X}^B_\mu\star\tilde{X}_B^\mu\right)^2-\frac{1}{2}\left(\frac{4\Omega^2}{1+\Omega^2}\tilde{X}^\mu_0\star\tilde{X}^0_\mu\right)^2\nonumber \\
&&+\frac{1}{2}\left(\frac{(1-\Omega^2)^2}{2}-\frac{(1-\Omega^2)^4}{6(1+\Omega^2)^2}\right)(F^A_{\mu\nu}\star F_A^{\mu\nu}+F^B_{\mu\nu}\star F_B^{\mu\nu})\nonumber \\
&&+\frac{1+\Omega^2}{2}\left(D_\mu\phi\star\overline{D^\mu\phi}+\overline{D_\mu\phi}\star D^\mu\phi\right)\Bigg\}(z)+\mathcal{O}(\chi_1)
\end{eqnarray}\normalsize
Or after some rearrangements:
\begin{eqnarray}
&&S(\mathcal{D}_A)= \frac{\theta^4\chi_{-4}}{8\Omega^4}+\frac{2\theta^2\chi_{-2}}{3\Omega^2}+\frac{52\chi_{0}}{45} \nonumber \\
&&+\frac{\chi_0}{2\pi^2(1+\Omega^2)^2} \int d^4z \Bigg\{\left(\frac{(1-\Omega^2)^2}{2}-\frac{(1-\Omega^2)^4}{6(1+\Omega^2)^2}\right)(F^A_{\mu\nu}\star F_A^{\mu\nu}+F^B_{\mu\nu}\star F_B^{\mu\nu})\nonumber \\
&&+\left(\bar{\phi}\star\phi+ \frac{4\Omega^2}{1+\Omega^2}\tilde{X}^A_\mu\star\tilde{X}_A^\mu-\frac{\chi_{-1}}{\chi_0}\right)^2\nonumber \\
&&+\left(\phi\star\bar{\phi}+\frac{4\Omega^2}{1+\Omega^2}\tilde{X}^B_\mu\star\tilde{X}_B^\mu-\frac{\chi_{-1}}{\chi_0}\right)^2\nonumber \\
&&-2\left(\frac{4\Omega^2}{1+\Omega^2}\tilde{X}^\mu_0\star\tilde{X}^0_\mu-\frac{\chi_{-1}}{\chi_0}\right)^2+ 2(1+\Omega^2)D_\mu\phi\star\overline{D^\mu\phi}\Bigg\}(z)+\mathcal{O}(\chi_1)\label{faction}
\end{eqnarray}
We notice that Higgs mechanism introduces an extension of the standard Higgs potential in the commutative case, in fact the Higgs scalar field $\phi$ and the $\tilde{X}_A^\mu$, $\tilde{X}_B^\mu$ fields are present together in the potential. In this way the  gauge field  takes part in the definition of the vacuum. Another important property of the action,  considering the $\tilde{X}_A^\mu$, $\tilde{X}_B^\mu$ as independent, is the invariance under the translations:
\begin{equation}
\phi(x)\to\phi(x+a),\ X^\mu_A(x)\to X_A^\mu(x+a), \  X^\mu_B(x)\to X^\mu_B(x+a) , \ X^\mu_0(x)\to X^\mu_0(x+a)
\end{equation} 
which in others $\phi^4$-renormalizable theory is broken.
Beside, the action is invariant under $U(1) \times U(1)$ transformations:
\begin{equation}
\phi \to u_A\star\phi\star\overline{u_B}, \ \tilde{X}\to u_A\star \tilde{X}^\mu_A\star \overline{u_A}, \ \tilde{X}^\mu_B\to u_B\star\tilde{X}^\mu_B\star \overline{u_B} \label{gauge}
\end{equation}
In field theory the ground state can be defined through the minimum of the action, the relevant part of the \eqref{faction} for the minimization is:
\begin{eqnarray}
&&S(\mathcal{D}_A)=  \nonumber \\
&& \int d^4z \Bigg\{\left(\frac{(1-\Omega^2)^2}{2}-\frac{(1-\Omega^2)^4}{6(1+\Omega^2)^2}\right)(F^A_{\mu\nu}\star F_A^{\mu\nu}+F^B_{\mu\nu}\star F_B^{\mu\nu})\nonumber \\
&&+\left(\bar{\phi}\star\phi+ \frac{4\Omega^2}{1+\Omega^2}\tilde{X}^A_\mu\star\tilde{X}_A^\mu-\frac{\chi_{-1}}{\chi_0}\right)^2\nonumber \\
&&+\left(\phi\star\bar{\phi}+ \frac{4\Omega^2}{1+\Omega^2}\tilde{X}^B_\mu\star\tilde{X}_B^\mu-\frac{\chi_{-1}}{\chi_0}\right)^2\nonumber \\
&& -2\left(\frac{4\Omega^2}{1+\Omega^2}\tilde{X}^\mu_0\star\tilde{X}^0_\mu -\frac{\chi_{-1}}{\chi_0}\right)^2+2(1+\Omega^2)D_\mu\phi\star\overline{D^\mu\phi}\Bigg\}(z)+\mathcal{O}(\chi_1) \label{faction1}
\end{eqnarray}
Where we have omitted the constant part and we have rescaled the coefficient in front of the integral.
Considering the fields $X_A^\mu$, $X_B^\mu$ as fields variables instead $A^\mu$, $B^\mu$ we can state that each terms of the action is semi-positive defined, so in order to find the minimum it is sufficient to minimize them separately.
There are the two possible minima for the field strength part and for the covariant derivative part: the trivial solution with $\phi$ and $\tilde{X}_A^\mu$, $\tilde{X}_B^\mu$ equal to the null fields and the solution with  $\phi$, $X_A^\mu$, $X_B^\mu$ proportional to the identity. In each cases both the field strength part and the covariant derivative part 	disappear. For the potential parts we have:
\begin{eqnarray*}
V_A=V_B=\left(\frac{\chi_{-1}}{\chi_0}\right)^2 &&  \textrm{for} \ \phi=\tilde{X}^\mu_A=\tilde{X}^\mu_B=0 \\
V_A=(\alpha^2_\phi+4\frac{4\Omega^2}{1+\Omega^2}\alpha^2_A-\frac{\chi_{-1}}{\chi_0})^2,  &&  \textrm{for} \ \phi=\alpha_\phi \textbf{I}, \ \tilde{X}^\mu_A=\alpha_A \textbf{I}^\mu, \  \tilde{X}^\mu_B=\alpha_B \textbf{I}^\mu \\
V_B=(\alpha^2_\phi+4\frac{4\Omega^2}{1+\Omega^2}\alpha^2_B-\frac{\chi_{-1}}{\chi_0})^2 &&
\end{eqnarray*} 
Referring to the second case and minimizing the potentials, the minimum seems to be  for 
\begin{equation}
\phi_0=\frac{\chi_{-1}}{\chi_0}\cos\alpha \textbf{I},\ \tilde{X}^\mu_{A0}=\tilde{X}^\mu_{B0}= \frac{1}{2}\sqrt{\frac{\chi_{-1}}{\chi_0}}\sqrt{ \frac{1+\Omega^2}{4\Omega^2}} \sin\alpha \textbf{I}^\mu \label{Vacuum}
\end{equation}
However, the previous position is not allowed because the identity does not belong to the algebra under consideration. In general the non-triviality of the vacuum makes very difficult to explicit the vacuum configuration of the system in \cite{Goursac1} A. de Goursac, J.C. Wallet, and R. Wulkenhaar, using the matrix base formalism, have found an expressions from vacuum solutions deriving them from the relevant solutions equations of motion.
Although, the  complexity  of the vacuum configuration makes the perturbative approach very complicated, in order to conduct some investigation in the next chapters will be consider a non-perturbative approach using a discretized matrix model of the action \eqref{faction1} obtained using a particular Moyal base in which 
the fields become matrices, the star product becomes the matrix multiplication and the integral turns in a matrix trace. In this setting the action reduces to  
\begin{eqnarray}
&&S(\mathcal{D}_A)=  \nonumber \\
&& \int d^4z \Bigg\{\left(\frac{(1-\Omega^2)^2}{2}-\frac{(1-\Omega^2)^4}{6(1+\Omega^2)^2}\right)(F^A_{\mu\nu}\star F_A^{\mu\nu}+F^B_{\mu\nu}\star F_B^{\mu\nu})\nonumber \\
&&+\left(\bar{\phi}\star\phi+\frac{4\Omega^2}{1+\Omega^2}\tilde{X}^A_\mu\star\tilde{X}_A^\mu-\frac{\chi_{-1}}{\chi_0}\right)^2 \nonumber \\
&&+\left(\phi\star\bar{\phi}+ \frac{4\Omega^2}{1+\Omega^2}\tilde{X}^B_\mu\star\tilde{X}_B^\mu-\frac{\chi_{-1}}{\chi_0}\right)^2\nonumber \\
&&+ 2(1+\Omega^2)D_\mu\phi\star\overline{D^\mu\phi}\Bigg\}(z)+\mathcal{O}(\chi_1) \label{S}
\end{eqnarray}
The omitted factor for the finite matrix model of size $N$	becomes constant so can be ignored. The minimum is obtained like before and formally is \eqref{Vacuum}  in this case the identity,  of course, belongs to the matrix space. It is interesting to notice that the vacuum of the finite model, due to the Higgs field,  is no longer invariant under the transformations \eqref{gauge}, but is invariant under a subgroup of $U(N)\times U(N)$:
\begin{equation}
u_A=\overline{u_B} \longrightarrow   u_A\star \textbf{I} \star\overline{u_B}=u_A\star \overline{u_A}=\textbf{I}
\end{equation}
Having discretized the model will be performed a Monte Carlo simulation studying some statistical quantity such the energy density, specific heat, varying the parameters $\Omega, \frac{\chi_{-1}}{\chi_0}, \alpha$ and gathering some informations on the various contributions of the fields to the action. The simulations are quite cumbersome due to the 
complexity of the action and to the number of independent matrix to handle. Anyway, using particular algorithm and some 	simplifications we are able to get an acceptable balance between the computation precision and the computation time.     

%% file: cap3.tex
\chapter{Introduction to numerical analysis}

\emph{In this chapter will be explained the basics of the Monte Carlo simulation, focusing ourself on the application of such simulation on the field theory. In the first part will be hinted the path integral formulation which is essential to connect the field theories to  statistical systems in which the Monte Carlo methods are born.  Then will be introduced the Metropolis algorithm used to produce a Markov chain. In order to resume all the previous concepts and to show an example of phase analysis will be presented an example of numerical simulation on the well know Ising model.}

\section{Path integrals and functional integrals}

The path integral formalism in quantum mechanic was introduced by R. P. Feynman as a generalization \cite{path,path1} 
and provides a powerful tool to study quantum field theories. The path integral is based on the superposition law, in fact one has to consider a superposition of all possible paths in order to compute the transition amplitude from an initial state at $t_0$ to a final at time $t_1$.
Considering the initial and final states written as $|\psi(x_0)\rangle$ at $t_0$
and $|\psi(x_1)\rangle$ at time $t_1$, the transition amplitude is determined taking the matrix element of the temporal evolution:
\begin{equation}
\langle\psi(x_1)\mid U(t_1,t_0)\mid \psi(x_0)\rangle \label{matrixele}
\end{equation}
where
\begin{equation}
U(t_1,t_0) = e^{-\frac{iH(t_1-t_0)}{\hslash}} \label{time-evolu0}
\end{equation}
is the time evolution and $H$ is the Hamiltonian of the system, which we assume to be time independent.
The time interval $(t_0,t_1)$ can be divided into $N$ subintervals of size $\epsilon=\tau_{i+1}-\tau_i$, with $t_0=\tau_0$ and $t_1=\tau_N$, in this way is possible to write the time evolution operator as:
\begin{equation}
e^{-\frac{iH(t_1-t_0)}{\hslash}}=e^{-\frac{i}{\hslash} H(\tau_N -\tau_{N-1}+\tau_{N-1}-\cdots-\tau_2 +\tau_2-\tau_1+\tau_1-\tau_0)}=
\left(e^{-\frac{iH\epsilon}{\hslash}}\right)^N
\end{equation}
In the case $H$ has the general form $H = H_0 + V$, with $H_0=\frac{p^2}{2m}$, we can write using Trotter formula
\begin{equation}
e^{-\frac{iH(t_1-t_0)}{\hslash}} \sim \left(e^{-iH_0\epsilon/\hslash}e^{-iV\epsilon/\hslash}\right)^N \label{expdec}
\end{equation}
for $\epsilon \to 0$ and supposing that $H_0$ and $V$ are semi-bounded.
In order to factorize \eqref{matrixele} is useful to consider a set of complete states 
 \begin{equation}
\int|\psi(\tilde{x}_i)\rangle \langle \psi(\tilde{x}_i)| \textrm{d}\tilde{x}_i = 1, \ \ \tilde{x}_1=x(\tau_i)
\end{equation}
and inserting it between each term $e^{-iH_0\epsilon/\hslash}e^{-iV\epsilon/\hslash}$.
As result the \eqref{matrixele} can be rewritten as $N$ terms product  
\small
\begin{eqnarray}
\langle\psi(x_1)\mid U(t_1,t_0)\mid\psi(x_0)\rangle &\sim &\int\textrm{d}\tilde{x}_{N-1} \int \textrm{d}\tilde{x}_{N-2}\cdots\int\textrm{d}\tilde{x}_1 \nonumber \\
& & \times\langle\phi(\tilde{x}_N)\mid e^{-iH_0\epsilon/\hslash}e^{-iV\epsilon/\hslash}\mid \psi(\tilde{x}_{N-1})\rangle  \nonumber \\
& &  \times\langle\phi(\tilde{x}_{N-1})\mid e^{-iH_0\epsilon/\hslash}e^{-iV\epsilon/\hslash}\mid \psi(\tilde{x}_{N-2})\rangle \nonumber \\ 
& & \vdots \nonumber \\
& & \times\langle\phi(\tilde{x}_0)\mid e^{-iH_0\epsilon/\hslash}e^{-iV\epsilon/\hslash}\mid \psi(\tilde{x}_1)\rangle \label{expfac}
\end{eqnarray}
\normalsize
Considering a conservative $V(x)$ and $H_0$ dependent only on $p$, it is possible to demonstrate that \cite{path1,path2} :
\begin{equation}
\langle\phi(x_{i+1})\mid e^{-iH_0\epsilon/\hslash}e^{-iV\epsilon/\hslash}\mid \psi(x_i)\rangle\sim\int \frac{\textrm{d}p_{i+1}}{2\pi\hslash} e^{\frac{i}{\hslash}\left(p_{i+1}(x_{i+1}-x_i )-\epsilon H(p_{i+1},\frac{1}{2}(x_{i+1}+x_i))\right)}\label{expdec2}
\end{equation}
In the previous statement the argument in the exponential can be written as
\begin{equation}
\frac{i\epsilon}{\hslash}\left(p_{i+1}\frac{x_{i+1}-x_i}{\epsilon}-H(p_{i+1},\frac{1}{2}(x_{i+1} + x_i)\right)
\end{equation}
It easy to recognize the action in discrete interval $(\tau_i,\tau_{i+1})$ like the Lagrangian in the interval 
times the duration of the interval. Using \eqref{expdec2} and \eqref{expfac} taking $N \to \infty$ we finally obtain:
\begin{equation}
 \langle\psi(x_1)\mid U(t_1,t_0)\mid\psi(x_0)\rangle=\int^{x(t_1)=x_1}_{x(t_0)=x_0}
e^{-iS[x]/\hslash}\textrm{D}x(t)\label{time-evolu} 
\end{equation}
where 
\begin{equation}
S[x]=\int^{t_1}_{t_0} L(x,\dot{x})\textrm{d}t,
\end{equation}
$S[x]$ and $L$ are the action and the Lagrangian of the system, $ \int_{x_0}^{x_1}\textrm{D}x(t)$
stands for the integral over all paths $x(t)$ starting from $x(t_0 ) = x_0$ to $x(t_1) = x_1$  and $\textrm{D}x(t)$ is the functional measure. The generalization \eqref{time-evolu} to quantum fields is  straightforward.  
For completeness, before introducing its expression,  it is important to remind that
in the path-integral methods, when Osterwalder-Schrader axioms hold \cite{path3,path4}, it is common to have the action with imaginary time in order to simplify the calculations.
Beside, there is a strong advantage of considering an imaginary time, in fact it allows
to establish a relation between statistical mechanic and quantum field theory.

The euclidean frame approach is based on the analytic continuation, which is a technique used in the domain of definition of a given analytic function in order to extend a real valued function into the complex plane. Our purpose is to start from real time $t$ to the imaginary time $\tilde{t}$, the so called the Euclidean time \cite{path2}.  In this way  we force the imaginary time and spatial coordinates to have the same signature. A simple application is the  D'Alembertian operator which in real time is given by:
\begin{equation}
\Box=\frac{\partial^2}{\partial t^2}-\frac{\partial^2}{\partial x_1^2}-\frac{\partial^2}{\partial x_2^2}-\frac{\partial^2}{\partial x_3^2}
\end{equation}
Applying the previous procedure we obtain: 
\begin{equation}
\triangle=-\frac{\partial^2}{\partial x_0^2}-\frac{\partial^2}{\partial x_1^2}-\frac{\partial^2}{\partial x_2^2}-\frac{\partial^2}{\partial x_3^2} 
\end{equation}
with $x_0 =\tilde{t}=it$. 
In quantum mechanic, when  we consider  the path integrals over all possible particle paths between two points, we treat space and time in different ways. Otherwise, in field theory, in order to treat
them in the same manner, we introduce a function of 4-dimensional space-time manifold $\bar{x}=(x,t)$, which we call field $\psi(\bar{x})$. In the case $\psi$ is a neutral scalar field, the field values are real, $\psi(\bar{x}) \in \mathbb{R}$. We define a configuration like a fixed field value in each space-time point $\bar{x}$, the configuration  plays the role of  particle path in quantum mechanic. The functional integral is over all field configurations
$\int \textrm{D}\psi$, the Lagrangian $L$ becomes a Lagrangian density $ L(\psi(x),\partial_\mu\psi(x))$ at each 
point $x$ and the action is obtained by an integral over the space-time volume $S =\int L d^4x$.
The generalization of \eqref{time-evolu} for quantum fields is given by $(\hslash=1)$:
\begin{equation}
 \langle\psi_1|T(\tilde{t}_1,\tilde{t}_0)|\psi_0\rangle = \int  e^{-S[\psi]}\textrm{D}\psi(x,\tilde{t}) \label{qft-stat}
\end{equation}
Beside the path integral there is also a second interpretation of the resulting functional integral, as 
a functional integral in statistical field theory. In this frame the Euclidean action is seen as the energy
functional of an analog statistical mechanical system with $k_B T = 1$ where $k_B$ is Boltzmann's constant. A statistical system in equilibrium is naturally described by \eqref{qft-stat}.

\subsection{Expectation values}
The expectation values of an observable $O$ it is defined as follows:
\begin{equation}
\langle O\rangle =\int  \frac{O(\psi)e^{-S[\psi]}}{Z}\textrm{D}\psi \label{expectation}  
\end{equation}
where
$$Z=\int e^{-S[\psi]}\textrm{D}\psi $$
is called  partition function.
The analytically  integration in \eqref{expectation}, in general, is impossible, indeed it involves all the possible
configurations in the functional space. But we can try, using numerical methods, to estimate the value in \eqref{expectation}. 
One of  the most popular approach is the Monte Carlo method, this method is founded of the assumption that we can estimate the
expectation values using representative sets of random configurations.
There are many way to generate a representative set, usually  are used random moves from a starting configuration
in order to explore the configuration space. In this thesis, to estimate the expectation values of the observables defined in the next chapter,  will be used a particular algorithm for the configurations generation called the Metropolis algorithm \cite{Metro}.

\section{Monte Carlo methods} 

Many physical and mathematical systems are often treated using Monte Carlo methods thanks to their computational algorithms which use a random (or pseudo-random) generated  sampling, these methods are most suited to calculation by a computer and tend to be used when it is hard or impossible to compute an exact result with a deterministic algorithm. These methods are widely used in mathematics; one of the classic  application is the evaluation of definite integrals, particularly multidimensional integrals with complicated boundary conditions. Beside, Monte Carlo methods have a great importance in statistical mechanic for example in the statistical systems with a large number of degrees of freedom, such as strongly coupled solids, disordered materials and fluids. 
A precise definition of Monte Carlo methods is hard to give, in fact  there is no single Monte Carlo method, instead the term describes a large and heavily used class of approaches and algorithms. However, these approaches tend to follow a particular procedure. The first step is to define a domain of possible inputs, generate inputs randomly from the domain using a certain specified probability distribution, then perform a deterministic computation using the inputs and at the end join the results of the individual computations into a final result. 
For the present thesis one possible definition  Monte Carlo could be: a method of approximating an expectation value by 
the sampled mean of a function of random variables, invoking the laws of large numbers.

We can formalize the previous ideas \cite{monte0} considering a random variable $x$
having probability function or probability density function $p_X (x)$ which is greater than zero
on a set $X$ of values. Then the expected value of a function $f$ of  the continues variable $x$ is
\begin{equation}
E(f(z)) = \int_{x \in X} f(x)p_X (x)dx  \label{mean}
\end{equation} 
For the discrete case :
\begin{equation}
E(f(X)) =\sum_{x \in X} f(x)p_X (x) \label{mean2}
\end{equation}
A Monte Carlo estimate of \eqref{mean} or \eqref{mean2} can be defined taking $N$ samples of $X$, $(x_1 ,\cdots , x_N )$ and computing the mean of $f(x)$ on the sample: 
\begin{equation}
\tilde{f}_N (x) =\frac{1}{N}\sum_{i=1}^N f(x_i) \label{mean3}
\end{equation}
$\tilde{f}_N (x)$ it is called Monte Carlo estimator of $E(f(X))$.
This estimations can be applied either when the generated variables are mutually independent or when they are
correlated one to another (for example if they are generated by an ergodic Markov chain). For simplicity we will
consider the random variables independent, but all can be extended to samples obtained from a Markov chains via
the weak law of large numbers. 
If $E(f(X))$ exists, then the weak law of large numbers tells us that for any arbitrarily small $\epsilon$:
\begin{equation}
\lim_{N\to\infty} P(|f_N(x)-E(f(x))|\geq\epsilon) = 0
\end{equation}
In other words, for $N$ large the probability that $\tilde{f}_N(x)$ deviates much from
$E(f(x))$ becomes small so we are justified to use \eqref{mean3} as estimator of $E(f(x))$. It is interesting to note that expectation value of the \eqref{mean3} is:
\begin{equation}
E(\tilde{f}_N(x))=E\left(\frac{1}{N}\sum_{i=1}^N f(x_i)\right)=\frac{1}{N}\sum_{i=1}^N E(f(x_i))= E(f(x))
\end{equation}
In this case we say that $f_N(x_i)$ is unbiased for $E(f(x))$.
The previous method turns to be very useful when it is  applied in all the situations in which the quantities of interest are 
formulated as expectations value, for example: probabilities, integrals, and summations.
\\
Let $Y$ be a random variable, the probability that $Y$ takes on some value in a set $X$
can be expressed as an expectation using the function:
\begin{equation}
P(Y\in X)= E(C_{(X)}(Y))
\end{equation}
where $C_{X}(Y)$ is the characteristic function of $X$ that takes the value 1 when $Y \in X$ otherwise is 0.
\\
For integrals we consider a problem  which is completely deterministic the integration of a function $f(x)$
from $a$ to $b$. This integral can be expressed as an expectation respect to a uniformly distributed continuous random 
variable $u$ between $a$ and $b$, with density function $P_u (u) = 1/(b-a)$. Rewriting the integral we obtain
\begin{equation}
(b-a)\int_b^a\frac{q(x)}{b-a}dx = (b-a)\int_b^a q(x)p_wu(x)dx =(b-a)E(q(u)) 
\end{equation} 
\\
A discrete sum is just the discrete version of the previous example; the sum of a function $q(x)$ over some numerable values of $x$ in a set $X$. Using a random variable $w$ which takes values in $X$ all with constant probability $P$ with $\sum_{w \in X}P = 1$ the  sum can be seen as the expectation:
\begin{equation}
\sum_{x \in X} q(x) =\frac{1}{P}\sum_{x \in A}q(x)P=\frac{1}{P}E(q(w))
\end{equation}
\\
The immediate consequence is that all probabilities, integrals, and summations can be
approximated by the Monte Carlo method. However, it is  very important to point out that there is no
a priori reason to use uniform distributions. 
As we have seen many quantity of interest can be formulated as an expectation approximate by a Monte
Carlo estimator, but it is not always so easy to actually have a Monte Carlo estimator that  can provide a
sufficient good estimation in a reasonable amount of computer time. For the same problems  various number of
Monte Carlo estimators can be constructed (essentially varying the probability distribution), of course some Monte Carlo estimators are more efficient
than others. In order to find the best  estimator we need to compute the variance, we look for the Monte Carlo estimator with smallest variance taking the amount of computational effort fixed. 
We have to compute the variance  of the Monte Carlo estimator of $E(f(x))$
with the random variable $f_N(x)$. The standard formulas for a random variables are 
\begin{eqnarray}
Var(\tilde{f}_N(X))&=&Var\left(\frac{1}{N}\sum_{i=1}^N f(X_i)\right)=\frac{Var(f(X))}{N} \nonumber \\
&=&\sum_{x\in X}[f(x) - E(f(X))]^2 P_X (x) 
\end{eqnarray}
if $x$ is discrete, and
\begin{eqnarray}
Var(\tilde{f}_N(X))&=&Var\left(\frac{1}{N}\sum_{i=1}^N f(X_i)\right)=\frac{Var(f(X))}{N}  \nonumber \\
&=&\int_{x\in X}[f(x) - E(f(X))]^2 P_X (x) 
\end{eqnarray}
if $x$ is continuous.  
There are numerous sophisticated algorithm used to obtain a better approximation reducing the variance, many of them uses a non-uniform probability distributions, in the next section we introduce a such kind of algorithm called Metropolis algorithm which is particularly useful for the systems with many degrees of freedom. 

\subsection{The Metropolis algorithm} 
\cite{Metro0} Studying systems with a great number of degrees of freedom it becomes clear that the number
of possible configurations becomes exceedingly  large very quickly. Even
for extremely simple binary models, in which each particle may exist in one of two possible
states, the number of configurations of the system grows extremely rapidly with N, the
number of particles. A very instructive and simple example is the Ising model of magnetic
spins, in which each spin-1/2 particle can be spin up or spin down  respect
to a fixed axis. In this way each of the $N$ particles has only two possible states, so the 
total number of discernible configurations is
$$2^N$$
In easy to notice that using a model of a square lattice of such spins taking a quite small scale of a macroscopic sample of matter with $32 \times 32 = 2^{10}$ spins, there are
$$2^{2^{10}} \approx 10^{308}$$
distinct configurations of this system. But remind the basic postulate of statistical mechanic, 
we know that an isolated system in equilibrium is equally likely to be found in any one of these configurations.
To compute the time-averaged properties we compute an ensemble average over all the accessible configurations.
If system is not isolated from its external environment, but may
exchange energy with it, the exchange is ruled by a temperature $T$, which
quantifies how the energy is shared with its surroundings. The Boltzmann
factor $e^{-E/T}$ is proportional to the probability that the system will be found in a particular
configuration at energy $E$ when the temperature of the environment is $T$ ($K_b=1$)
\begin{equation}
P \propto e^{-E/T} \label{metro}
\end{equation}
Higher $T$ means
that the environment tends to give energy to the system otherwise in the small the temperature case  the
environment tends to gather energy 	bringing the system in a lower energy state. 
In view the identification of \eqref{qft-stat} as description of a statistical system in equilibrium $(k_bT=1)$, explained in the previous section, we can repeat the same treatment in the case we have 
\begin{equation}
P \propto e^{-S} \label{metro1}
\end{equation}
In the study of a such model, unless we can treat it analytically, it is very hard to try investigating each configuration, 
or to average over all of them. Usually when we handle an interacting system, we can almost never compute the sums analytically, so we have no choice but to seek an approximation either analytically or numerically. One choice is to use the a Monte Carlo approach with  the Metropolis algorithm in order to limit the huge number of possible configurations. The Metropolis algorithm  is based on the notion of detailed balance that describes equilibrium 
for systems whose configurations have probability proportional to the Boltzmann
factor. Essentially we sample the space of possible configurations in a way that agrees with \eqref{metro1}. 
Beside we complete the exploration  by taking in account the possible transitions between close configurations.
Consider two configurations $A$ and $B$, each of which occurs with probability proportional
to \eqref{metro1} and computing the ratio of the probability of the configurations we obtain 
\begin{equation}
\frac{P(A)}{P(B)}=\frac{e^{-S_A}}{e^{-S_B}}= e^{-(S_A-S_B)} \label{metro2}
\end{equation}
The notable thing about considering the ratio is that it tuns a relative probabilities involving
an unknown proportionality constant (the inverse of the partition function) into a
pure number. Metropolis et al. \cite{Metro} have proposed a practical algorithm to achieve the
relative probability of \eqref{metro2}. In a simulation the procedure is the following:
\begin{enumerate}
\item  Beginning from a configuration $A$, with known $S_A$, make a  change (typically small) in the configuration to obtain a new configuration $B$.
\item   Compute $S_B$ (usually do not differs so much from $S_A$).
\item  If $S_B < S_A$, accept the new configuration, since it has lower energy according to \eqref{metro1}.
\item Otherwise the number $e^{-(S_B-S_A)}$ is compared to a random number, $\textrm{ran} \in
[0,1]$; if $e^{-(S_B-S_A)} >\textrm{ran}$ the configuration $B$ is accepted, in the other
case it is rejected. 
\item If the configuration $B$ is accepted go to point 1 using $B$ as new starting configuration, in case of rejection go to point 1 trying another configuration $B$. 
\end{enumerate}
Following these prescriptions, we will sample points in the space of all possible configurations 
with probability proportional to the exponential factor consistently with the theory
of statistical mechanic equilibrium. We can approximate expectation value like \eqref{mean3} by summing them
along the path generated by Metropolis procedure.
Some hard parts about implementing the Metropolis algorithm are the first step,  how to decide the staring configuration and how
to generate useful new configurations, the first issue will been solved using the thermalization graphs.

\subsection{Phase analysis}
\cite{monte1} In order to show and test advanced analysis tools for stochastically generated data it
is convenient to work with very simple models, in fact a very few exact analytical results are available 
for comparison. On the other hand, one should always be prepared that a really complex system may add further
complications which are not present in the simple test cases. Nevertheless, to introduce the phase transitions
we go back to the well known Ising model whose partition function is defined by the Hamiltonian $H_I$:
\begin{equation}
  H_I = -\sum_{(i,j)} \sigma_i \sigma_j \ ,\ \sigma_i = \pm 1  \ ,\  Z_I(\beta)=\sum_{\sigma_i }e^{-\beta H_I}
\end{equation}
where $\beta = J/k_B T$ is the inverse temperature in natural units, the spins $\sigma_i$ take place on the
sites $i$ of a $n$-dimensional cubic lattice of volume $V = L^n$ and the symbol $(ij)$ indicates
that the lattice sum is on all nearest pairs assuming periodic
boundary conditions. In two dimensions and with no external field this model has been solved
exactly even on finite lattices. Unfortunately for the three dimensional model is not so easy and an exact solution
is not available. Instead, there is an huge amount of very precise  Monte Carlo simulations.
The standard observables are the internal energy density $e = E/V$, with $E =
−d\textrm{ln}Z_I /d\beta ≡ \langle H_I\rangle$, the specific heat
\begin{equation}
C/k_B = \frac{de}{d(k_BT)} = \beta^2\left( \frac{\langle H^2_I \rangle - \langle H_I\rangle^2}{V}\right)
\end{equation}
the magnetization density
\begin{equation}
m = M/V = \langle|\mu|\rangle \ , \ \mu= \sum_i\sigma_i/V 
\end{equation}
and the susceptibility
\begin{equation}
\chi = \beta V \left( \langle \mu^2 \rangle - \langle|\mu|\rangle^2 \right)
\end{equation}
Beside, the spin-spin correlation function can be defined as
\begin{equation}
G(x_i - x_j) =\langle \sigma_i \sigma_j\rangle 
\end{equation}
Can be showed that at large distances, the spin-spin correlator $G(x)$ decays exponentially $G(x) \sim e^{-|x| / \xi}$, so we can define the so called spatial correlation length $\xi$ as
\begin{equation}
\xi= − \lim_{|x|\to\infty} \frac{|x|}{\textrm{ln} G(x)}
\end{equation}
The theory of phase transitions is a large subject, here
we shall confine ourself to those properties that are important for understanding the basic
concepts for the data analysis.
Most phase transitions in nature are of first order and are characterized by
discontinuities in the order parameter  $\Delta m$ of the magnetization $m$ in Fig. 1
or in the energy like latent heat $\Delta e$ or both, at the transition point $T_0$. This is connected to the
coexistence of two or more phases  at $T_0$. In contrast to a second-order
transition, the correlation length in the coexisting pure phases is finite. Therefore, the
specific heat, the susceptibility and also the autocorrelation time do not diverge in the pure
phases. However, can be present superimposed delta function like singularities associated with
the jumps of $e$ and $m$. The standard example exhibiting a first-order phase transition is the $q$-state Potts
model defined by the Hamiltonian
\begin{equation}
H_P = -\sum_{(i,j)}\delta_{\sigma_i \sigma_j} \ , \ \sigma_i \in 1,\cdots,q 
\end{equation}
where $\delta_{ij}$ is the Kronecker symbol. This model in two dimension is exactly solved and exhibits a
first order transition at $\beta_t = \textrm{log}(1 + \sqrt{q})$ for all $q \geq 5$ and for $q \leq 4$ the
transition is of second order, including the Ising case $(q = 2)$ and the special percolation
limit $(q = 1)$. In three dimension there are  many numerical evidences that suggest that for $q \geq 3$ the transition is
of first order.

\begin{figure}[htb]
\begin{center}
\includegraphics[scale=0.5]{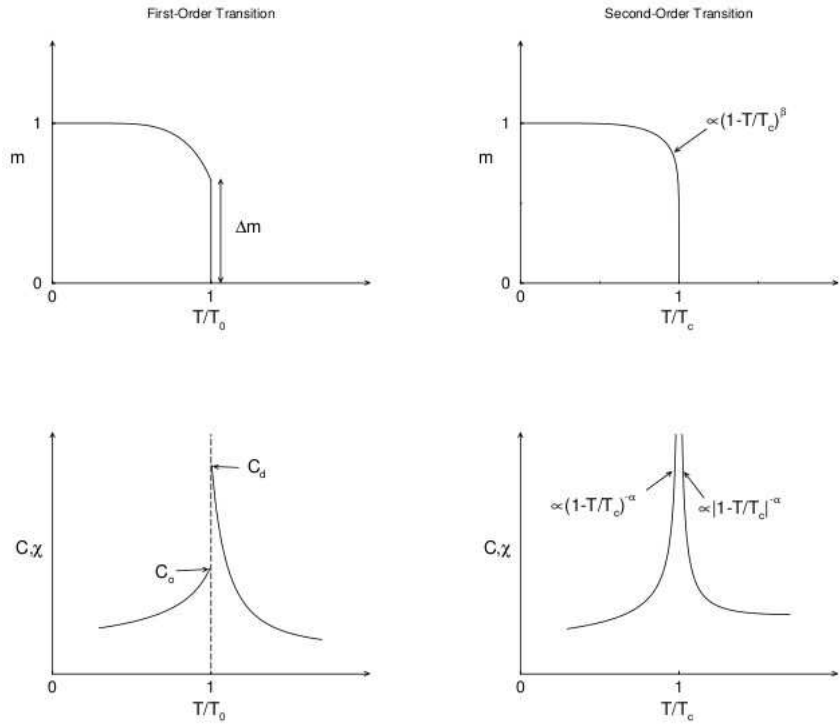}
\end{center}
\caption{\sl The characteristic behavior of the magnetization, $m$, specific heat $C$ and susceptibility, $\chi$ densities  at first
and second order phase transitions.}\label{Figure}\end{figure}

The characteristic feature of second-order phase transitions is a divergent spatial correlation
length $\xi$ at the transition point $\beta_c$. This causes a scale invariance, the core of
renormalization group treatments \cite{monte} and is the origin of the important concept of universality. 
At $\beta_c$  are expected fluctuations on all length scales, implying 
singularities in statistical functions such as the correlation length
\begin{equation}
\xi = \xi_0^{+,-} t^{-\nu}+\cdots ,
\end{equation}
where $t \equiv |1-T/T_c|$ and the $ \cdots$ indicate higher order  corrections.
In the previous expansion are defined the critical exponent $\nu$ and the critical amplitudes
$\xi_0^{+,-}$  where $\pm$ symbols denotes the high and low temperature side of the transition. 
Singularities are present in the specific heat, magnetization and susceptibility too fig.\ref{Figure} \cite{monte1}. They define, respectively, the
critical exponents $\alpha,\beta$ and $\gamma$, can be proved that these coefficients are related each other through scaling
relations and only two of them are independent.

\begin{figure}[htb]
\begin{center}
\includegraphics[scale=0.5]{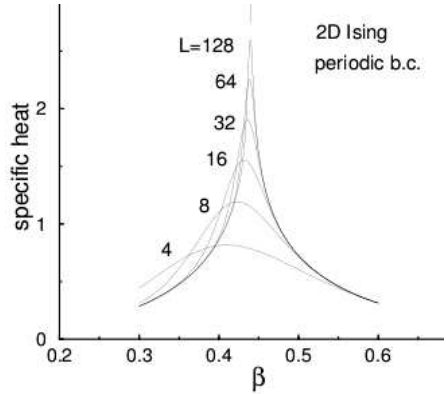}
\end{center}
\caption{\sl Finite size scaling behavior of the specific heat density of the $2D$ Ising model on $L \times L$ lattices close to the
 critical point $\beta_c$ .}\label{Figure 0}\end{figure}

When  the spins are updated with an Metropolis-Monte Carlo process, \cite{monte1} the information 
on the updated state of the spins has to propagate over the correlation volume before
one obtains a new truly independent configuration. The number of update steps required is measured by the autocorrelation time $\tau$ (its definition is given in appendix) which
close to $\beta_c$  behaves according to
\begin{equation}
\tau\propto\xi^z \propto t^{-\nu z} 
\end{equation}
Here we have introduced the independent dynamical critical exponent $z$ but depends on
the used update algorithm. For the Metropolis 
we expect that the updated information performs a random walk
in configuration space, requiring on the average $\xi^2$ steps to propagate over a distance proportional 
to $\xi$ ,therefore we expect a dynamical critical exponent of $ z \sim 2$. 
Of course in systems characterized by finite size, as in any numerical simulation, every statistical function like the correlation length cannot
diverge and  the divergences in all other quantities occurring in the infinite model are rounded and can be shifted. The example of
the specific heat density for the two dimensional Ising model  is illustrated in fig.\ref{Figure 0}. The finiteness  affects  the scaling formulas 
and  $\xi$ is then substituted  by the linear size of the system $L$. If we write
\begin{equation}
t \propto \xi^{-1/\nu} \to L^{-1/\nu}  
\end{equation}
it easy to understood  that the thermodynamic scaling laws  $\chi \propto t^{-\gamma}$ ,
for finite models, has to be replaced by finite-size scaling (FFS)   $\chi \propto L^{\gamma/\nu} $.
In particular, \cite{monte} we obtain for the autocorrelation time a FSS of the form
\begin{equation}
\tau \propto L^z  
\end{equation}

Even for first order phase transition in a finite systems the singularities can be split in  narrow peaks with  height proportional to the characteristic volume $V$ and  width  proportional to $V^{-1}$. The system is now capable to  exist in a  mixed phase configurations and jump from one phase into the other, however 
the probability  of the appearing  of this mixed configurations is smoothed by the Boltzmann factor:
$\textrm{exp}(2\sigma L^{n-1})$. The double-peak structure of the energy and magnetization densities that can occur for a first-order phase transition too and can be explained  by the previous argumentations. For large system sizes it may require many Monte Carlo steps before the systems explore  	enough phase space.
In fact, the autocorrelation time associated with this transitions is  the inverse of
the suppression factor:
\begin{equation}
\tau \propto \textrm{exp}(2\sigma L^{n-1}) 
\end{equation}
Since in this case the autocorrelations grows exponentially with the system size, this behavior
is  usually  called supercritical slowing down.

%% file: cap4.tex
\chapter{Preparing for numerics}
\emph{ In this chapter will be set up all the ingredients required to 	implement a Monte Carlo simulation of the model treated in the chapter two. The first part will be devoted to the discretization scheme, introducing a Moyal base in which   the model 
described by the action \eqref{S} on Moyal plane will be discretized.
In the second part, in order to define the  measurements of the conducted Monte Carlo  simulation,  will be applied some arguments treated in the third chapter outlining the expectation values, statistical quantities and some order parameters}.

\section{Discretization of the action}
The first step across the numerical analysis is to apply a discretization scheme. Various schemes can be used like lattice approximation, but the nature of star product due to its oscillator exponential,  makes the lattice approach not suitable without adaptations. We will use  another discretization scheme in which our fields are approximated  by finite matrices and the star product becomes the standard matrix multiplication.
Our  model is Euclidean, as  remarked in chapter 3, the main advantage of working in this
formalism is that we can  establish a direct connection to statistical physic
and the functional integral converges with a relatively modest statistic.
The action to discretize is
\footnotesize
\begin{eqnarray}
&S(\phi,\tilde{X}_A,\tilde{X}_B)&=\int d^4z\Bigg\{\left(\frac{\left(1-\Omega^2\right)^2}{2}-\frac{\left(1-\Omega^2\right)^4}{6\left(1+\Omega^2\right)^2}\right)\left(F^A_{\mu\nu}\star F^{\mu\nu}_A +F^B_{\mu\nu}\star F^{\mu\nu}_B\right) \nonumber \\ 
& &+ \left(\phi\star\bar{\phi}+\frac{4\Omega^2}{1+\Omega^2}\tilde{X}_A^\mu\star\tilde{X}_{A\mu}-\frac{\chi_{-1}}{\chi_0}\right)^2\nonumber \\
& & +\left(\bar{\phi}\star\phi +\frac{4\Omega^2}{1+\Omega^2}\tilde{X}_B^\mu\star\tilde{X}_{B\mu}-\frac{\chi_{-1}}{\chi_0} \right)^2  \nonumber \\
& &+2(1+\Omega^2)D_\mu\phi\star\overline{D_\mu\phi}\Bigg\}(z)+\mathcal{O}(\chi_1)
\end{eqnarray}
\normalsize
Using the identity $D_\mu\phi =\phi\star\tilde{X}_{B\mu}-\tilde{X}_{A\mu}\star\phi$  we can recast the action in the following form:
\footnotesize
\begin{eqnarray}
&S(\phi,\tilde{X}_A,\tilde{X}_B)&= \int d^4z\Bigg\{\left(\frac{\left(1-\Omega^2\right)^2}{2}-\frac{\left(1-\Omega^2\right)^4}{6\left(1+\Omega^2\right)^2}\right)\Big(\left[\tilde{X}_{A\mu},\tilde{X}_{A\nu}\right]_\star\left[\tilde{X}_{A}^\mu,\tilde{X}_{A}^\nu\right]_\star  \nonumber \\
& &+\left[\tilde{X}_{B\mu},\tilde{X}_{B\nu}\right]_\star\left[\tilde{X}_{B}^\mu,\tilde{X}_{B}^\nu\right]_\star\Big)
+ \left(\phi\star\bar{\phi}+\frac{4\Omega^2}{1+\Omega^2}\tilde{X}_A^\mu\star\tilde{X}_{A\mu}-\frac{\chi_{-1}}{\chi_0}\right)^2 \nonumber \\
& & +\left(\bar{\phi}\star\phi +\frac{4\Omega^2}{1+\Omega^2}\tilde{X}_B^\mu\star\tilde{X}_{B\mu}-\frac{\chi_{-1}}{\chi_0} \right)^2  \nonumber \\
& &+2(1+\Omega^2)\left(\phi\star\tilde{X}_{B\mu}-\tilde{X}_{A\mu}\star\phi\right)
\left(\bar{\phi}\star\tilde{X}_{A}^\mu-\tilde{X}_{B}^\mu\star\bar{\phi}\right)\Bigg\}(z) \nonumber \\
& &+\mathcal{O}(\chi_1)\label{S0}
\end{eqnarray}\normalsize
As a first approach to the numerical simulation and forced by limited computation resource,  we will consider the Monte Carlo simulation of the previous action around  its minimum and the simulation will take $\sqrt{\frac{\chi_{-1}}{\chi_0}} $  as a positive parameter. In this setting  the behavior of the simulations will be identical for the negative case and avoiding $\frac{\chi_{-1}}{\chi_0}$ to be negative we  have not any problems about the thermalization. In order to define the previous action around the minimum we translate the fields $\phi$, $\tilde{X}_{A\mu}$, $\tilde{X}_{B\mu}$ using the following translated fields:
\begin{eqnarray}
\phi &=& \psi +\sqrt{\frac{\chi_{-1}}{\chi_0}}\cos\alpha\textbf{I} \ \\
\tilde{X}_{A\mu}&=& Y_{A\mu} +\frac{1}{2}\sqrt{\frac{\chi_{-1}}{\chi_0}}\sqrt{\frac{2\Omega^2}{(1+\Omega^2)}}\textbf{I}_\mu\sin\alpha  \\
\tilde{X}_{B\mu}&=& Y_{B\mu} +\frac{1}{2}\sqrt{\frac{\chi_{-1}}{\chi_0}}\sqrt{\frac{2\Omega^2}{(1+\Omega^2)}}\textbf{I}_\mu\sin\alpha  
\end{eqnarray}
Substituting the previous fields into \eqref{S0} we get a positive action with minimum in zero:

\begin{eqnarray}
&S(\psi,Y_A,Y_B)&= \int d^4z\Bigg\{D\Big(\left[Y_{A\mu},Y_{A\nu}\right]_\star\left[Y_{A}^\mu,Y_{A}^\nu\right]_\star 
+ \left[Y_{B\mu},Y_{B\nu}\right]_\star\left[Y_{B}^\mu,Y_{B}^\nu\right]_\star\Big) \nonumber \\
& &+ \left(\psi\star\bar{\psi}+\mu\cos\alpha(\psi+\bar{\psi})+ CY_A^\mu\star Y_{A\mu} +\mu \textbf{I}^\mu Y_{A\mu}\sin\alpha \right)^2  \nonumber \\
& &+ \left(\bar{\psi}\star\psi +\mu\cos\alpha(\psi+\bar{\psi})+ CY_B^\mu\star Y_{B\mu} +\mu \textbf{I}^\mu Y_{B\mu}\sin\alpha  \right)^2 \nonumber \\ \nonumber \\
& &+2(1+\Omega^2)\left((Y_{B\mu}-Y_{A\mu})\mu\cos\alpha +\psi\star Y_{B\mu}-Y_{A\mu}\star\psi\right)\nonumber \\
& &\star\left((Y_{A}^\mu-Y_{B}^\mu)\mu\cos\alpha+\bar{\psi}\star Y_{A}^\mu -Y_{B}^\mu\star\bar{\psi}\right)\Bigg\}(z)  +\mathcal{O}(\chi_1)
\label{Sf} \end{eqnarray}

Where for simplicity we put:
\begin{eqnarray}
C=\frac{1+\Omega^2}{4\Omega^2},& D=\frac{\left(1-\Omega^2\right)^2}{2}-\frac{\left(1-\Omega^2\right)^4}{6\left(1+\Omega^2\right)^2} , & \frac{\chi_{-1}}{\chi_0}=\mu^2
\end{eqnarray}
Recalling the first chapter we consider a  Moyal base in which our fields became infinite matrices and the integral becomes a double trace, we will discretize the action introducing a cut off and then we will show how to simplify the discretization in  order to conduct a feasible numerical simulations. 

\subsection{Discretization by Moyal base}
The following treatment is mainly taken from \cite{moyal-triple,bondia2} as introduction to the Moyal base which will be used later.
We can define on the algebra $\mathbb{R}_\Theta^2$  a natural basis of eigenfunctions $f_{mn}$ of the harmonic
oscillator, where $m, n \in \mathbb{N}$. If we define the Hamiltonian  
\begin{equation}
H = H_1 + H_2 + \cdots  + H_N,
\end{equation}
with
\begin{equation}
H_l := \frac{1}{2}(x^2_l + x^2_{l+N} ) \ \textrm{for} \ l = 1, \cdots , N
\end{equation}
then the $f_{mn}$ diagonalize these harmonic oscillator Hamiltonians:
\begin{equation}
H \star f_{mn} = \theta(m + \frac{1}{2} )f_{mn},\ f_{mn} \star H_l = \theta(n + \frac{1}{2} )f_{mn}  \label{moy-hamil}
\end{equation}
In order to derive $f_{mn}$ we note that  the Gaussian function
\begin{equation}
f_0(x) = 2e^{-\frac{1}{\theta}(x_1^2 +x_2^2 )}
\end{equation}
is an idempotent,
\begin{equation}
(f_0 \star f_0 )(x) = 4\int d^2y\int \frac{d^2k}{(2\pi)^2} e^{-\frac{1}{\theta}(2x^2+y^2 +2x\cdot y+x\cdot \theta\cdot k+\frac{1}{4}\theta^2_1 k^2)+ik\cdot y}=f_0(x) \label{idemp}
\end{equation}
We can define the graded creation and annihilation operators
\begin{eqnarray}
a = \frac{1}{\sqrt{2}}(x_1+ix_2), & & \frac{\partial}{\partial a} = \frac{1}{\sqrt{2} }(\partial_1 - i\partial_2) \nonumber \\
\bar{a} = \frac{1}{\sqrt{2}}(x_1-ix_2), & & \frac{\partial}{\partial\bar{a}} = \frac{1}{\sqrt{2} }(\partial_1 + i\partial_2)
\end{eqnarray}
Computing the left and right star multiplication of this graded operator with any $f \in \mathbb{R}_\Theta^2$ we obtain
\begin{eqnarray}
(a\star f)=a(x)f(x)+\frac{\theta}{2}\frac{\partial f}{\partial\bar{a}}(x), & & (f\star a)=a(x)f(x)-\frac{\theta}{2}\frac{\partial f}{\partial\bar{a}}(x) \nonumber \\
(\bar{a}\star f)=\bar{a}(x)f(x)-\frac{\theta}{2}\frac{\partial f}{\partial a}(x), & & (f\star\bar{a})=\bar{a}(x)f(x)+\frac{\theta}{2}\frac{\partial f}{\partial a}(x) \nonumber \\
\end{eqnarray}
The multiple application of the star multiplication for $a$ and $\bar{a}$ to $f_0$ leads to
\begin{equation}
\bar{a}^{\star n}\star f_0 = 2^m \bar{a}^mf_0, \  f_0\star a^{\star n} = 2^na^nf_0 
\end{equation}
and
\begin{eqnarray}                            
 a \star \bar{a}^{\star m} \star f_0 &=& \Bigg\{ \begin{array}{cc}
 m\theta( \bar{a}^{\star(m-1)}\star f_0 ) & \textrm{for} \  m \geq 1 \\
0 & \textrm{for} \  m=0 \end{array}\nonumber \\
 f_0 \star a^{\star n} \star \bar{a} &=& \Bigg\{ \begin{array}{cc}
 n\theta( f_0\star a^{\star(m-1) }) & \textrm{for} \ m \geq 1 \\
0 & \textrm{for} \  m=0\end{array}\label{moy-prop}
 \end{eqnarray}
with $a^{\star n} = a \star a \star \cdots \star a$ and similarly for $\bar{a}^{\star n}=\bar{a} \star \bar{a} \star \cdots \star \bar{a}$, where in the left side appear $n$ factors. Finally, can be proved by induction the following relation for $f_{mn}$:
\begin{eqnarray}
f_{mn}&=&\frac{1}{\sqrt{n!m!\theta^{m+n}_1}} \bar{a}^{\star m}\star f_0 \star a^{\star n} \\
      &=& \frac{1}{\sqrt{n!m!\theta^{m+n}_1}}\sum^{\textrm{min}(m,n)}_{k=0} (-1)^k \binom{m}{k}\binom{n}{k} k!  2^{m+n-2k}\theta^k_1 \bar{a}^{m-k} a^{n-k}f_0 \nonumber
\end{eqnarray}
From the equations \eqref{idemp} and \eqref{moy-prop} we obtain the $\star$-multiplication rule: 
\begin{equation}
(f_{mn} \star f_{kl} )(x) = \delta_{nk} f_{ml} (x) \label{star-rule}
\end{equation}
The previous multiplication rule  associates the $\star$-product between $f_{ml}$ with the ordinary matrix product:
\begin{eqnarray}
a(x) = \sum^\infty_{m,n=0} a_{mn} f_{mn}(x), & & b(x) = \sum^\infty_{m,n=0} b_{mn} f_{mn}(x) \nonumber \\
(a \star b)(x) =\sum^\infty_{m,n=0} (ab)_{mn} f_{mn} (x), & & (ab)_{mn}=\sum^\infty_{k=0}a_{mk}b_{kn}\label{matrix-prod}
\end{eqnarray}
In this base we can write any elements of $ \mathbb{R}_\Theta^2$ but we have to require the rapid decay \cite{bondia2} of the sequences of coefficients $\{a_{mn}\}$  appearing in the \eqref{matrix-prod}:
\begin{equation}
\sum^\infty_{m,n=0} a_{mn} f_{mn}(x) \in \mathbb{R}_\Theta^2, \ \textrm{if}, \ \sum^\infty_{m,n=0} \left((2m+1)^{2k} (2n+1)^{2k} |a_{mn}|^2 \right)^\frac{1}{2} < \infty \ \textrm{for all}  \ k .
\end{equation}
Using  again \eqref{idemp}, the trace property of the integral and \eqref{moy-prop} we find this useful property:
\begin{eqnarray}
\int d^2 x f_{mn}(x)&=&\frac{1}{\sqrt{n!m!\theta^{m+n}_1}}\int d^2x (\bar{a}^{\star n} \star f_0 \star f_0 \star a^{\star n}(x)) = \delta_{mn} \int d^2x  f_0 \nonumber \\
&=&2\pi\theta\delta_{mn}
\end{eqnarray}
The eigenfunctions $f_{nm}$ can be expressed with the help of
Laguerre functions \cite{bondia2,phi4-non,moy-base}:
\begin{equation}
f_{mn}(\rho,\varphi)= 2(-1)^m \sqrt{\frac{m!}{n!}} e^{i\varphi(n-m)} \left(\sqrt{\frac{2}{\theta}}\rho\right)^{n-m} e^{-\frac{\rho^2}{\theta}}L_{m}^{n-m}\left(\frac{2}{\theta}\rho^2\right) \label{mb}
\end{equation}
Our fields can be expanded in this base as:
\begin{equation}
X^\mu(x) =\sum_{m_i,n_i \in  \mathbb{N}} X^\mu_{\di{m_1n_1}{m_2n_2}} f_{m_1n_1}(x_0,x_1)f_{m_2n_2}(x_2,x_3 )\label{X-exp}
\end{equation}
and
\begin{equation}
\psi(x) =\sum_{m_i,n_i \in  \mathbb{N}} \psi_{\di{m_1n_1}{m_2n_2}} f_{m_1n_1}(x_0,x_1)f_{m_2n_2}(x_2,x_3 )\label{psi-exp}
\end{equation}
Using this base we can forget the Moyal product in this way the model becomes to 9-matrix model. In view of \eqref{matrix-prod}
a $\star$-product between two fields using \eqref{star-rule} can be written as
\begin{eqnarray}
\Psi(x)\star\Phi(x)&=&\sum_{m_i,n_i,k_1,l_1\in  \mathbb{N}}\Psi_{\di{m_1n_1}{m_2n_2}}\Phi_{\di{k_1l_1}{k_2l_2}} f_{m_1n_1}(x_0,x_1)\star f_{k_1l_1}(x_0,x_1) \nonumber \\
&\times & f_{m_2n_2}(x_2,x_3)\star f_{k_2l_2}(x_2,x_3) \nonumber \\
&=& \sum_{m_i,n_i,k_1,l_1\in  \mathbb{N}}\Psi_{\di{m_1n_1}{m_2n_2}}\Phi_{\di{k_1l_1}{k_2l_2}}\delta_{n_1k_1}\delta_{n_2k_2}f_{m_1l_1}(x_0,x_1)f_{m_2l_2}(x_2,x_3)\nonumber \\
&=& \sum_{m_i,l_1\in  \mathbb{N}}\Psi\Phi_{\di{m_1l_1}{m_2l_2}} f_{m_1l_1}(x_0,x_1)f_{m_2l_2}(x_2,x_3)
\end{eqnarray}
\normalsize
where 
\begin{equation}
\Psi\Phi_{\di{m_1l_1}{m_2l_2}}=\sum_{n_1,n_2 \in  \mathbb{N}}\Psi_{\di{m_1n_1}{m_2n_2}}\Phi_{\di{n_1l_1}{n_2l_2}}
\end{equation}
So the star product became a "double" matrix multiplication, the action, the equations of field and all treatments can be conducted on the infinite matrices instead directly on the continues fields. Beside, for finite matrices, the $\mathbb{N}^2$-indexed double sequences
  can be written as tensor products of ordinary matrices, 
\begin{equation}
X_{\genfrac{}{}{0pt}{}{m_1n_1}{m_2n_2}}= \sum_{i=1}^K 
X^i_{m_1n_1} \otimes X^i_{m_2n_2} \;.
\label{tensorproduct}
\end{equation}
Since the matrix product and trace also
factor into these independent components, the action factors into 
$S=\sum_{i=1}^K
S(\psi^{1i},Y_A^{1i},Y_B^{1i})S(\psi^{2i},Y_A^{2i},Y_B^{2i})$. 
Then, regarding all 
$\psi^{1i},Y_A^{1i}$, $Y_B^{1i},\psi^{2i},Y_A^{2i},Y_B^{2i}$
as random variables over which to integrate in the partition function,
the partition function factors, too:
\begin{align}
&\int
\mathcal{D}(\psi^{11},Y_A^{11},Y_B^{11},\psi^{21},Y_A^{21},Y_B^{21})
\cdots 
\mathcal{D}(\psi^{1K},Y_A^{1K},Y_B^{1K},\psi^{2K},Y_A^{2K},Y_B^{2K})
\;e^{-S}
\nonumber
\\
&=\bigg(
\int \mathcal{D}(\psi^{1i},Y_A^{1i},Y_B^{1i},\psi^{2i},Y_A^{2i},Y_B^{2i})
\;e^{-S(\psi^{1i},Y_A^{1i},Y_B^{1i})\cdot S(\psi^{2i},Y_A^{2i},Y_B^{2i})}
\bigg)^K\;.
\end{align}
We may therefore restrict ourselves to $K=1$ and we can split the model in  2+2 dimensions, in this way the calculus will be performed just on standard matrices.
This decomposition makes the  numerical treatment of 2 dimensional case simpler respect the complete one, indeed  will be examined both cases. Using this base now our problem is reduced to a infinite matrix problem but is not enough to be handled numerically, we have to perform a truncation in order to obtain finite matrices, this truncation will consist in a maximum $m,n<N$ in the expansion \eqref{psi-exp}-\eqref{X-exp}. It is easy to verify that this kind of approximation corresponds in a cut in energy, in fact recalling the definition \eqref{moy-hamil} of  $f_{mn}$ we have:
\begin{equation}
\{H, f_{mn}\}_\star= \theta(m+n+\frac{1}{2} )f_{mn}(\varphi,\rho)
\end{equation}
Beside, the functions $f_{mn}$ with $m, n < N$ induce a cut-off  in position space and momentum space, indeed  the function 
\begin{equation}
L_m^\alpha(z)z^{\alpha/2} e^{-z/2}
\end{equation}
with $z=\frac{2}{\theta}\rho^2$, is rapidly decreasing beyond the last maximum $(z^\alpha_m )_{max}$. 
Numerically  can be found $(z^\alpha_m )_{max} < 2\alpha + 4m$ and thus the radial cut-off
\begin{equation}
\rho_{max} \sim \sqrt{2\theta}  \ ,\ \textrm{for} \ m, n < N 
\end{equation}
In addition, for $p_1 = −p \sin \psi$, $p_2 = p \cos\psi $ we obtain \cite{moybase}  using \eqref{mb} , 
\footnotesize
\begin{eqnarray}
\tilde{f}_{mn}(p,\psi)&=&\int_0^\infty\rho d\rho\int_0^{2\pi} e^{ip\rho\sin(\varphi-\psi)}f_{mn}(\rho,\varphi) \nonumber \\
&=&4\pi(-1)^n \sqrt{\frac{m!}{n!}} e^{i\psi(n-m)}\int_0^\infty d\rho\left(\sqrt{\frac{\theta}{2}}p\right)^{n-m} L_{m}^{n-m} \Big(\frac{\theta}{2}p^2\Big)J_{n-m}(\rho p)e^{-\frac{\rho^2}{\theta} }\nonumber \\
&=&2\pi\theta\sqrt{\frac{m!}{n!}}e^{i(\phi+\pi)(n-m)}\left(\sqrt{\frac{\theta}{2}}p\right)^{n-m}L_{m}^{n-m} \Big(\frac{\theta}{2}p^2\Big)e^{-\frac{\theta}{4}p^2 }
\end{eqnarray}
\normalsize
In the end we obtain
\begin{equation}
p_{max} \sim \sqrt{\frac{8N}{\theta}}  \,\ \textrm{for} \ m, n < N 
\end{equation}
Summarizing, to operate the discretization we have the following correspondences:
\begin{eqnarray}
\phi(x) \in \mathbb{R}_\Theta^4  &\rightarrow & \hat{\phi} \in \mathbb{M}_N \\
Y^A_\mu(x) \in \mathbb{R}_\Theta^4 &\rightarrow & \hat{Y}^A_\mu \in  \mathbb{M}_N \\
Y^B_\mu(x) \in \mathbb{R}_\Theta^4 &\rightarrow & \hat{Y}^B_\mu \in  \mathbb{M}_N \\
 \int a(x)dx &\rightarrow & \textrm{Tr}(\hat{a}) 
\end{eqnarray}
After truncating the representative matrices is convenient to operate  another substitution \cite{Goursac1}:
\begin{eqnarray}
 Z_0=\hat{Y}^A_0+i\hat{Y}^A_1, & \bar{Z}_0=\hat{Y}^A_0-i\hat{Y}^A_1  \nonumber \\
 Z_1=\hat{Y}^B_0+i\hat{Y}^B_1, & \bar{Z}_1=\hat{Y}^B_0-i\hat{Y}^B_1 \nonumber \\
 Z_2=\hat{Y}^A_2+i\hat{Y}^A_2, & \bar{Z}_2=\hat{Y}^A_2-i\hat{Y}^A_3  \nonumber \\
 Z_3=\hat{Y}^B_2+i\hat{Y}^B_3, & \bar{Z}_3=\hat{Y}^B_2-i\hat{Y}^B_3  \label{Z-sub}
 \end{eqnarray}
Implementing the previous replacements for the two dimensional model, thus using only $Z_0$, $Z_1$, in eq. \eqref{Sf}  we arrive at
\begin{equation}\\
S_2=\operatorname{Tr}\left(\mathcal{L}_F+\mathcal{L}_{V_0}+\mathcal{L}_{V_1}+\mathcal{L}_{D_0}\bar{\mathcal{L}}_{D_0} +\mathcal{L}_{D_1}\bar{\mathcal{L}}_{D_1}\right) \label{S2}
\end{equation}
with
\footnotesize
\begin{eqnarray}
\mathcal{L}_{2F}&=&\frac{D}{2}\left(\left[\bar{Z}_0,Z_0\right]^2 +\left[\bar{Z}_1,Z_1\right]^2\right)\nonumber \\
\mathcal{L}_{2V_0}&=&\left(\psi\bar{\psi}+\mu\cos\alpha(\psi+\bar{\psi})+ \frac{1}{2}\left\{\bar{Z}_0,Z_0\right\} +\frac{\mu\sin\alpha}{\sqrt{2C}}((-1+i)Z_0+(1+i)\bar{Z}_0)\right)^2    \nonumber \\
\mathcal{L}_{2V_1}&=&\left(\bar{\psi}\psi +\mu\cos\alpha(\psi+\bar{\psi})+ \frac{1}{2}\left\{\bar{Z}_1,Z_1\right\} +\frac{\mu\sin\alpha}{\sqrt{2C}}((-1+i)Z_1+(1+i)\bar{Z}_1)\right)^2    \nonumber \\
\mathcal{L}_{2D_0}&=&  \sqrt{2(1+\Omega^2)}\left(\mu\cos\alpha(Z_1+\bar{Z}_1-Z_0-\bar{Z}_0 ) + \psi(Z_1+\bar{Z}_1)-(Z_0+\bar{Z}_0)\psi\right)\nonumber \\
\mathcal{L}_{2D_1}&=&  \sqrt{2(1+\Omega^2)}\left(\mu\cos\alpha(Z_1-\bar{Z}_1-Z_0+\bar{Z}_0 ) + \psi(Z_1-\bar{Z}_1)-(Z_0-\bar{Z}_0)\psi\right) \nonumber 
\end{eqnarray}
\normalsize
Where for simplicity we have omitted the hat on the matrices and the bars stand for the hermitian conjugate. With this discretized action \eqref{S2}, the scalar $\psi$ and the gauge fields $Z_0$, $Z_1$ become 3-complex matrix instead to have one complex matrix for the scalar and four real matrices for the $Y_{A\mu}$, $Y_{B\mu}$ fields. In this setting the algorithm for Monte Carlo simulation becomes simpler especially with the update algorithm used here.
In  the four dimensional model it is convenient to use the \eqref{Z-sub} too, in this case there are some additional terms. In particular appear the crossed commutators between the $Z_i$ in the Y-M part:
\begin{equation}
S_4=\operatorname{Tr}\left(\mathcal{L}_F+\mathcal{L}_{V_0}+\mathcal{L}_{V_1}+\mathcal{L}_{D_0}\bar{\mathcal{L}}_{D_0} +\mathcal{L}_{D_1}\bar{\mathcal{L}}_{D_1}+\mathcal{L}_{D_2}\bar{\mathcal{L}}_{D_2}+\mathcal{L}_{D_3}\bar{\mathcal{L}}_{D_3}\right) \label{S4}
\end{equation}
With
\footnotesize
\begin{eqnarray}
\mathcal{L}_{4F}&=&\frac{D}{2}\Big(\left[\bar{Z}_0,Z_0\right]^2 +\left[\bar{Z}_1,Z_1\right]^2 + 
\frac{1}{4}\Big(\left[Z_0+\bar{Z}_0,Z_2-\bar{Z}_2\right]^2-\left[Z_0+\bar{Z}_0,Z_2+\bar{Z}_2\right]^2\nonumber \\
&+& \left[Z_0-\bar{Z}_0,Z_2+\bar{Z}_2\right]^2- \left[Z_0-\bar{Z}_0,Z_2-\bar{Z}_2\right]^2 -\left[Z_1+\bar{Z}_1,Z_3+\bar{Z}_3\right]^2\nonumber \\
&+& \left[Z_1+\bar{Z}_1,Z_3-\bar{Z}_3\right]^2 +\left[Z_1-\bar{Z}_1,Z_3+\bar{Z}_3\right]^2 -\left[Z_1-\bar{Z}_1,Z_3-\bar{Z}_3\right]^2\Big)\Big)\nonumber \\
\mathcal{L}_{4V_0}&=&\big(\psi\bar{\psi}+\mu\cos\alpha(\psi+\bar{\psi})+ \frac{1}{2}\left(\left\{\bar{Z}_0,Z_0\right\} +\left\{\bar{Z}_2,Z_2\right\}\right)\nonumber \\
&+&\frac{\mu\sin\alpha}{2\sqrt{C}}((-1+i)(Z_0+Z_2)+(1+i)(\bar{Z}_0+\bar{Z}_2))\big)^2    \nonumber \\
\mathcal{L}_{4V_1}&=&\big(\bar{\psi}\psi+\mu\cos\alpha(\psi+\bar{\psi})+ \frac{1}{2}\left(\left\{\bar{Z}_1,Z_1\right\} +\left\{\bar{Z}_3,Z_3\right\}\right)\nonumber \\
&+& \frac{\mu\sin\alpha}{2\sqrt{C}}((-1+i)(Z_1+Z_3)+(1+i)(\bar{Z}_1+\bar{Z}_3))\big)^2    \nonumber \\
\mathcal{L}_{4D_0}&=&  \sqrt{2(1+\Omega^2)}\left(\mu\cos\alpha(Z_1+\bar{Z}_1-Z_0-\bar{Z}_0 ) + \psi(Z_1+\bar{Z}_1)-(Z_0+\bar{Z}_0)\psi\right)\nonumber \\
\mathcal{L}_{4D_1}&=&  \sqrt{2(1+\Omega^2)}\left(\mu\cos\alpha(Z_1-\bar{Z}_1-Z_0+\bar{Z}_0 ) + \psi(Z_1-\bar{Z}_1)-(Z_0-\bar{Z}_0)\psi\right)\nonumber \\
\mathcal{L}_{4D_2}&=&  \sqrt{2(1+\Omega^2)}\left(\mu\cos\alpha(Z_3+\bar{Z}_3-Z_2-\bar{Z}_2 ) + \psi(Z_3+\bar{Z}_3)-(Z_2+\bar{Z}_2)\psi\right)\nonumber \\
\mathcal{L}_{4D_3}&=&  \sqrt{2(1+\Omega^2)}\left(\mu\cos\alpha(Z_3-\bar{Z}_1-Z_2+\bar{Z}_2 ) + \psi(Z_3-\bar{Z}_3)-(Z_2-\bar{Z}_2)\psi\right)\nonumber \\
\nonumber 
\end{eqnarray}
\normalsize

In this case the action \eqref{S2} becomes 5 complex matrix model instead eight real matrices $Y_{A\mu}$, $Y_{B\mu}$ and one complex matrix $\psi$. This form  may seem cumbersome but it still  more comfortable for numerical simulations. The next step is to define the estimator for the average values of interest and to develop some numerical parameters in order to analyze  the numerical results.  

\section{Definition of the observables}
Calling $(\psi,Z_i)$  a configuration of the fields (where $i=1,2$ for the two dimensional case and $i=1,\cdots,4$ for the four dimensional one), the probability  to encounter this configuration is given by
\begin{equation}
P[(\psi,Z_i)] =\frac{e^{-S[(\psi,Z_i)]}}{\mathcal{Z}}
\end{equation}
where $S[(\psi,Z_i)]$ is the action \eqref{S2} or \eqref{S4} of the system evaluated in the particular configuration and $\mathcal{Z}$ is the partition function:
\begin{equation}
\mathcal{Z}=D[(\psi,Z_i)] e^{-S[(\psi,Z_i)]}
\end{equation}
$D [(\psi,Z_i)]$ denotes the integration over all field configurations. The expectation
value of the observable $O$ is defined by the expression:
\begin{equation}
\langle O\rangle=\int D[(\psi,Z_i)] \frac{e^{-S[(\psi,Z_i)]}O[(\psi,Z_i)]}{\mathcal{Z}}
\end{equation}
Following  Monte Carlo methods, will be produced a  sequence of configurations
$\{(\psi,Z_i)_j \}, j = 1, 2,\cdots,T_{MC}$  and evaluated the average of the observables over that set
of configurations. The sequences of configurations obtained,  a Monte Carlo chain,  are  representations of 
the configuration space at the given parameters. 
In this frame the expectation value is approximated as
\begin{equation}
\langle O\rangle \approx \frac{1}{T_{MC}}\sum_{j=1}^{T_{MC}}O_j
\end{equation}
where $O_j$ is the value of the observable $O$ evaluated in the $j$-sampled configuration,
$(\psi,Z_i)_j$, $O_i= O[(\psi,Z_i)_j]$.

\subsection{Energy and specific heat}
The internal energy is defined as:
\begin{equation}
E(\Omega,\mu,\alpha)= \langle S\rangle
\end{equation}
and the specific heat takes the form
\begin{equation}
C(\Omega,\mu,\alpha)= \langle S^2\rangle - \langle S\rangle^2
\end{equation}
These terms correspond to the usual definitions for energy
\begin{equation}
E(\Omega,\mu,\alpha) = -\frac{1}{\mathcal{Z}}\frac{\partial\mathcal{Z}}{\partial\beta}
\end{equation}
and specific heat
\begin{equation}
C(\Omega,\mu,\alpha) =\frac{\partial E}{\partial\beta}
\end{equation} where $\mathcal{Z}$ is the partition function. 
It is very useful to compute separately the average values of the four contributions:
\begin{eqnarray}
S_F(\psi,Z_i)&=&  \operatorname{Tr} \mathcal{L}_F \label{sf}\\
S_{V_0}(\psi,Z_i)&=& \operatorname{Tr} \mathcal{L}_{V_0} \\
S_{V_1}(\psi,Z_i)&=& \operatorname{Tr} \mathcal{L}_{V_1}  \\
S_D(\psi,Z_i)&=& \operatorname{Tr}\left(\mathcal{L}_{D_j}\bar{\mathcal{L}}_{D^j}\right) \label{sd} 
\end{eqnarray}
Where $i,j=1,2$ or $i,j=1,\cdots,4$. The corresponding expectation values to \eqref{sf}-\eqref{sd} are
\begin{eqnarray}
F(\Omega,\mu,\alpha)&=& \langle S_F \rangle \\
V_0(\Omega,\mu,\alpha)&=& \langle S_{V_1} \rangle \\
V_1(\Omega,\mu,\alpha)&=& \langle S_{V_1} \rangle \\
D(\Omega,\mu,\alpha)&=& \langle S_D \rangle \\
V&=&V_0+V_1
\end{eqnarray}

\subsection{Order parameters}
The previous quantities are not enough if we want to measure the various contributions of different modes of the fields 
to the configuration $\psi,Z_i$. Therefore we need  some control parameters usually called order parameters. As a first idea we can think about a quantity related to the norms of the fields for example the sums $\sum_{nm} |\psi_{nm}|^2$, $\sum_{nm} |Z_{inm}|^2$ of all squared entries of our matrices, this quantity is called the full power of the field \cite{order-par,order-par1} and it can be computed as  the trace of the square:
\begin{eqnarray}
\varphi^2_a &=& \operatorname{Tr}(|\psi|^2) \label{vara} \\
Z^2_{ia} &=& \operatorname{Tr}(|Z_i|^2) 
\end{eqnarray}
This parameter turns to be very useful  to find the region where the configurations are considered disordered for which we expect to have a random distribution of the modes  $\langle\varphi^2_a\rangle \sim 0$, $\langle Z^2_{ia}\rangle \sim 0 $ and the so called ordered phase in which we have  
 $\langle\varphi^2_a\rangle >> 0$, $\langle Z^2_{ia}\rangle >> 0$. The full power of the field can be used to determine for which independent parameters  there is a transition between the disordered and ordered regime.
However $\langle\varphi_a\rangle$ alone  is not a real order parameter because does not distinguish contributions from the different modes but we can use it as a reference to  define the quantities:
\begin{eqnarray}
\varphi^2_0 &=& \sum^N_{n=0} |a_{nn}|^2  \nonumber \\
Z^2_{i0} &=& \sum^N_{n=0} |z_{inn}|^2 \label{var0}
\end{eqnarray} 
Referring to the base \eqref{mb} it is easy to see that  such parameters are connected to the pure spherical contribution.  This quantity will be used to analyze the spherical contribution to the full power of the field.  
We can define the corresponding susceptibility  as:
\begin{eqnarray}
\chi_{\varphi_0}&=& \langle \varphi_0^2\rangle - \langle \varphi_0\rangle^2\nonumber \\
\chi_{Z_{i0}}&=& \langle Z^2_{i0}\rangle - \langle Z_{i0}\rangle^2
\end{eqnarray}
We can generalize the previous quantity defining some parameters $\varphi_l$ in such a way they form a decomposition of the full power of the fields. 
\begin{equation}
\varphi^2_a=\varphi^2_0+\sum_l \varphi^2_l , \  Z_{ia}^2=Z_{i0}^2+\sum_l Z_{il}^2
\end{equation}
Following this prescription the other quantity for $l>0$ can be  defined as:
\begin{equation}
\varphi^2_l =\sum^l_{n,m=0} |a_{nm}(1-\delta_{nm})|^2 , \  Z_{il}^2 =\sum^l_{n,m=0} |z_{lnm}(1-\delta_{nm})|^2  \label{varl}
\end{equation}
If the contribution is dominated from  the spherical symmetric parameter we expect to have  $\langle\varphi^2_a\rangle \sim \langle\varphi^2_0\rangle$, $\langle Z_{ia}^2\rangle \sim \langle Z^2_{i0}\rangle$. In this regime for the uniform ordered region we have  $\langle\varphi^2_0\rangle \sim 0$, $\langle Z_{i0}^2\rangle \sim 0$ and in the disordered phase we expect $\langle\varphi^2_l\rangle \sim 0$, $\langle Z_{il}^2\rangle \sim 0$ for all $l$. With \eqref{varl}  we can define the order parameters $\varphi^2_1$, $Z_{i1}^2$ as a particular case
$l = 1$ we have
\begin{equation}
\varphi^2_1 =|a_{10}|^2+ |a_{01}|^2, \  Z_{i1}^2 =|z_{i10}|^2+ |z_{i01}|^2
\end{equation}
and its susceptibility, 
\begin{equation}
\chi_{\varphi_1}=\langle \varphi_1^2\rangle - \langle \varphi_1\rangle^2, \ \chi_{Z_{i1}}= \langle Z_{i1}^2\rangle - \langle Z_{i1}\rangle^2
\end{equation}
In the next simulations we be evaluated  the quantities related to $l = 0$ and to $l = 1$ as representative of
those contribution where the rotational symmetry is broken.
 A region characterized by $\langle\varphi^2_1\rangle >> 0$, $\langle Z_{i1}^2\rangle >> 0$ is called \cite{order-par}  non-uniform, such region feature a large contributions in which the spherical symmetry is broken. Using higher $l$ in \eqref{varl} we can analyze  the contributions of the remaining modes, but turns out that the measurements 
of the first two modes are  enough to characterize the behavior of the system. 

%% file: cap5.tex
\chapter{Numerical results }
\emph{This last chapter is devoted to present the numerical results of the 4-dimensional and 2-dimensional model constructed and approximated in the previous chapters. In the first part will be discussed the behaviors of the defined observables of the full 4-dimensional model varying the parameters. Meanwhile in the second part will be showed the result for the 2-dimensional model for the same values of parameters and compared  with the full model results.}

\section{Four dimensional case}
Now we are ready to  discuss the results of the Monte Carlo simulation on the approximated spectral model described in the chapter 2 and discretized in the previous chapter. As a first approach to Monte Carlo simulation of such kind of model, we use some restrictions on the parameters. Will be used the spectral action around its minimum \eqref{S4} in order to simplify the calculation, in this frame in the action  is symmetric under the transformation $\mu\to-\mu$ so  will be used the condition $\mu\geq0 $ and $\mu^2\geq0$. Will be explored the range $\mu\in[0,3]$, this interval was chosen in such a way to show a particular behavior of the system for fixed $\Omega$. The parameter $\Omega$  appears only with its square and is defined as a real parameter, therefore for the $\Omega$ too we require $\Omega\geq0$. Beside, if we refer to the scalar model, is possible to prove \cite{phi4-non} using the Langmann-Szabo duality \cite{L-S-dual}, that the model can be fully described varying $\Omega$ in the range $[0,1]$, for higher $\Omega$ the system can be remap inside the previous interval. In the present model the L-S duality does not hold any more, but forced by limited resource, we conjecture that the interval $\Omega\in[0,1]$ still enough to describe the system. The last parameter to consider is  $\alpha$,  it is connected to the choice of the vacuum state, of course, the range of $\alpha$ is $[0,2\pi]$. The study of the system varying this parameter is quite important; from a theoretical point of view, because is related to the vacuum invariance and for numerical reasons too. In fact, in the action of both 4-dimensional and 2-dimensional model appear some contributions  $\sim(\sin\alpha)/\Omega$ which seem to diverge for $\Omega=0$, numerically we have verified that is an eliminable divergence and the curves of the observables can be extended in zero by continuity. Nevertheless, this factor slow down the computations due to its high value close to $\Omega=0$. Studying the dependence on $\alpha$ we can conclude that in the limit $N\to\infty$ the observables are independent from $\alpha$, therefore for our purposes $\alpha$ will be fixed equal to zero avoiding the annoying terms. In general for each observable  are computed the graphs for $N=$5, 10, 15, 20 matrix size.

\subsection{Varying $\alpha$ }
We start  looking at the variation of energy density and the full power of the fields density for fixed $\mu$ and $\Omega$, varying $\alpha \in [0,2\pi]$. As representative here will be presented the graphs for $\mu=1$, $\Omega=\{1,0.5\}$ but we obtain the same behaviors for any other choice of the parameters allowed in the range considered.
\begin{figure}[htb]
\begin{center}
\includegraphics[scale=0.45]{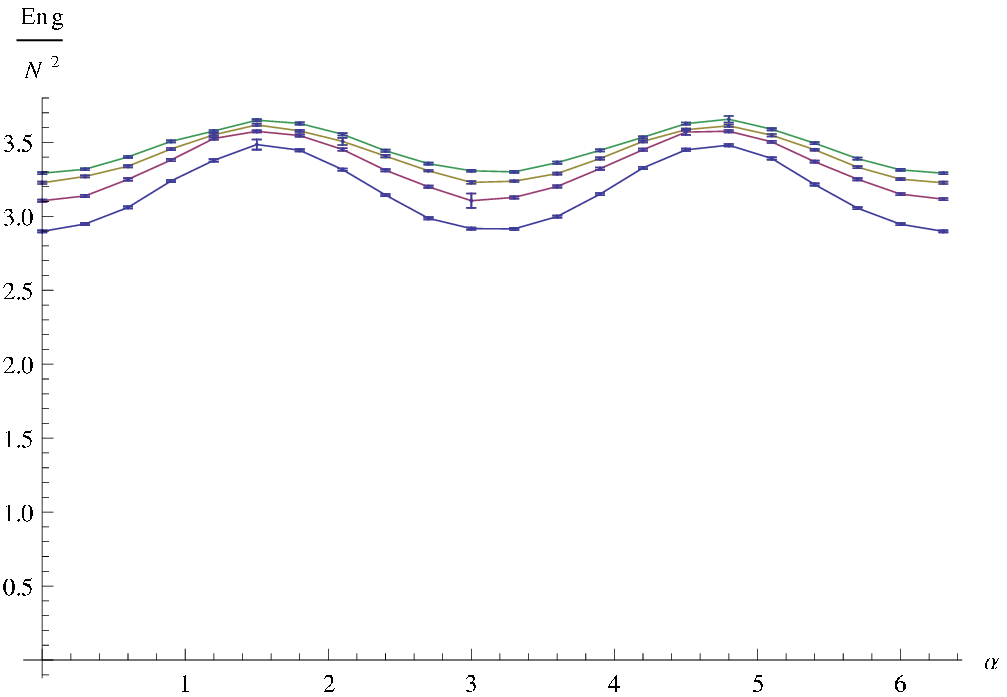}
\includegraphics[scale=0.45]{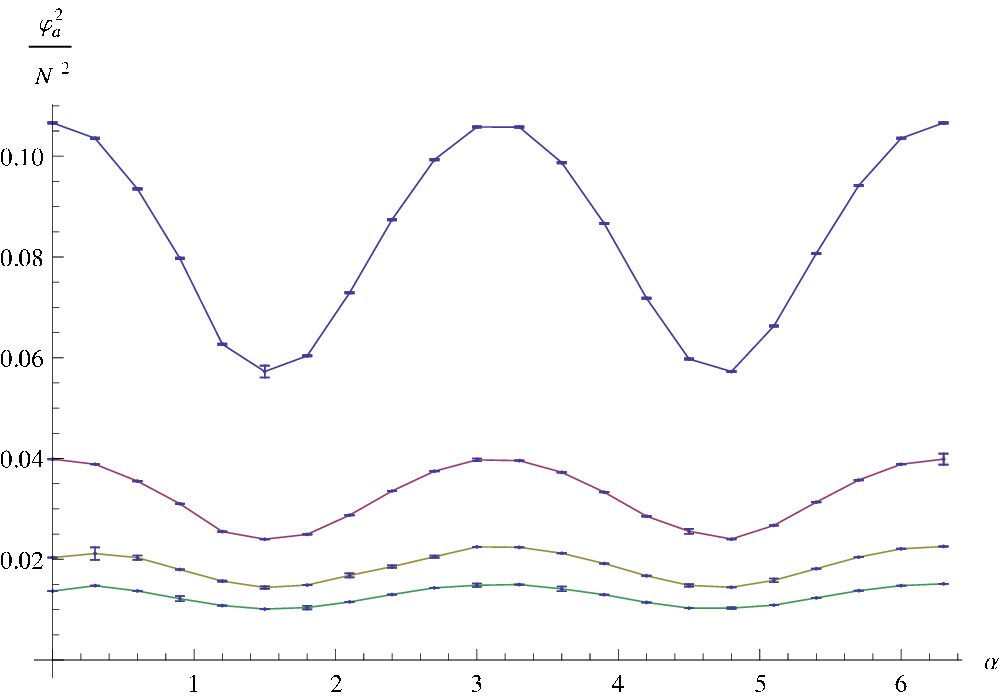}
\includegraphics[scale=0.45]{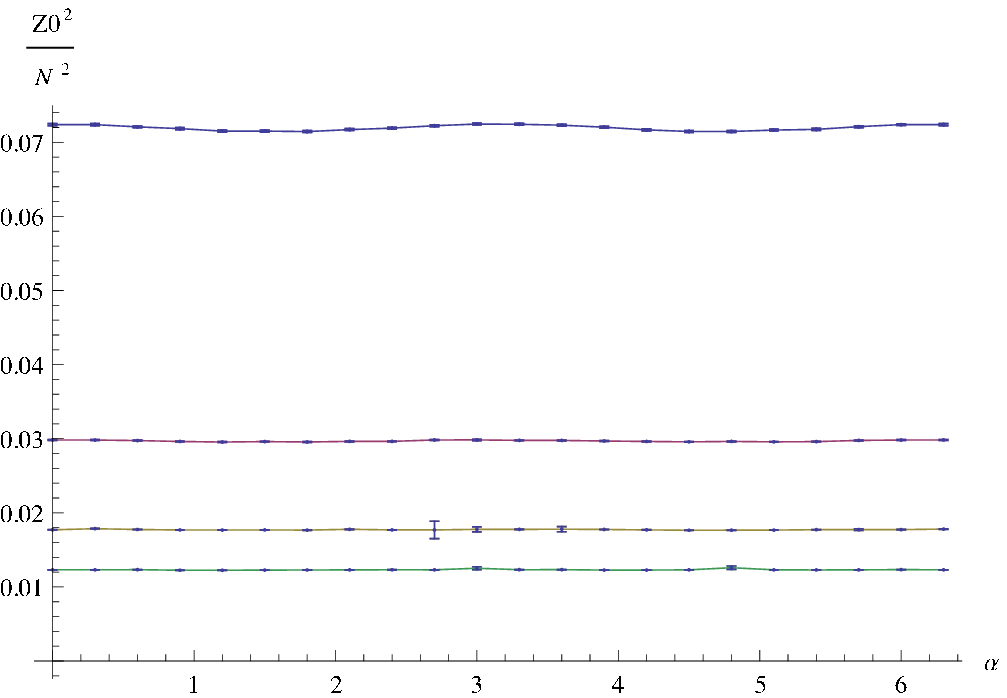}
\end{center}
\caption{\footnotesize Total energy density and full power of the fields density for $\varphi^2_a$, $Z_{0a}^2$ (from the left to the right) fixing $\mu=1$, $\Omega=1$, varying $\alpha$ and $N$. $N=5$ (blue), $N=10$ (purple), $N=15$ (brown), $N=20$ (green).  \normalsize}\label{Figure 1}\end{figure}
All tree graphs show an oscillating behavior of the values, this oscillation is present in all other quantities measured. The amplitude of this oscillation becomes smaller and smaller increasing the size of the matrix and this is true for all the quantity measured. The same  trend  is described in fig.\ref{Figure 2} in which  position of the maximum are different but the amplitudes become smaller increasing $N$ 

\begin{figure}[htb]
\begin{center}
\includegraphics[scale=0.45]{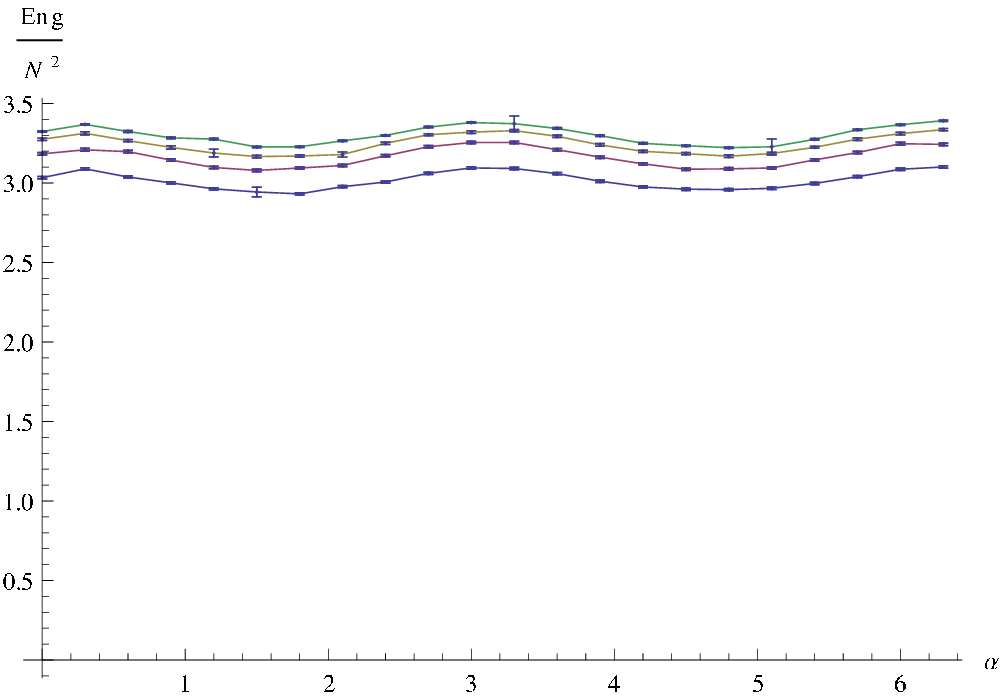}
\includegraphics[scale=0.45]{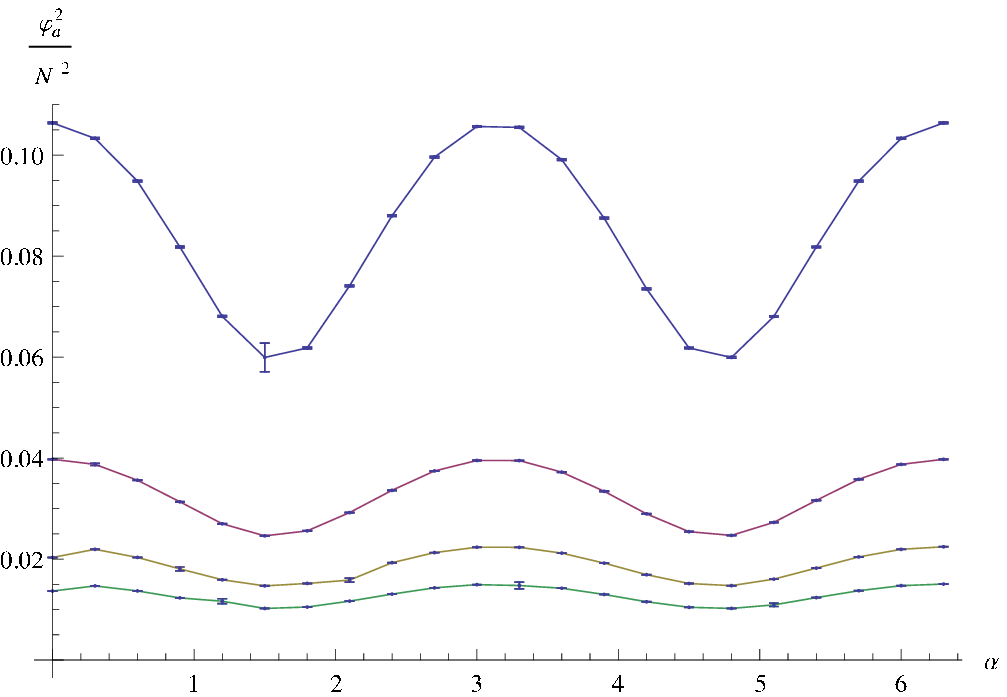}
\includegraphics[scale=0.45]{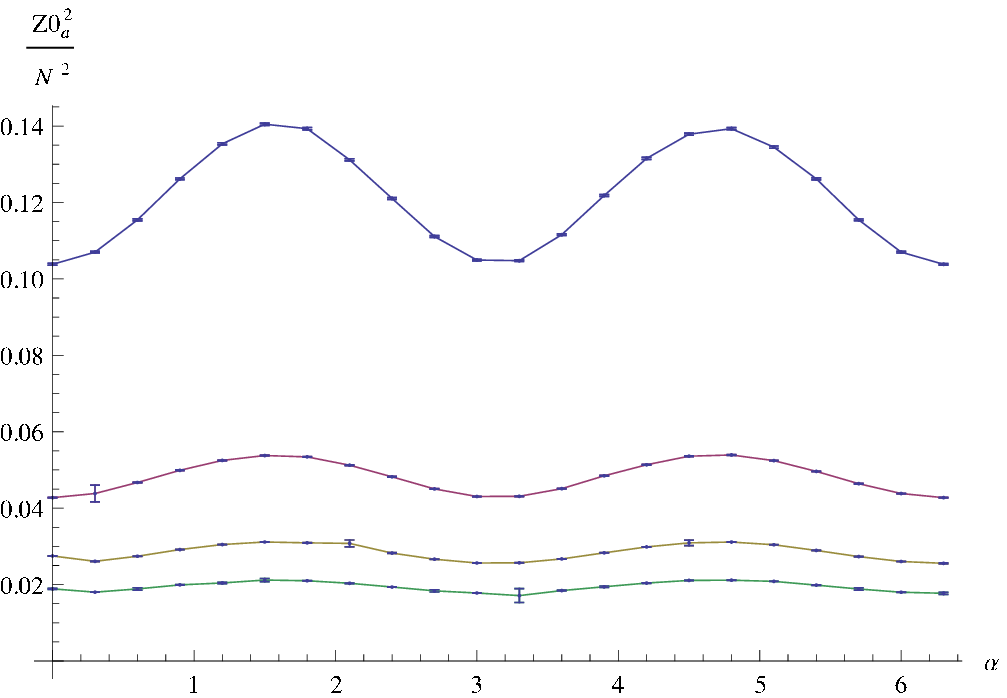}
\end{center}
\caption{\footnotesize Total energy density and full power of the fields density for $\varphi^2_a$, $Z_{0a}^2$  (from the left to the right) fixing $\mu=1$, $\Omega=0.5$, varying $\alpha$ and $N$. \normalsize}\label{Figure 2}\end{figure}
This results allow us to consider $\alpha=0$ for all next graphs, since we are interested in the behavior of the system for $N \to \infty $. This occurrence simplify all the next simulations thanks to the vanishing the of terms  $\sim(\sin\alpha)/\Omega $  appearing in the discretized action. Beside, such results induce us to reckon the parameter $\alpha$ as connected to the remaining invariance of the vacuum state for the exact model. 

\subsection{Varying $\Omega$ }
Now we will analyze three cases in which $\mu$ is fixed to $0,1,3$, $\alpha$ is zero and we vary $\Omega\in[0,1]$.  Will be showed the results of simulation
for matrix size of 5, 10, 15, 20. In the rest of this chapter we ignore for the computation of
$E,D,V,F$ the prefactor $(1+\Omega^2)$. In this way we focus our
attention to the integral as the source of possible phase transitions. 
\begin{figure}[htb]
\begin{center}
\includegraphics[scale=0.4]{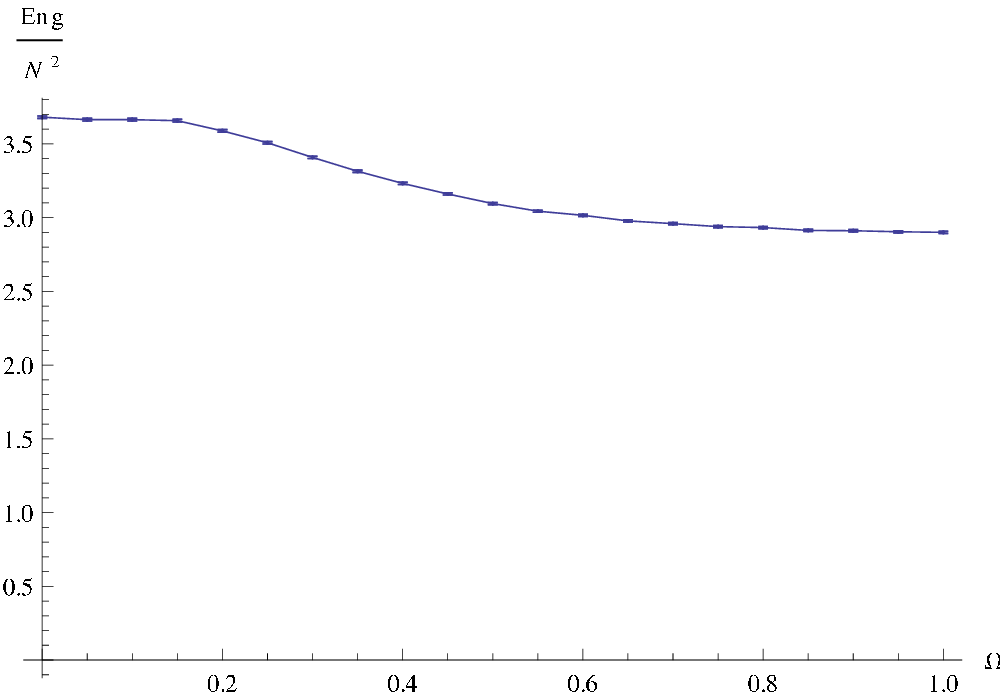}
\includegraphics[scale=0.4]{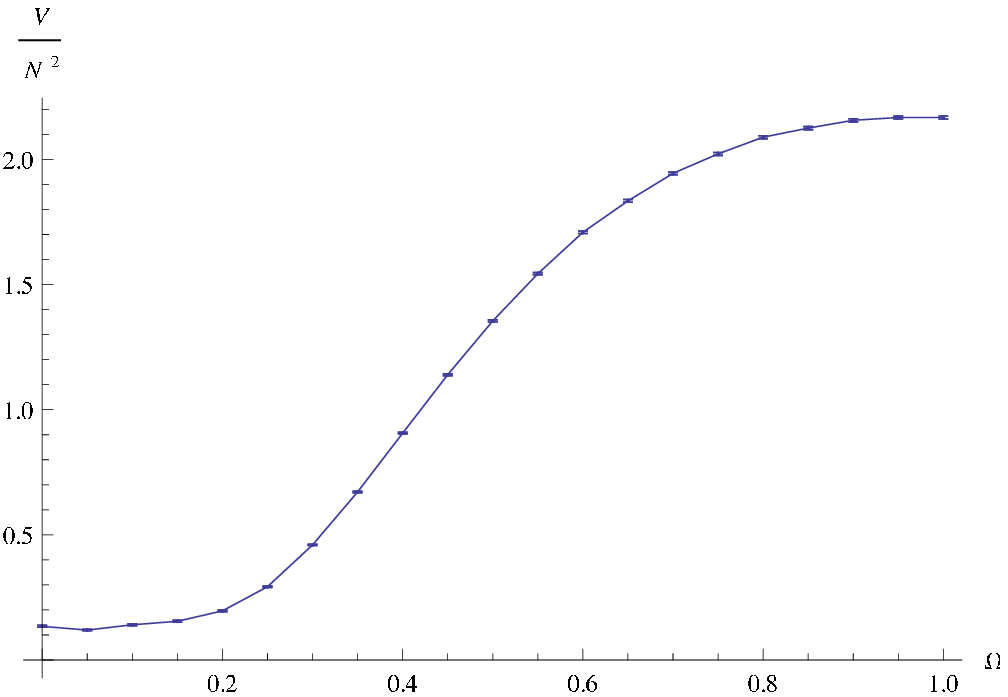}
\includegraphics[scale=0.4]{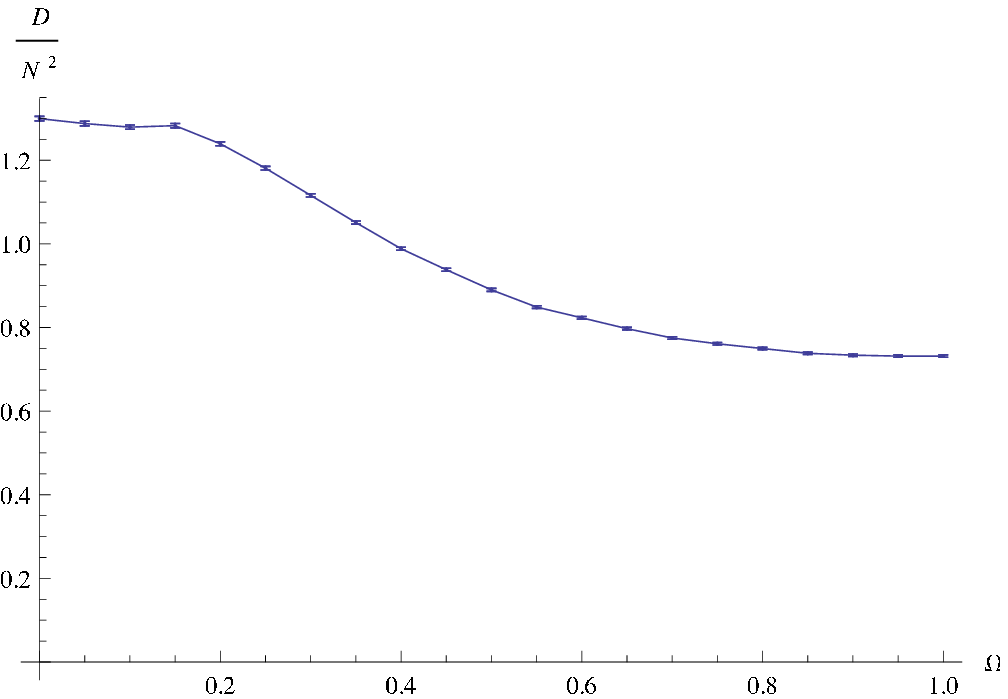}
\includegraphics[scale=0.4]{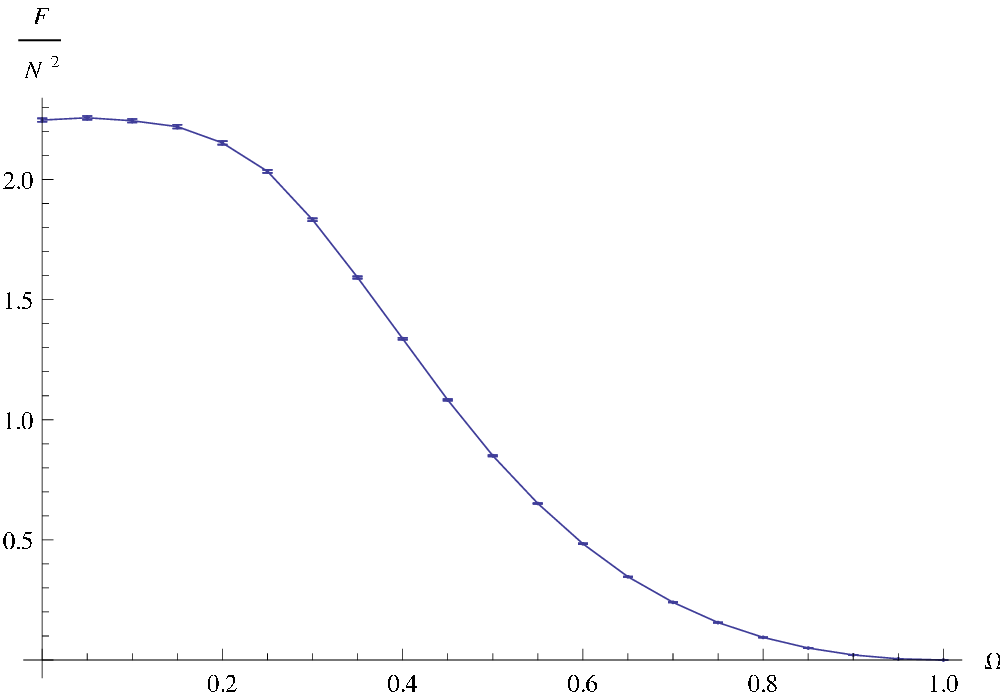}
\end{center}
\caption{\footnotesize Total energy density and the various contributions for $\mu=1$. From the left to the right the density for: E, V, D, F. \normalsize}\label{Figure 3}\end{figure}
The graphs in fig.\ref{Figure 3} show the total energy density $\langle S \rangle / N^2$ and the various contributions: the potential $V / N^2$, the Yang-Mills part $F / N^2$ and the covariant derivative part $D / N^2$, for $\mu=1$  $\alpha=0$ defined in the previous chapter. There is no evident discontinuity or peak, apart a small deviation visible in the origin, but we can interpret it  as a finite volume effect. 
Increasing the size of the matrices the curves and the apparent discontinuity close to zero become smoother fig.\ref{Figure 4}.   \newpage
\begin{figure}[htb]
\begin{center}
\includegraphics[scale=0.4]{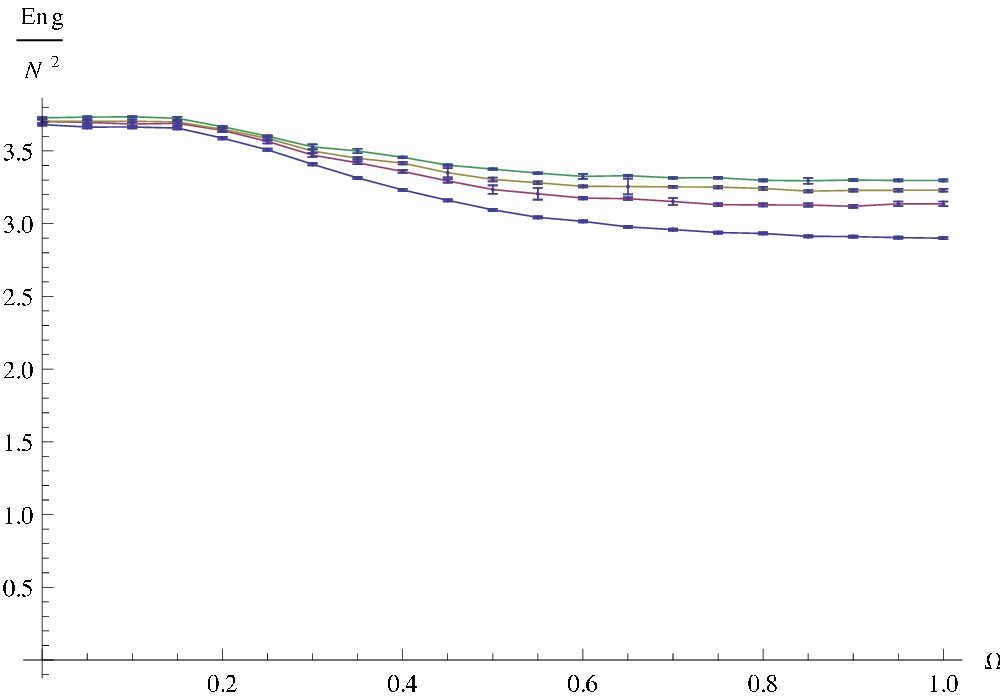}
\includegraphics[scale=0.4]{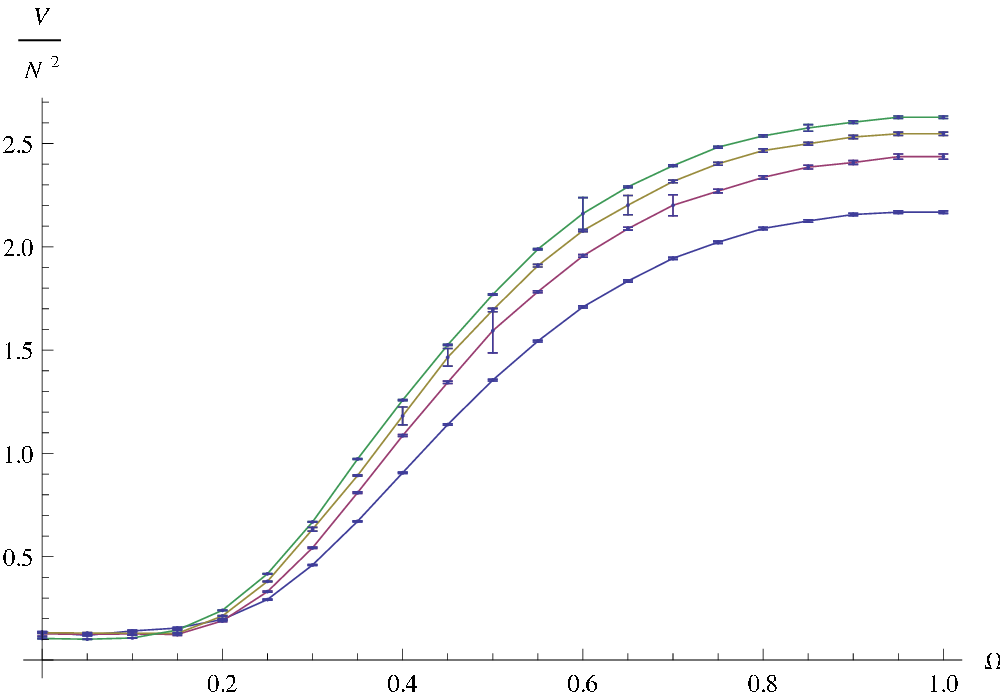}
\includegraphics[scale=0.4]{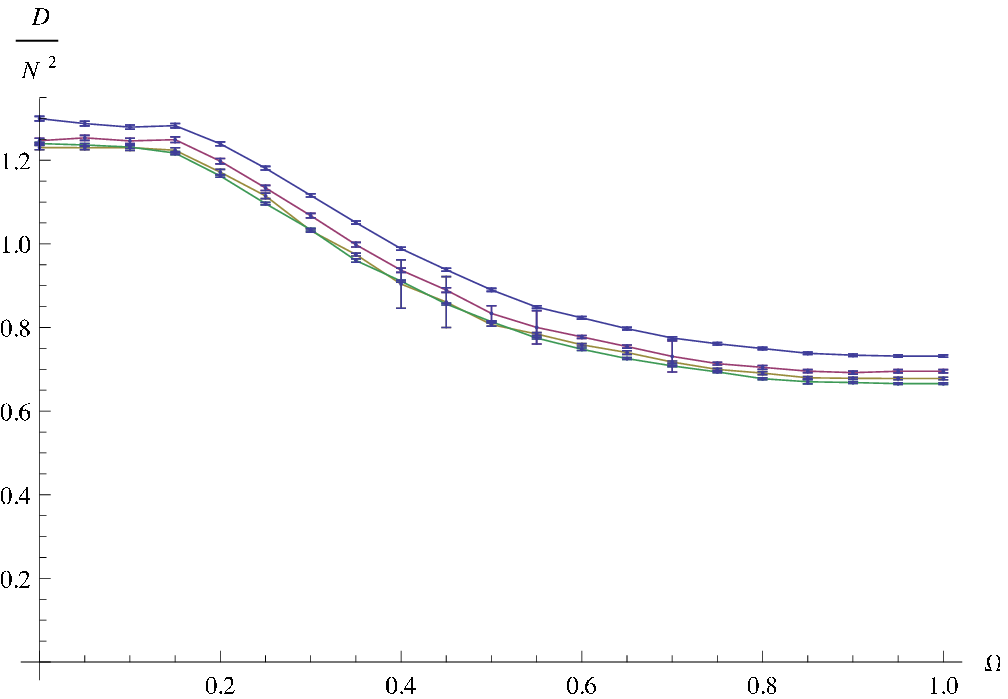}
\includegraphics[scale=0.4]{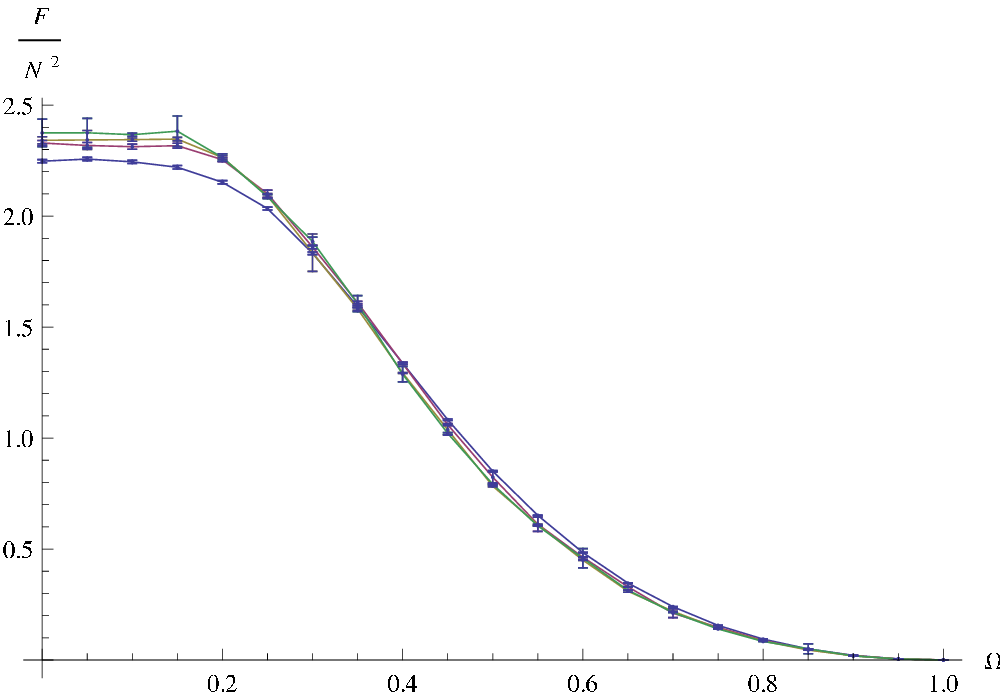}
\end{center}
\caption{\footnotesize Total energy density and the various contributions  for $\mu=1$, $\alpha=0$ varying $\Omega$ and $N$. From the left to the right  $E$, $V$, $D$, $F$ with $N=5$ (blue), $N=10$ (purple), $N=15$ (brown), $N=20$ (green).\normalsize}\label{Figure 4}\end{figure}
Comparing the energy density and the various contributions fig.\ref{Figure 5} we notice that the contributions between $F$ and $V$ balance each other and the total energy follows the slope of $D$, this  behavior continues increasing the size of the matrices fig.\ref{Figure 5}. \\
\begin{figure}[htb]
\begin{center}
\includegraphics[scale=0.55]{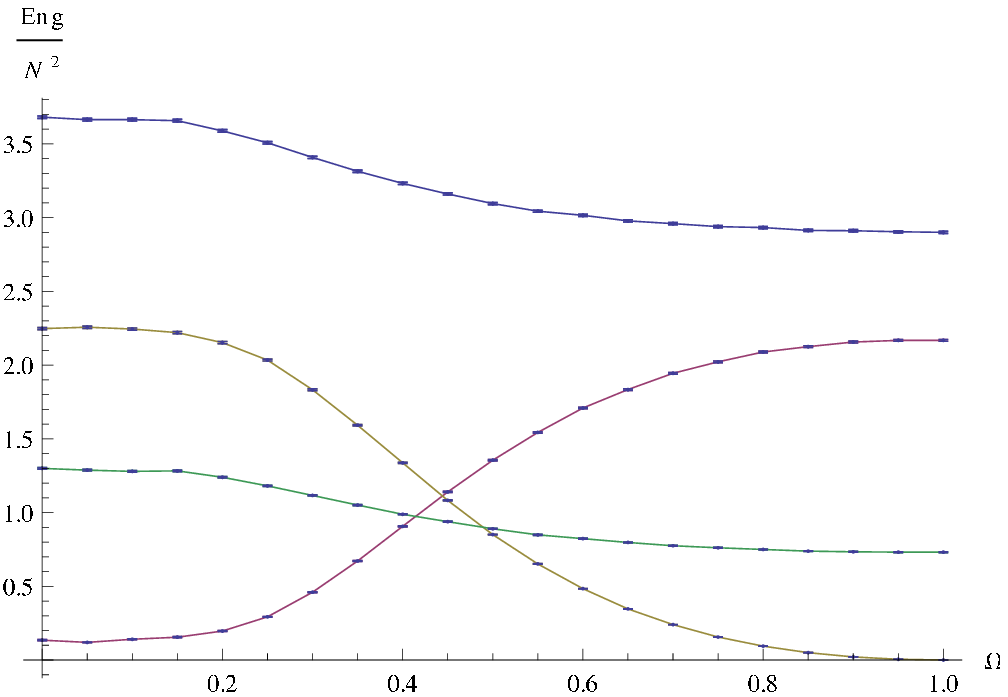}
\includegraphics[scale=0.55]{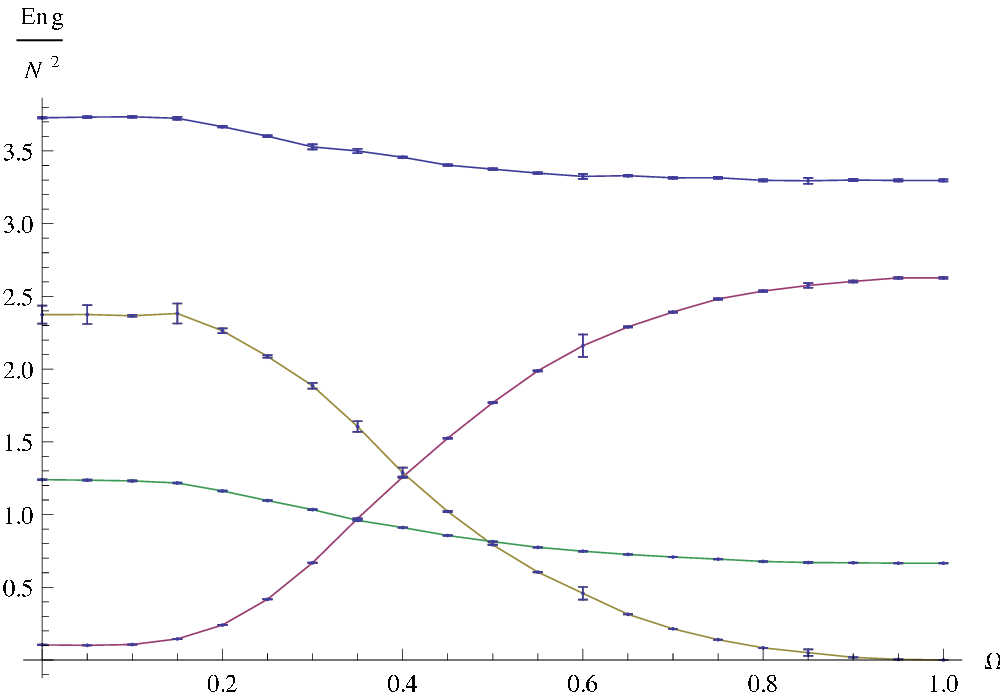}
\end{center}
\caption{\footnotesize Comparison of the total energy density and the various contributions for $\mu=1$, $\alpha=0$. $E$ (blue), $F$ (brown), $D$ (green), $V$ (purple). With $N=5$ (left) and $N=20$ (right).  \normalsize}\label{Figure 5}
\end{figure}\newpage
\begin{figure}[htb]
\begin{center}
\includegraphics[scale=0.8]{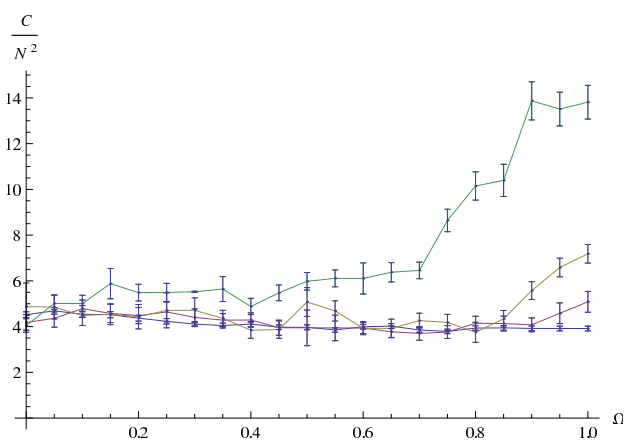}
\end{center}
\caption{\footnotesize Specific heat for $\mu=1$.\normalsize}\label{Figure 6}
\end{figure}
The specific heat density defined as $(\langle S \rangle^2 -\langle S^2 \rangle)/N^2$ shows fig.\ref{Figure 6} a peak in $\Omega=1$, this peak increases as $N$ increase. This behavior is typical of a phase transition and the peak is not clear for small $N$ due to the finite volume effect. 
In order to gain some informations on the composition  of the fields we look at the order parameters defined in the previous chapter. Starting from $\psi$ field, in the figure \ref{Figure 6} it is showed the graphs for $\varphi_a^2 $, $\varphi^2_0 $  and $\varphi^2_1 $ for $N=5$. \\
 \begin{figure}[htb]
\begin{center}
\includegraphics[scale=0.45]{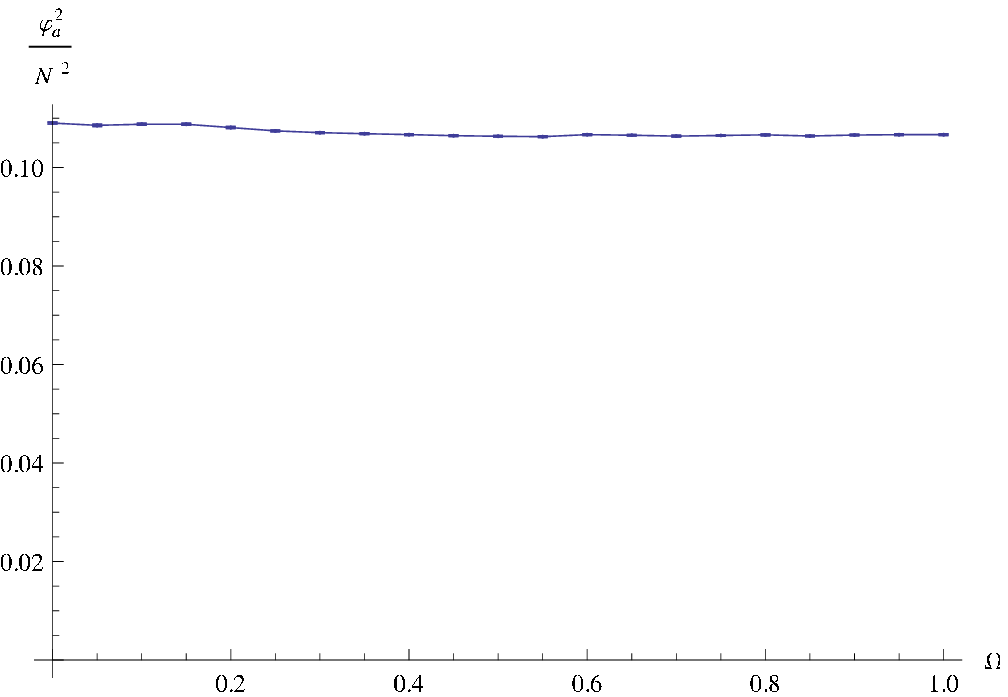}
\includegraphics[scale=0.45]{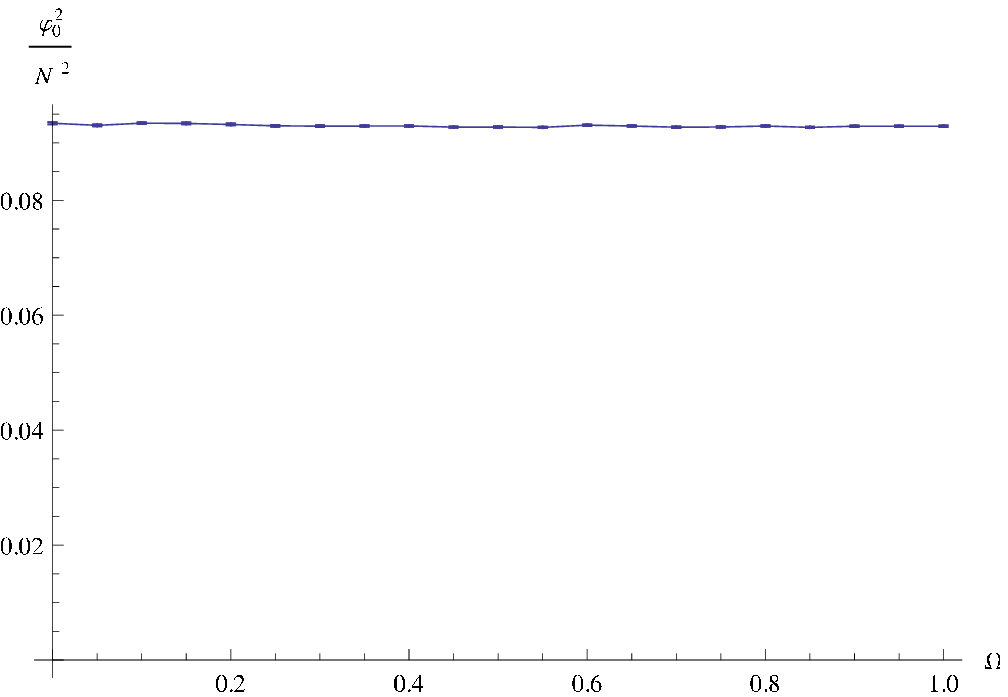}
\includegraphics[scale=0.45]{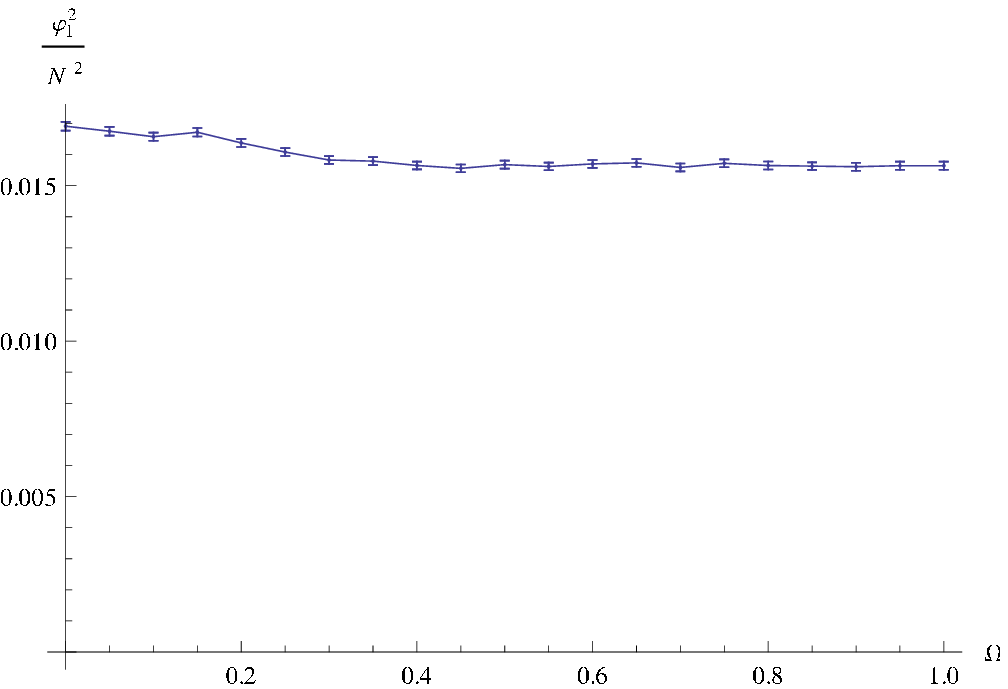}
\end{center}
\caption{\footnotesize From the left to the right $\varphi_a^2 $, $\varphi^2_0 $  and $\varphi^2_1 $ for $\mu=1$ varying $\Omega$ \normalsize }\label{Figure 7}\end{figure}
The three values  $\varphi_a^2 $, $\varphi^2_0 $  and $\varphi^2_1 $ seem essentially constant, comparing the three graphs fig.\ref{Figure 7} it is easy to see the dominance of the spherical contribution $\varphi^2_0 $ to the full power of the field.   \\
\begin{figure}[htb]
\begin{center}
\includegraphics[scale=0.55]{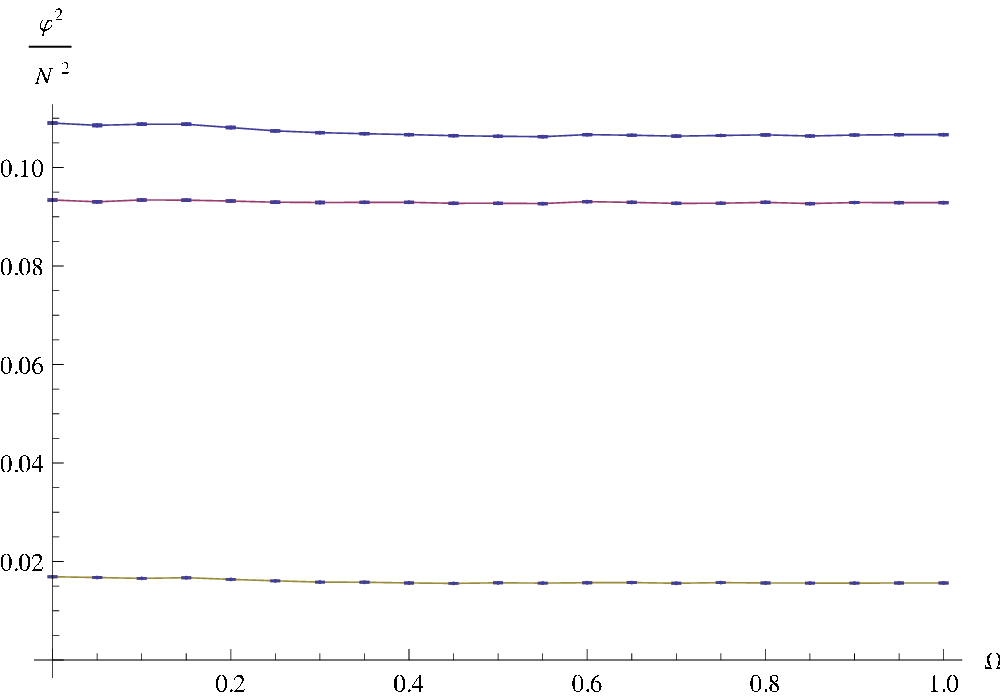}
\includegraphics[scale=0.55]{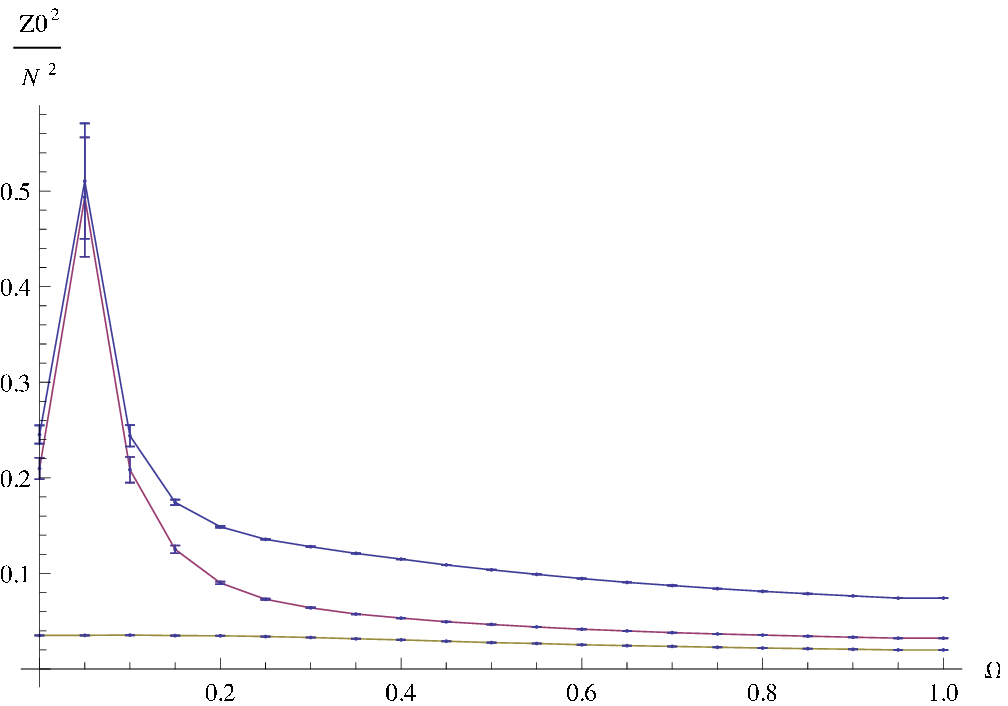}
\end{center}
\caption{\footnotesize On the left comparison of  $\varphi_a^2 $ (blue), $\varphi^2_0 $ (purple)  and $\varphi^2_1 $ (green) density. On the right comparison of $Z_{0a}^2$ (blue), $Z_{00}^2$ (purple) and $Z_{01}^2$ (green) density.\normalsize}\label{Figure 8}
\end{figure}\newpage
The behavior of the $Z_0$ fields is different, referring to figure \ref{Figure 8} where are showed the quantity $Z_{0a}^2 $, $Z_{00}^2$  and $Z_{01}^2 $ over $N^2$, the spherical contribution becomes dominant approaching to $\Omega=0$ starting from a zone in which the contribution of $Z_{00}^2$ and $Z_{01}^2 $ are comparable. 
For brevity  will not be showed only the graphs for $Z_{0a}^2 $, $Z_{00}^2$  and $Z_{01}^2 $ but taking in account the statistical errors the other $Z_i$ related graphs appear compatible to the $Z_0$ case. The dependence of the previous quantities on $N$ are showed in the following graphs fig.\ref{Figure 9}. 
\begin{figure}[htb]
\begin{center}
\includegraphics[scale=0.4]{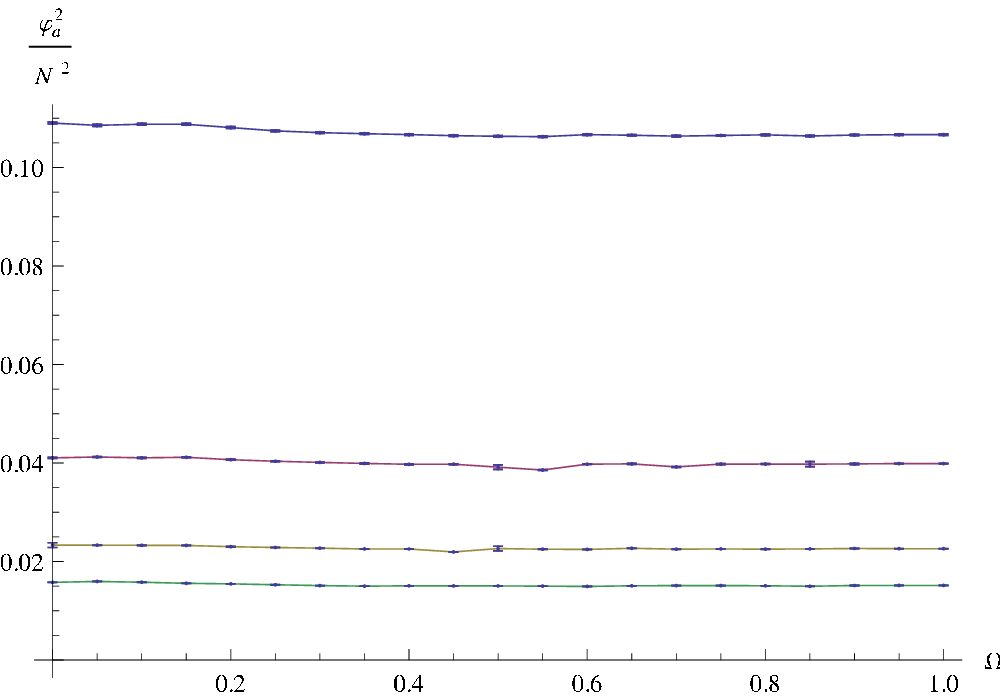}
\includegraphics[scale=0.4]{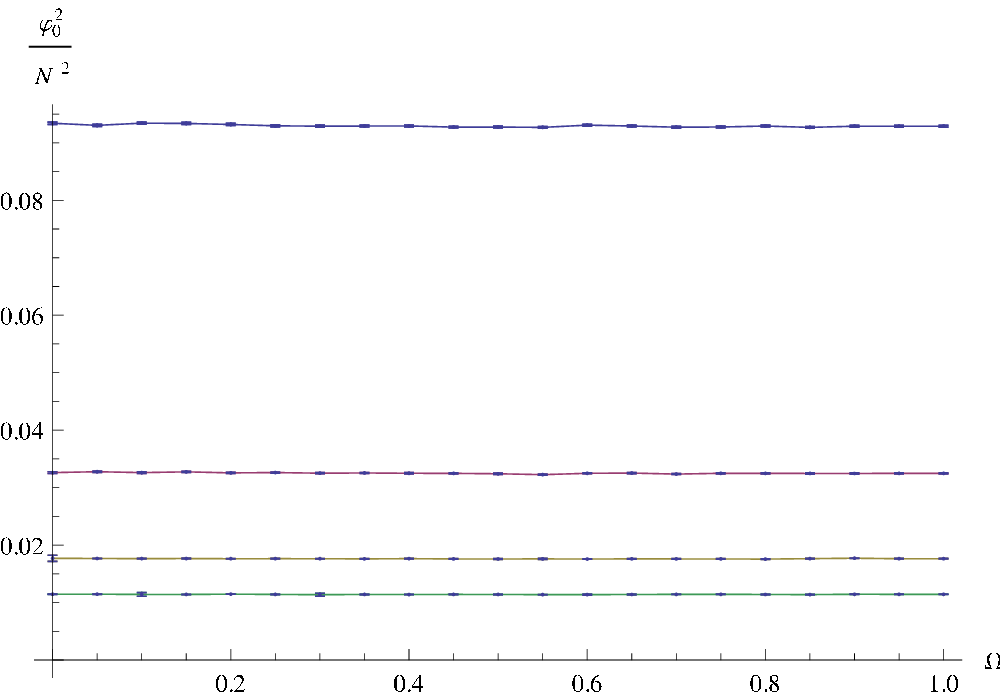}
\includegraphics[scale=0.4]{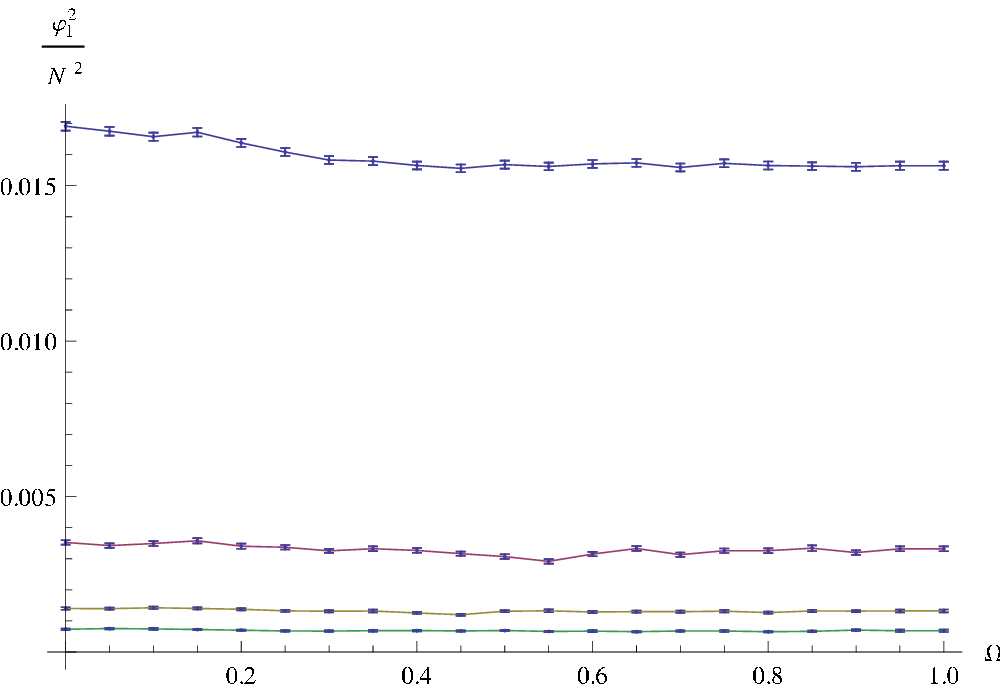}
\includegraphics[scale=0.4]{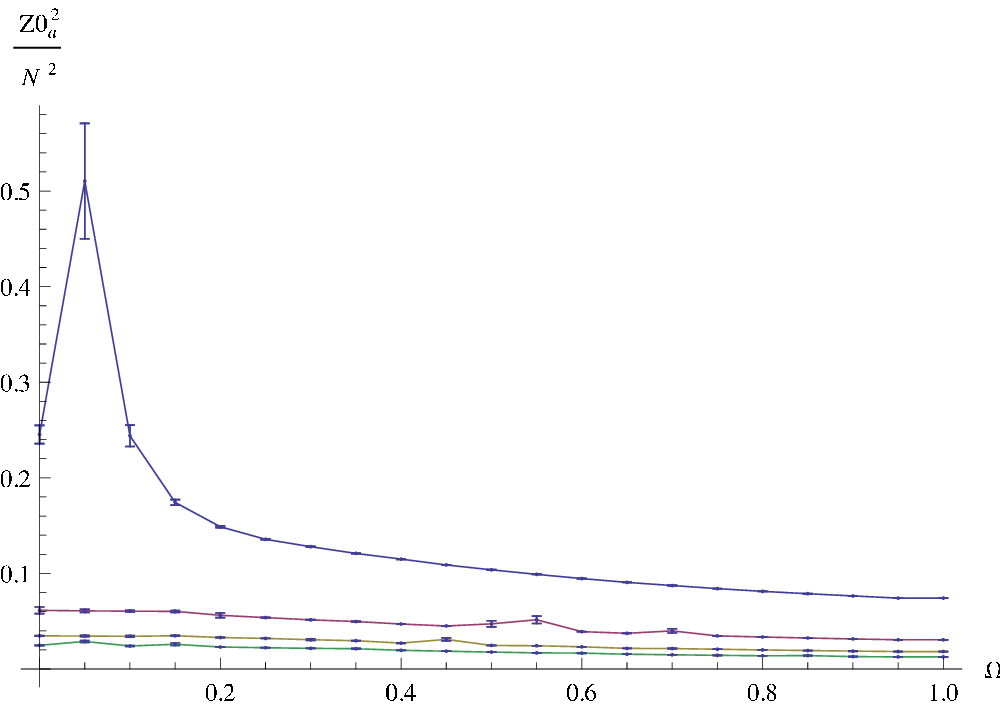}
\includegraphics[scale=0.4]{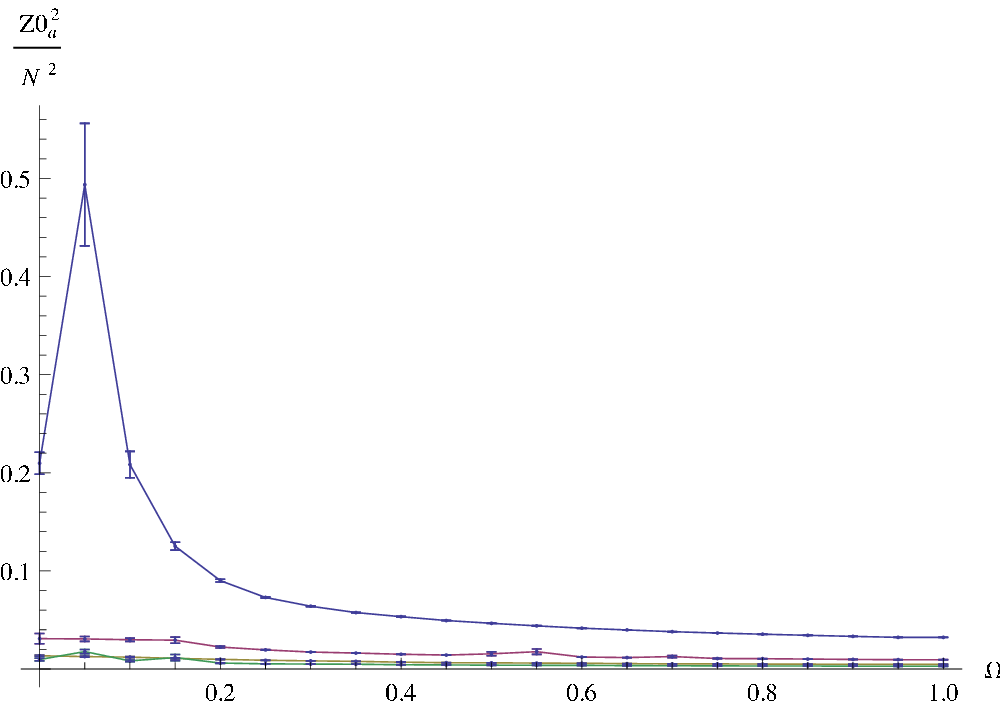}
\includegraphics[scale=0.4]{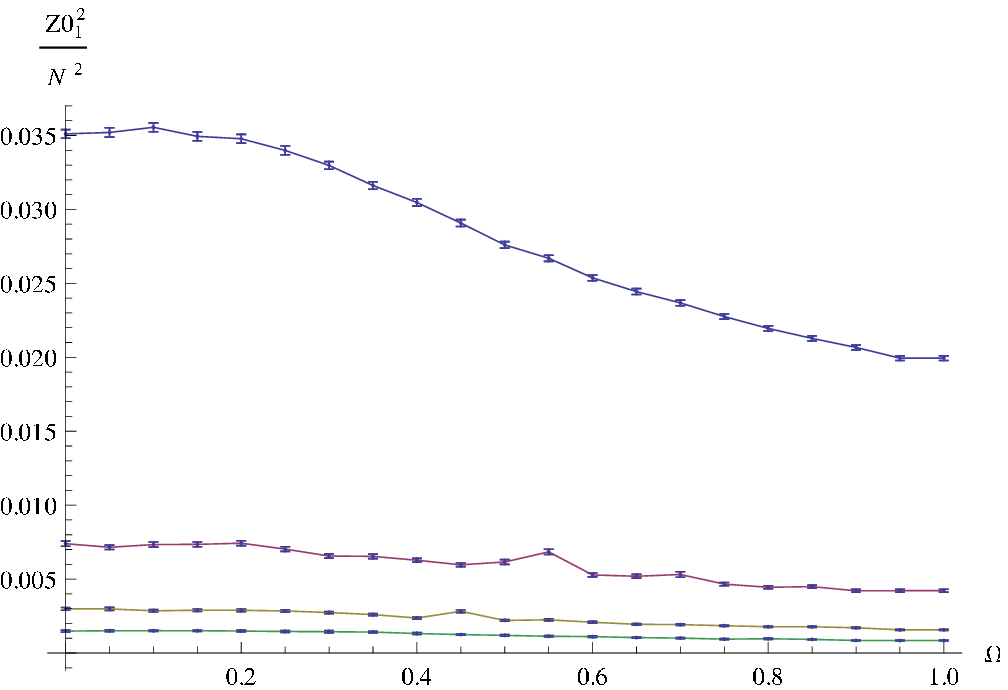}
\end{center}
\caption{\footnotesize Starting from the up left corner and from the left to the right the 	densities for $\varphi_a^2 $, $\varphi^2_0 $, $\varphi^2_1 $, $Z_{0a}^2$, $Z_{00}^2 $ and $Z_{01}^2$ for $\mu=1$ varying $\Omega$ and $N$. \normalsize}\label{Figure 9}
\end{figure}The values of the quantity of all the previous parameters decreases with $N$  but the dominance of the $\varphi_0$ on the total power of the field is independent by $N$. The peak  related  to $Z_0$ decrease, but if look at the single  graph for $N=20$ the spherical contribution approaching the point $\Omega=0$ features a peak.

\begin{figure}[htb]
\begin{center}
\includegraphics[scale=0.4]{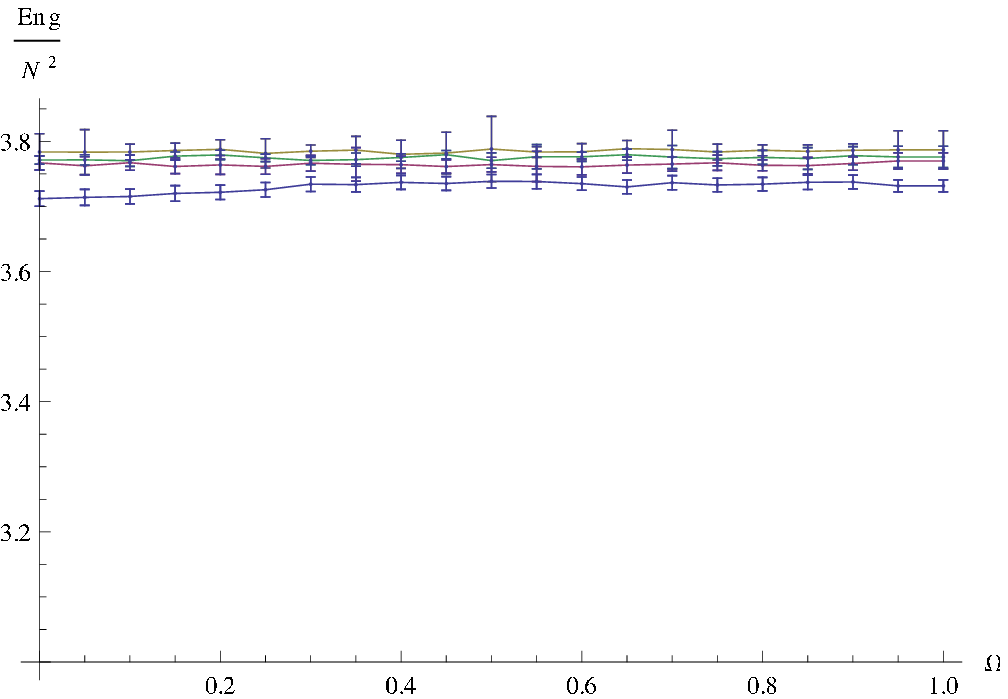}
\includegraphics[scale=0.4]{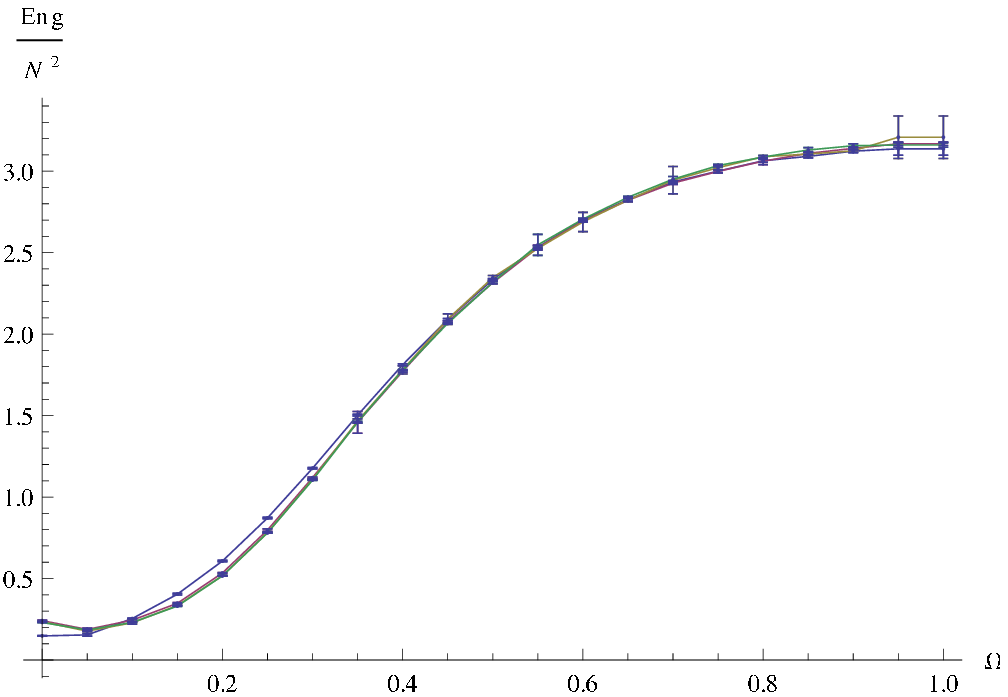}
\includegraphics[scale=0.4]{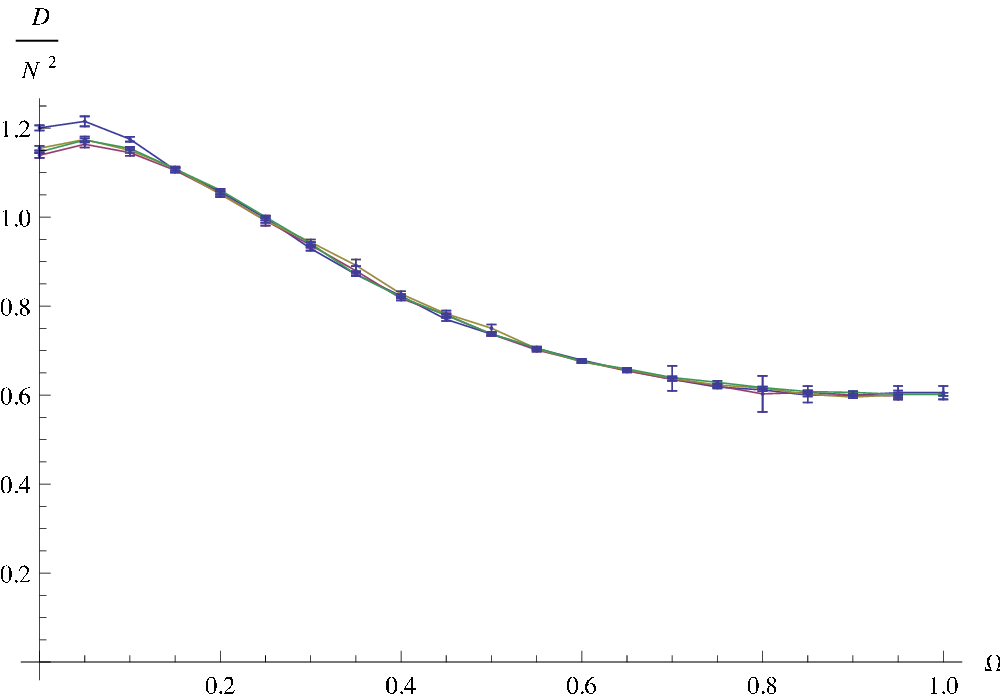}
\includegraphics[scale=0.4]{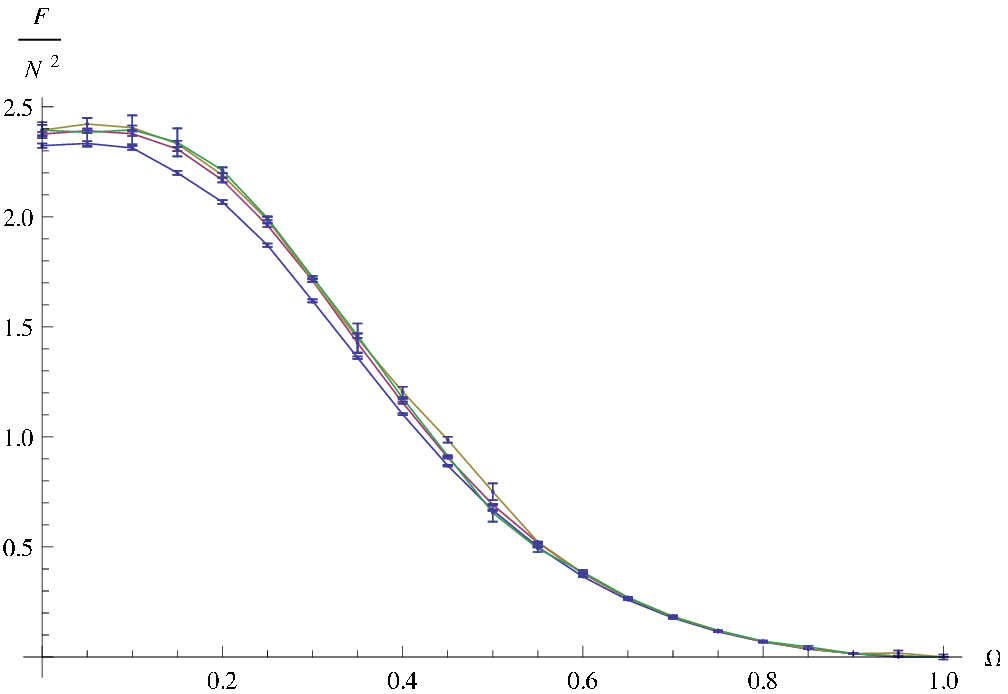}
\includegraphics[scale=0.45]{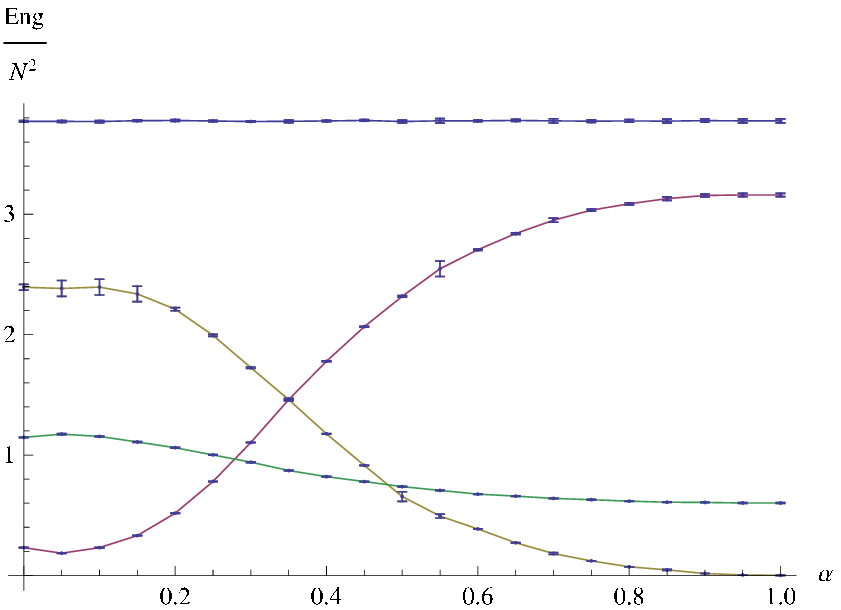}
\end{center}
\caption{\footnotesize Total energy density, various contributions and the comparison among them for $\mu=0$ varying $\Omega$ and $N$. From the left to the right  $E$, $V$, $D$, $F$ and comparison.\normalsize}\label{Figure 10}\end{figure}

Now we will analyze the  model for $\mu=0$; fig.\ref{Figure 10} shows the  graphs for total energy density and the contributions $V$, $D$, $F$. The slope of the total energy density seems to be constant. The $D$ contribution and the $F$ do not balance each other like in the previous case, but all the three contributions balance among them self to produce a constant sum. 
\begin{figure}[htb]
\begin{center}
\includegraphics[scale=0.6]{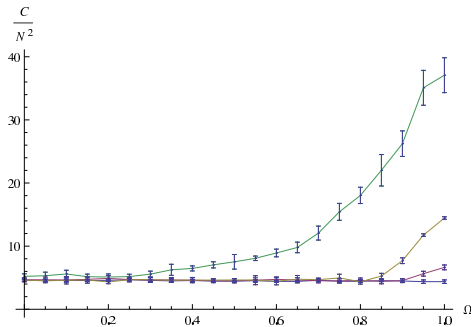}
\end{center}
\caption{\footnotesize Specific heat density for $\mu=0$ varying $\Omega$ and $N$.\normalsize}\label{Figure 11}
\end{figure}
The specific heat density shows fig.\ref{Figure 11} again the peak in $\Omega=1$ as $N$ increase.
For the other 	quantities  $\varphi_a^2 $, $\varphi^2_0 $, $\varphi^2_1 $ and  $Z_{0a}^2 $, $Z_{00}^2$,  $Z_{01}^2$ we have the same  behavior fig.\ref{Figure 14} of $\mu=1$ case, except for the oscillation appearing in the $Z_{0a}^2 $, $Z_{00}^2$ graphs close to zero, anyway it appears only for $N=5$.\newpage
\begin{figure}[htb]
\begin{center}
\includegraphics[scale=0.35]{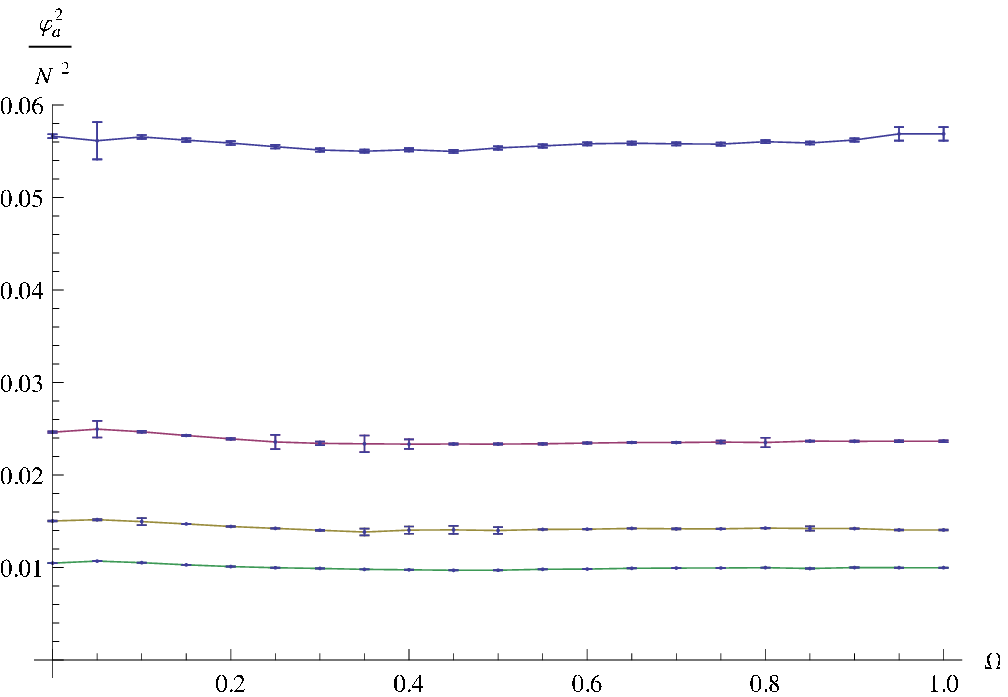}
\includegraphics[scale=0.35]{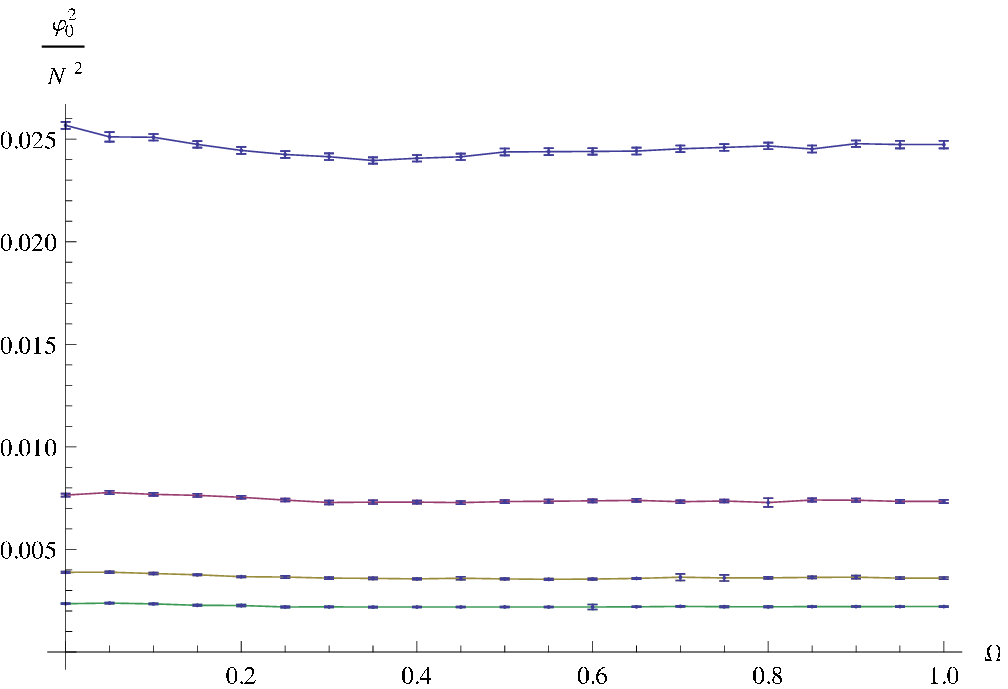}
\includegraphics[scale=0.35]{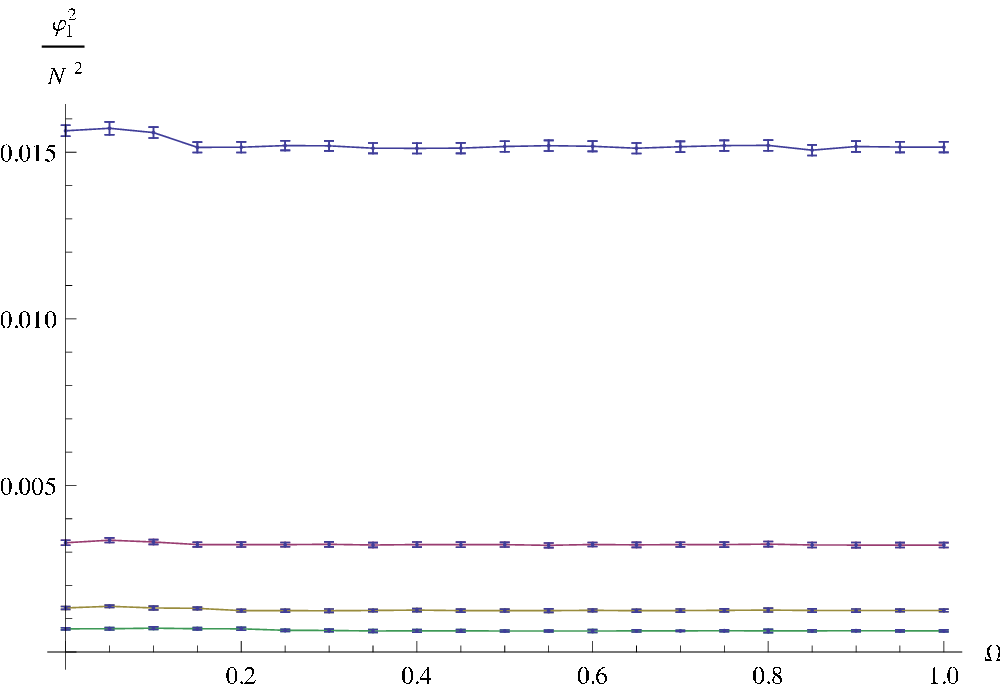}
\includegraphics[scale=0.35]{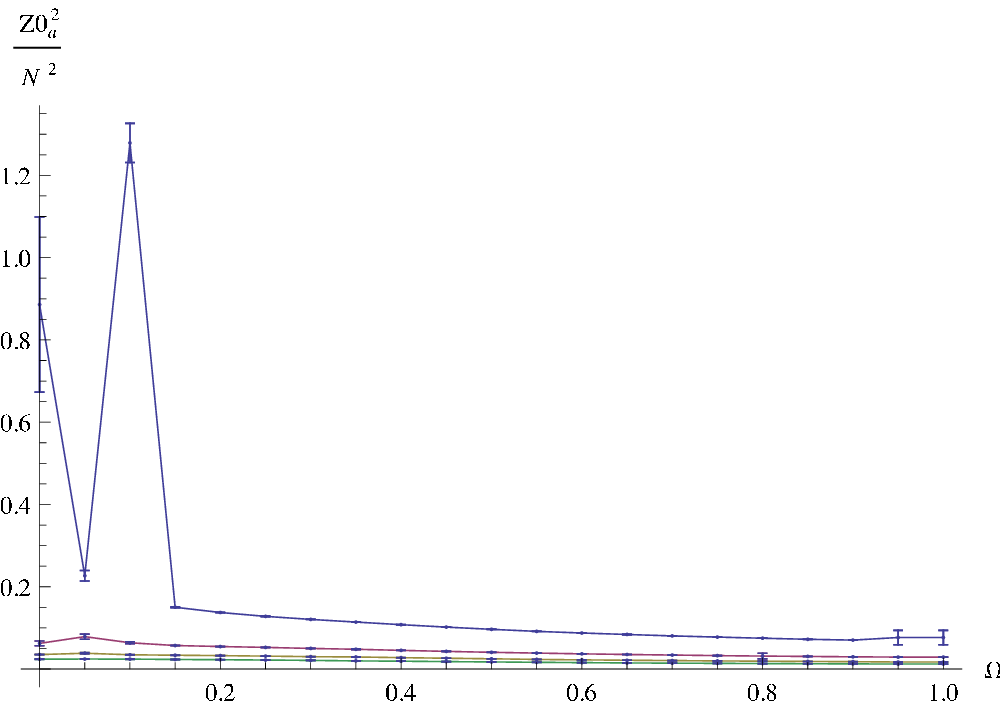}
\includegraphics[scale=0.35]{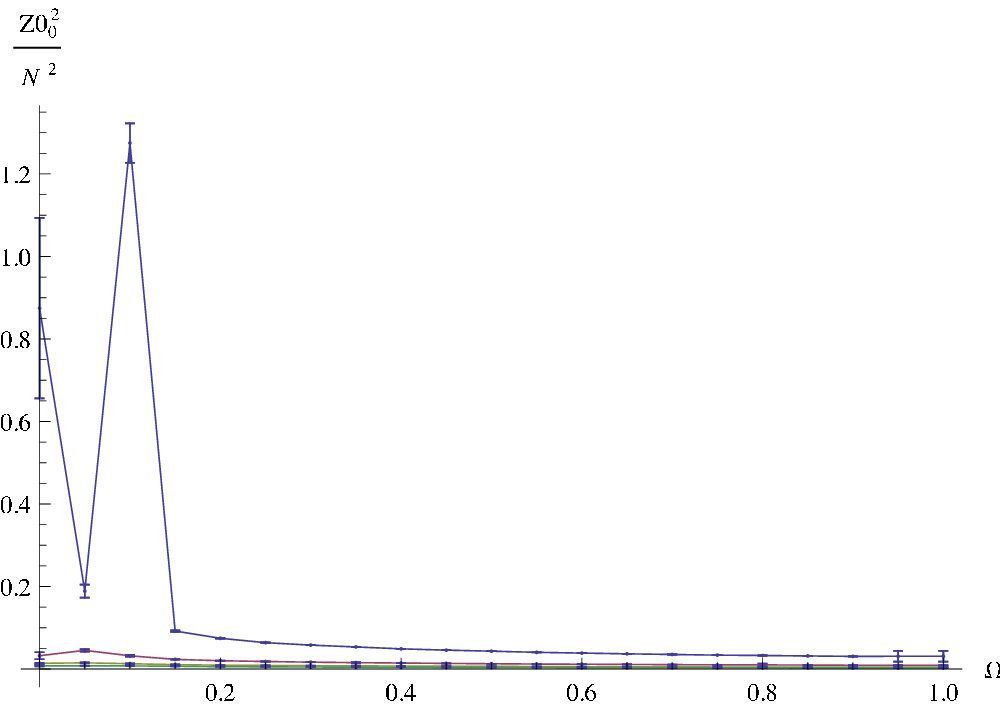}
\includegraphics[scale=0.35]{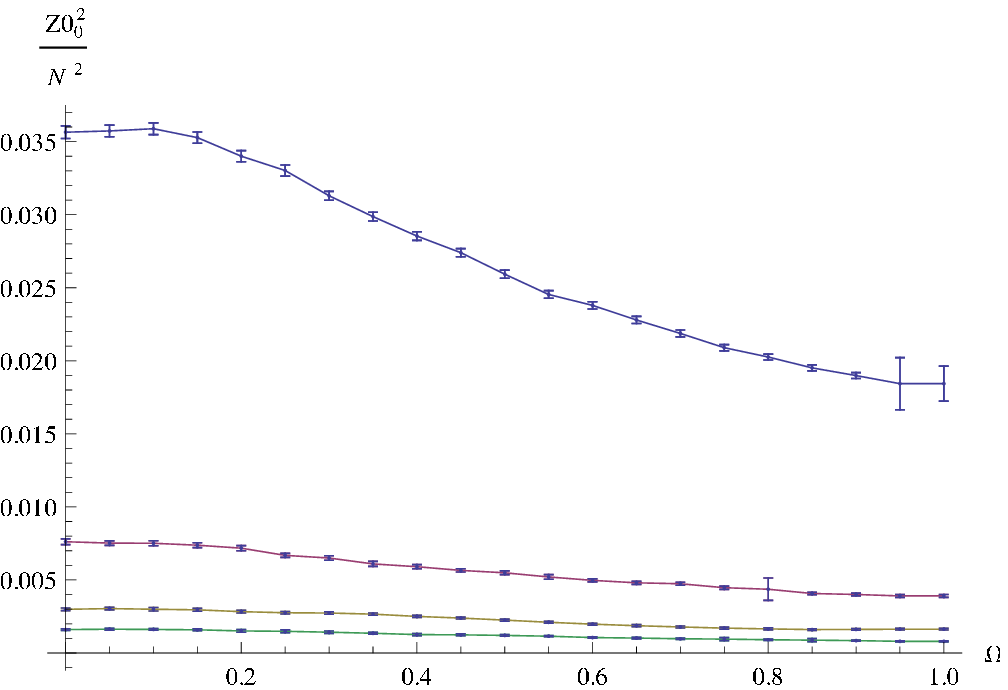}
\end{center}
\caption{\footnotesize Starting from the up left corner and from the left to the right the 	densities for $\varphi_a^2 $, $\varphi^2_0 $, $\varphi^2_1 $, $Z_{0a}^2$, $Z_{00}^2 $ and $Z_{01}^2$ for $\mu=0$ varying $\Omega$ and $N$. }\label{Figure 12}
\end{figure}
A complete different response of the system is described in the graphs for $\mu=3$, as we can see from fig.\ref{Figure 13}. The slope of total energy density  is very similar to the $F$ component instead $D$. Beside, appears a maximum around  $\Omega=0.4$ for $N\to\infty$. This dramatic change of the graphs might be interpreted as consequence of a phase transition ruled  by the parameter $\mu$, actually in the next section we will find a peak in the specific heat density for some fixed $\Omega$ and varying $\mu\in[0,3]$. 
\begin{figure}[htb]
\begin{center}
\includegraphics[scale=0.45]{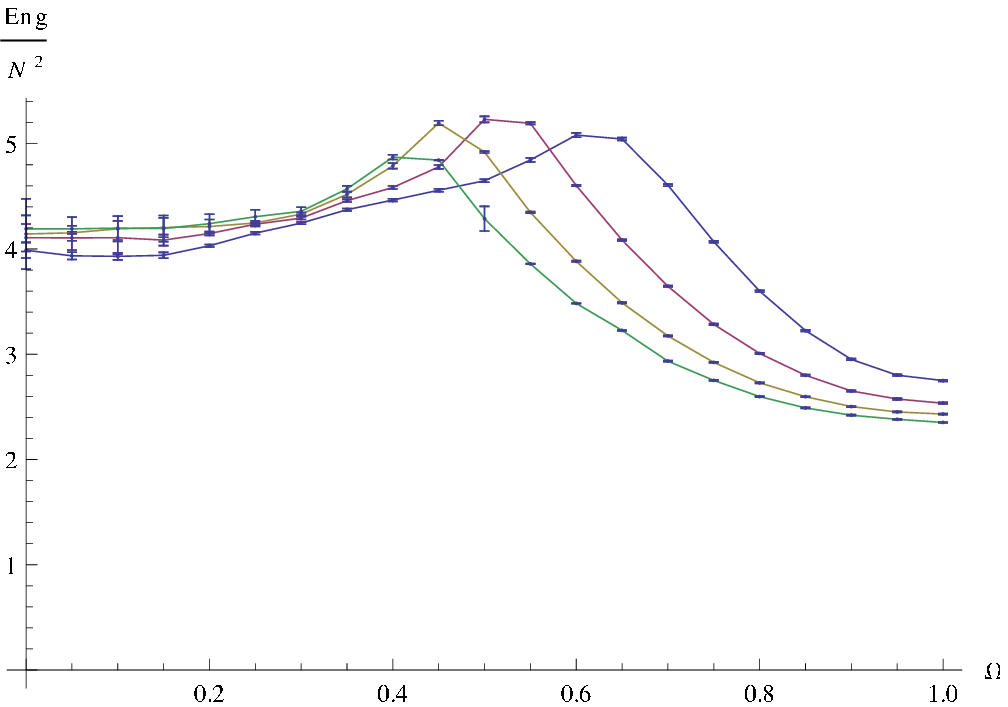}
\includegraphics[scale=0.45]{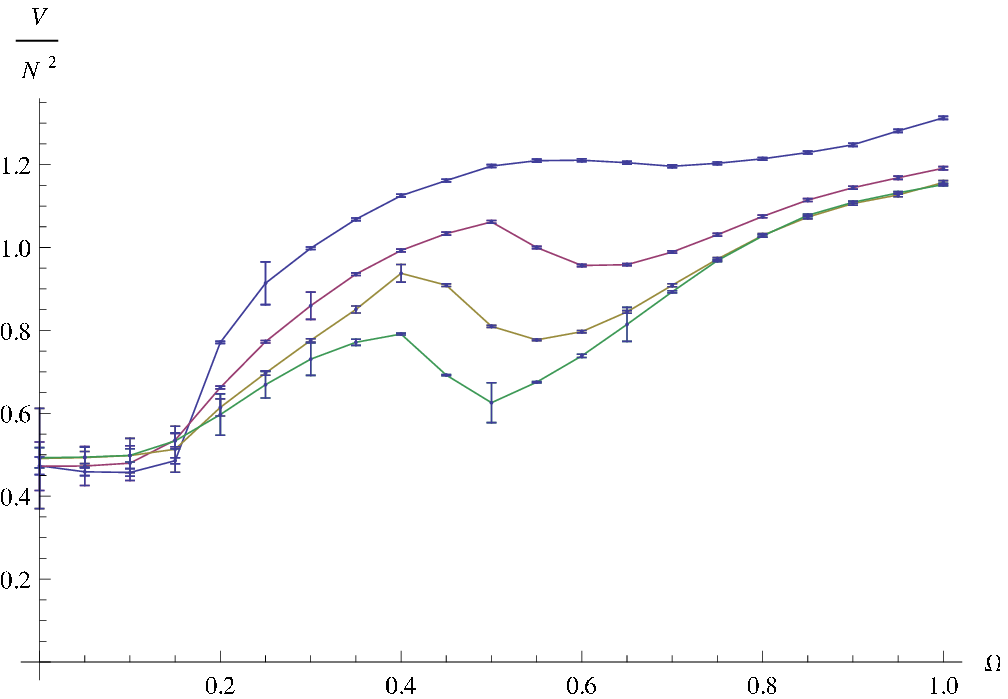}
\includegraphics[scale=0.45]{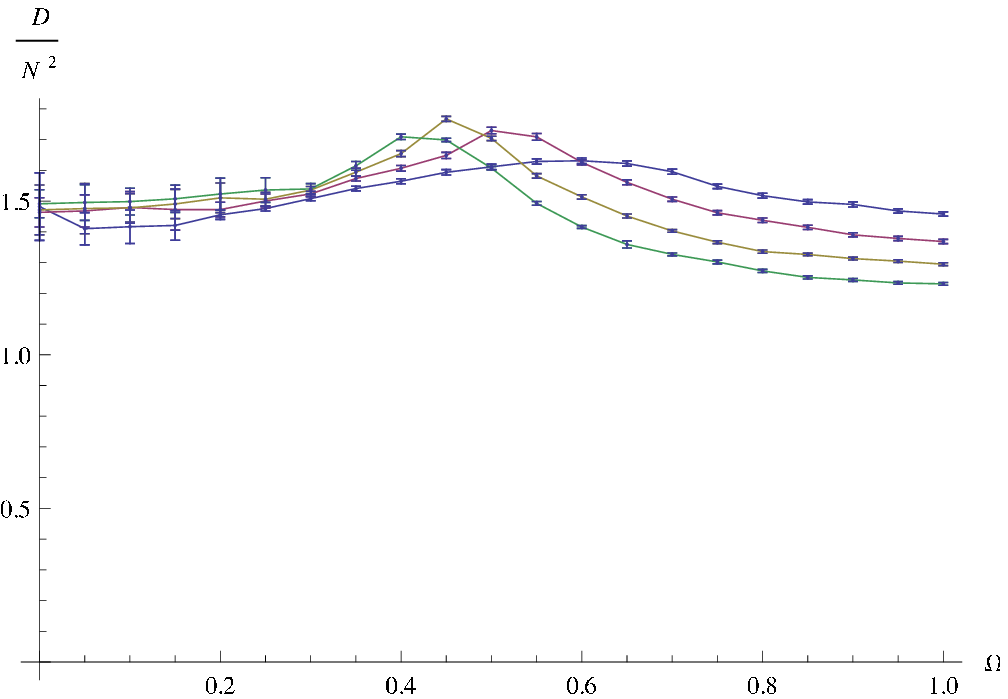}
\includegraphics[scale=0.45]{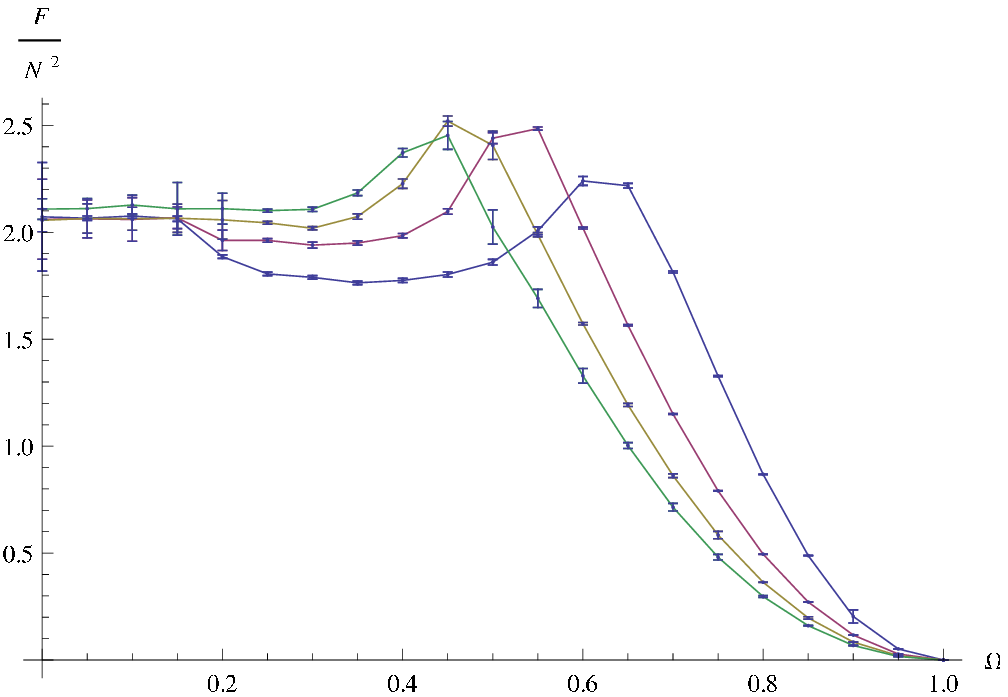}
\includegraphics[scale=0.45]{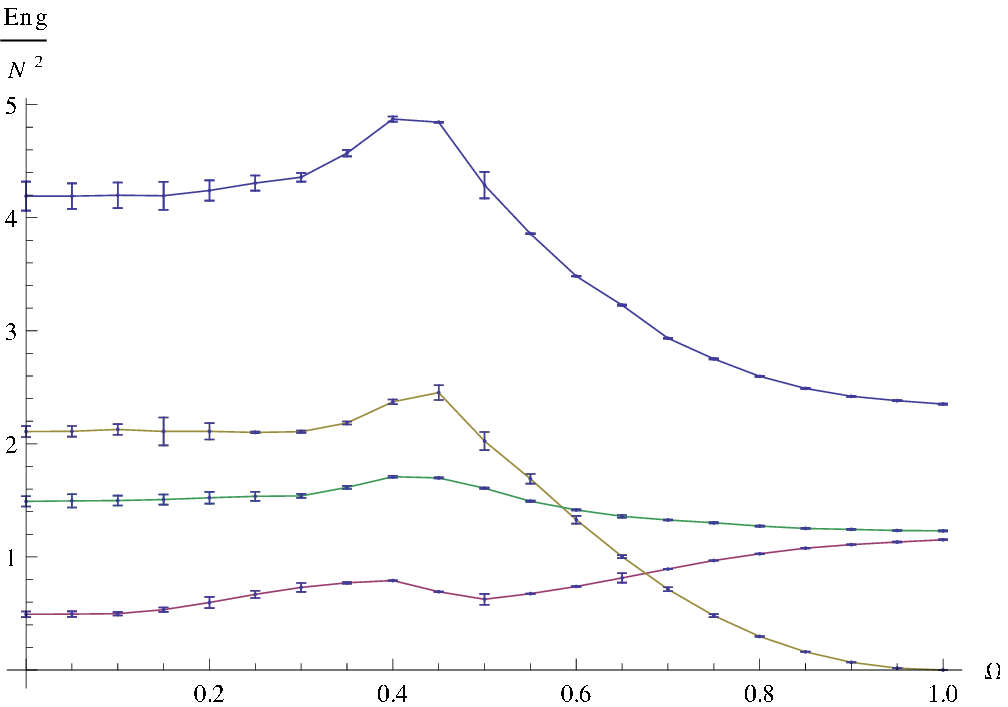}
\end{center}
\caption{\footnotesize Total energy density, various contributions and the comparison among them for $\mu=3$ varying $\Omega$ and $N$. From the left to the right  $E$, $V$, $D$, $F$ and comparison. \normalsize}\label{Figure 13}\end{figure}\newpage
Specific heat density displays fig.\ref{Figure 14} this change too, in fact instead the peak in $\Omega=1$, it  appears in the opposite side of the studied interval  in $\Omega=0$. This peak too, due to its grows  increasing $N$ indicates a phase transition. 
\begin{figure}[htb]
\begin{center}
\includegraphics[scale=.6]{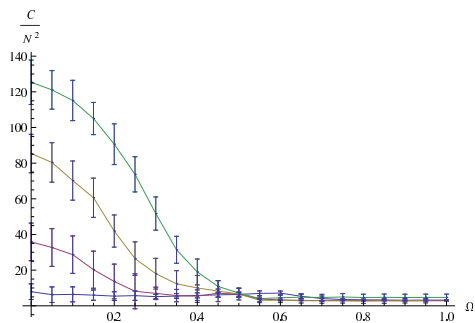}
\end{center}
\caption{\footnotesize Specific heat density for $\mu=3$ varying $\Omega$ and $N$.\normalsize}\label{Figure 14}
\end{figure}
\begin{figure}[htb]
\begin{center}
\includegraphics[scale=0.35]{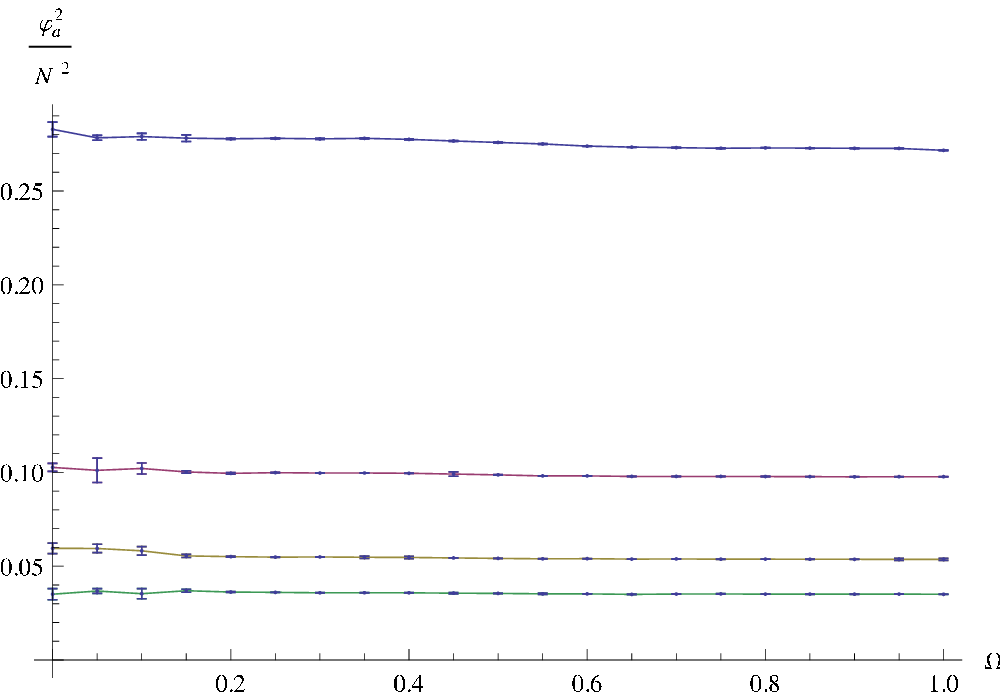}
\includegraphics[scale=0.35]{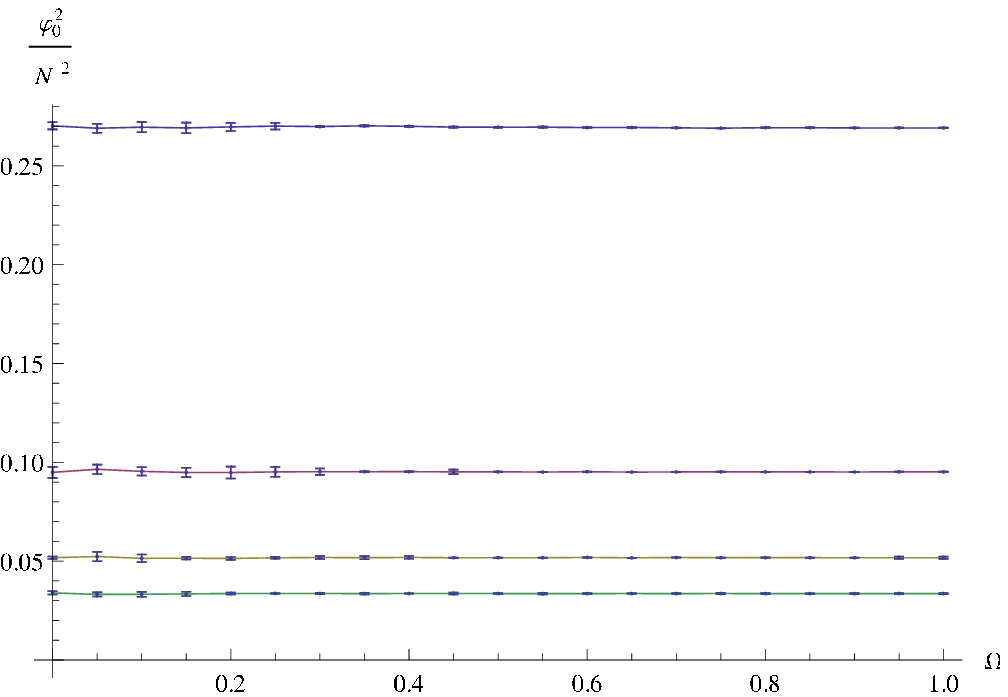}
\includegraphics[scale=0.35]{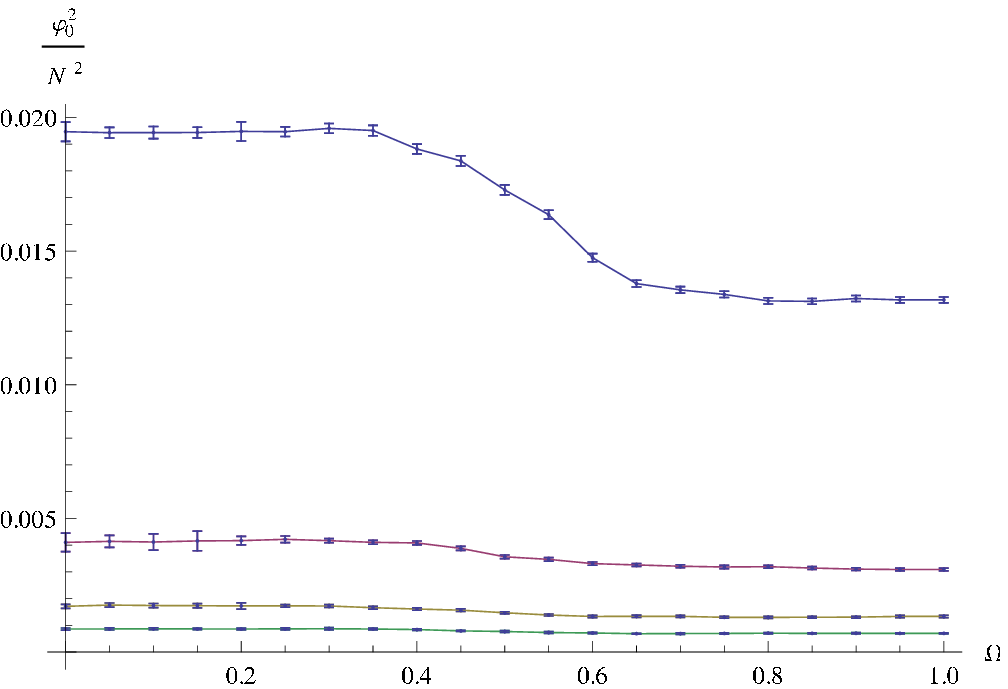}
\includegraphics[scale=0.35]{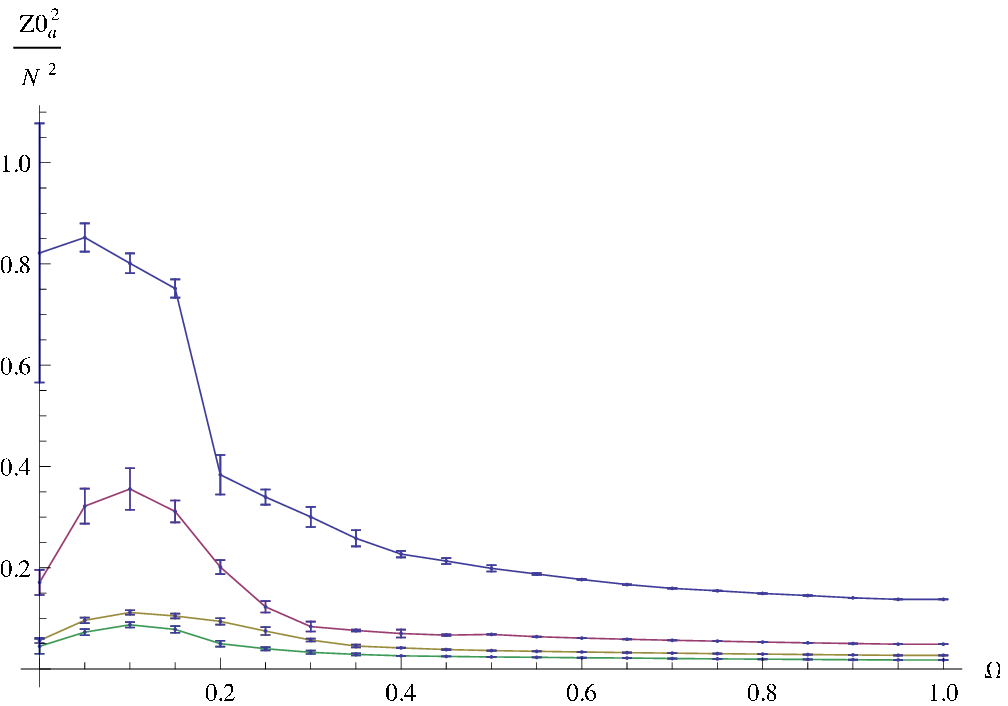}
\includegraphics[scale=0.35]{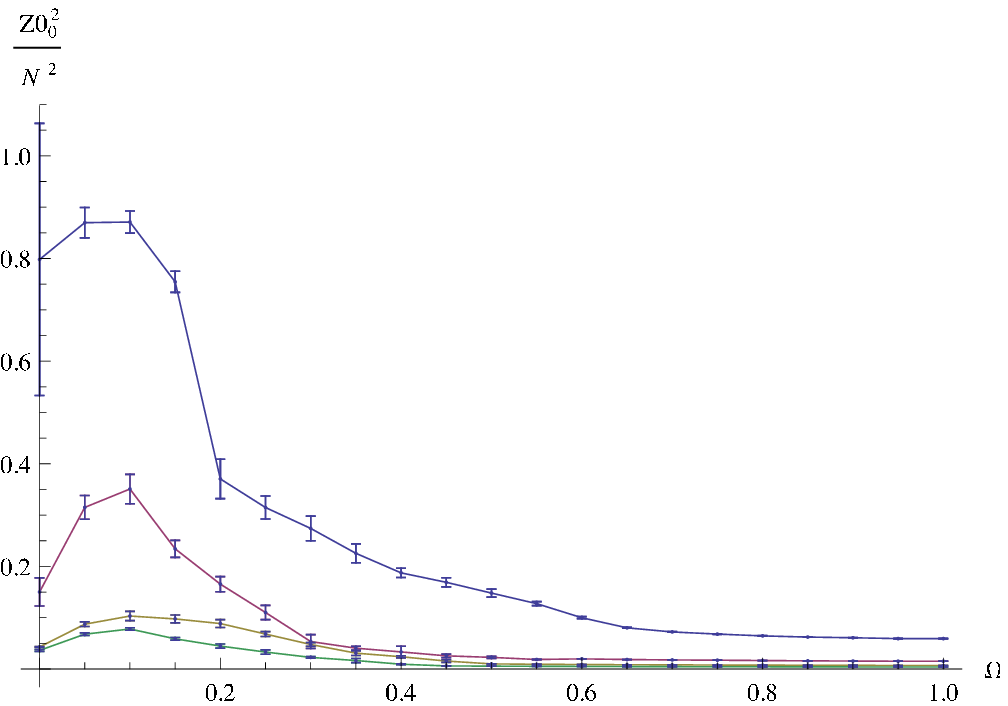}
\includegraphics[scale=0.35]{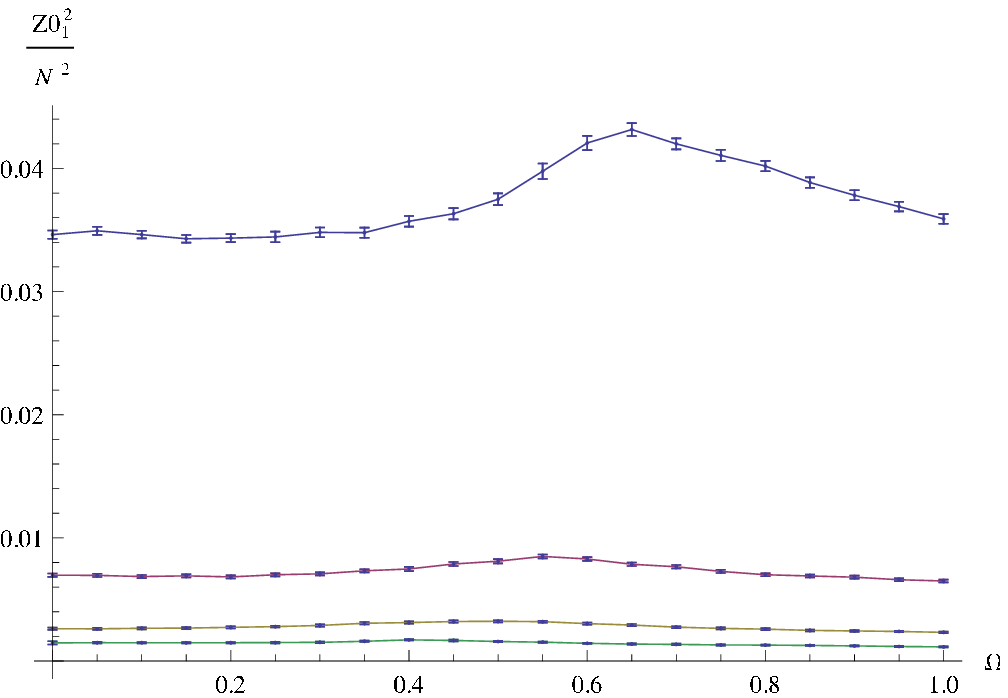}
\end{center}
\caption{\footnotesize Starting from the up left corner and from the left to the right the 	densities for $\varphi_a^2 $, $\varphi^2_0 $, $\varphi^2_1 $, $Z_{0a}^2$, $Z_{00}^2 $ and $Z_{01}^2$ for $\mu=3$ varying $\Omega$ and $N$.\normalsize}\label{Figure 15}
\end{figure}
The fig.\ref{Figure 15} describes the behavior of the order parameters 	densities $\varphi_a^2 $, $\varphi^2_0 $, $\varphi^2_1 $ and $Z_{0a}^2$, $Z_{00}^2 $, $Z_{01}^2$, they have a similar aspect to the previous relative graphs. For the $\psi$ field the spherical contribution remains dominant, beside  in the $\varphi^2_1 $ graph appears a deviation from the constant slope this deviation is evident for $N=5$ but still there for higher $N$. The order parameters for $Z_0$ display a peak close to the origin without oscillations even for $N=5$. This maximum for higher $N$ does not move closer to the origin, in other words, this shift in not due to the finite volume effect. Even for  $Z^2_{01} $ graph appears a deviation from the constant slope, a small peak which 	becomes shifted and smoother for higher $N$ 

\subsection{Varying $\mu$ }
\begin{figure}[htb]
\begin{center}
\includegraphics[scale=0.45]{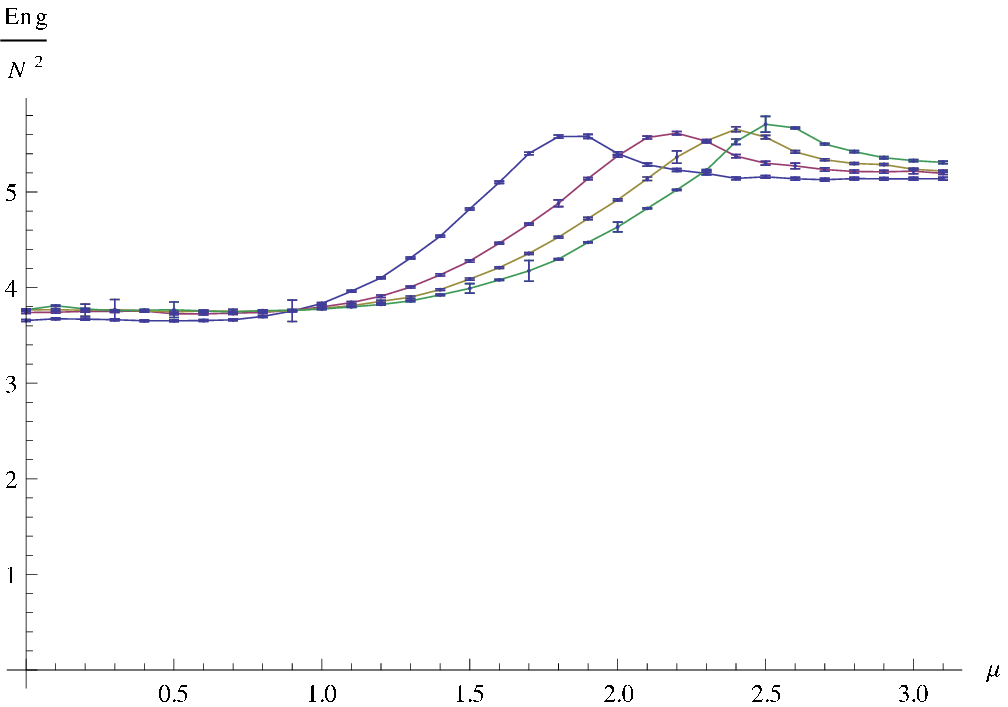}
\includegraphics[scale=0.45]{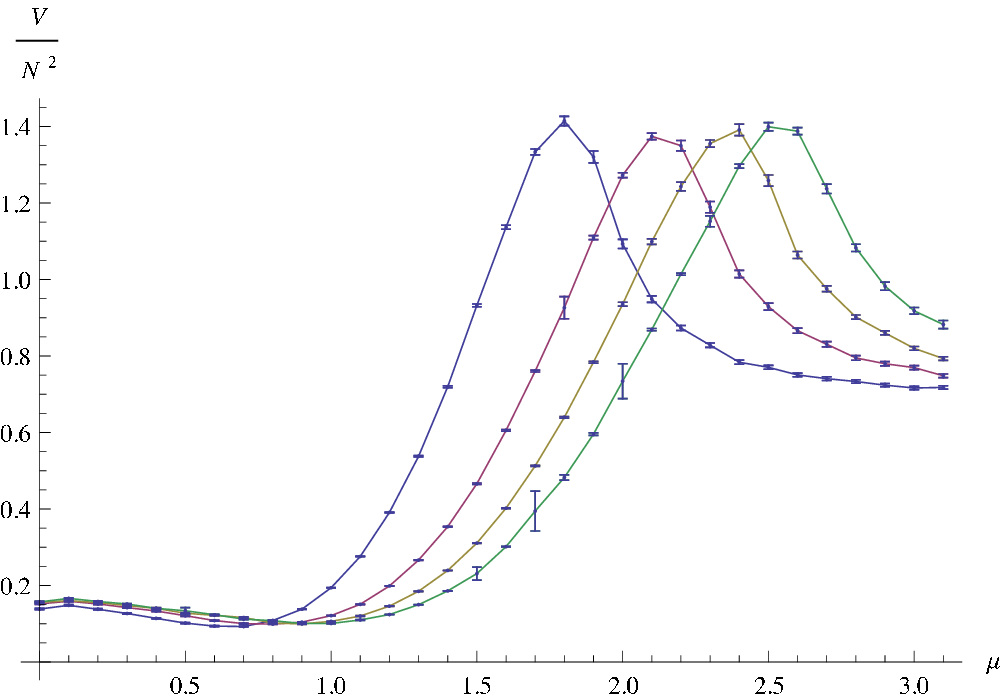}
\includegraphics[scale=0.45]{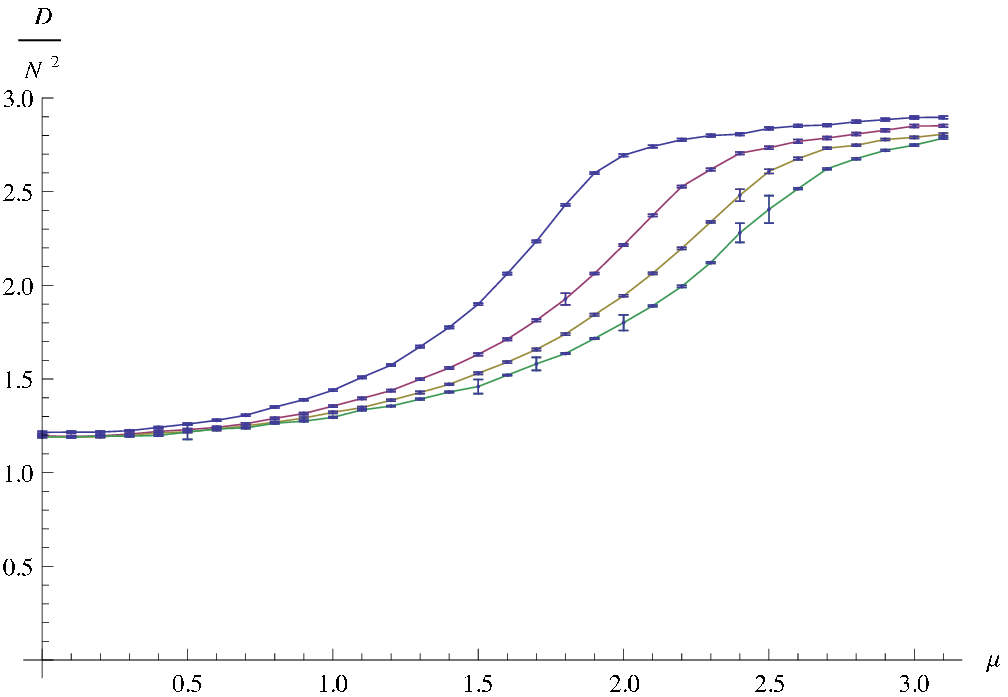}
\includegraphics[scale=0.45]{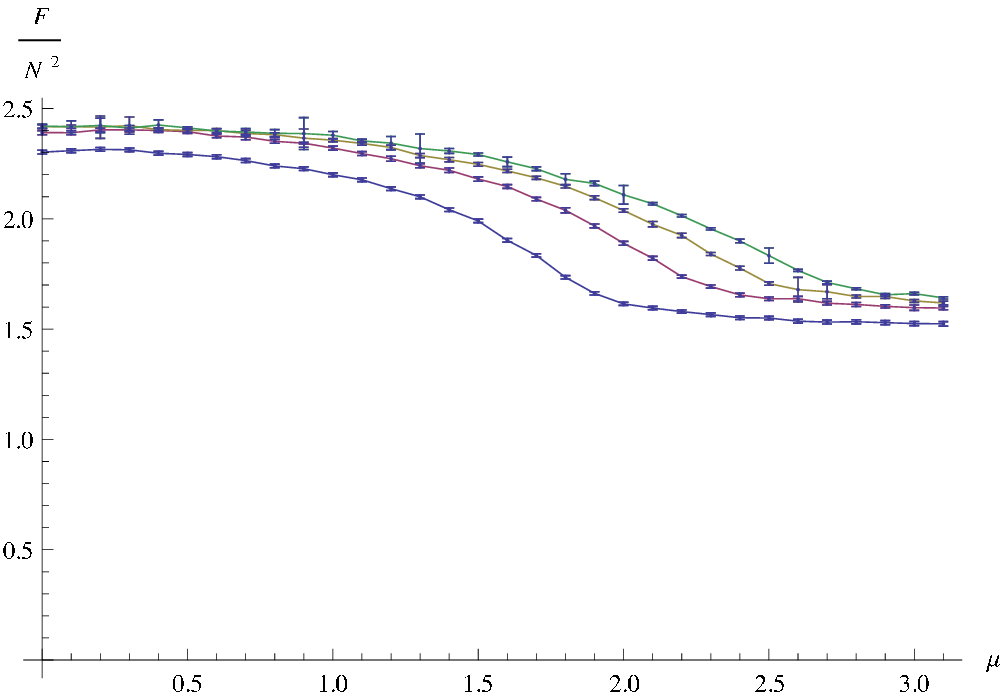}
\includegraphics[scale=0.45]{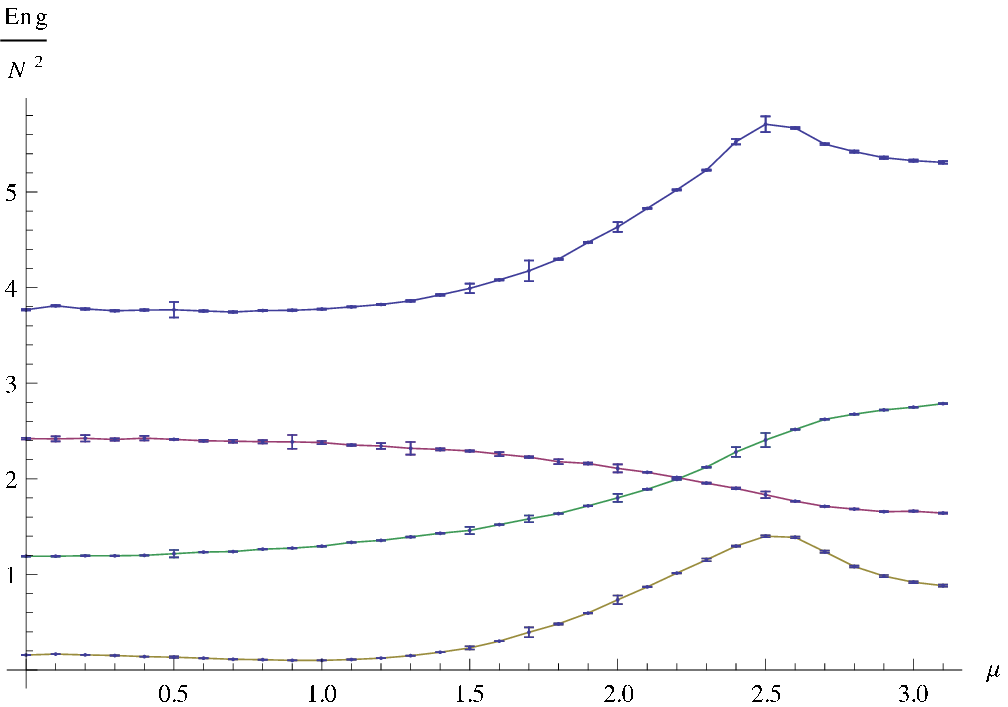}
\end{center}
\caption{\footnotesize The total energy density and the various contributions for $\Omega=0$ varying $\mu$ and $N$. From the left to the right  $E$, $V$, $D$, $F$ an comparison with $N=5$ (blue), $N=10$ (purple), $N=15$ (brown), $N=20$ (green). For the 	comparison: $E$ (blue), $V$ (purple), $D$ (brown), $F$ (green).\normalsize}\label{Figure 16}\end{figure}

In this section is analyzed  the response of the system varying $\mu\in[0,3]$ where $\Omega$ is fixed to $0,0.5,1$ and $\alpha$ is always zero. We start displaying the graphs fig.\ref{Figure 16} of the total energy density and of various contributions for $\Omega=0$. There is no evident discontinuity but appears a peak in the total energy density around $\mu\approx 2.5$ for $N=20$. Comparing all the contributions is easy to notice that the slope of the total energy is dictated by the curve $V$ of the potential part.    
\begin{figure}[htb]
\begin{center}
\includegraphics[scale=0.6]{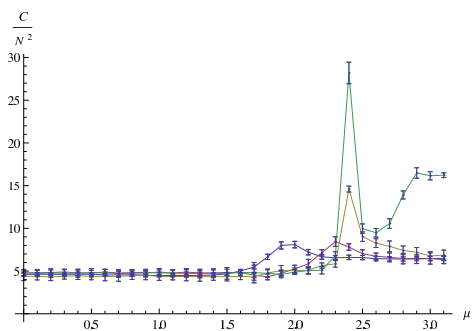}
\end{center}
\caption{\footnotesize Specific heat density for $\Omega=0$ varying $\mu$ and $N$.\normalsize}\label{Figure 17}
\end{figure}

\begin{figure}[htb]
\begin{center}
\includegraphics[scale=0.4]{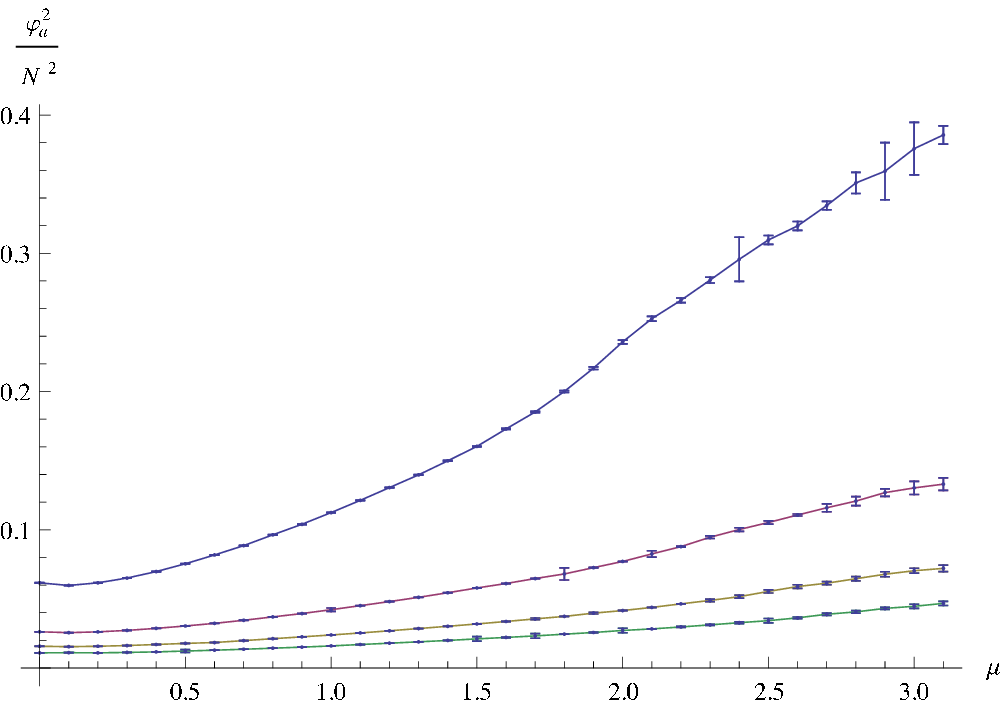}
\includegraphics[scale=0.4]{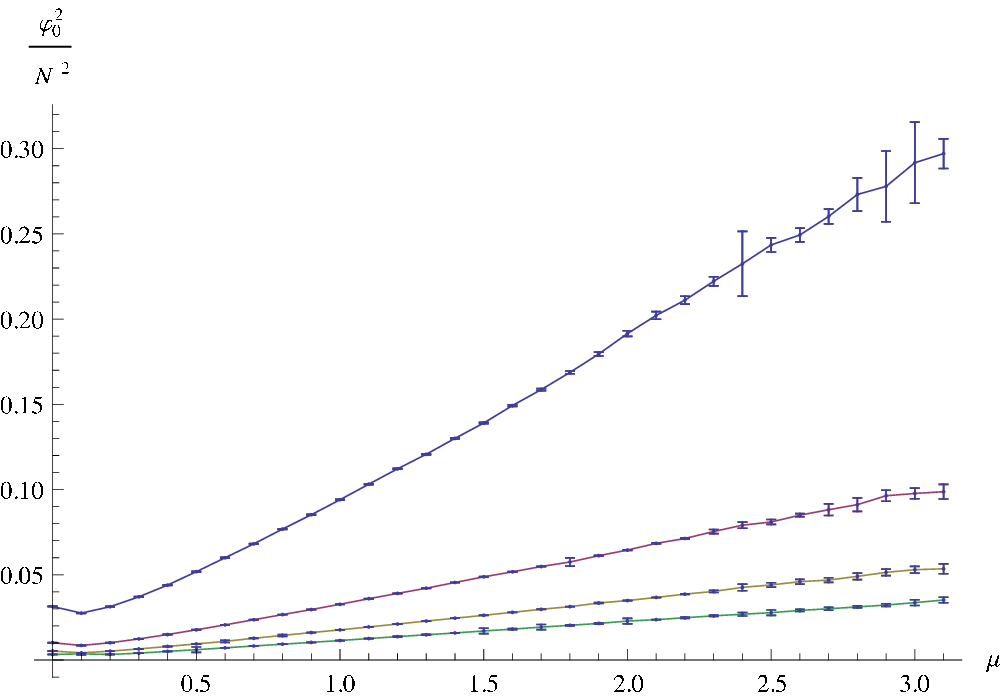}
\includegraphics[scale=0.4]{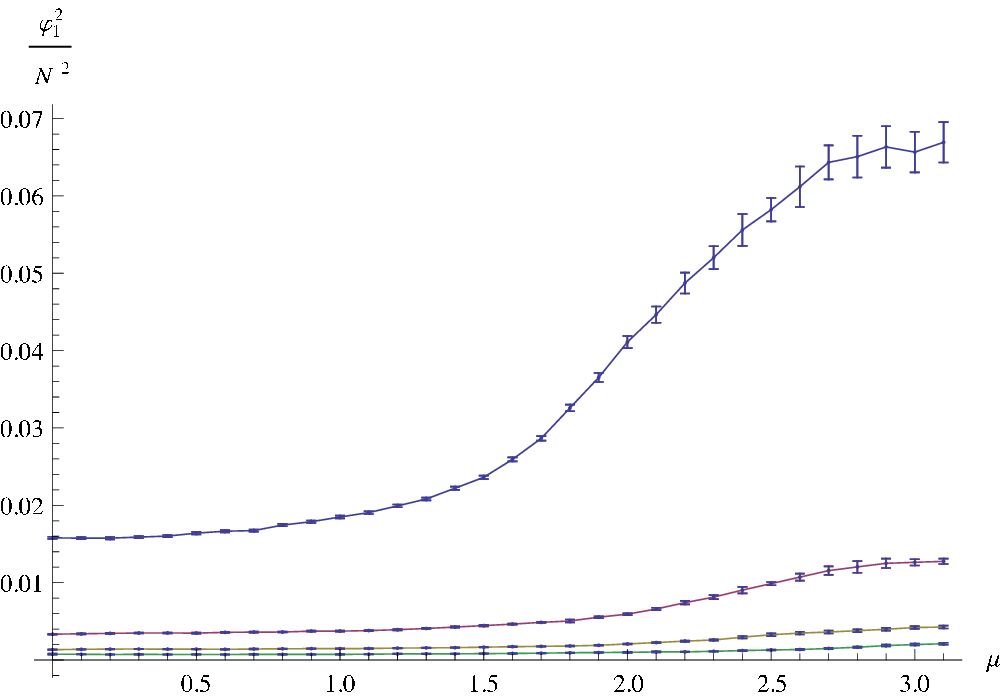}
\includegraphics[scale=0.4]{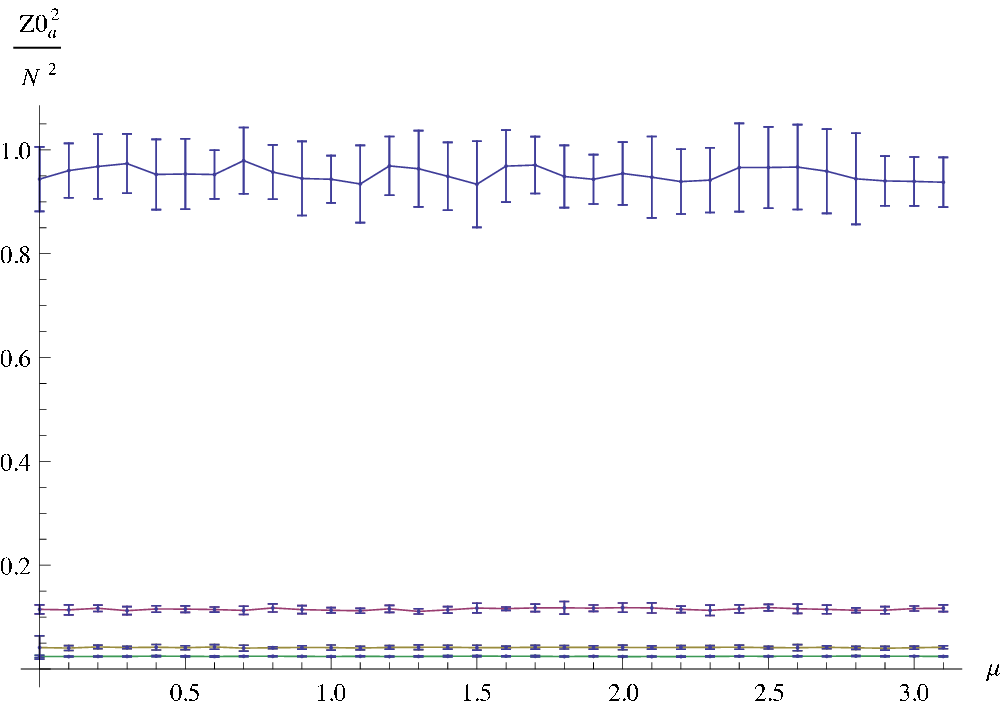}
\includegraphics[scale=0.4]{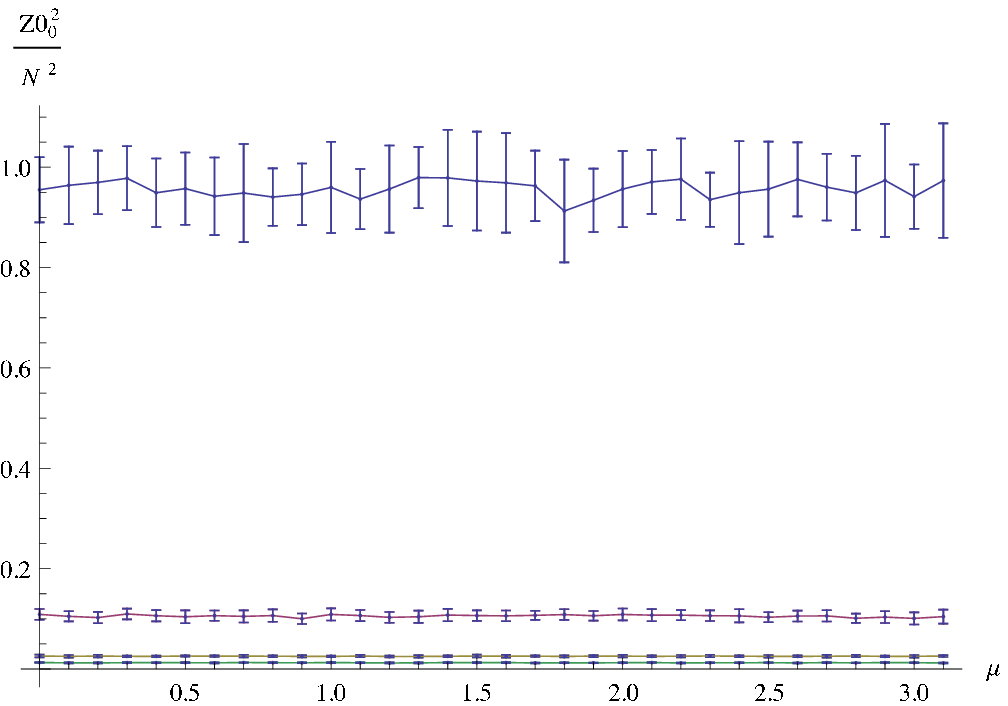}
\includegraphics[scale=0.4]{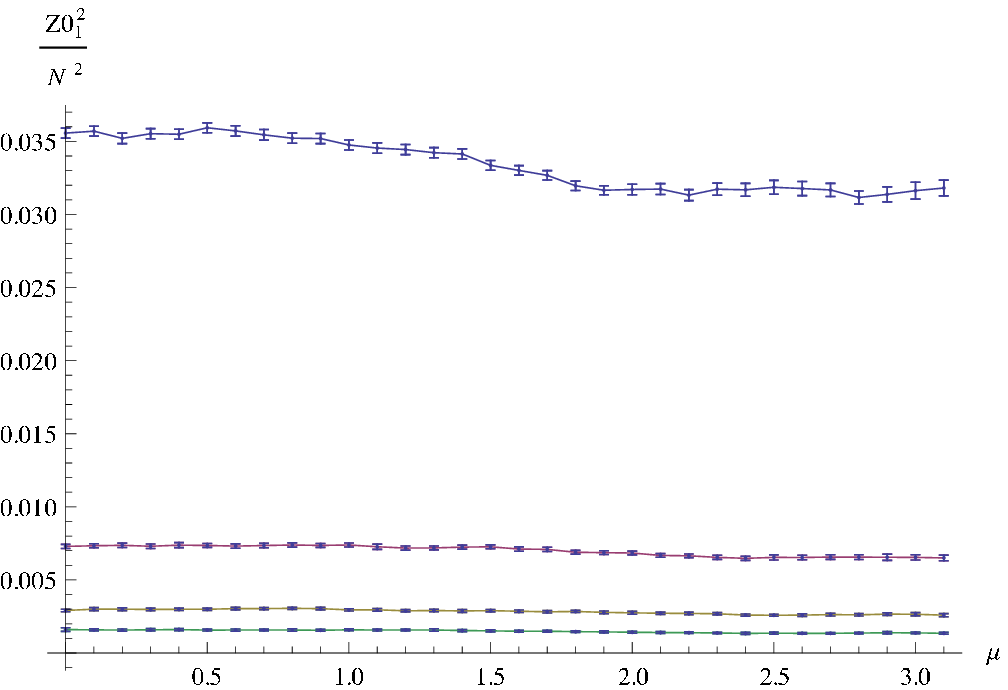}
\end{center}
\caption{\footnotesize Starting from the up left corner and from the left to the right the 	densities for $\varphi_a^2 $, $\varphi^2_0 $, $\varphi^2_1 $, $Z_{0a}^2$, $Z_{00}^2 $ and $Z_{01}^2$ for $\Omega=0$ varying $\mu$ and $N$. \normalsize}\label{Figure 18}
\end{figure}

As mentioned before, the specific heat density fig.\ref{Figure 17} features a peak  around $\mu\approx2.5$ for $N=20$ and again, due to this behavior as $N$ increase, we could relate this peak to a phase transition.
The plots for the quantities $\varphi_a^2 $ and $\varphi^2_0 $ denote a strong dependence on $\mu$, in particular the slope of $\varphi^2_0 $ seems mostly linear, for $\varphi^2_1 $
 we notice  a similar behavior but the slope is no longer linear. Comparing the three graphs fig.\ref{Figure 18} we deduce that close to the origin
 the non spherical contribution is bigger the spherical  one, increasing $\mu$ this situation capsizes and $\varphi^2_0 $ becomes dominant respect 
$\varphi^2_1 $. 
The behavior of the $Z_0$ fields is quite different, referring to figure \ref{Figure 18}, the spherical contribution is always dominant for the all interval $\mu\in[0,3]$. The curves for $Z_{0a}^2$, $Z_{00}^2 $ are compatible to the constant slope, for $Z_{01}^2 $ we have  the same dependence on $\mu$ in particular there is a smooth descending step, however this step becomes smoother for bigger $N$. It is behooves to say that due to some cancellations effects the statistical errors are quite big and they can hide some dependence, anyway this results tell us about the dependence of the order parameter for $Z_i$ and in general of the system, on the two  choice $\Omega=0$ or $\Omega\neq0$.

\begin{figure}[htb]
\begin{center}
\includegraphics[scale=0.4]{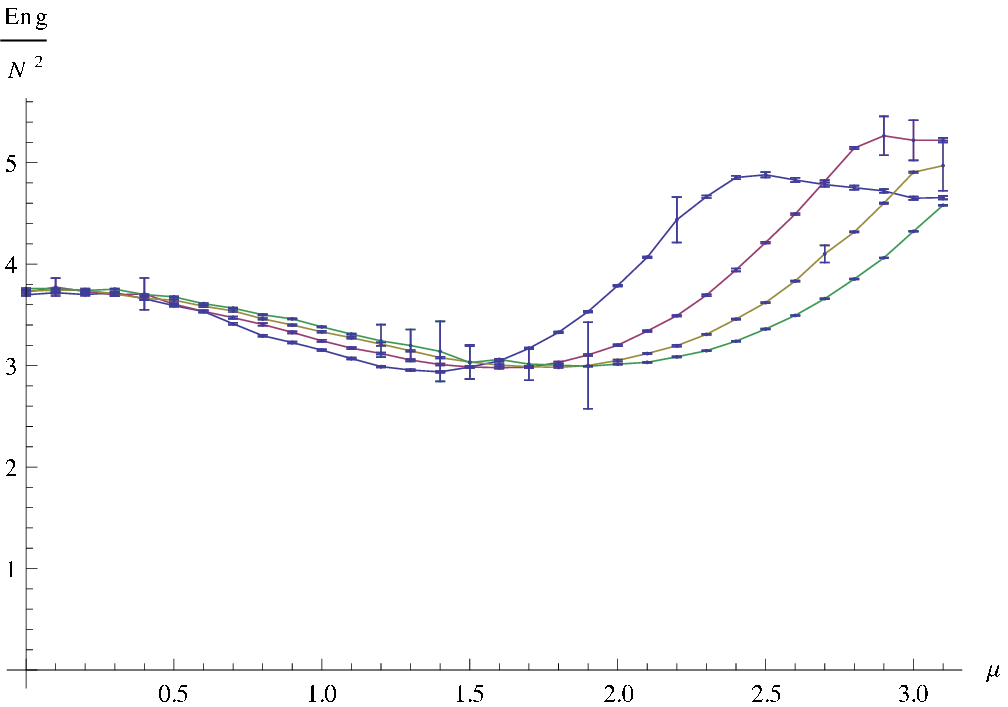}
\includegraphics[scale=0.4]{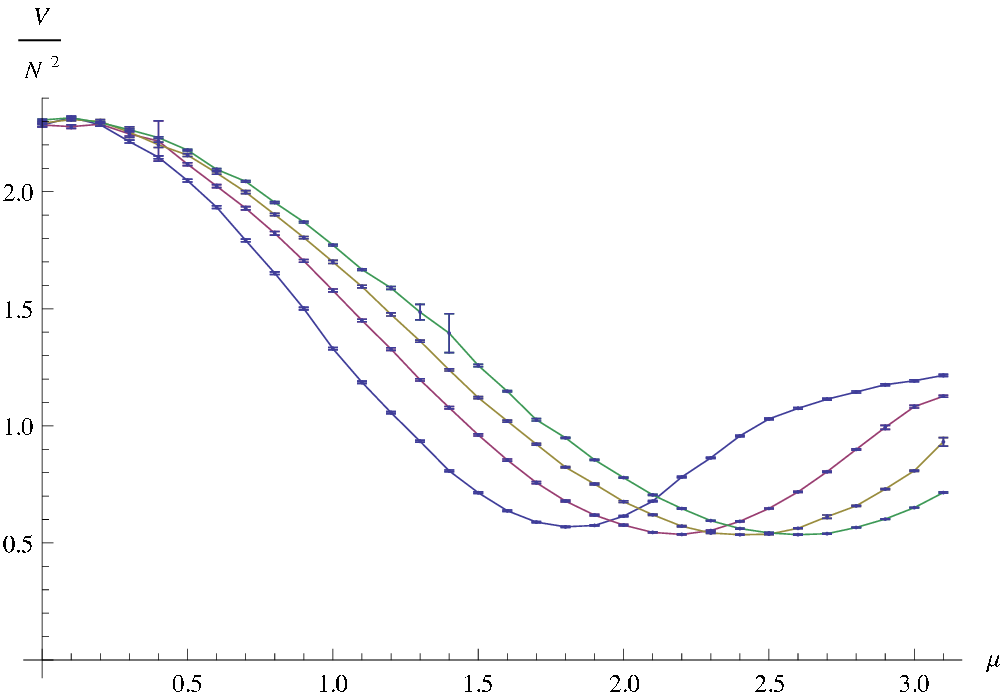}
\includegraphics[scale=0.4]{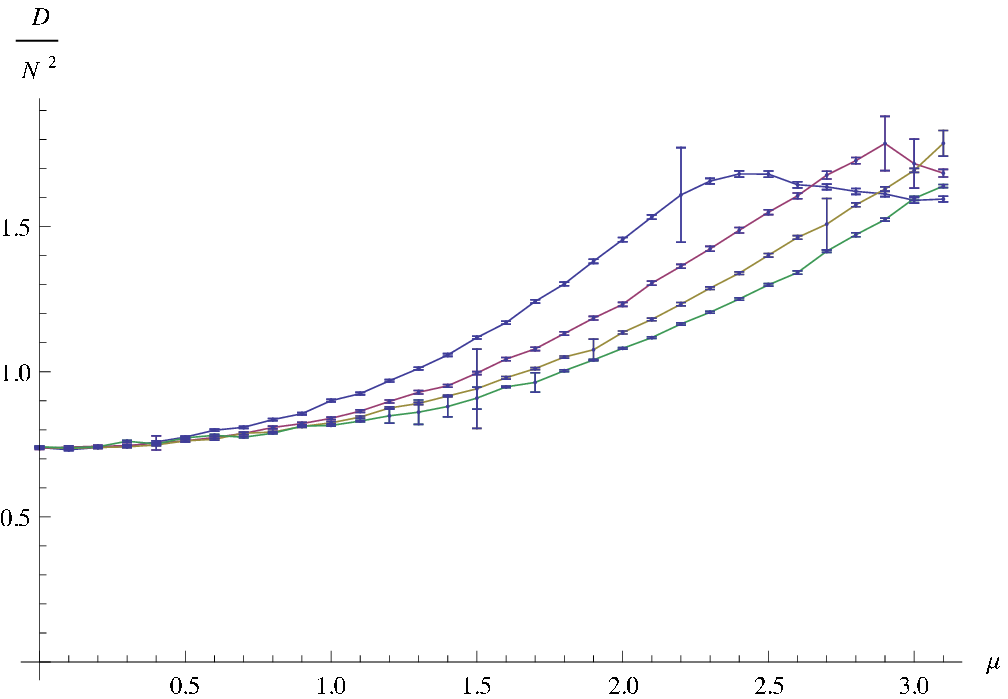}
\includegraphics[scale=0.4]{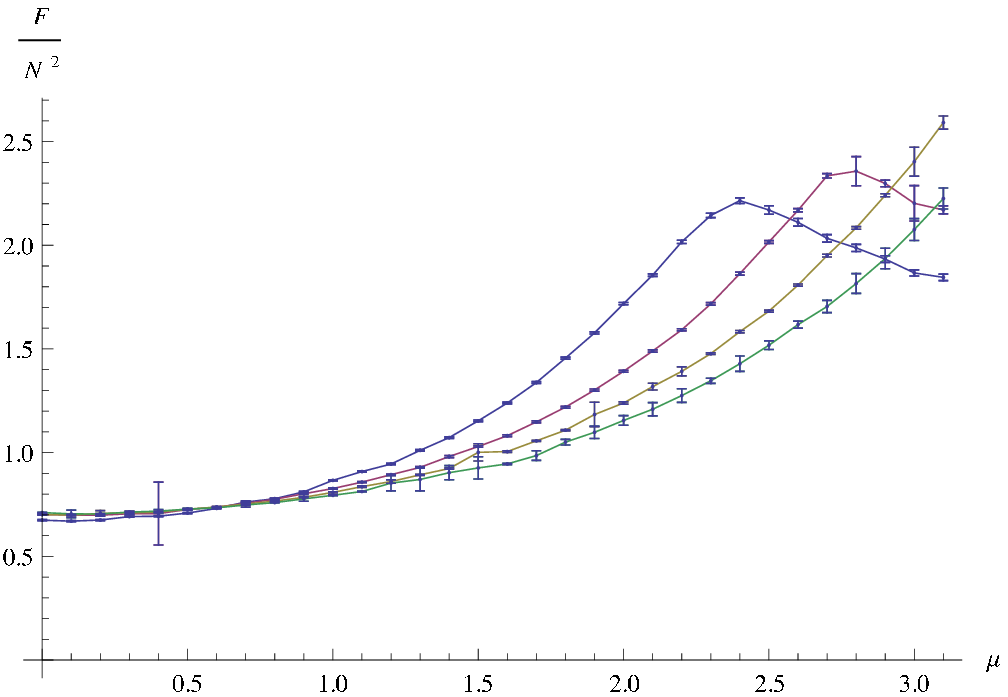}
\end{center}
\caption{\footnotesize Total energy density and contributions for $\Omega=0.5$ varying $\mu$ and $N$. From the left to the right  $E$, $V$, $D$, $F$.\normalsize}\label{Figure 19}\end{figure}

\begin{figure}[htb]
\begin{center}
\includegraphics[scale=0.7]{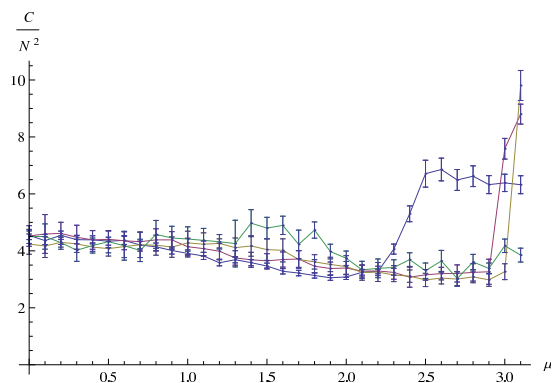}
\end{center}
\caption{\footnotesize Specific heat density for $\Omega=0.5$ varying $\mu$ and $N$.\normalsize}\label{Figure 20}
\end{figure}

Now we will analyze the  model for $\Omega=0.5$; as fig.\ref{Figure 19} shows the graphs have a different slope comparing to the previous case, the maximum of total energy density follow the  one of the $V$ component. If we focus ourself only on the total energy graph and we compare it with the one for $\Omega=0$, we notice a shift of the maximum for each $N$. In particular in fig.\ref{Figure 19} some maximum are moved outside the studied interval. 
We can find this shift very clearly looking at specific heat density graph fig.\ref{Figure 20}, we find again the peak as $N$ increase but  it is shifted around $\mu\approx 3.3$. 
The  graphs fig.\ref{Figure 23} for $\varphi_a^2 $, $\varphi^2_0 $ have the same 
behavior of $\Omega=0.5$ case, excluding some fluctuations close to the origin for $N=5$ due to the finite volume effect, $\varphi^2_1 $ graph displays an almost constant curve. However, close to the origin, the spherical contribution and the first non spherical one are comparable.
\begin{figure}[htb]
\begin{center}
\includegraphics[scale=0.35]{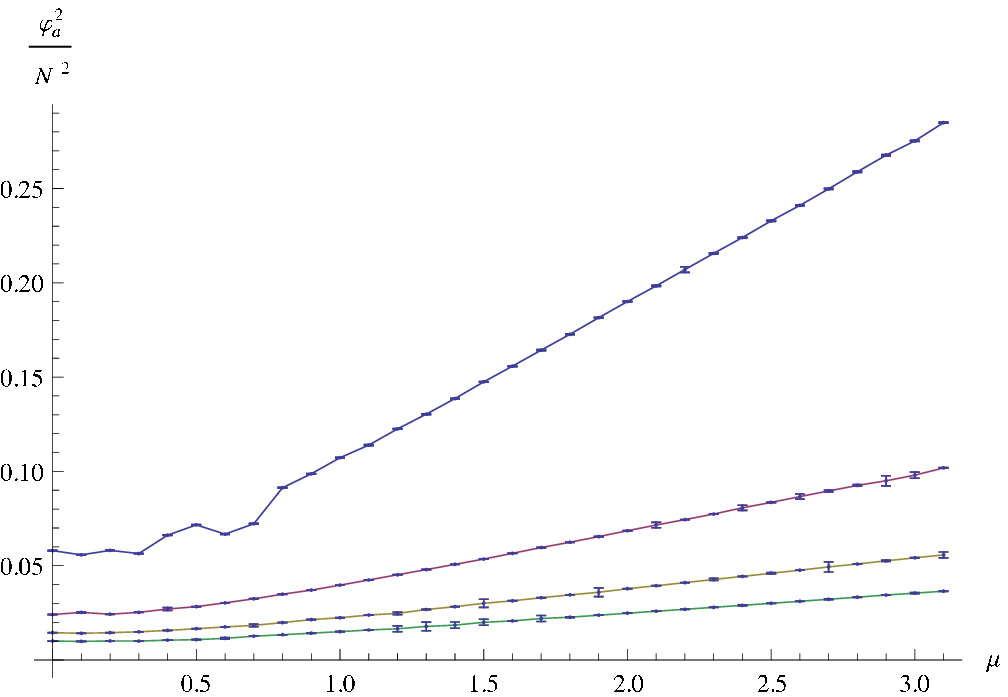}
\includegraphics[scale=0.35]{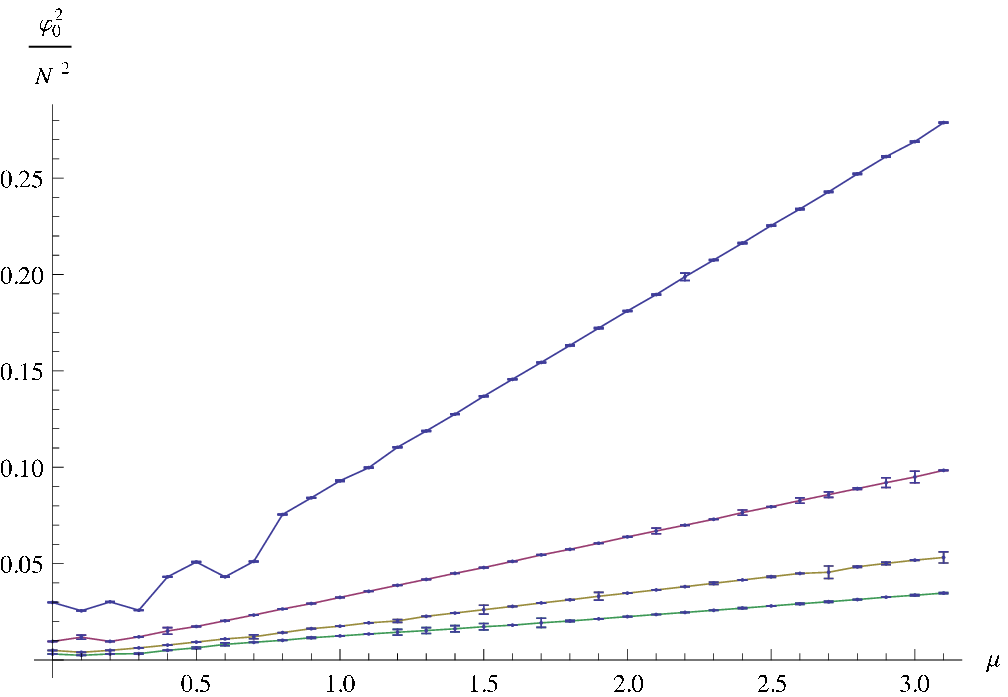}
\includegraphics[scale=0.35]{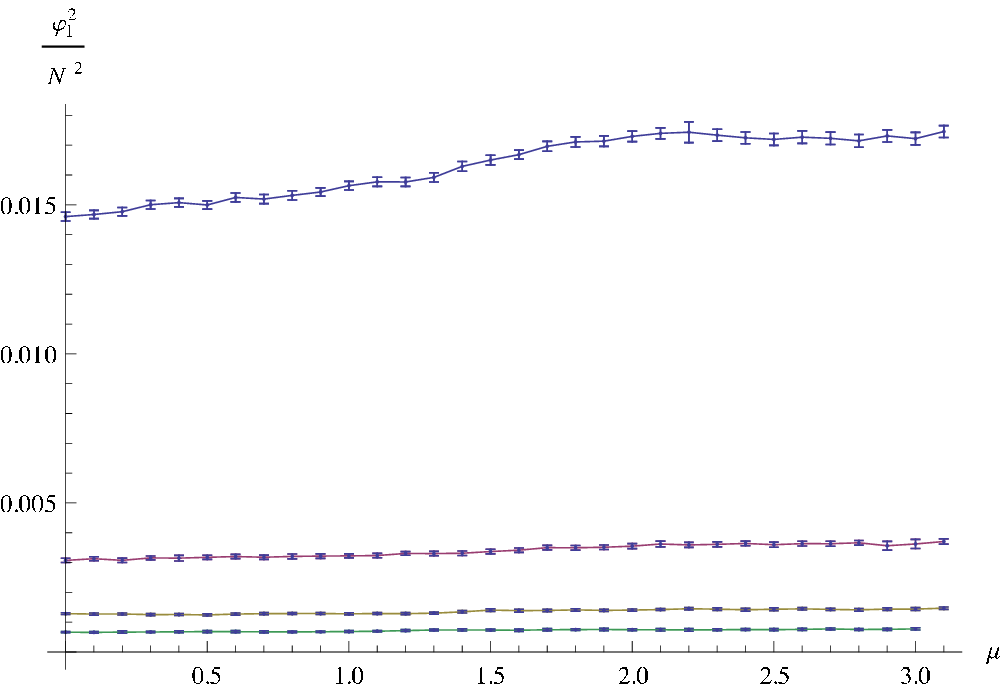}
\includegraphics[scale=0.35]{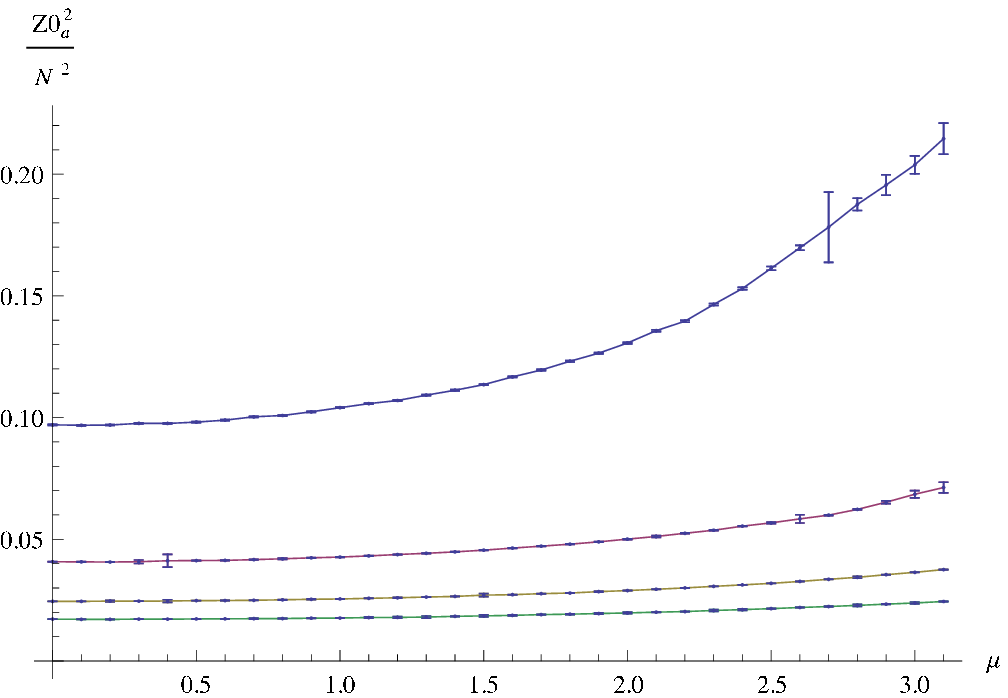}
\includegraphics[scale=0.35]{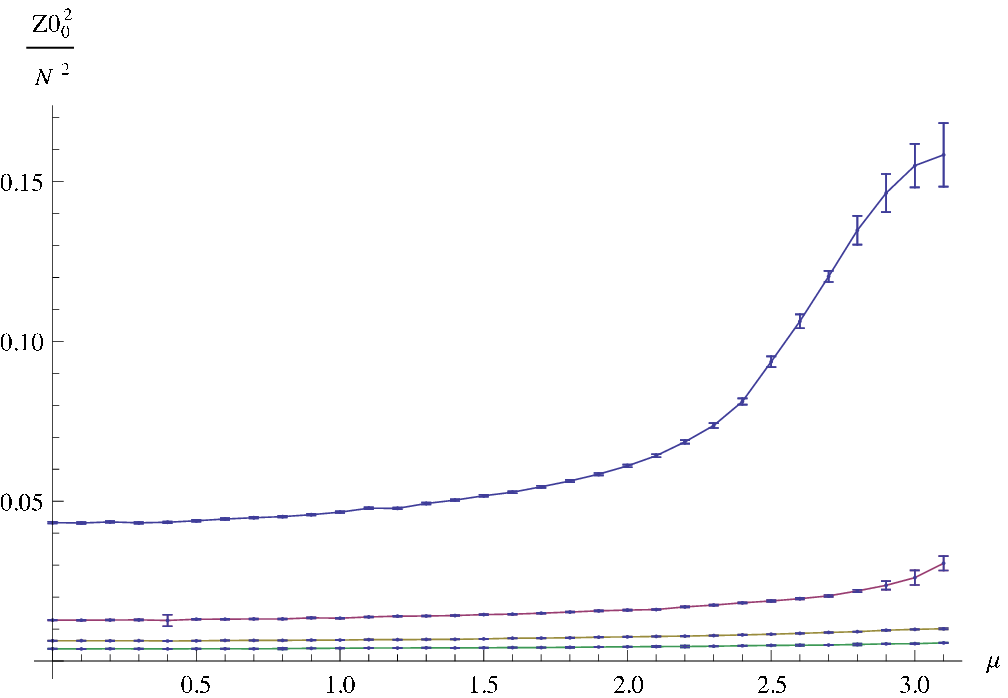}
\includegraphics[scale=0.35]{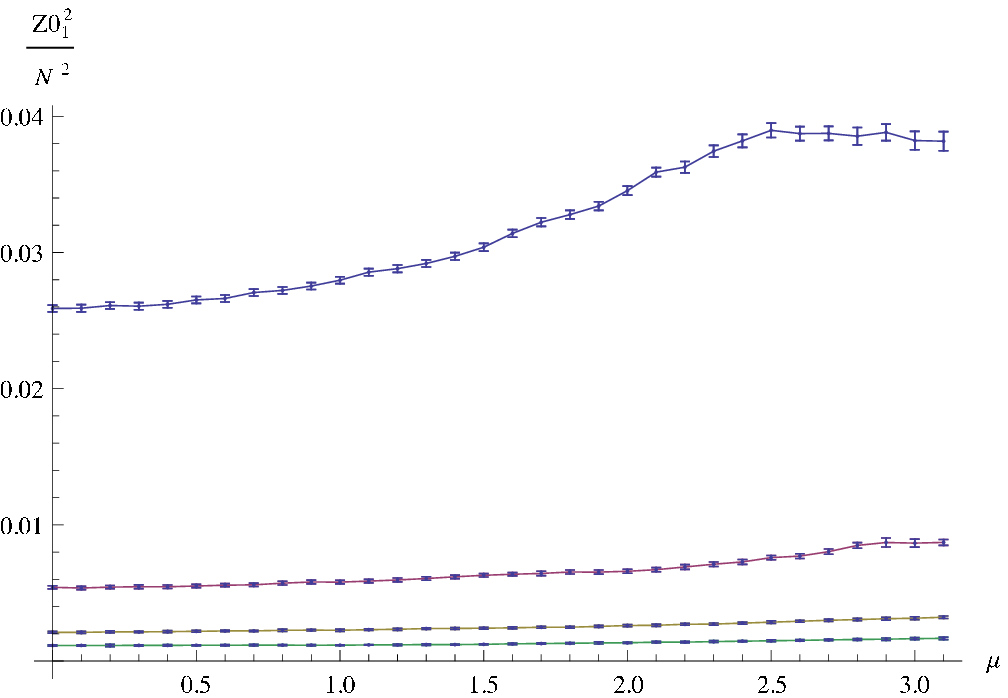}
\end{center}
\caption{\footnotesize Starting from the up left corner and from the left to the right the 	densities for $\varphi_a^2 $, $\varphi^2_0 $, $\varphi^2_1 $, $Z_{0a}^2$, $Z_{00}^2 $ and $Z_{01}^2$ for $\Omega=0.5$ varying $\mu$ and $N$.\normalsize}\label{Figure 21}
\end{figure}
The introduction of $\Omega\neq 0$ creates, in the $Z_0$ fields order parameters fig.\ref{Figure 21}, a dependence similar to the graphs for the $\psi$; the full power of the field density and the spherical contribution are no more constant and they grow increasing $\mu$. Even in this case the spherical contribution is always dominant excluding the region around $\mu=0$. 

\newpage
The last set of graphs for the 4-dimensional model are obtained fixing  \newline $\Omega=1$, due to the vanishing of prefactor in front of the Yang-Mills part of the action  the $F$ contribution is always zero. The following diagrams for the energy and contributions show the absence of the previous peak and comparing again them with former graphs they seem a sort dilatation. 
\begin{figure}[htb]
\begin{center}
\includegraphics[scale=0.4]{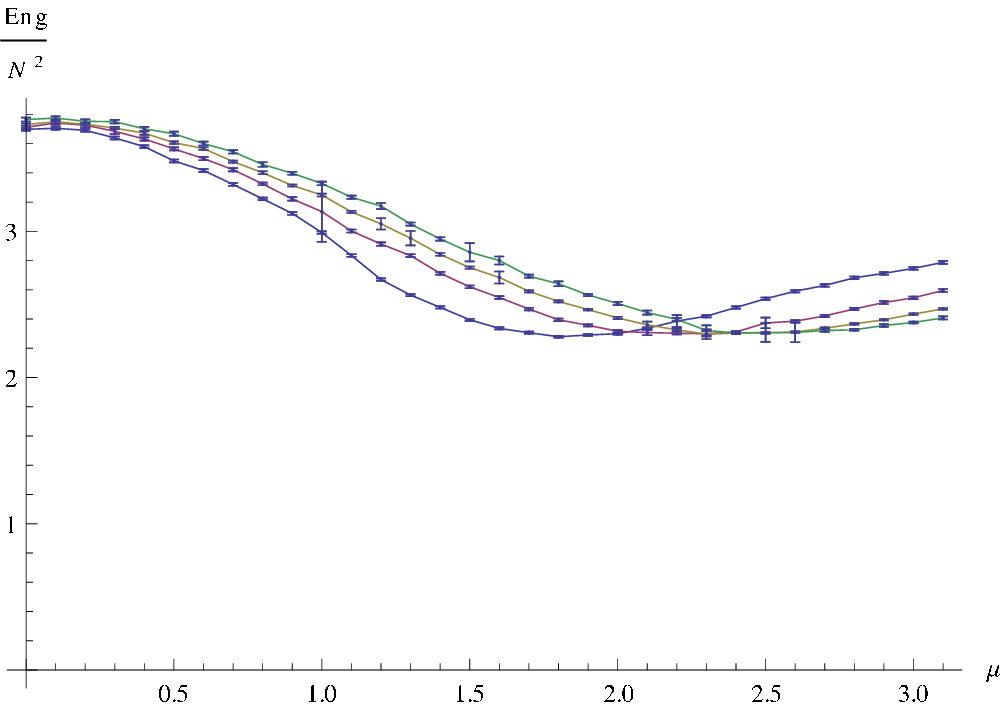}
\includegraphics[scale=0.4]{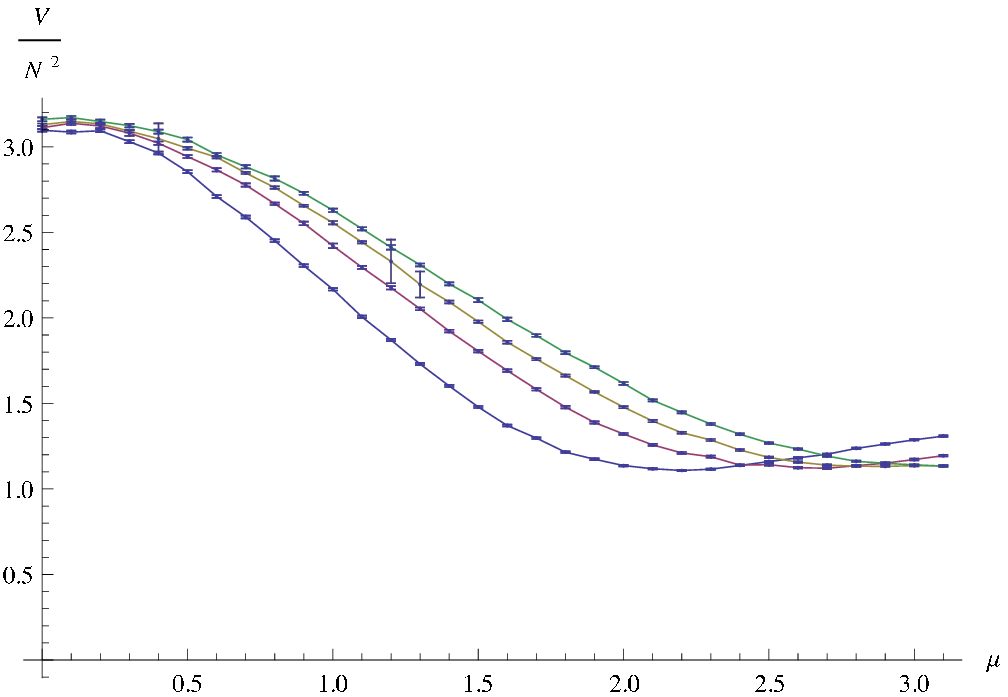}
\includegraphics[scale=0.4]{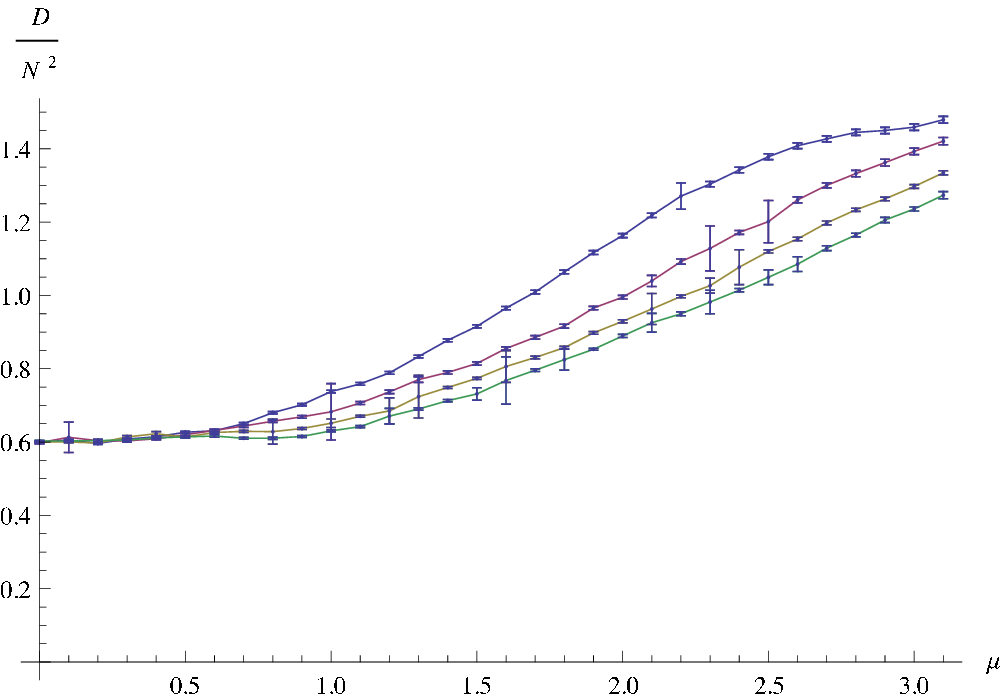}
\end{center}
\caption{\footnotesize Total energy density and the various contributions  for $\Omega=1$ varying $\mu$ and $N$. From the left to the right  $E$, $V$, $D$. \normalsize}\label{Figure 22}\end{figure}
The specific heat density does not show the peak in zero any more fig.\ref{Figure 23} and the curves does not show any particular point as $N$ increase, actually the peak can be found for higher $\mu$.  
\begin{figure}[htb]
\begin{center}
\includegraphics[scale=.7]{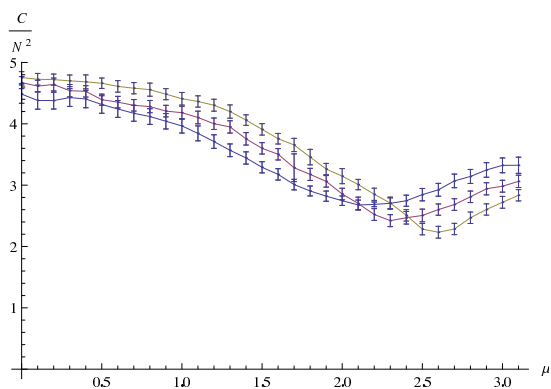}
\end{center}
\caption{\footnotesize Specific heat density for $\Omega=1$ varying $\mu$ and $N$.\normalsize}\label{Figure 23}
\end{figure}
\begin{figure}
\begin{center}
\includegraphics[scale=0.35]{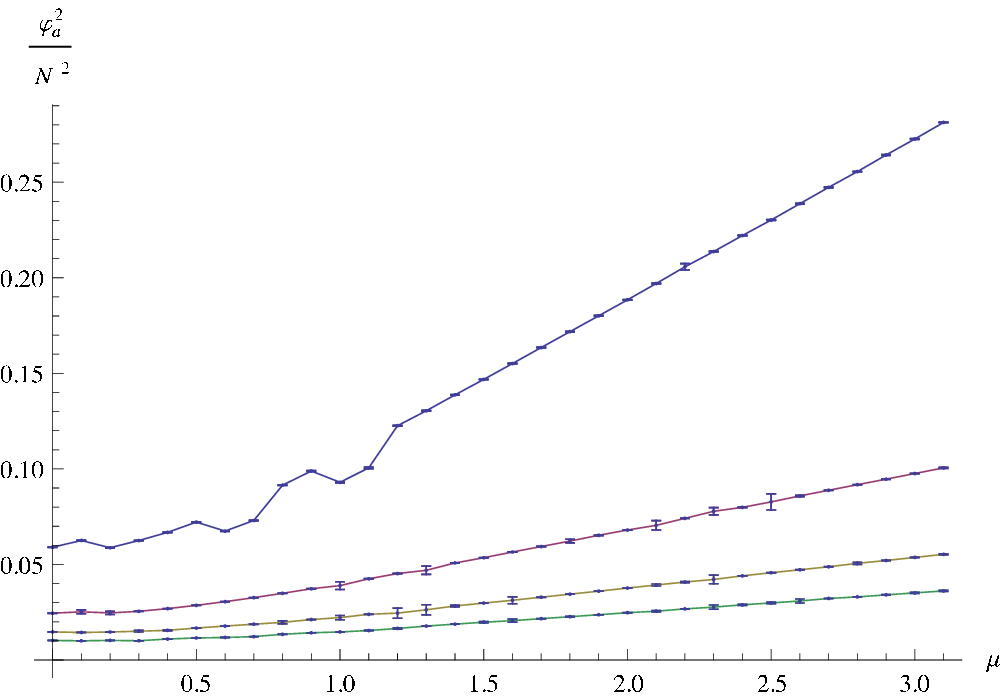}
\includegraphics[scale=0.35]{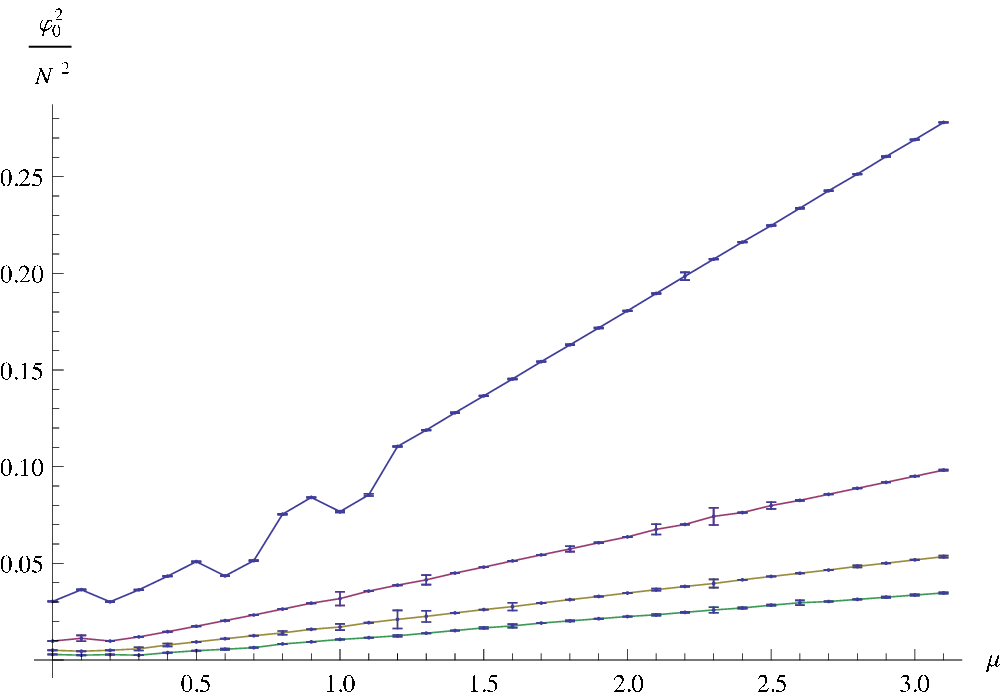}
\includegraphics[scale=0.35]{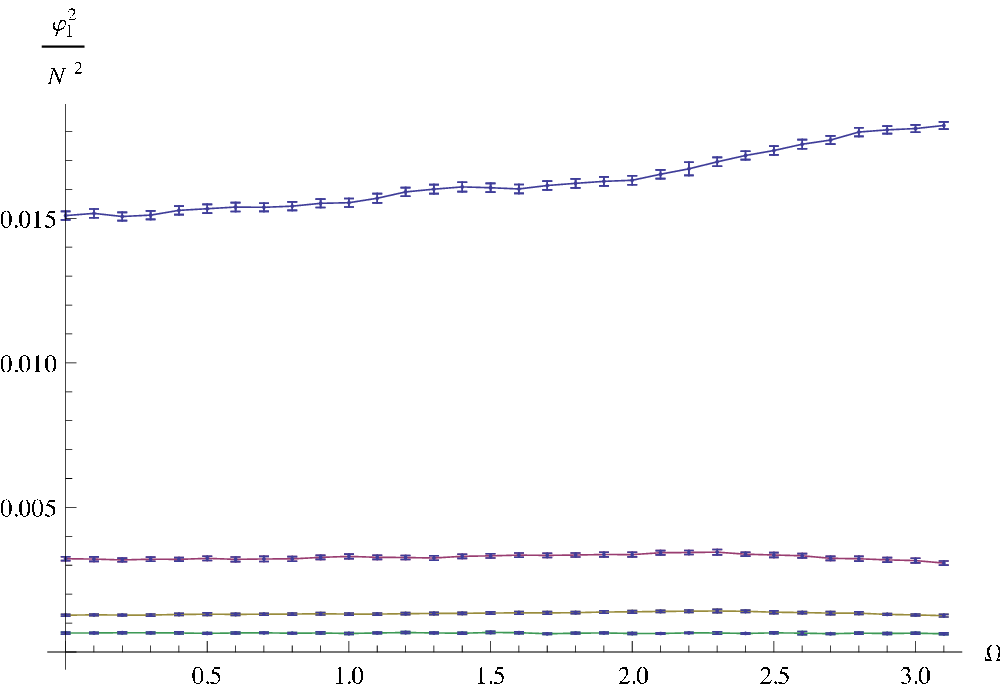}
\includegraphics[scale=0.35]{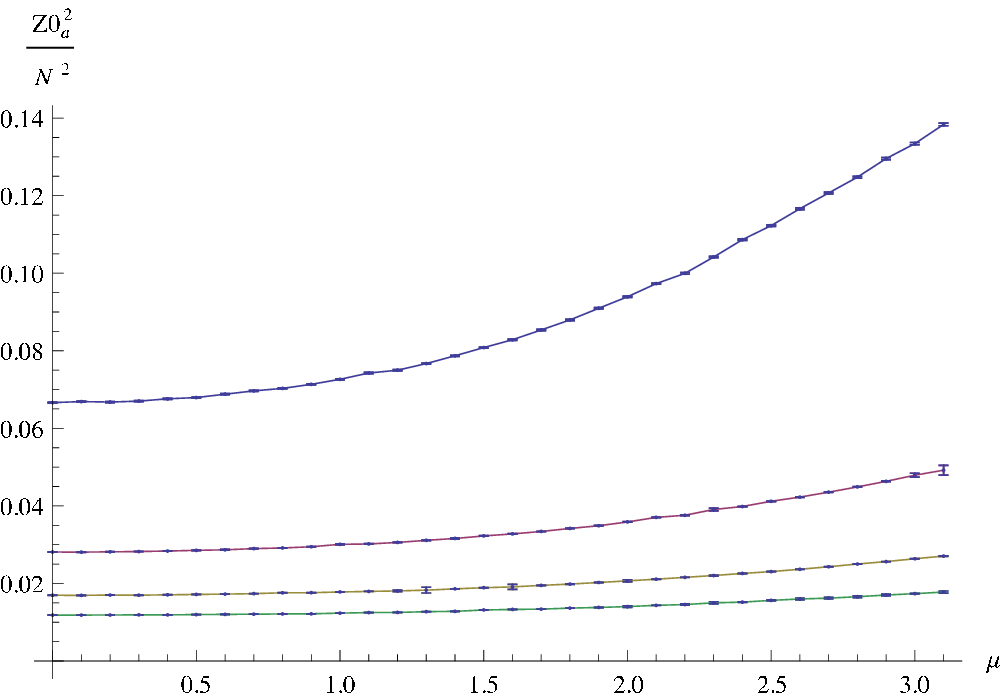}
\includegraphics[scale=0.35]{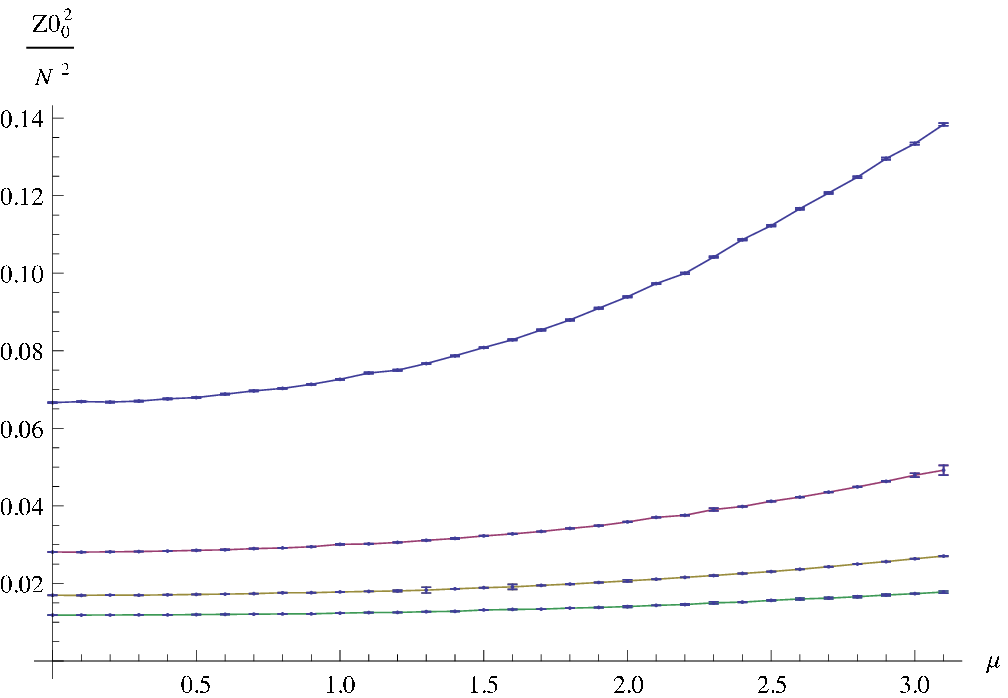}
\includegraphics[scale=0.35]{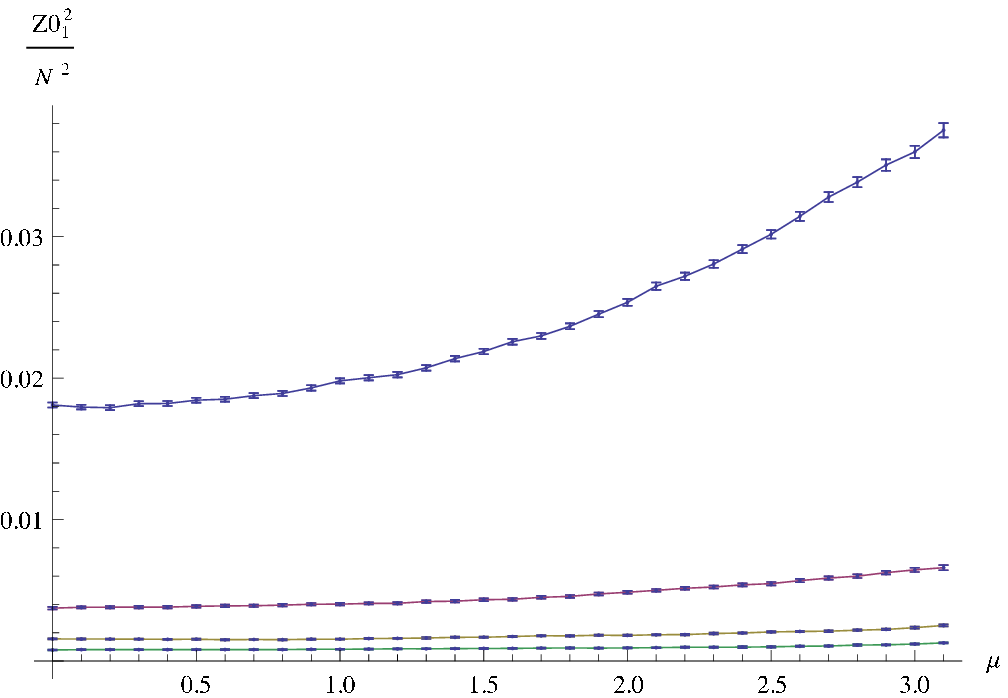}
\end{center}
\caption{\footnotesize Starting from the up left corner and from the left to the right the 	densities for $\varphi_a^2 $, $\varphi^2_0 $, $\varphi^2_1 $, $Z_{0a}^2$, $Z_{00}^2 $ and $Z_{01}^2$ for $\Omega=1$ varying $\mu$ and $N$.\normalsize}\label{Figure 24}
\end{figure}
At last in fig.\ref{Figure 24} we found a behavior of the order parameters density for $Z_0$ and $\psi$ fields similar to the former graphs for $\Omega=0.5$ and they are compatible with a dilatation of the previous diagrams.
\newpage
\section{Two dimensional case}
The main aim of this chapter was to show the numerical result for the 4-dimensional model, beside it is very interesting to analyze the 2-dimensional model (which was used as a first test of the program) described by the discretized action \eqref{S2} and note the differences with the full model. Will be used the same set of parameters used for the full model and due to the same dependence on $\alpha$ all the simulation  will be conducted with $\alpha=0$. 
 
\subsection{Varying $\Omega$ }
\begin{figure}[htb]
\begin{center}
\includegraphics[scale=0.45]{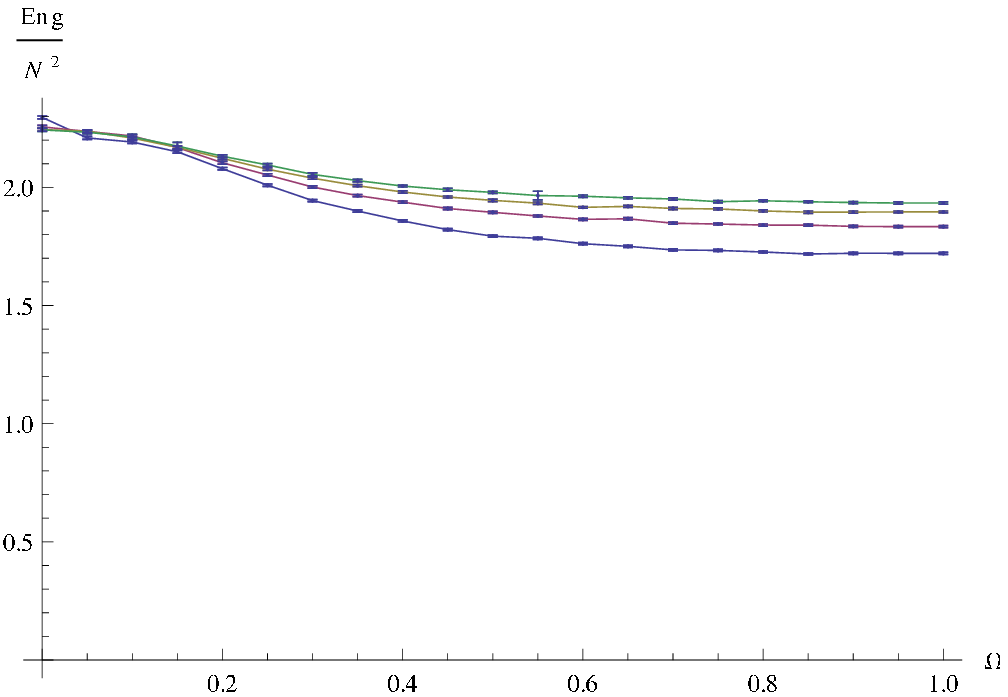}
\includegraphics[scale=0.45]{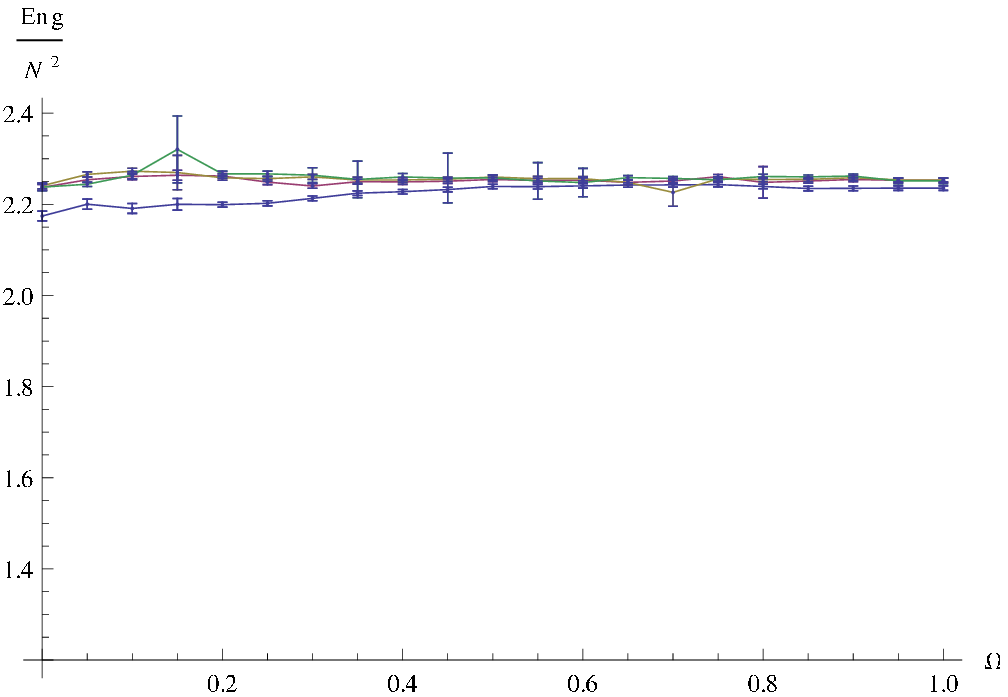}
\includegraphics[scale=0.45]{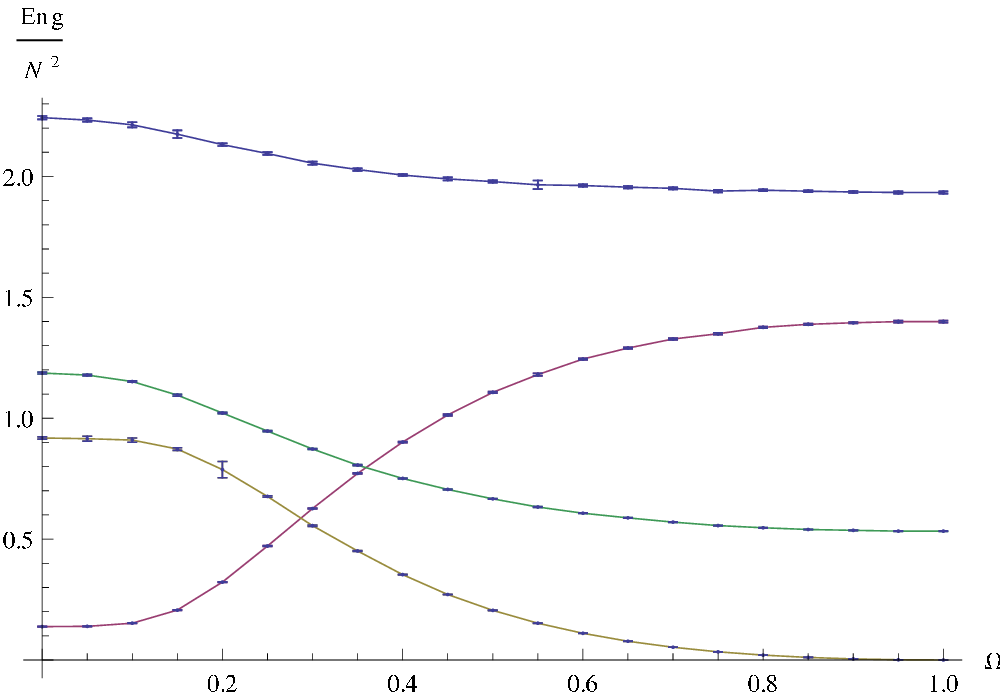}
\includegraphics[scale=0.45]{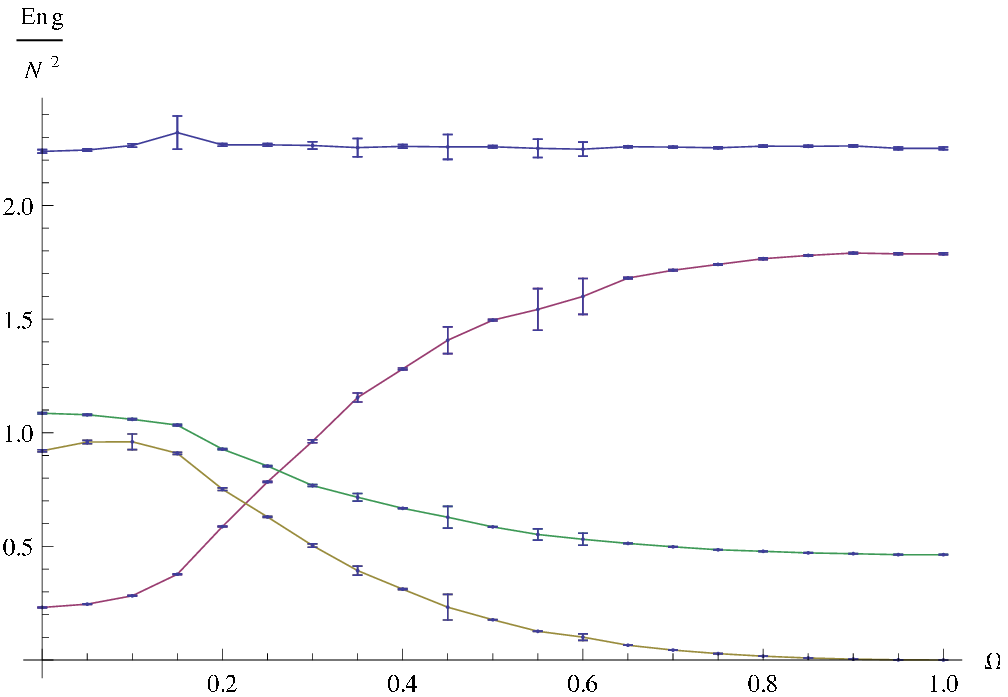}
\end{center}
\caption{\footnotesize Total energy density and contributions for $\mu=1$ (left side) and  $\mu=0$ (right side) varying $\Omega$ and $N$. Where $E$ (blue), $V$ (purple), $D$ (brown), $F$ (green).\normalsize}\label{Figure 25}\end{figure}

\begin{figure}[htb]
\begin{center}
\includegraphics[scale=0.5]{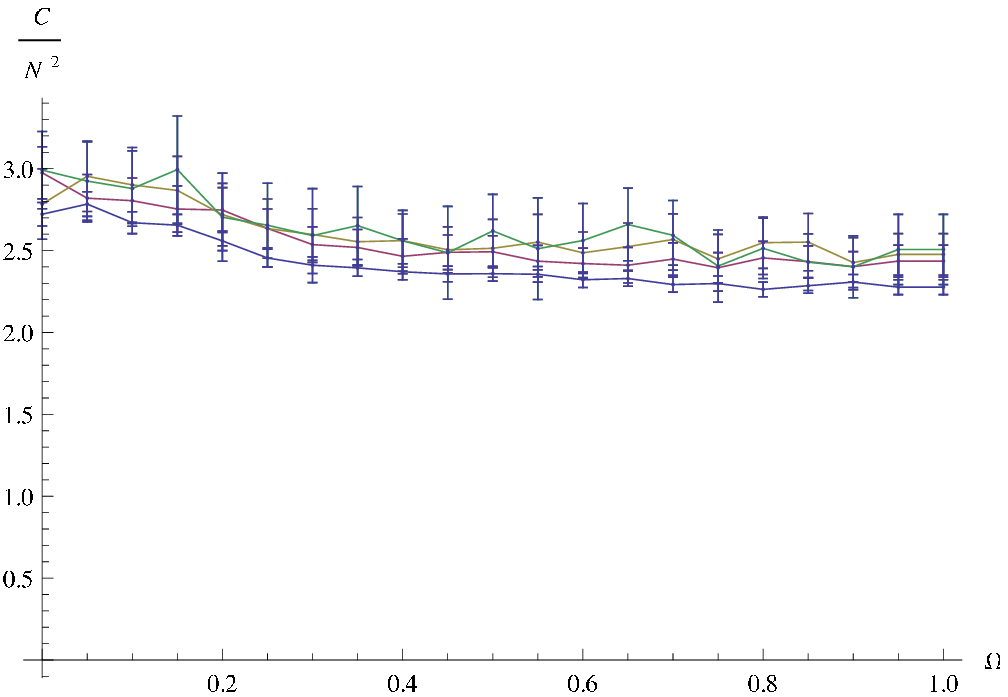}
\includegraphics[scale=0.5]{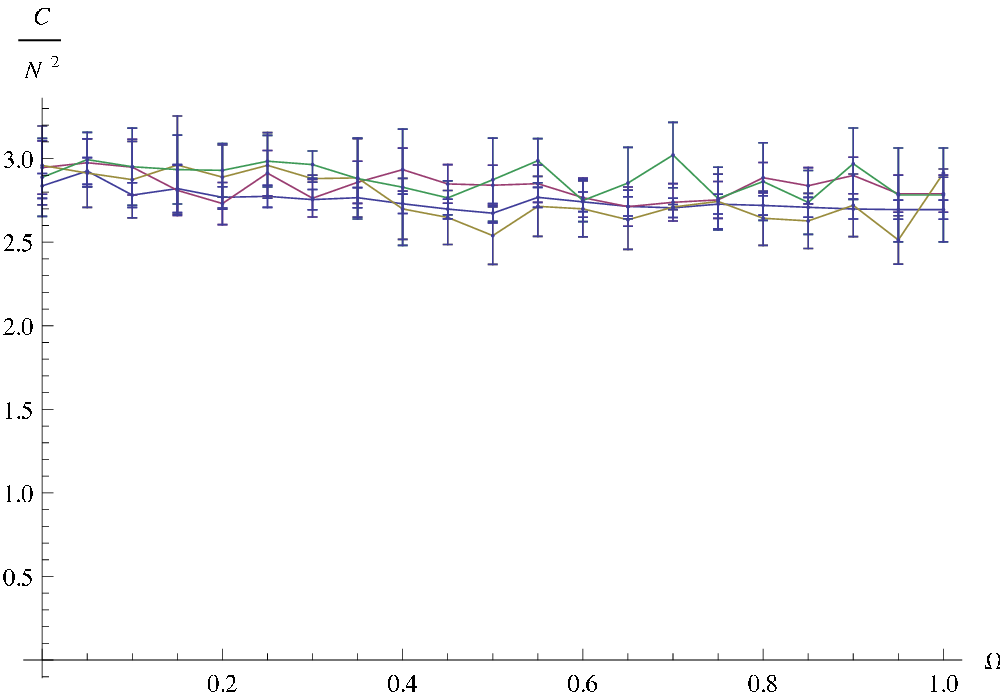}
\end{center}
\caption{\footnotesize Specific heat densities for $\mu=1$ (left) and $\mu=0$ (right) varying $\Omega$ and $N$.\normalsize}\label{Figure 26}
\end{figure}

\begin{figure}[htb]
\begin{center}
\includegraphics[scale=0.45]{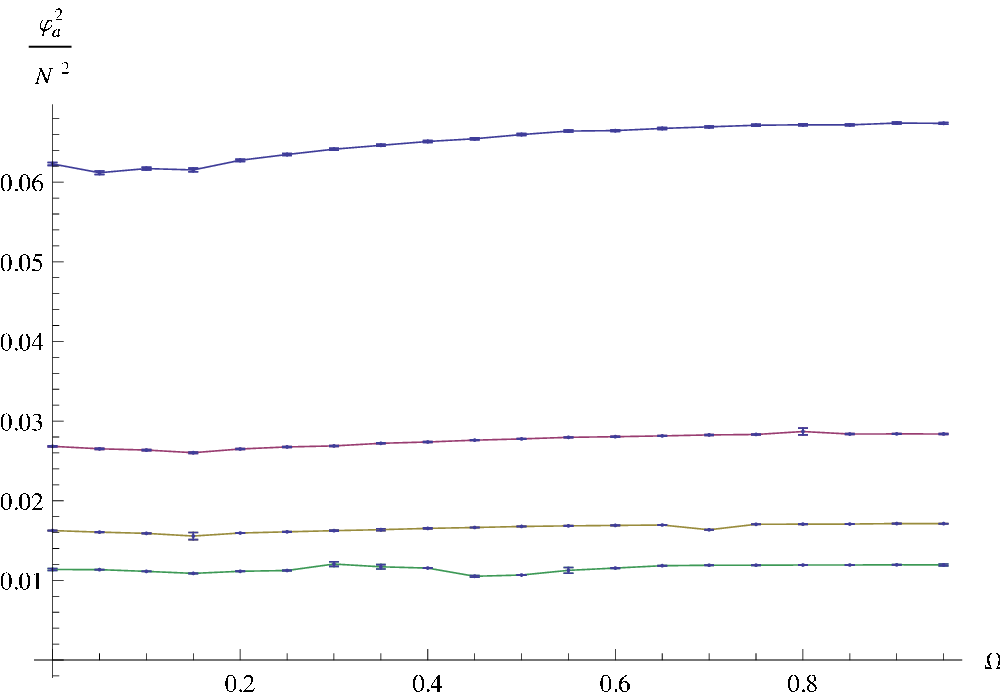}
\includegraphics[scale=0.45]{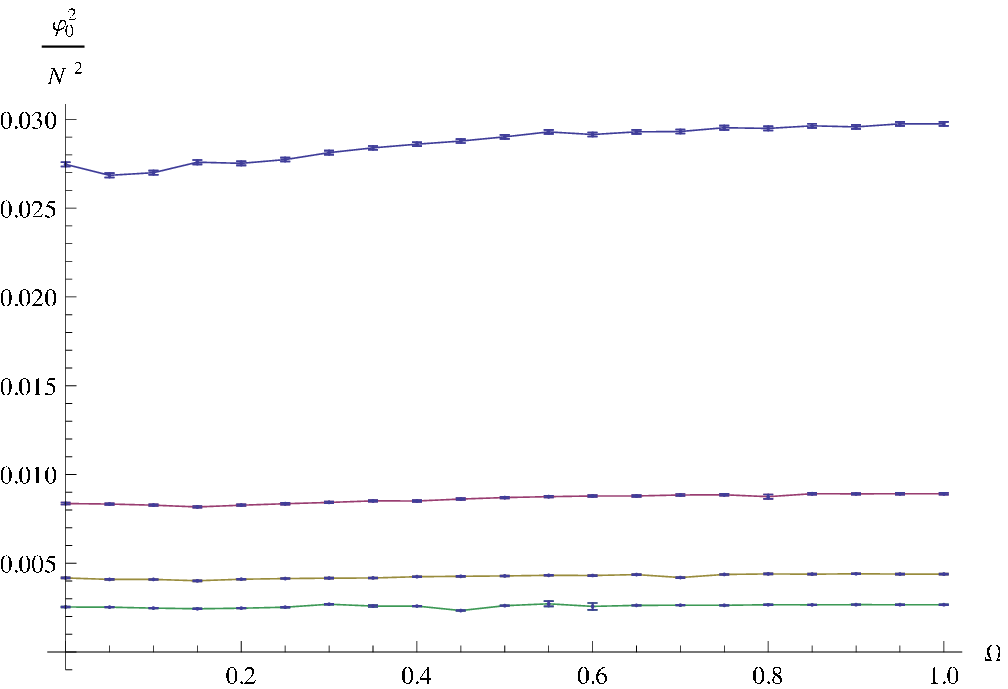}
\includegraphics[scale=0.45]{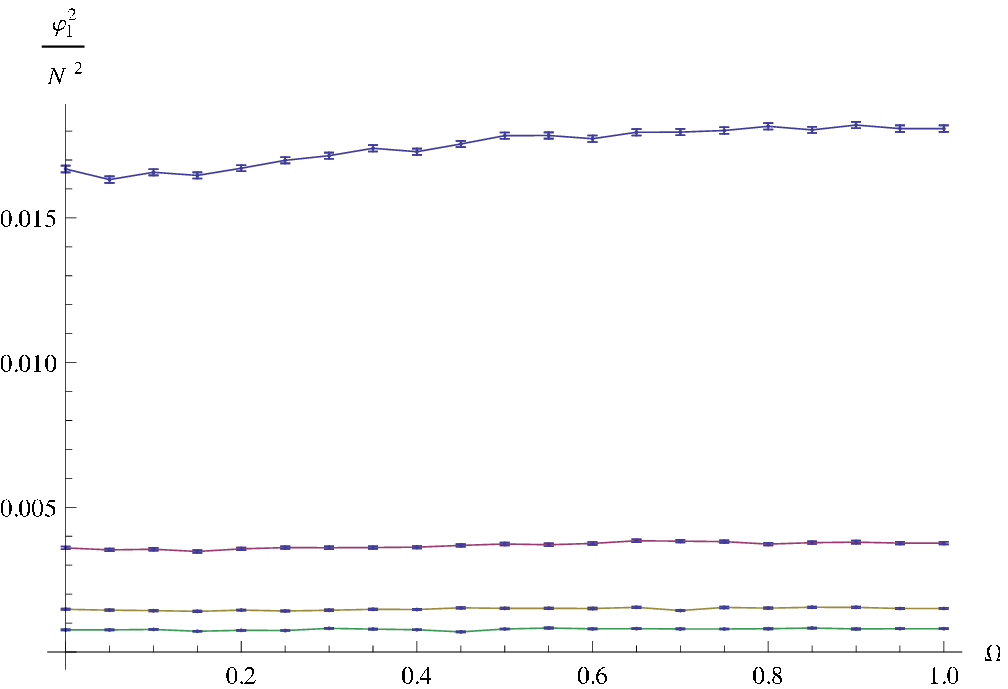}
\includegraphics[scale=0.45]{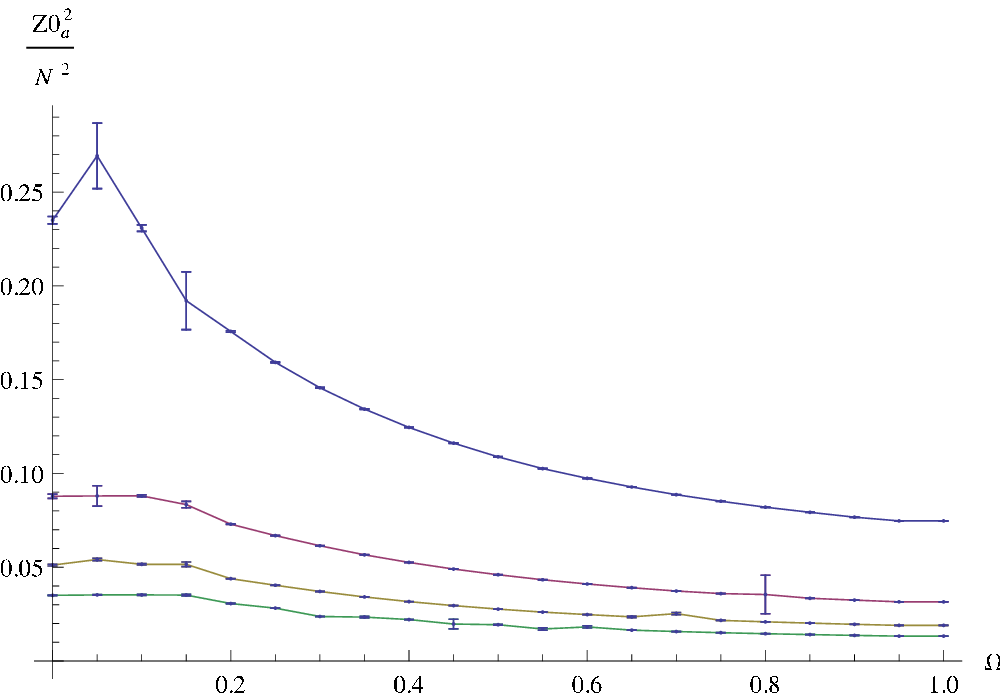}
\includegraphics[scale=0.45]{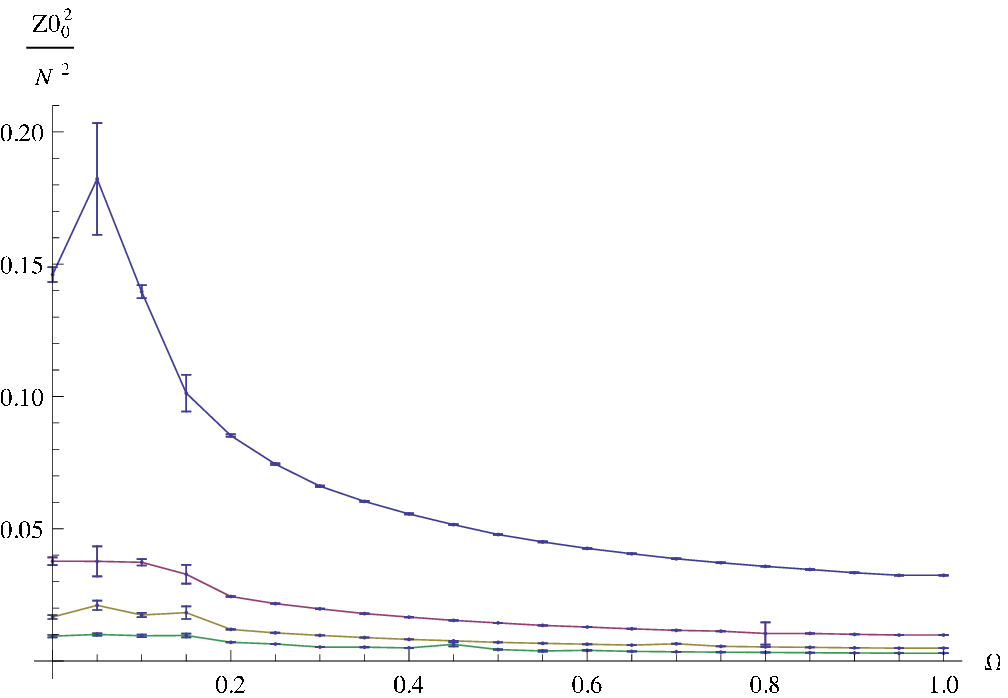}
\includegraphics[scale=0.45]{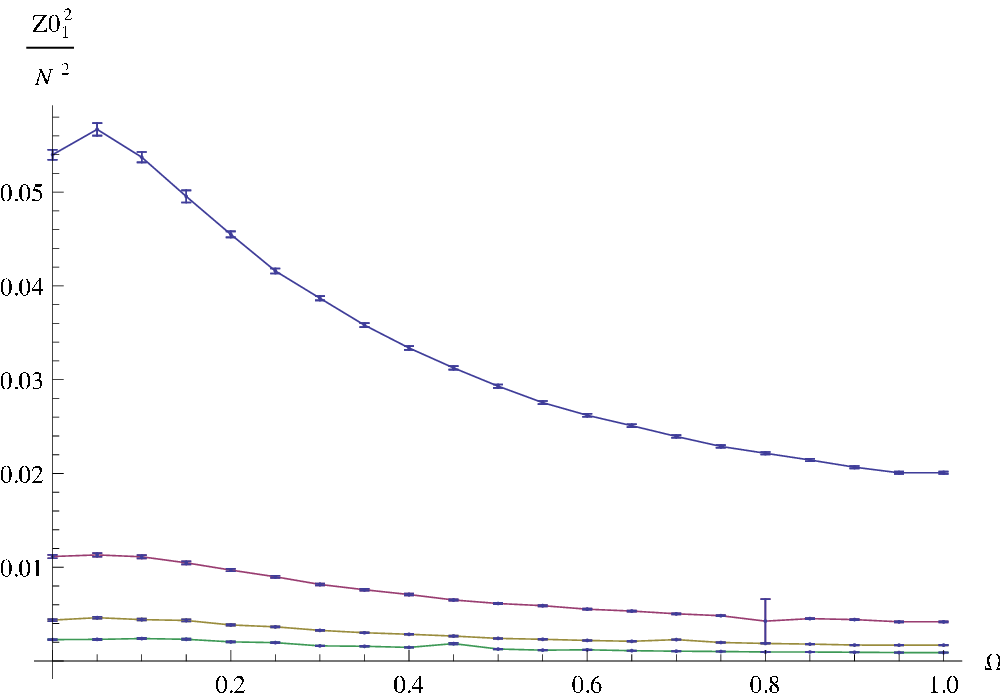}
\end{center}
\caption{\footnotesize Starting from the up left corner and from the left to the right the 	densities for $\varphi_a^2 $, $\varphi^2_0 $, $\varphi^2_1 $, $Z_{0a}^2$, $Z_{00}^2 $ and $Z_{01}^2$ for $\mu=0.5$ varying $\Omega$ and $N$.}\label{Figure 27}
\end{figure}

The behavior of the energy density for $\mu=\{1,0\}$ is the same of the full model, in the graphs fig \ref{Figure 25} are plotted  the total energy density and the comparison of the various contributions for $\mu=\{1,0\}$. As we can see there is no evident difference from the 4-dimensional case; even in this case the total energy  for $\mu=1 $ follows the slope of the $D$ contribution and the total energy for $\mu=0$ is constant. Even the various contributions  have the same trend of the full model; for $\mu=1$ the $V$ and $F$ have an ``opposite`` slope  and for $\mu=0$
they balance each other in order to obtain a constant sum.
The specific heat density  shows fig.\ref{Figure 26} some differences:  for $\mu=0$ the peak  is no more located around $\Omega=1$ but it appears, not so clearly, around zero  and the specific heat for $\mu=0$ does not show any particular point. 
In the figure fig.\ref{Figure 27} are showed the graphs for $\varphi_a^2 $, $\varphi^2_0 $  and $\varphi^2_1 $ and $Z_{0a}^2 $, $Z_{00}^2$  and $Z_{01}^2 $ for $\mu=1$.
For brevity are  showed the graphs only for $Z_0$, however  even for the  2-dim model taking in account the statistical errors they are compatible with  the $Z_1$ field related graphs.
In this results there are no significant changes with the 4-dim model, indeed  comparing the three graphs fig.\ref{Figure 27} it easy to see the dominance of the spherical contribution $\varphi^2_0 $ to the full power of the field and for the $Z_0$ fields the spherical contribution becomes dominant approaching to $\Omega=0$ starting from a zone in which the contributions of $Z_{00}^2$ and $Z_{01}^2 $ are comparable. Beside, we notice a decrease of values of the quantiles related to $\psi$  with $N$,  nevertheless the dominance of the $\varphi_0$ on the total power of the field remains independently by $N$. The peak appearing in $Z_{0a}^2$ and $Z_{00}^2 $  decrease, but if look at the single graph for fixed $N$ the spherical contribution approaching the point $\Omega=0$ still dominant. The graphs of the previous quantities for $\mu=0$  show the same slope as the full model so will be not presented here. 

\begin{figure}[htb]
\begin{center}
\includegraphics[scale=0.45]{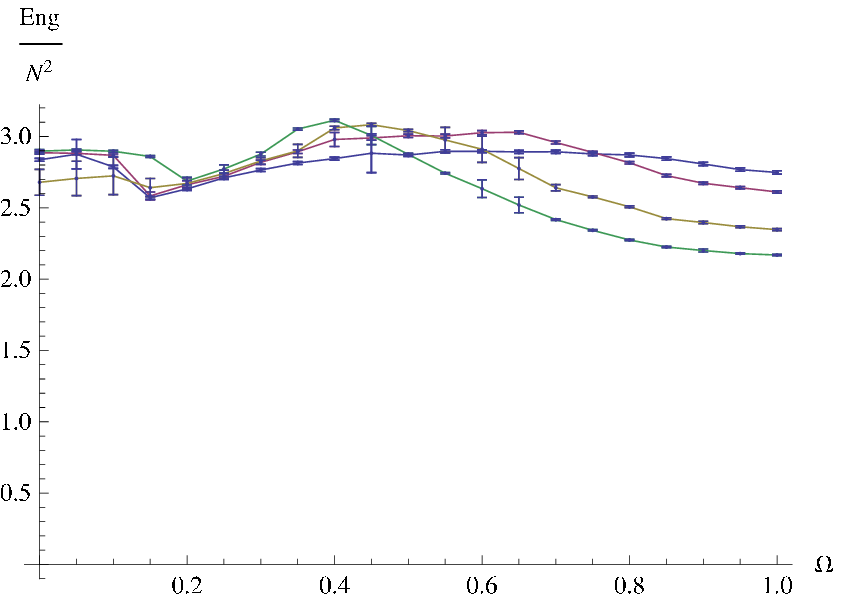}
\includegraphics[scale=0.45]{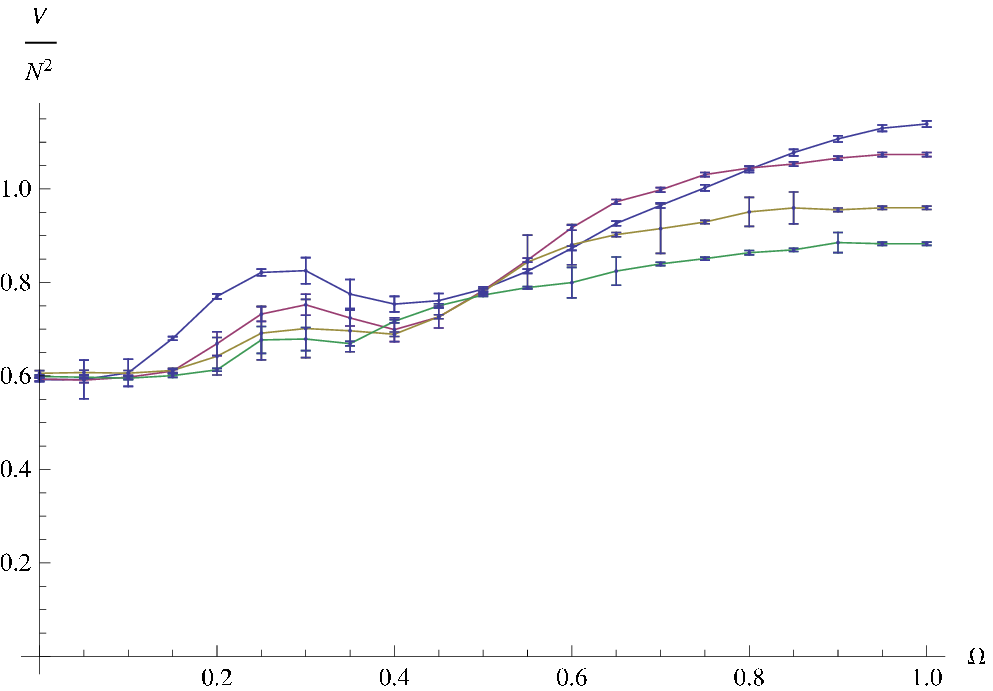}
\includegraphics[scale=0.45]{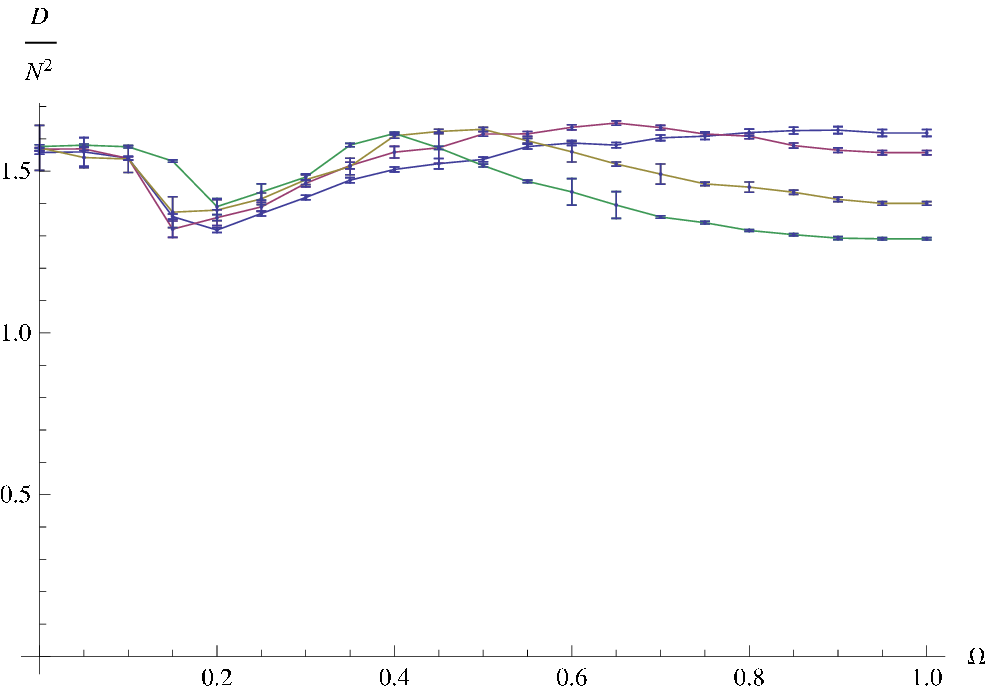}
\includegraphics[scale=0.45]{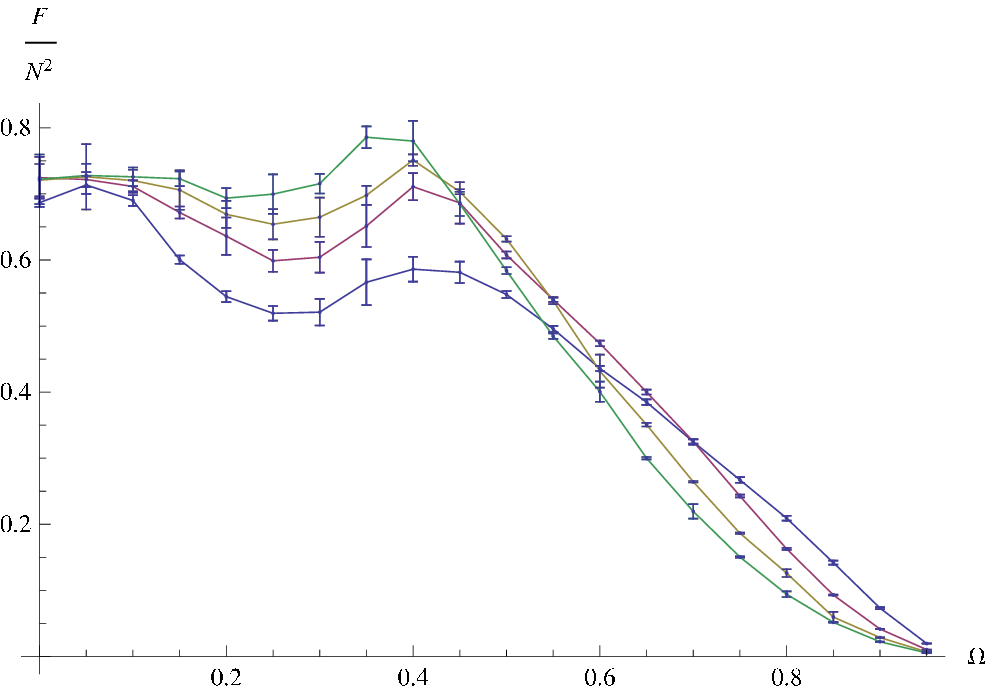}
\includegraphics[scale=0.46]{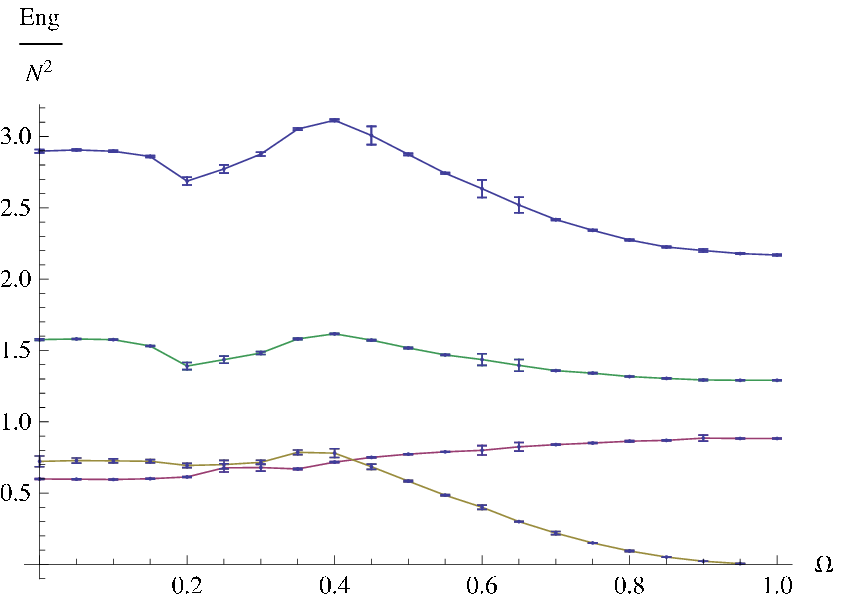}
\end{center}
\caption{\footnotesize  Total energy density and contributions  for $\mu=1$ varying $\Omega$ and $N$. From the left to the right  $E$, $V$, $D$, $F$ and comparison.\normalsize}\label{Figure 28}\end{figure}

A radical change emerges for the case $\mu=3$ as in the full model, but the plots of the energy density are very different fig.\ref{Figure 28}. From the comparison of the various contributions with the total energy density for $N=20$, we  deduce that the total energy density seems to follow the $D$ contribution instead $F$ as in the 4-dim model. This can be justified by the presence of additional terms in the full model coming from the Y-M part. \newpage
\begin{figure}[htb]
\begin{center}
\includegraphics[scale=0.7]{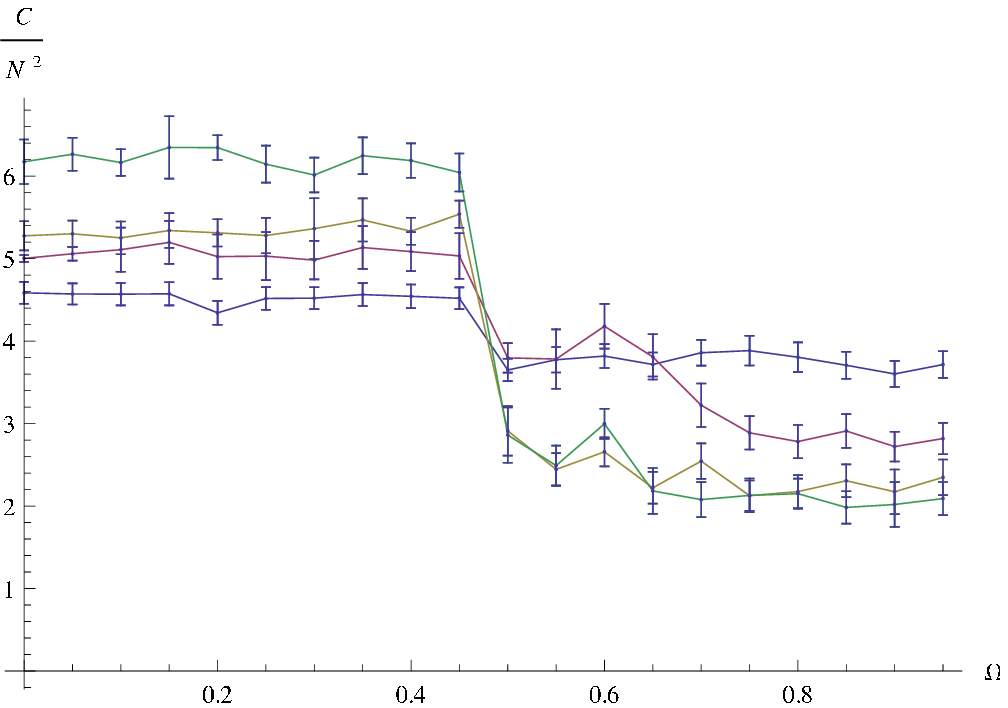}
\end{center}
\caption{\footnotesize Specific heat density for $\mu=3$  varying $\Omega$ and $N$.\normalsize}\label{Figure 29}
\end{figure}
\begin{figure}[htb]
\begin{center}
\includegraphics[scale=0.5]{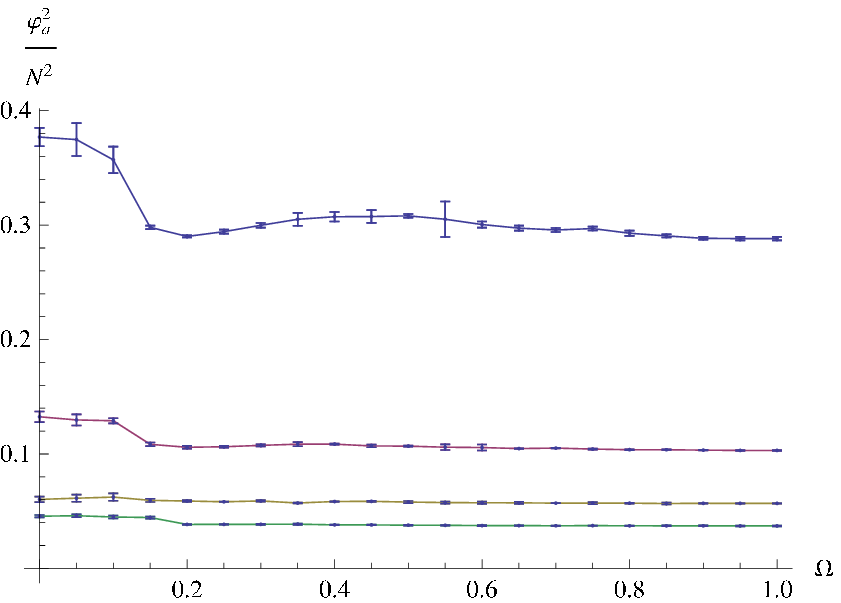}
\includegraphics[scale=0.5]{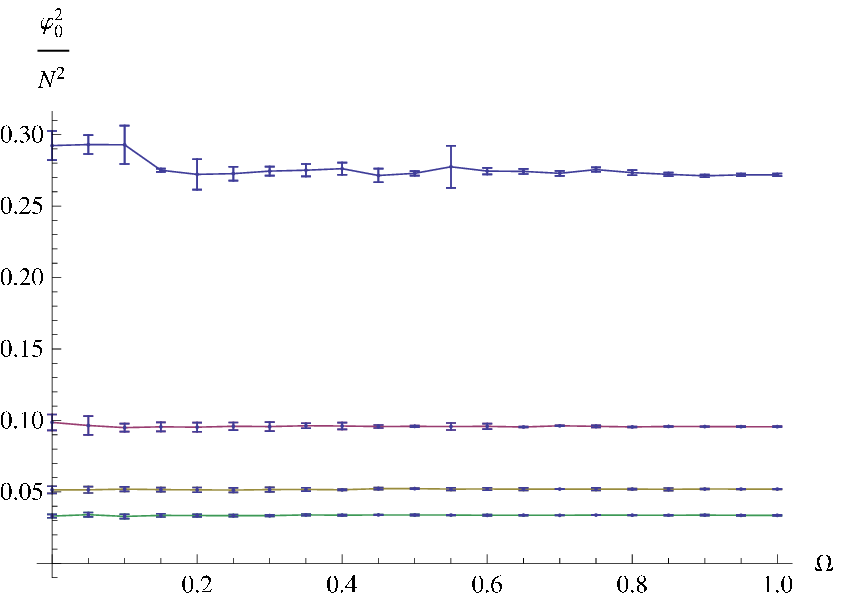}
\includegraphics[scale=0.5]{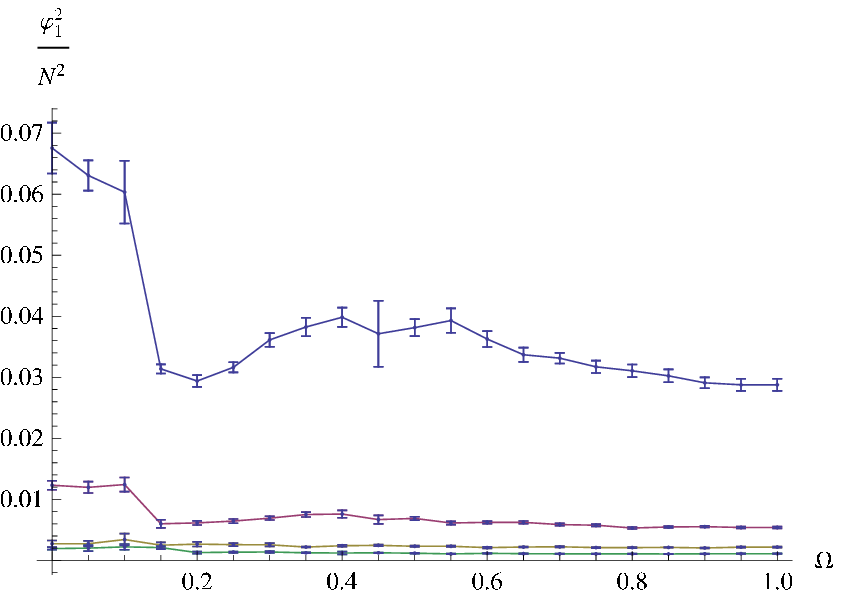}
\includegraphics[scale=0.5]{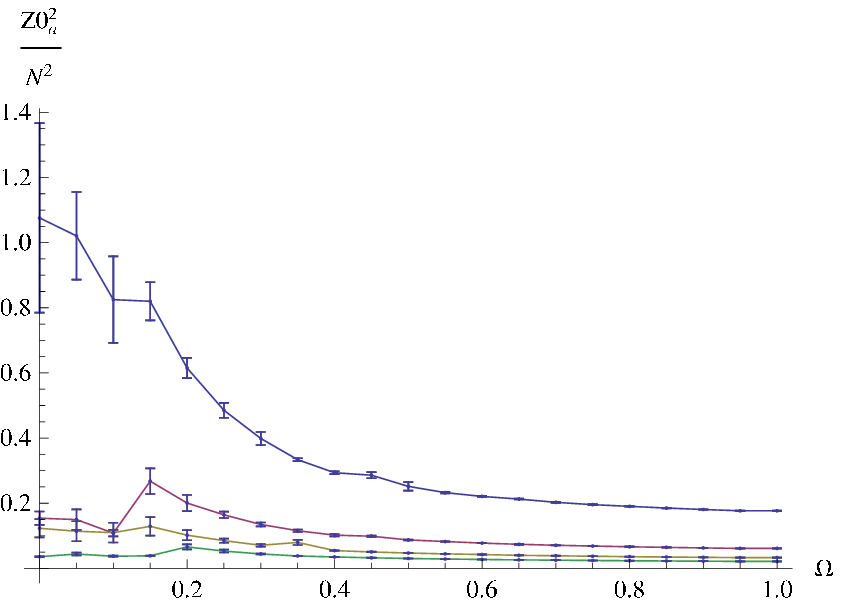}
\includegraphics[scale=0.5]{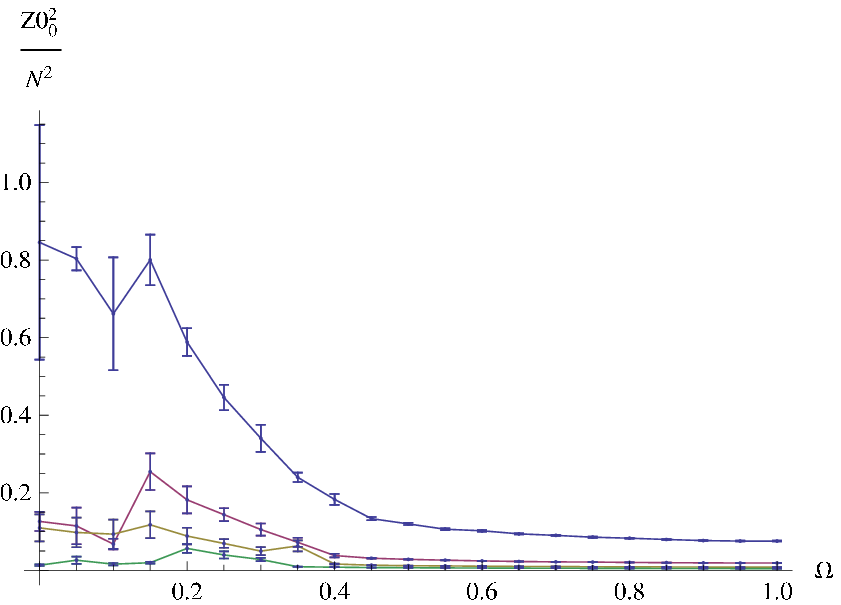}
\includegraphics[scale=0.5]{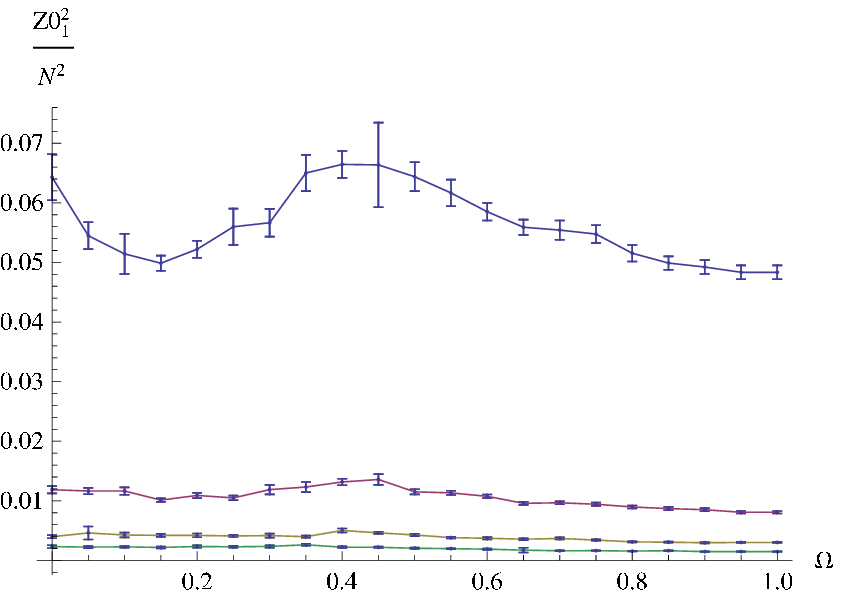}
\end{center}
\caption{\footnotesize Starting from the up left corner and from the left to the right the 	densities for $\varphi_a^2 $, $\varphi^2_0 $, $\varphi^2_1 $, $Z_{0a}^2$, $Z_{00}^2 $ and $Z_{01}^2$ for $\mu=3$ varying $\Omega$ and $N$.\normalsize}\label{Figure 30}
\end{figure}In this case the specific heat density does not shows fig.\ref{Figure 29} the peak for $\Omega=1$, but we discover a step around $\mu\approx 0.5$ this seems to becomes sharper as $N$ increase. Regarding the order parameters $\varphi_a^2 $, $\varphi^2_0 $, $\varphi^2_1 $ and $Z_{0a}^2$  $Z_{00}^2 $, $Z_{01}^2$ they have similar behavior of the full model with some more evident oscillations around the origin. 

\subsection{Varying $\mu$ }
\begin{figure}[htb]
\begin{center}
\includegraphics[scale=0.45]{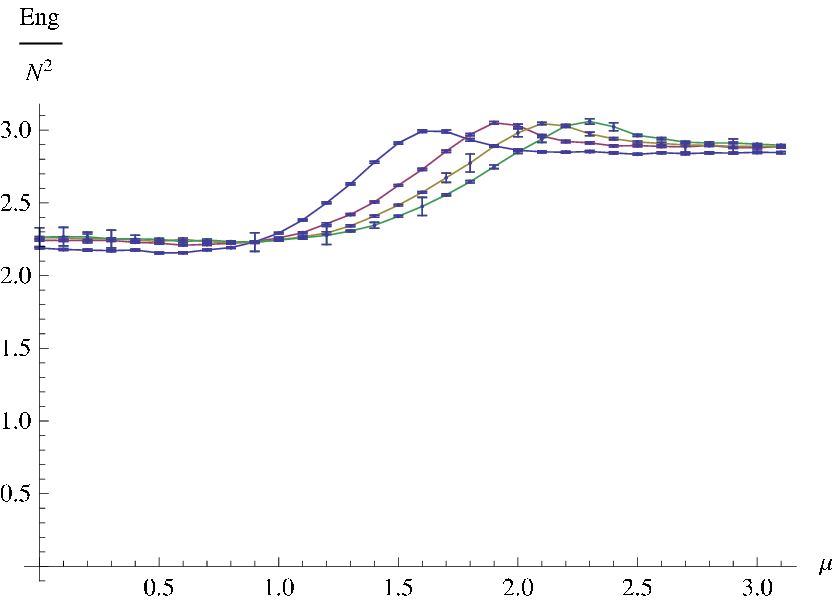}
\includegraphics[scale=0.45]{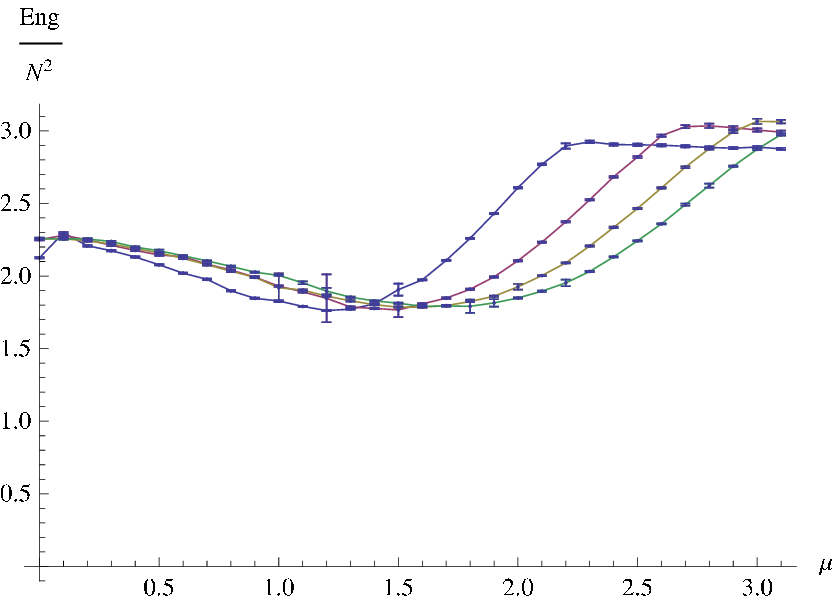}
\includegraphics[scale=0.45]{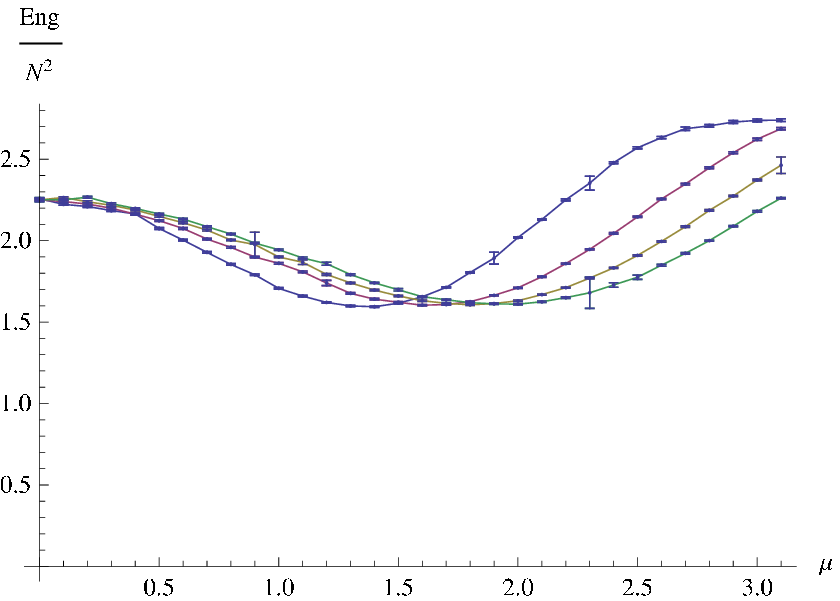}
\includegraphics[scale=0.45]{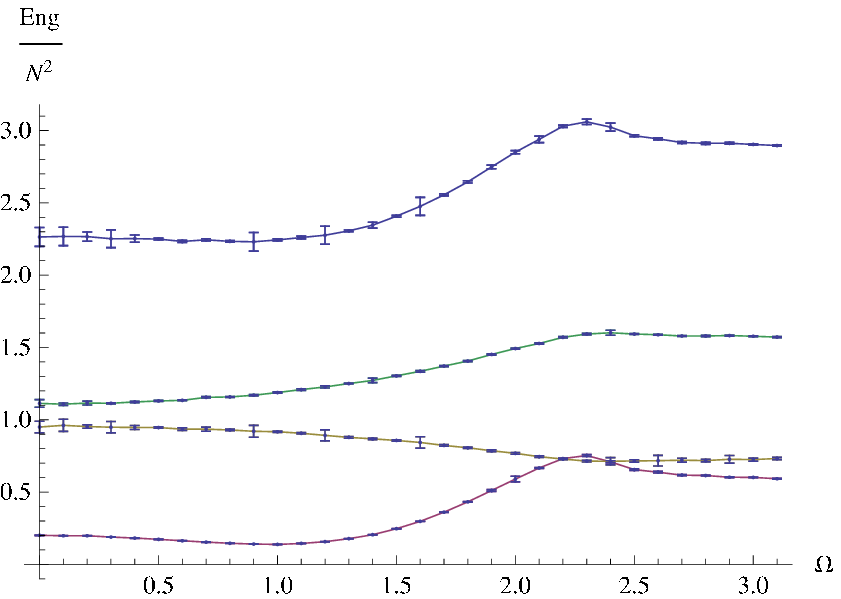}
\includegraphics[scale=0.45]{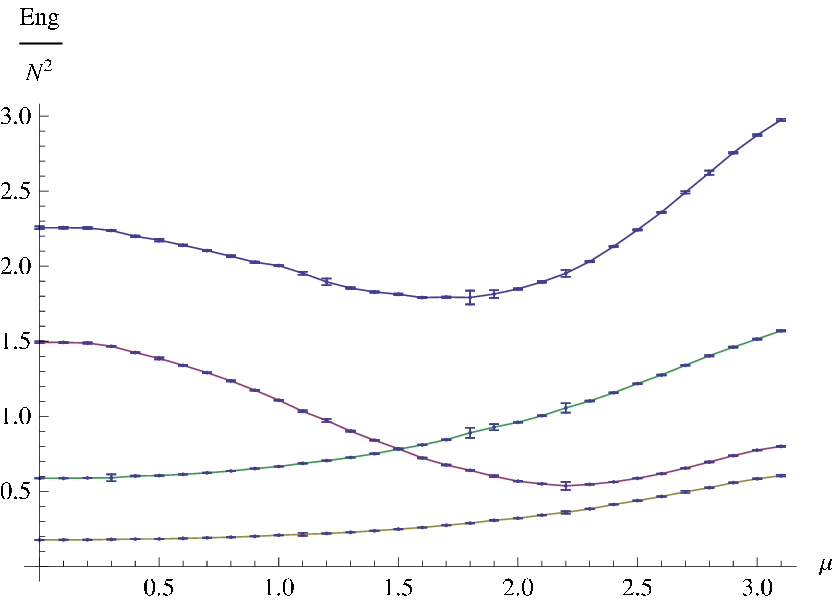}
\includegraphics[scale=0.45]{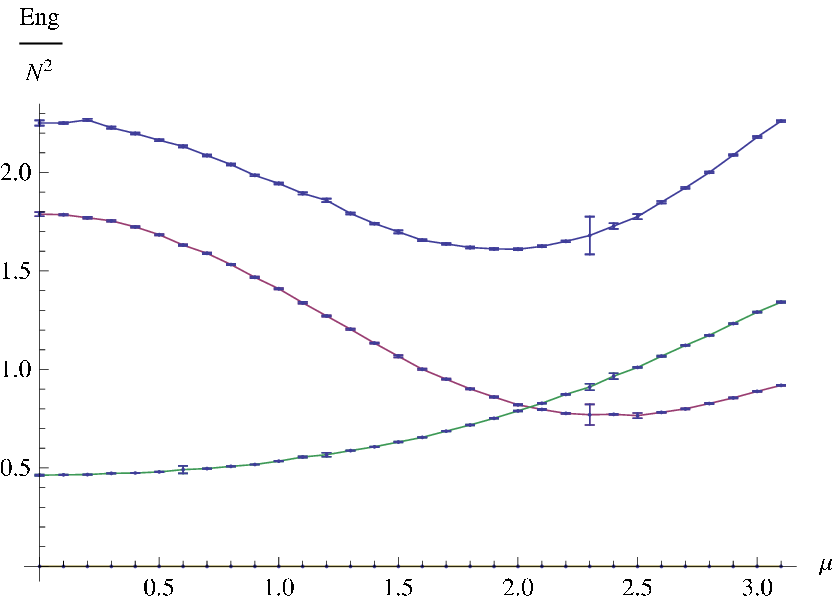}
\end{center}
\caption{\footnotesize Total energy density and contributions  for $\mu=1,\alpha=0$ varying $\Omega$ and $N$. From the left to the right  $E$, $V$, $D$, $F$ and comparison with $N=5$ (blue), $N=10$ (purple), $N=15$ (brown), $N=20$ (green).\normalsize}\label{Figure 31}\end{figure}

The resemblance between the two models is  more evident when we compare the plots obtained fixing $\Omega=\{0,0.5,1\}$ and varying $\mu$. The graphs in fig.\ref{Figure 31} show the total energy density and the various contributions  for $\Omega=\{0,0.5,1\}$. It is very clear the similarity with the full model; for $\Omega=0$ appears a peak  in the total energy density around $\mu\approx 2.3$ for $N=20$, for $\Omega=0.5$ and $\Omega=1$ the peaks are shifted. The various contributions follows the same behavior of the full model where,  in the case $\Omega=1$, the $F$ component is absent due to the vanishing of the prefactor in front of the Y-M part of the action. 
\begin{figure}[htb]
\begin{center}
\includegraphics[scale=0.5]{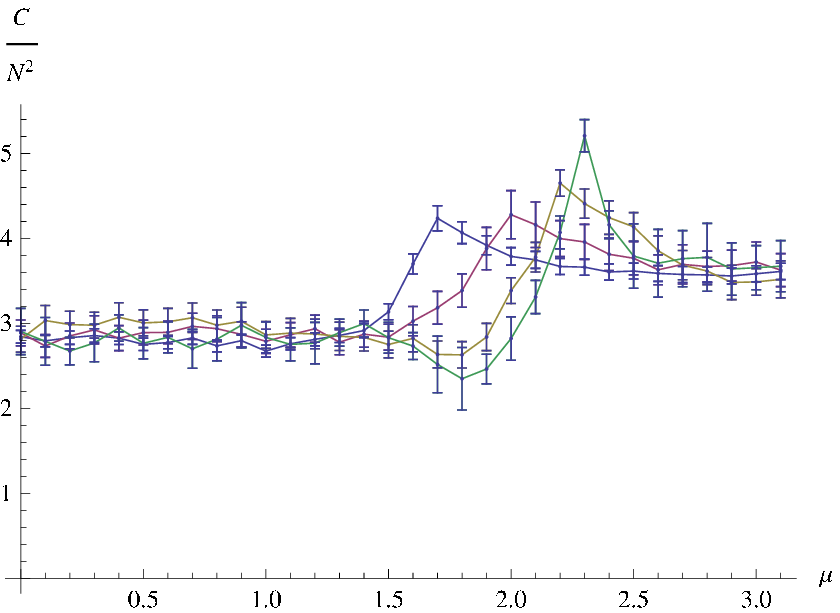}
\includegraphics[scale=0.5]{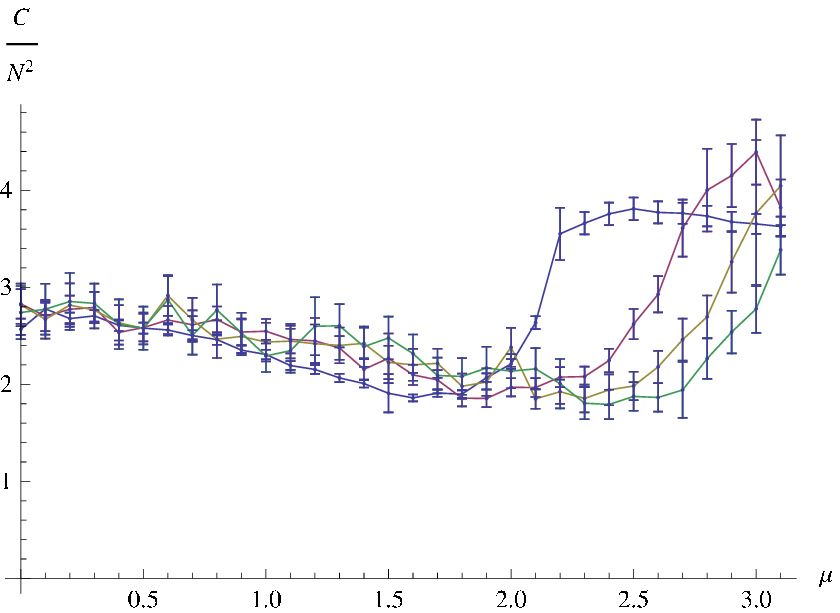}
\includegraphics[scale=0.5]{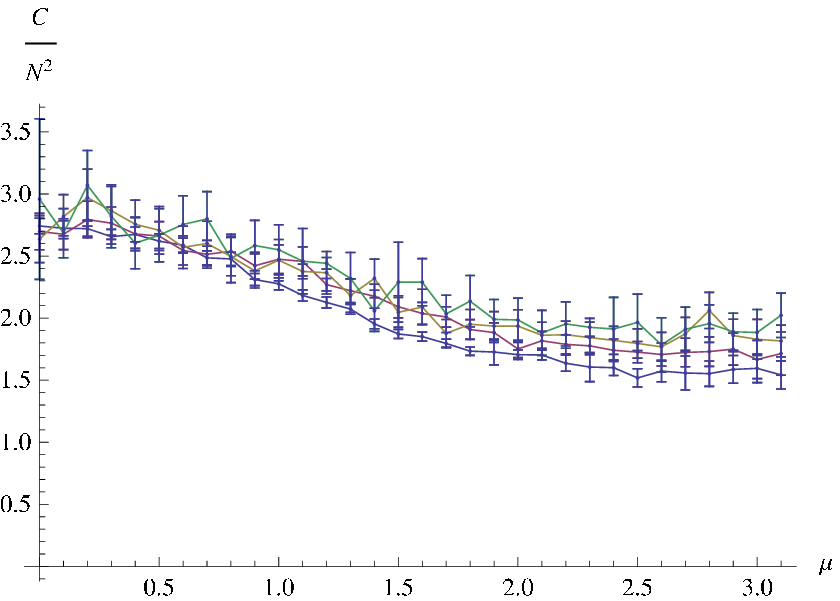}
\end{center}
\caption{\footnotesize Specific heat density for $\Omega=0$ (left) and $\Omega=0.5$ (center), $\Omega=1$ (right) varying $\mu$ and $N$.\normalsize}\label{Figure 32}
\end{figure}\newpage
The resemblance persists in the specific heat density too, in fig.\ref{Figure 32} the peaks are located as the 4-dim model and for $\Omega=1$  does not appear any peak. \\

\begin{figure}[htb]
\begin{center}
\includegraphics[scale=0.5]{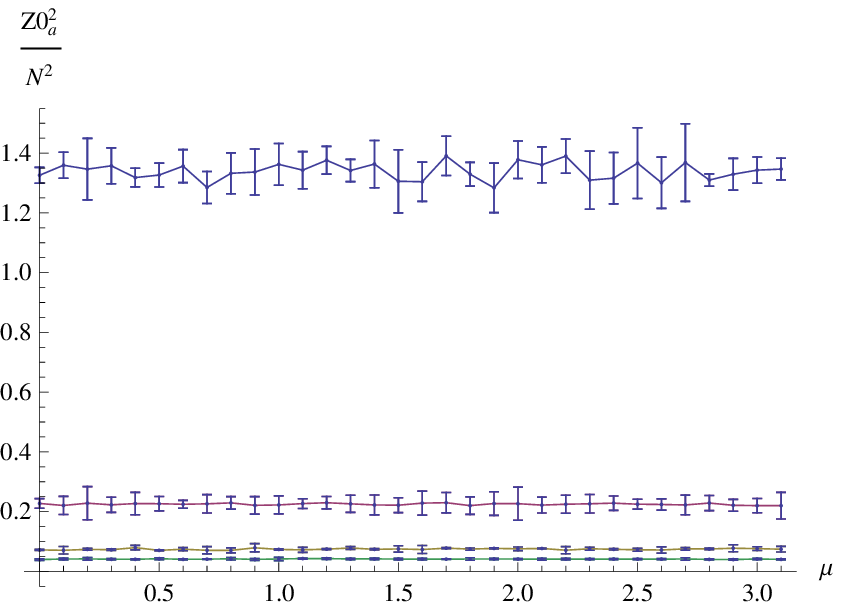}
\includegraphics[scale=0.5]{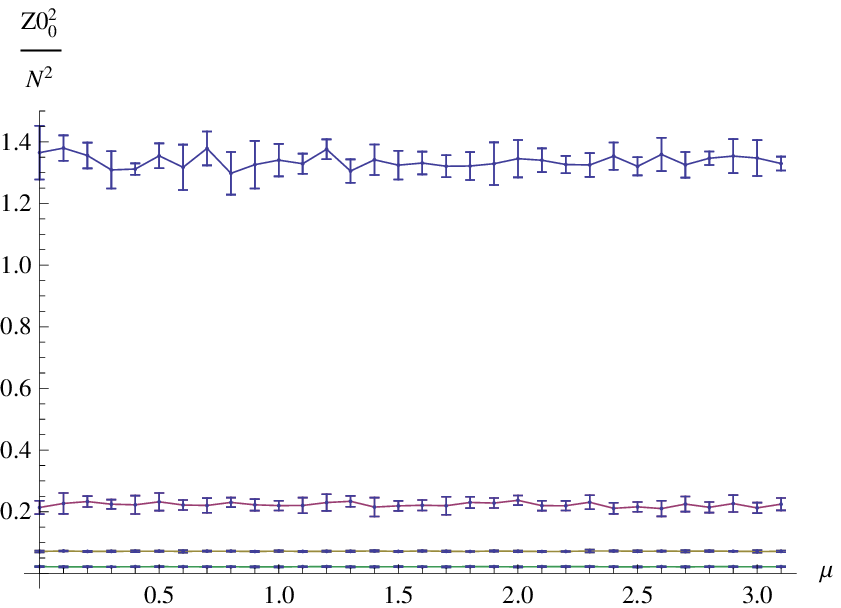}
\includegraphics[scale=0.5]{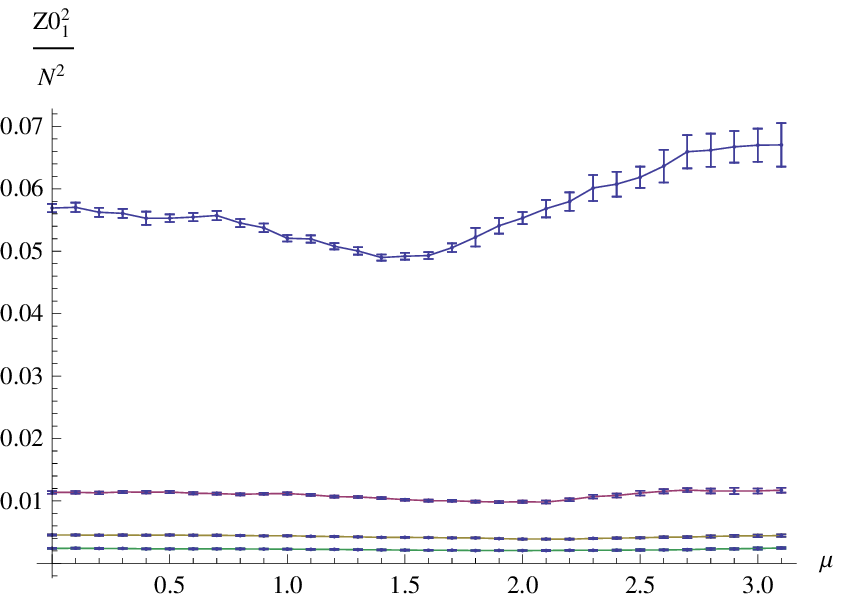}
\end{center}\caption{\footnotesize From the left to the right the 	densities for $Z_{0a}^2$, $Z_{00}^2 $ and $Z_{01}^2$ for $\Omega=0$ varying $\mu$ and $N$.\normalsize}\label{Figure 34}
\end{figure}At last in fig.\ref{Figure 34} is displayed the behavior of the order parameters density of $Z_0$ for $\Omega=0$, even in the 2-dimensional case the slopes of $Z_{0a}^2$, $Z_{00}^2 $ densities are constant within the errors bars. The other  orders parameters  are identical to the full model related graphs and are compatible with a dilatation of the previous diagrams for $\Omega=0.5$.

%% file: concl.tex
\chapter*{Conclusions and prospectives}

In this thesis we have presented a study of a spectral action model constructed in order to extend to the Yang-Mills theories the non-commutative  Wulkenhaar-Grosse model.
The main aim of this work was to test a first Monte Carlo approach based on a non-perturbative regularization method  alternative to the standard
lattice regularization, which in our case is not suitable due to the oscillator factor of the Moyal product. We have performed Monte Carlo simulations and obtained the values of the defined observables varying the parameters of the system. Despite the complexity of the approximated spectral action considered here we were able to obtain some reliable numerical results, we can conclude that a numerical approach of this kind of model using the matrix Moyal base seems feasible. The specific heat density shows various peaks indicating  phase transitions, in particular studying the behaviors for some fixed $\mu$  we found a peak around $\Omega=0$ for $\mu=\{0,1\}$ and peak in $\Omega=1$ for $ \mu=3$, beside we notice a huge change in the energy density and in its contributions between the cases $\mu=\{0,1\}$ and $\mu=3$. Others peaks in the specific heat density can be found varying $\mu$ and fixing $\Omega$, the graphs show that increasing $\Omega$  the peak in the specific heat, starting from  $\mu\approx 2.4$ for $\Omega=0$,  is moved towards higher $\mu$. The order parameters introduced show a strong dependence on the occurrence of $\Omega=0$ or $\Omega\neq0$. Referring to the fixed $\mu$ graphs we found a peak in the spherical contribution for the gauge fields $Z_i$, we can interpreted this slope as a sort of symmetry breaking introduced by $\Omega\neq0$. Additionally, varying $\mu$ and fixing $\Omega$ the other parameters display an increasing slope with  $\mu$ for all fields and all situation but one;  the graphs of  the order parameters concerning  $Z_{0a}$, $Z_{00}$  for $\Omega=0$ show a constant behavior. As an aside result we have ran the same computations on the simpler 2-dimensional model, from the comparison despite many similarities in the results varying $\mu$, we note a complete different trend of the specific heat varying $\Omega$, the peaks present in the full model does not appear any more in the 2-dimensions one.  From this comparison we deduce that the dimension is a crucial parameter of the model. About the prospective, there is a lot of work to do; the  natural next steps in the numerical study of this model, could be the computation of the transition curves in order to separate the phase regions and classify them using eventually some additional  order parameters. Our treatment, forced by limited resource, was conducted conjecturing  that the system can fully described varying $\Omega$ in the range $[0,1]$ but since the L-S duality does not hold any more in our case, will be very useful to extend this range. Actually, the computed graphs does not show any periodicity in  $ \Omega\in[0,1]$ so there are some hints to infer that the range $[0,1]$ is not enough, however only a direct computation will  clarify this point.
At last, in analogy with the matrix model will be very interesting to do not require any more the condition $\mu^2>0$, in order to implement this change we have to conduct the calculation no more around the minimum of the actions. The problems that can occur are related the thermalization process, which can be much more problematic due to the eventual emerging of local minimum.
The expansion of the parameters space, together with the classification of the different phase regions allows us to compare our model with the results of the simulation conducted on the fuzzy spaces, looking in particular about the occurrence of the so called non-uniformly ordered phase which is connected with the UV/IR mixing. Since we have constructed our model starting from a renormalizable one this study is very desirable.

%% file: appt.tex
\chapter{Axioms  non-compact  geometries} 

\emph{In this appendix will be introduced the definition of non-compact spectral triple and the modified set of axioms required for such spectral triple. In the last part of the chapter will be hinted a more rigorous treatment of the spectral triple used in chapter 3, extending it in the setting of non-compact triple.}

\section{General remarks}
A compact spectral triple is a collection $(\mathcal{A},\mathcal{D},\mathcal{H})$ where $\mathcal{A}$ is a unital pre-$C^*$-algebra,
$\mathcal{H}$ is a Hilbert space on which $\mathcal{A}$ is represented  by bounded operators and $\mathcal{D}$ is a
Dirac selfadjoint operator on $\mathcal{A}$, with  resolvent $R_\mathcal{D}(\lambda)=(\mathcal{D}-\lambda)^{-1}$  compact and  the commutator $[\mathcal{D}, a]$ is bounded on $\mathcal{H}$ for every $a \in \mathcal{A}$.
In the non-compact case this definition is modified to allow $\mathcal{A}$ to be no longer unital.  
The usual way to construct a non-commutative spin manifolds, for the unital case, is to  start from an operatorial version of ordinary spin geometry  that can be generalized  to a non-commutative manifold. This generalization is obtained using a reconstruction theorem in order to recover all of the structure from the abstract geometry over a suitable  algebra \cite{bondia,Connes2,Rennie}. However, there is no proper direct reconstruction theorem for  non-compact Riemanian
spin manifolds, so the question of axioms for non-unital spectral triples is  guided by the resemblance to the compact case,  requiring  a collection of   axioms for not
necessarily compact non-commutative manifolds. Of course, such  list of conditions should be compatible
with the previous axiomatic framework and be fulfilled by non-compact commutative manifolds.
Modified systems of axioms for non-compact spectral triples has been formulated
by A. Rennie \cite{Rennie1} for the commutative case and by Gayral et al. \cite{moyal-triple}
in a fully non-commutative framework on the Moyal plane. 
A very important aspect for these formulations  is the introduction of various  different 
types of non-commutative dimension \cite{Carey}. Indeed, the non-compact spectral triple analyzed  will be characterized
by different metric and KO-dimensions. When $\mathcal{A}$ is not unital, we need to choose a preferred unitization  $\mathcal{B}$. For compact or unital spectral triples it is further required that the operator $\mathcal{D}$ has a compact resolvent, or in other words, $(\mathcal{D}-\lambda)^{-1}$ must belong to $K(\mathcal{H})$ for $\lambda \notin sp\mathcal{\mathcal{D}}$, where $K(\mathcal{H})$ denote the compact operators on $\mathcal{H}$. Consequently $\mathcal{D}$ must have discrete spectrum of finite multiplicity. Since this is clearly not the case of Dirac operator ${\not}\mathcal{D}$ on $ \mathbb{R}^d$, in the nonunital case we only demand \cite{Connes2} that $a(\mathcal{D}-\lambda)^{-1}$ is compact for $a \in \mathcal{A}$. This condition ensures that the spectral
triple $(\mathcal{A},\mathcal{D},\mathcal{H})$ corresponds to a well-defined $d$-homology class \cite{Higson}
for an, in general, non-compact non-commutative geometry.
Now we analyze some modified conditions which spectral triples are required to satisfy. 
We use a regularity condition \cite{non-com-tri}  in terms of the operator $\langle \mathcal{D} \rangle =(\mathcal{D}^2+1)^{\frac{1}{2}}$, in this way we avoid the hypothetical zero modes of the Dirac operator. Can be shown that the previous regularity condition coincides with the original definition of Connes and Moscovici \cite{local-index}.  
The condition of finiteness, in the unital case, is that the smooth domain $\mathcal{H}^\infty$ of $\mathcal{D}$ in $\mathcal{H}$ can be finitely generated by a projective left module over the unital algebra $\mathcal{A}$, or $\mathcal{H}^\infty \sim \mathcal{A}^mp$ for some projector $p$ in $M_m(\mathcal{A})$ with a suitable $m$.
In the case of $\mathbb{R}^d$, using the commutative or the Moyal product, the module of
smooth spinors is free, since the triviality of the spinor bundle. 

When $\mathcal{A}$ is nonunital, in order to obtain a projective $\mathcal{A}$-module, we have to find the projector $p$ in a matrix algebra over the preferred compactification  $\mathcal{B}$, therefore this condition for the nonunital case should then require that $\mathcal{H}^\infty$ densely contains a pullback of a finite projective $\mathcal{B}$-module to $\mathcal{A}$. The reality axiom is the same as the  compact case, thus is the
existence of an antilinear conjugation operator $J$ on $\mathcal{H}$ such that a $ \to Ja^{*}J^{-1}$, which define an another representation of $\mathcal{A}$ on $\mathcal{H}$ commuting with the previous one. The required algebraic properties of the real structure $J$ are the same  as the compact case. Even for the first order condition, there is no need to modify the axiom in the non-compact case, we have: 
\begin{equation}
 [[\mathcal{D}, a], Jb^{*}J^{-1}] = 0, \ \textrm{for all} \ a, b \in \mathcal{A}. \label{h-cycle}
\end{equation}
It is convenient formulate the orientability condition in the compact case as the property that the spectral triple $(\mathcal{A},\mathcal{H},\mathcal{D})$ contains an algebraic version of
a volume $d$-form. 
This algebraic version consists of an Hochschild $d$-cycle $c$, which is a sum of terms of the
form $(a_0\otimes b^o_0 )\otimes a_1 \otimes \cdots \otimes a_d$ satisfying  the cycle property $bc = 0$. Defining
 $b_0^o$  as  an element of the opposite algebra $\mathcal{A}^o$ with $b_0 \in \mathcal{A}$, 
we can represent the Hochschild $d$-cycle by bounded operators:
\begin{equation}
\pi_\mathcal{D}((a_0 \otimes b^o_0 ) \otimes a_1 \otimes \cdots \otimes a_d)= a_0Jb^{*}_0 J^{-1} [\mathcal{D}, a_1] \cdots [\mathcal{D}, a_d].
\end{equation}
on which we impose the orientation $\pi_\mathcal{D}(c) = \chi$, where $\chi$ is the given $\mathbb{Z}_2$-grading operator
on $\mathcal{H}$. We just use $\chi = 1$ if $d$ is odd and  in the even case, for ordinary spinors, we use
\begin{equation}
\chi = (-i)^m\gamma^1\gamma^2\cdots\gamma^{2m}
\end{equation}
For nonunital algebras, we ask the same condition  taking the cycle over the
unitization $ \mathcal{B}$ rather than $\mathcal{A}$ itself.

\section{Modified conditions for nonunital spectral triples}
By a real non-compact spectral triple of dimension $k$ we connote the data \\
$(\mathcal{A},\mathcal{B},\mathcal{H},\mathcal{D}, J, \chi),$ where $\mathcal{A}$ is an algebra, in general nonunital, acting faithfully via a representation 
denoted by $\pi$ on the Hilbert space $\mathcal{H}$, $\mathcal{B}$ is a preferred unitization of $\mathcal{A}$ acting on the same Hilbert
space and $\mathcal{D}$ is an unbounded selfadjoint operator on $\mathcal{H}$ such that $[\mathcal{D}, a]$ for each $a$ in $\mathcal{B}$ extends
to a bounded operator on $\mathcal{H}$.
In addition, $J$ and $\chi$ are respectively an antiunitary and a selfadjoint operator, such that
$\chi=1$ when $k$ is odd and otherwise $\chi^2 = 1$, $\chi a = a\chi$ for $a \in \mathcal{A}$ and $\mathcal{D}\chi = -\chi \mathcal{D}$. This objects are asked to satisfy the
conditions below.

\subsubsection*{0.Regularity}
The bounded operator , $\pi(a) $ and $[\mathcal{D},\pi(a)]$ for all $a  \in \mathcal{B}$ belongs to $\bigcap_{n=1}^\infty \textrm{Dom}(\delta^n) $, with $\delta(T)=[\langle \mathcal{D}\rangle,T]$ and $\langle \mathcal{D}\rangle=(\mathcal{D}^2+1)^{\frac{1}{2}} $
\subsubsection*{1.Compactness}
The operator $\pi(a)(\mathcal{D}-\lambda)^{-1}$ is compact for all $a \in \mathcal{A}$ and $\lambda \in sp(\mathcal{D})$, where $sp(\mathcal{D})$ is the spectrum of $\mathcal{\mathcal{D}}$.
\subsubsection*{2.Spectral dimension}
For any element $b$ of the algebra $\Psi_0(\mathcal{A})$ generated by $\delta^n(\pi(\mathcal{A}))$ and $\delta^n([D,\pi(\mathcal{A})])$,
the function $\zeta_b(z)=\textrm{Tr}(b\langle \mathcal{D} \rangle^{-z})$ is well defined, holomorphic for $\mathfrak{R}(z)$ large and analytically continues to $\mathbb{C}/\textrm{Sd}$, where $\textrm{Sd}\subset \mathbb{C} $  is a discrete set called dimension spectrum.
The dimension spectrum is  said simple  if all the poles are simple, finite if there is a $k \in \mathbb{N}$  such that all the poles are of the order  at most $k$ and if it is finite.

\subsubsection*{3. Metric dimension}
The metric dimension is defined as $d=sup\{\mathfrak{R}(z),z\in \textrm{Sd} \} $, the operator $\pi(a)\langle \mathcal{D} \rangle^{-d}$ belongs to the  Dixmier class $ \mathcal{L}^1_\omega(\mathcal{H})$ for any $a \in \mathcal{A} $. For any Dixmier trace the map $a \in \mathcal{A}_+ \to Tr_\omega(\pi(a)\langle \mathcal{D} \rangle^{-d}) $  in non vanishing.
\subsubsection*{4.Finiteness}
The algebra $\mathcal{A}$ and its preferred unitization $\mathcal{B}$ are pre-$C^{*}$-algebras. Each one is a $\star$-subalgebra of some $C^*$-algebra, stable under holomorphic functional calculus. The space of smooth spinors 
\begin{equation}
\mathcal{H}^\infty=\bigcap_{n=0}^\infty \mathcal{H}^n, \ \textrm{where} \ \mathcal{H}^n=\textrm{Dom}(\mathcal{D}^n)
\end{equation}
completed with the norm $||\xi ||^n=||\xi||^2 + ||\mathcal{D}^n \xi ||^2$, is a finitely generated projective $\mathcal{A}$-module $p\mathcal{A}^m$ for some $m\in \mathbb{N}$ and some projector $p=p^2=p^*\in M_m(\mathcal{B})$. The scalar product $(\cdot,\cdot)$ on $\mathcal{H}^\infty$ can be obtained composing the  Dixmier trace and the induced structure $\langle\cdot ,\cdot\rangle:\mathcal{H}^\infty\times \mathcal{H}^\infty\to \mathcal{A}$
\begin{equation}
\textrm{Tr}_\omega(\langle\xi,\eta\rangle \langle \mathcal{D} \rangle^{-d})=(\xi,\eta), \ \xi,\eta\in \mathcal{H}^\infty
\end{equation}
\subsubsection*{5.Reality}
There is an antiunitary operator $J$ on $\mathcal{H}$, such that: 
\begin{equation}
 [\pi(a), J\pi(b^{*})J^{-1}] = 0, \ \textrm{for all} \  a, b \in  \mathcal{B} 
\end{equation}
$b \to Jb^{*}J^{-1}$ is a commuting representation on $\mathcal{H}$ of the opposite algebra $\mathcal{A}^o$. Beside,
the operator is required to satisfy $J =\epsilon $ and $J\mathcal{D} = \epsilon^\prime \mathcal{D}J$ and also for the even case $J\chi = \epsilon^{\prime\prime}\chi J$ , where $\epsilon,\epsilon^\prime, \epsilon^{\prime\prime} \in \{-1,1\}$ given as a function of $k_O \in \mathbb{Z}_8 $. 
This function is represented by the table:
\begin{equation}
\begin{array}{|c|c|c|c|c|c|c|c|c|}
\hline 
k_O \ \textrm{mod 8}& 0& 1& 2& 3& 4 & 5 & 6 & 7\\
\hline 
\epsilon &1& 1&-1&-1&-1 & -1& 1 & 1 \\
\hline 
\epsilon^\prime &1&-1& 1& 1& 1& -1 & 1 & 1   \\
\hline  
\epsilon^{\prime\prime} &1 & &-1 & & 1 & &-1 &  \\
\hline
\end{array}\end{equation}
This $k_O$ is usually referred as the KO dimension of the spectral triple.
\subsubsection*{6.First order}
The bounded operators $[\mathcal{D},\pi(a)]$ also commute with the opposite algebra representation:
\begin{equation}
[[\mathcal{D}, \pi(a)], J\pi(b^{*})J^{-1}] = 0, \ \textrm{for all} \ a, b \in  \mathcal{B}.
\end{equation}
\subsubsection*{7.Orientation}
There is an Hochschild $d$-cycle $c$ on  $\mathcal{B}$, with values in  $\mathcal{B} \otimes  \mathcal{B}^{\prime o}$. Such  $d$-cycle is a finite sum of terms like $(a\otimes b^o)\otimes a_1 \otimes\cdots\otimes a_k$, which can be naturally represented by operators on $\mathcal{H}$  using $\pi_\mathcal{D}(c)$ in formula  \eqref{h-cycle}. The volume form $\pi_\mathcal{D}(c)$ must solve the equations:
\begin{eqnarray}
\pi_\mathcal{D}(c) &=& \chi \ \textrm{even case} \\ 
\pi_\mathcal{D}(c) &=& 1 \ \textrm{odd case} 
\end{eqnarray}
A geometry is called connected or irreducible if the only operators commuting with
$\mathcal{A}$ and $\mathcal{D}$ are the scalars.

Poincar\'e duality, for a non-compact orientable manifold $M$, is usually expressed as the isomorphism
between the compactly supported de Rham cohomology and the homology of $M$
mediated by the fundamental class [M]. We shall actually leave aside the final condition
of Poincar\'e duality, since it is not central in the present work.

\section{Non-compact harmonic Moyal case}
Referring to the spectral triple $(\mathcal{A}_4,\mathcal{D}_4,\mathcal{H}_4)$ defined in the chapter 3 it is possible to extend the previous triple to a non-compact spectral triple  of metric dimension 4 and KO dimension 0. In this section will be just listed the theorems about compactness, boundness, regularity and dimension remanding to \cite{non-com-tri} for the complete proofs. 

\subsubsection*{Compactness and Boundness}
As first step we define a preferred unitisation $\mathcal{B}_4$ of $\mathcal{A}_4$ as the space of smooth bounded function on $\mathbb{R}^4$ with all partial derivatives bounded. The Moyal product  extend to $\mathcal{B}_4$ and  $\mathcal{A}_4\subset \mathcal{B}_4$ is not a dense two-side essential  ideal \cite{moyal-triple}. The algebra $\mathcal{B}_4$ contains the 	plane waves $e^{ikx}$ and constant functions but no other non-constant polynomials. Can be proved \cite{moyal-triple}  that $\mathcal{B}_4$ is contained in the $C^*$-algebra:
\begin{equation}
 A_4=\{T\in S^\prime(\mathbb{R}^4) \ : \ T \star f \in L^2(\mathbb{R}^4) \ \textrm{for all} \ f \in L^2(\mathbb{R}^4) \}
\end{equation}
completed using operator norm. Defining the action $\pi=L_\star $ of $\mathcal{B}_4$ on $\mathcal{H}_4$ as the componentwise left star multiplication:
\begin{equation}
\star:\mathcal{A}_4\times\mathcal{H}_4 \to\mathcal{H}_4, \ \ (f,\psi)\to L_\star(f)\psi=f\star\psi
\end{equation}
For $f \in \mathcal{B}_4$ the operator $\pi(f)$ is  bounded. The commutator $[\mathcal{D}_4,\pi(f)]$ is computed  using the relation \eqref{d4-com}
\begin{equation}
 \mathcal{D}_4(f\star\psi)-f\star(\mathcal{D}_4\psi)= (i(\Gamma^\mu+\Omega\Gamma^{\mu+4})(\partial_\mu f))\star\psi
\end{equation}
with $f\in B_4$ and $\psi \in \textrm{Dom}(\mathcal{D}_4)\bigcap\textrm{Dom}(\mathcal{D}_4) \pi(f) $. We deduce that for $\partial_\mu f \in \mathcal{B}_4$ the commutator is extended to a bounded operator.
The compactness is verified taking in account that $(\mathcal{D}_4^2+1)^{-1}$ is already a compact operator on $\mathcal{H}_4$, as  bounded operator $K$ is compact if and only if $KK^*$ is compact too. In particular, we have that the operator $(\mathcal{D}_4+i)^{-1} $ is compact, beside the definition  resolvent equation implies that $(\mathcal{D}_4+\lambda)^{-1}$ is compact for any $\lambda$ in the resolvent set of $\mathcal{D}_4$. Whence $\pi(f)(\mathcal{D}_4+\lambda)^{-1}$ is compact for $f\in \mathcal{B}_4$. 

\subsubsection*{Regularity and dimension}
As in the compact case it is useful for the non-compact too, to define the unbounded operators $R$ and $L$ as
\begin{equation}
R(T)=[\mathcal{D}_4^2,T]\langle \mathcal{D}_4 \rangle^{-1}, \  L(T)=\langle \mathcal{D}_4 \rangle^{-1}[\mathcal{D}_4^2,T]
\end{equation}
Can be proved that they satisfy $\bigcap_{n=1}^\infty \textrm{Dom} \delta^n=\bigcap_{m,n=1}^\infty \textrm{Dom} R^nL^m$. 
Using this derivation can be demonstrated \cite{non-com-tri} the following proposition:
\begin{prop}
Both $\pi(f)$ and $[\mathcal{D}_4,\pi(f)]$ belong for any $ f \in \mathcal{B}_4$ to $\bigcap_{n=1}^\infty \textrm{Dom} \delta^n $, where $\delta^n$ are defined by the axiom 0.
\end{prop}
For the  dimension spectrum we have the following theorem:
\begin{thm}
The spectral triple $(\mathcal{A}_4,\mathcal{D}_4,\mathcal{H}_4)$ has dimension spectrum Sd$=4-N$ and therefore metric dimension 4.
\end{thm}

%% file: appn.tex
\chapter{Error estimation and update algorithm} 

\emph{In this appendix will be introduced the treatment of the statistical error in Monte Carlo simulation  and will be discussed  three  methods to estimate the statistical errors. In the end of the section will be discussed some issues concerning the estimation of the autocorrelation time and the implemented update algorithm.}\\ \\
We start defining the mean $O$ over a subset of $\{O_i\}^{T_{MC}}_{i=1}$ of $T$ elements as a simple arithmetic mean over the Markov chain:
\begin{equation}
\bar{O}=\frac{1}{T}\sum_{i=1}^T O_i
\end{equation}
where $T_{MC}$ are total number of measurements of the time series. 
The expectation value of $O$ is an ordinary number, meanwhile the estimator 
$\bar{O}$ is a random number fluctuating around the theoretically
expected value. For numerical simulations the estimation would require to repeat the simulation several times, forced by the limit of computing  power,  the fluctuations of the mean value are not probed directly. The usual way is to 
estimates its variance:
\begin{equation}
\sigma^2_{\bar{O}} = \langle(\bar{O} - \langle \bar{O}\rangle)^2 \rangle = \langle \bar{O}^2\rangle-\langle \bar{O}\rangle^2 \label{sigm-gen}
\end{equation}
using  the distribution of the individual measurements of $O$. In case of uncorrelated subsequent $T_{MC}$ measurements the previous formula reduces to:
\begin{equation}
\sigma^2_{\bar{O}}= \frac{\sigma^2_ O}{T_{MC}}\label{sigm-unc}
\end{equation}
$ \sigma^2_O$ is the variance of the individual measurements. Explicitly the \eqref{sigm-unc} expression is:
\begin{equation}
\sigma_{\bar{O}} =\sqrt{\frac{\frac{1}{T}\Sigma_{i=1}^T (O_i-\bar{O})}{T-1}} =\sqrt{ \frac{1}{T-1}\left( \overline{O^2}-(\bar{O})^2\right) }\label{sigm-unc1}
\end{equation}
To formulate the  \eqref{sigm-gen} we  assume time-translation invariance
over the Markov chain and the  equilibrium of simulation (values taken after the thermalization) for any distribution $P(O)$ of the $O$. 
It is usual to refer to $\sigma_{\bar{O}} $ as the variance of the mean, it is the squared width of this distribution and 
is used as the squared error quoted together with the  value $\bar{O}$. In the case of a
Gaussian distribution  we have a probability of about 68\%  that  all identical simulations  
will produce a mean value in the interval $[\bar{O}-\sigma_{\bar{O}}, \bar{O}+ \sigma_{\bar{O}}]$. If  we use  $2\sigma$ as the error the percentage become about 95\% and for   $3\sigma$ interval the probability is  higher than 99.7\%.

If the samples are not  statistically independent \eqref{sigm-unc} becomes an  under-estimation of
the error. The correct expression for $\sigma_{\bar{O}}$ is \cite{monte}:
\begin{equation}
\sigma_{\bar{O}}=\sqrt{\frac{1+\frac{2\tau}{\Delta T}}{T-1}\left(\overline{O^2}-(\bar{O})^2\right)}\label{sig-cor} 
\end{equation}
where $\Delta T$ is the Monte Carlo time interval ($T =T_{MC}/\Delta T$) and $\tau$ is called autocorrelation time.  
The autocorrelation time is defined as \cite{monte1}:
\begin{equation}
\tau=\frac{1}{2}+\sum_{k=1}^{T} A(k)\left(1-\frac{k}{T}\right) \label{cor-t} 
\end{equation}
with
\begin{equation}
A(k) =\sum_{i=1}^{i+k\leq T}\frac{\langle O_i O_{i+k}\rangle - \langle O_i\rangle \langle O_i\rangle }{\langle O_i\rangle - \langle O_i\rangle \langle O_i\rangle}\label{cor-t1} 
\end{equation}
For large $T$ and $2\tau$, equation \eqref{sig-cor}  turns into:
\begin{equation}
\sigma_{\bar{O}}=\sqrt{ \frac{2\tau}{T_{MC}}\left(\overline{O^2}-(\bar{O})^2\right)}\label{sig-cor1}
\end{equation}
The equation \eqref{sig-cor} states that due to temporal correlations of the measurements
the real statistical error $\sigma_{\bar{O}}$ of the Monte Carlo estimator $\bar{O}$ is multiplied by a factor of $ \sqrt{2\tau}$. 
We can reformulate the correlated statistical error looking to the uncorrelated case as:
\begin{equation}
\sigma^\prime_ {\bar{O}}=\sqrt{\frac{\sigma^2_ O}{T_{eff}} }
\end{equation} 
but now with a effective statistic parameter
\begin{equation}
T_{eff} = \frac{T_{MC}}{2\tau}\leq T_{MC}
\end{equation}
The effective statistic, for correlated data, is smaller of the number total number of  measurements and are approximately uncorrelated  for $2\tau$ iterations. To obtain an acceptable error estimation (especially for such quantity like specific heat and susceptibility) a simulation has to provide an estimation of autocorrelation times too.

\section{Methods to estimate the error}

\subsection{Binning method}

In many cases the absence of autocorrelations time  factor on statistical errors  bring an   
great underestimation, beside the calculus  of autocorrelation time usually involves a big amount of computation time. 
The  statistical error  underestimation can occur  even for completely uncorrelated data
in time,  in fact  arises the problem of the error estimation for quantities that are not
directly measured in the simulation but are derived as a general combination of direct measures, but in this case can be use the error propagation. An good balance, in some situation, can be found using some alternative methods to estimate the statistical errors; the binning analysis is much more easy to compute but is less accurate of the direct computation.
This method consist to split the original correlated measurements vector $\{O_i\}^{T_{MC}}_{i=1}$ into smaller  non-overlapping $T_B$ blocks of length $k$ such that $ T_{MC}= T_B k$ and compute the block average $O_{B,n}$ of the $n$-th block
\begin{equation}
O_{B,n} =\frac{1}{k}\sum^k_{i=1} O_{(n-1)k+i}
\end{equation}
with $n = 1, \cdots,T_B$. The mean value over all block satisfies the identity $ \bar{O}_B = \bar{O}$. For block length
$k$  large enough $(k \ll \tau)$ the blocks become approximately uncorrelated, therefore the variance
can be computed according to the  uncorrelated estimator \eqref{sigm-unc1}. 
Resuming, we have for the squared statistical error of the mean value:
\begin{equation}
\sigma^2_{\bar{O}} = \frac{\sigma^2_B}{T_B} = \frac{1}{T_B(T_B-1)}\sum^{T_B}_{n=1} (O_{B,n} - \bar{O}_B )^2
\end{equation}
To obtain an acceptable result we must probe various partition of $T_B$ and select the one
which maximize the statistical error.
The main disadvantage of this method is the strongly dependence of the error on the choice of the partition $T_B$

\subsection{Jackknife estimation}
This approach \cite{monte4,monte5} is a variation of the binning method and can be viewed as a re-sampling. 
This method too concerns to separate the vector of measurements into $T_B$ blocks but instead consider  small blocks of length $k$, in the Jackknife method we build  $T_B$ large Jackknife blocks $O_{J,n}$
containing all measurements but omitting the data contained in the  previous $n$-th binning block:
\begin{equation}
O_{J,n}=\frac{T_{MC}\bar{O} - kO_{B,n}}{T_{MC}-K}
\end{equation}
with $n = 1, \cdots,T_B$. The Jackknife blocks consists of $T_{MC} - k$ data, in this way the $n$-th block contains almost every
measurements of the complete vector. This allows to use this method even to estimate the errors in non-linear combinations of measured variables,  the bias is comparable to that of the total data.  
This choice of the $T_B$ Jackknife blocks makes them trivially correlated due to the common data 
coming from the same original data in $T_B - 1$ different Jackknife blocks. However, this correlation is
a pure re-sampling  and  it is not connected to time correlation.  
As a consequence, it does not magnify the error but the Jackknife block variance $\sigma_J$ will be 
smaller than the variance computed using binning method. 
This reduction can be corrected by multiplying $\sigma^2_J$ with a factor $(T_B - 1)^2$, obtaining:
\begin{equation}
\sigma^2_{\bar{O}}=\frac{T_B-1}{T_B}\sum^{T_B}_{n=1} (O_{J,n}-\bar{O}_J)^2 
\end{equation}

\section{Numerical estimation of autocorrelation time} 

A third option, the so called Sokal-Madras method \cite{monte3}, is an hybrid approach founded on the estimation of the autocorrelation time using  \eqref{cor-t}-\eqref{cor-t1}. In the case that the autocorrelation 
time satisfies $\tau<0.5$ the samples are not correlated thus, to estimate the error, is used the standard uncorrelated error given by \eqref{sigm-unc1} otherwise error is given by \eqref{cor-t1}. 

An estimation of the autocorrelation time  is  important  not only for the error estimation, but also for the computation 
of static quantities themselves. In general, it is very hard to have  a priori estimation and  a direct  numerical 
analysis is usually  too time consuming. An estimator $ \hat{A}(k)$ for the autocorrelation function is obtained by replacing
in \eqref{cor-t1} the expectation values  by mean values.
Can be shown that increasing $k$ the variance of $\hat{A}(k)$ grows quickly, 
 to achieve an rough estimation of  $\tau$ and thus a correct error estimate, it is useful to record the partial autocorrelation time estimator
\begin{equation}
\tau^\prime(k_{max}) =\frac{1}{2}+\sum^{k_{max}}_{K=1} \hat{A}(k) 
\end{equation}
In the limit of large $k_{max}$ the previous sum tents to $\tau$,  unfortunately  its statistical error
increases rapidly. An usual  procedure, used to find a reasonable balance between systematic and statistical errors, is to determine a self-consistently upper limit $k_{max}$  by cutting off the summation for $k_{max} > 6\tau^\prime(k_{max} )$. 

In this thesis  were compared the errors obtained using  the previous three methods.
Referring to the basic observables like energy density $E$ and the order parameters like $\varphi_i$, $Z_{ i\mu}$, we have found that the errors were compatible among the three methods. For quantities like the specific heat the errors were no more compatible, therefore  was applied the more accurate Sokal-Madras method using the described cut  for the estimation of the correlation time.

\section{The update algorithm} 
The model considered in this works involves five complex independent matrices for the 4-dimensional case and  3 complex independent
for the 2-dimensional one, beside  the approximated actions are quite complex and contain terms up the fourth power. The usual update algorithm, used to generate a new configuration, expects to change all the entries of the independent matrices: 
\begin{equation}
\psi_{ij} \longrightarrow  \psi_{ij} + a_{0ij}, \ \ Z_{kij}\longrightarrow Z_{kij}+a_{kij}, \ \textrm{with} \  k=0,\cdots 4, \ i,j=1,\cdots N 
\end{equation}
where $N$ is the matrix size and $a_{kij}$ are random complex numbers in which the real and imaginary parts vary in $[-2,2]$. The amplitude of the random interval was numerically determined to obtain stable results for the observables for larger interval.     
The new proposed configurations are judged by the metropolis algorithm, which requires to recompute the action for the new proposed configuration. Using an optimized algorithm, this operation involves for each Metropolis check, about $16N^3$ or $6N^3$ (taking in account the cross products between the independent matrices present in the actions) as a leading order of number of operations respectively for the 4-dimensional and 2-dimensional case. In order to reduce the computation time, instead update all entries we have changed just one matrix coefficient for each  Monte Carlo step, the coefficient to update is not chosen randomly but following an order used to vary all the coefficients after $5N^2$ or $3N^2$ steps. In this way to execute the Metropolis check we can compute the variation of the action respect the previous configuration with a leading order of operations proportional to $N$. The drawback of this approach are manly two: the complication of the code and the behavior of the correlation time in $N$. In fact two configurations will share at least one coefficient until $5N^2$ or $3N^2$ Monte Carlo steps, this correlations introduce an increasing of the correlation time increasing $N$ and presumably $\tau$ behaves like $ 5N^2, \ 3N^2$. However, using this
optimization, together with the already showed theoretical simplifications and the implementation of parallel computing, we were able to execute our simulation  for $N=20$ with   sufficient precision and in reasonable time despite our limited computation resource.    